\documentclass[12pt,a4paper]{article}

\usepackage{natbib}
\usepackage[T1]{fontenc}
\usepackage{graphics}
\usepackage{graphicx}
\usepackage{booktabs,caption}
\usepackage{threeparttable}
\usepackage{adjustbox}
\usepackage[format=plain, labelsep=period, labelfont=bf, font=footnotesize,singlelinecheck=false,skip=5pt]{caption}
\usepackage{tabularx}
\usepackage{array}
\newcolumntype{H}{>{\setbox0=\hbox\bgroup}c<{\egroup}@{}}
\usepackage{amsthm}
\usepackage{lscape}
\usepackage{appendix}
\usepackage[bookmarks]{hyperref}
\usepackage{subcaption}
\usepackage{amsmath}

\usepackage{multirow}
\usepackage{setspace}
\usepackage{amssymb}
\usepackage{textcomp}
\usepackage{amsfonts}
\usepackage{soul}
\usepackage{bbm}
\usepackage{multirow}
\usepackage{multicol}
\usepackage{verbatim}
\usepackage{mdwlist}
\usepackage{color,soul}
\usepackage{eurosym}
\usepackage{rotating}
\usepackage{lmodern}
\usepackage{color}
\usepackage{longtable}
\usepackage{etoolbox}

\usepackage{epigraph}
\usepackage{xargs}
\usepackage[left=2cm,right=2cm,bottom=2cm,top=2cm]{geometry}

\makeatletter

\newcommand*{\@rowstyle}{}
\newcommand*{\rowstyle}[1]{
	\gdef\@rowstyle{#1}%
	\@rowstyle\ignorespaces%
}
\newcolumntype{=}{
	>{\gdef\@rowstyle{}}%
}
\newcolumntype{+}{
	>{\@rowstyle}%
}
\makeatother

\title{The attachment of adult women to the Italian labour market in the shadow of COVID-19\thanks{We thank participants to the 2022 EALE and 2022 AIEL conferences, seminar participants at the University of Modena and Reggio Emilia, and one anonymous referee for useful comments. The usual disclaimers apply.}}

\author{Davide Fiaschi\thanks{University of Pisa, Dipartimento di Economia e Management, University of Pisa, REMARC and Centro DAGUM, Via Ridolfi 10, 56124 Pisa (Italy), Phone: +39 0502216208, Email: davide.fiaschi@unipi.it. }  \and Cristina Tealdi\thanks{Corresponding author. Edinburgh Business School, Heriot-Watt University, EH14 4AS Edinburgh (UK) and IZA Institute of Labor, Phone: +44 0131 4513803, Email: c.tealdi@hw.ac.uk.}
}

\date{\today}

\begin{document}
	
	\maketitle
	
	\begin{abstract}
		We investigate the attachment to the labour market of women in their 30s, who are combining career and family choices, through their reactions to an exogenous, and potentially symmetric shock, such as the COVID-19 pandemic. We find that in Italy a large number of females with small children, living in the North, left permanent (and temporary) employment and became inactive in 2020. Despite the short period of observation after the burst of the pandemic, the identified impacts appear large and persistent, particularly with respect to the males of the same age. We argue that this evidence is ascribable to specific regional socio-cultural factors, which foreshadow a potential long-term detrimental impact on female labour force participation.
	\end{abstract}
	
	\noindent \textbf{Keywords}: female labour force participation, labour market flows, transition probabilities, childcare, socio-cultural factors.

	\noindent \textbf{JEL Classification}: E24, J21, J82.
	
%
	\doublespacing

	\clearpage
	
	\section{Introduction}
	
	The compatibility of labour market participation and childcare still represents a serious challenge for women. While in the past having a career and having a family were deemed as mutually exclusive choices, i.e, opting for one would imply giving up on the other, today, most women in developed countries aim to have both a family and a successful career \citep{doepke2022new}. The combination of career and family has led to a substantial change in the life cycle of women's labour force participation for the most recent cohorts, which looks relatively high and fairly flat, similar to that of men's \citep{Fernandez2004}, but  with a lower and squishier middle in their 30s \citep{goldin2020productivity}. In this paper, we use the burst of COVID-19 pandemic, which appears as an exogenous and potentially symmetric shock, to analyse the factors affecting the attachment to the labour market of women, in particular  those between the age of 30 and 39.
	
	For decades policymakers and researchers have been interested in the determinants of female labour force participation \citep{goldin2002power}. Labour force participation increased significantly during the last century; many reasons have been identified behind this phenomenon, ranging from the diffusion of the contraceptive pills, the electricity revolution, the relative change in returns to experience compared with the male's, the decrease in the gender wage gap, and the discovery of the infant formula \citep{jones2015married, goldin2002power, albanesi2016gender, greenwood2005engines, olivetti2006changes}. The levelling off of female labour force participation in the late 1990s has been attributed to the lack of appropriate policies to support families \citep{Blau2013}, the diffusion of long-hour occupations \citep{goldin2021career} and the marriage patterns of women to highly educated and high-income husbands \citep{albanesi2021gendered}. The recent `sagging middle' life cycle seems to be ascribable to  the lack or ineffectiveness of ``family-friendly'' policies  \citep{Blau2013, olivetti2017economic} and the rise in ``greedy jobs'', which are increasingly dominant among highly educated workers \citep{goldin2021career}. The literature points to a number of factors which could facilitate the combination of career and childcare responsibilities, among which: (i) policies in support of families, such as longer maternal and paternal leaves, reduction of childcare costs, increased availability of pre-school opportunities  \citep{GIVORD201599,BARUA2014129,COMPTON201472,HUEBENER2020510,andresen2019child,BETTENDORF2015112}; (ii) favourable social norms, such as the equal gender division of housework and childcare \citep{myong2021social, del2020women}; and (iii) labour markets which are sufficiently flexible \citep{del2002effect, da2006fertility}. However, the overlapping of various policies in support of families, heterogeneous socio-cultural factors, and region-specific economic factors makes it difficult to quantify the contribution of each single determinant to the labour force participation of women \citep{cascio2015effectiveness, olivetti2017economic}.
	
	We investigate the attachment of women to the Italian labour market, specifically those in the 30-39 age category, in the period 2013-2020, i.e., before and during the COVID-19 pandemic, by estimating transition probabilities to and from inactivity.
	The COVID-19 shock appears well-suited for our aim as it was (i) unexpected and (ii) potentially symmetric, thus fit for the analysis of its impact as purified by potential inter-temporal effects linked to the business cycle and long-term planning of individuals.\footnote{Appendix \ref{appsec:italianPolicies} for a detailed description of timing and  management of the COVID-19 crisis in Italy.}
	Moreover, the Italian labour market displays large heterogeneities, with vast and persistent regional disparities:  industrial activities are mostly concentrated in the North, while agriculture and commerce are prevalent in Southern regions (Tables \ref{apptab:workersBySector}-\ref{apptab:workersBySectorPercentage} in the Appendix and \citealp{OECDItaly}). Italy also ranks among the weakest of OECD countries regarding job quantity, defined as employment, unemployment and underemployment \citep{OECD2018}, reflecting persistently large gender employment gaps, with a remarkably low female labour force participation in the South \citep{agovino2019local}. Specifically, while in the North the labour force participation of 30-39 females ranges between 75\% and 80\%, in line with the US rate \citep[Fig.1]{goldin2020productivity}, in the South it is as low as 52\%, despite a comparable total fertility rate of 1.3 in the North and 1.25 in the South \citep{ISTAT}.
	
	Withstanding these different conditions, the COVID-19 shock  had asymmetric effects across categories of individuals by age, gender and geographical location, but their size and persistence over time are more surprising.\footnote{The literature on the asymmetric effects of the pandemic on different categories of individuals is large, see, e.g., \citet{caselli2021mobility,chetty2020economic, fiaschi2022young, alon2021mancession, albanesi2021gendered,fabrizio2021covid, zamarro2020gender, shibata2020distributional,adams2020inequality,casarico2020heterogeneous,hupkau2020work,bluedorn2021gender, dang2021gender}.}
	In particular, we find large flows of females in their 30s with small children who live in the North of Italy, moving from both permanent and temporary employment to inactivity during the whole pandemic year of 2020. In the South, instead, where female labour force participation was already very low, we do not find any evidence of such phenomenon. Despite the data availability being limited to the four quarters of 2020, the identified effects appear large and persistent, in particular if compared to males in the same age cohort.
	We argue that while in the North the combination of economic, social and cultural factors allows for an easier combination of career and family, this exposes women in this area to a high risk, i.e., the female labour market attachment is weaker. Such explanation seems to be strongly anchored to the different regional female socio-cultural factors.
	
	The rest of the paper is organized as follows. Section \ref{sec:Data} illustrates the data and the methodology used. Section \ref{sec:ItalianLabourMarket} provides evidence of large heterogeneity in the Italian labour market before the COVID-19 pandemic, while Section \ref{sec:Post} analyses the determinants of the changes in female labour force participation due to the pandemic. Finally, Section \ref{sec:concludingRemarks} concludes the paper and discusses potential explanations for our findings.
	
	\section{Data and methodology\label{sec:Data}}
	
	We use Italian quarterly longitudinal labour force data as provided by the Italian Institute of Statistics (ISTAT) for the period 2013 (quarter I) to 2020 (quarter IV).\footnote{Data for the period 2013 (quarter I) to 2020 (quarter IV) are available upon request at \url{https://www.istat.it/it/archivio/185540}. All codes and datasets used in the analysis are available at \url{https://people.unipi.it/davide_fiaschi/ricerca/}.} The Italian Labour Force Survey (LFS) follows a simple rotating sample design where households participate for two consecutive quarters, exit for the following two quarters, and come back in the sample for other two consecutive quarters. As a result, 50\% of the households, interviewed in a quarter, are re-interviewed after three months, 50\% after twelve months, 25\% after nine and fifteen months. This rotation scheme allows to obtain 3 months longitudinal data, which include almost 50\% of the original sample.
	
	The longitudinal feature of these data is essential for achieving a complete picture of significant economic phenomena of labour market mobility. Per each individual who has been interviewed, we observe a large number of individual and labour market characteristics at the time of the interview and three months before.
	On average, approximately 70,000 individuals are interviewed each quarter, of which 45,000 are part of the working age population. The average quarterly inflow of younger individuals in the working age population is 0.3\%, while the average quarterly outflow of older individuals from the working age population is 0.4\%, which implies a (almost) constant working age population within quarters.
	
	The dynamics of the labour market is described by Markov Chains with discrete states in discrete time. Our dataset allows to consider quarters as unit of time and to define seven labour market states: \textit{permanent} (PE), \textit{temporary} (TE), \textit{self-employment} (SE), \textit{unemployment} (U), the \textit{furlough scheme} (FS), \textit{education} (EDU) and \textit{inactivity} (INACT).  The dynamics are  represented through a \textit{Transition Probability Matrix} (TPM), which shows both persistence in each labour market state and the probability of transition from one state to another, and fully characterizes the dynamics of the shares of the whole population in each state. In particular, the shares of individuals in different states provide a picture of the long-term trends, as they take longer to react to shocks, while the transition probabilities inform about the sudden impact of the (pandemic) shock. We compute the labour market flows by calculating the quarter-on-quarter transitions made by individuals between labour market states. In the analysis we take the first quarter of 2020, which marks the time of the initial spread of the virus, as the period when the dynamics of the Italian labour market is expected to change. The inferential analysis on the shares and transition probabilities is computed via bootstrap using 1000 draws from the original sample.
	
	Important data limitations are to be mentioned. First, the point-in-time measurement of the worker's labour market state fails to capture transitions within the period (quarter). For instance, if an employed worker becomes unemployed and finds a new job within a quarter, we do not observe those transitions in our data. Second, the available data stop at quarter IV of 2020, while it would be desirable to have data also for 2021 to explore the further persistence of pandemic shock. Third, we do not have information on the household composition (as we only observe the household size), individual social attitudes and beliefs, and parents' characteristics. For this reason, in the last part of our analysis, we use data from the European Labour Force, which contains detailed information on the number and age of children, and the European Value Survey, from which we draw information on socio-cultural factors of individuals at regional level. Finally, the short longitudinal span of the data, having longitudinal observations only in two consecutive quarters,  constrains our analysis to be based on a Markovian process of order one.

\section{The female labour market before the COVID-19 pandemic}\label{sec:ItalianLabourMarket}

\begin{figure}[!htbp]
	\caption{Share of inactive individuals aged 30-39 in the North and South of Italy.}
	\label{fig:shares3039text}
	\centering
	\begin{subfigure}[t]{0.49\textwidth}
		\centering
		\includegraphics[width=\linewidth]{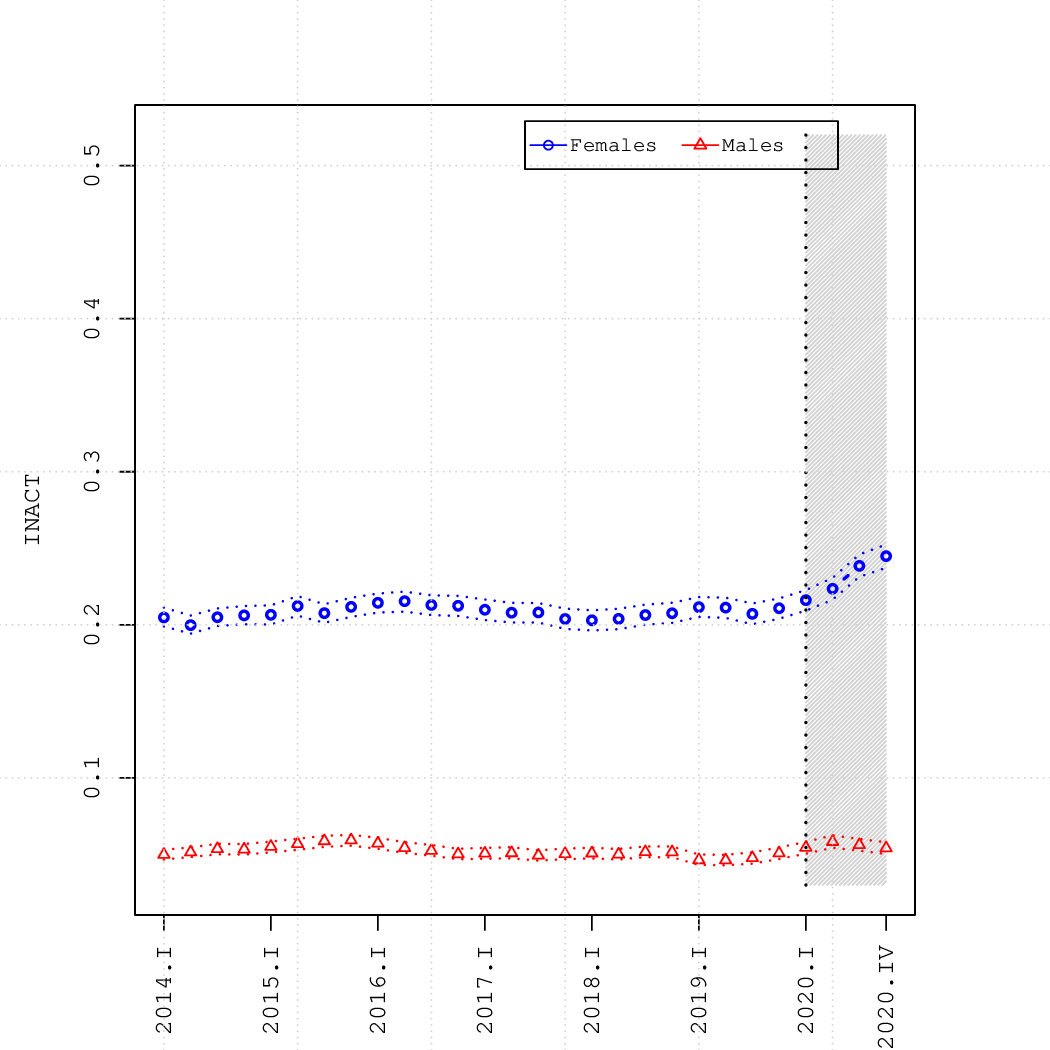}
		\caption{Inactive - North.}
		\label{fig:transProbFromEDUtoPE_I}
	\end{subfigure}
	\begin{subfigure}[t]{0.49\textwidth}
		\centering
		\includegraphics[width=\linewidth]{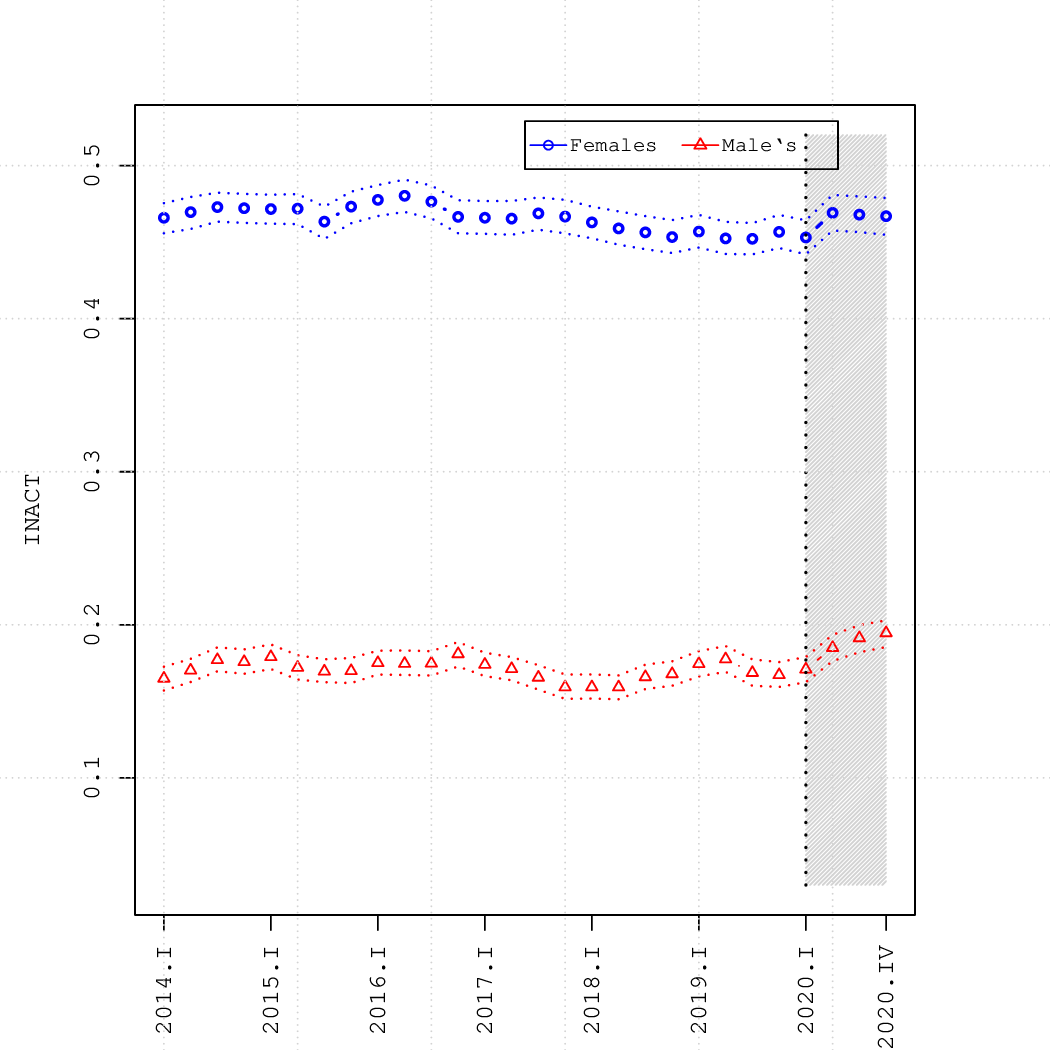}
		\caption{Inactive  - South.}
		\label{fig:transProbFromEDUtoU_I}
	\end{subfigure}
	\vspace{0.2cm}
	\caption*{\scriptsize{\textit{Note}: Confidence intervals at 90\% are computed using 1000 bootstraps. The gray area identifies the COVID period. \textit{Source}: LFS 3-month longitudinal data as provided by the Italian Institute of Statistics (ISTAT).}}
\end{figure}

The Italian labour market pre-COVID-19 presented specific characteristics, which we deem crucial to understand the impact of the pandemic shock on different categories of individuals. In particular, women participate much less to the labour market compared to men, but women in the South also participate much less to the labour market compared to women in the North. As reported in Figure \ref{fig:shares3039text}, before the pandemic, in the North, 20\% of females were inactive compared to 5\% of males; on the contrary, in the South, 45\% of females and 15\% of males were not participating to labour market. 

These patterns have their roots at the early stages of the lives of women and men. Figure \ref{appfig:shares1519text} in Appendix \ref{appsec:dynamicsprecovidIN} shows that already the 15-19 age cohort presents the first signs of heterogeneity by geographical area. While we do not observe significant differences by gender, the share of inactive individuals in the North ranges around 3.5\% and 4.5\%, while in the South it is already as high as 9\%. The gender gap starts appearing among the 20-24 age cohort, particularly in the North, where the female inactive share is between 10\% and 12\% compared to a range of 8\% to 10\% for males. In the South, the gender gap is less relevant, but the average inactive share is as high as 22\%. The gap by gender and geographical location becomes significant among the 25-29 age cohort: 20\% of females in the North are inactive, compared to 7\% of males and 35\% of females in the South are inactive compared to 20\% of males.

The bleeding of women into the inactivity state almost stops with the 25-29 age cohort in the North, and the share of inactive women stabilizes around 20\% among older cohorts. A similar pattern is observed for males in both geographical areas. However, the share of inactivity continues to grow for women in the South: it is as high as 45\% among the 30-39 age cohort and 50\% among the 40-49 age cohort. Hence, the decision to leave the labour force for many women starts very early in life in the South, and it continues as they grow older and across different stages of life. In the North, instead, the decision to participate to the labour force for women mainly happens in the early stage of life, below the age of 30. 
Interestingly, the median age of the first-child birth is slightly higher than 30 in both geographical locations, with minimal differences across regions of the North and the South (Table \ref{apptab:birthagebycitiz} in Appendix \ref{appsec:dynamicsprecovid}). This evidence, paired with the  statistics on inactivity, suggests that the birth of a child does not affect significantly the labour market participation of women in the North, while it represents an important determinant in the South. 
 9
The literature has already identified significant gender and geographical differences as structural features of the Italian labour market \citep{bertola2003structure}. Women, on average, are found to have a lower attachment to the labour force, together with a lower commitment to the labour market compared to men \citep{schiattarella2018old}. The North-South divide characterizes many dimensions of the economic and cultural life in Italy, but it is particularly  striking in women's work. Women in Southern Italy are comparatively more likely not to work and not to return to the labour market after marriage or childbearing. On average, 30\% of  Italian mothers in employment stop working to care for children or other relatives, and of these only about 12\% go back to work at some point in life \citep{SOAS2021gender}. The latter is still much lower in the Italian South, due to the predominant role of the male breadwinner model \citep{baussola2014disadvantaged,picchio2021if,pacelli2013labor}.

This evidence points to the presence of a strong geographical heterogeneity in the labour market even before the COVID-19 pandemic, with female inactivity rates being much higher in the South across all age cohorts, withstanding similar fertility rates.

\section{Female labour market participation during the COVID-19 pandemic} \label{sec:Post}

The impact of the COVID-19 pandemic on the labour market of females aged 30-39 has been heterogeneous in the North and South of Italy.
Table \ref{tab:sharechangesqIVtext} reports the difference between the inactive share in quarter IV of 2020 and the same quarter one year before for males and females across age categories and geographical areas.\footnote{In Appendix \ref{app:dynamicssharesduring} we report the full set of statistics for the seven labour states and the same statistics comparing the shares of individuals by age group in quarter III of 2020 with the same quarter one year before.}
In quarter IV of 2020, females in the North of Italy had a larger presence in the inactive  state across all age categories, although the increase was particularly large for the 20-24 and 30-39 age cohorts. Only for the 25-29 age cohort, the share of inactive females  increased across geographical locations, but more in the South.\footnote{Although the focus of our analysis is on females in older age categories, we provide in Appendix \ref{app:2529women} some explanations for this result.}  
By comparison, the shares of males living in the North of Italy in the inactive  state increased in quarter IV of 2020 across all age groups (except for the 30-39 age cohort) with respect to the same quarter one year before.  Similar patterns are displayed by males living in the South: across all age categories the share of males in the inactive  state is higher. Nevertheless, the magnitude of the changes is  smaller for males, compared to females, within the same age category.

\begin{table}[htbp] \centering 
	\caption{Changes in the inactive shares between quarter IV of 2019 and quarter IV of 2020 by category of individuals.}  
	\label{tab:sharechangesqIVtext} 
	\scriptsize{
		\begin{tabular}{@{\extracolsep{5pt}} ccccc} 
			\hline \hline
			&\multicolumn{2}{c}{Females}&\multicolumn{2}{c}{Males}\\
			\hline \\[-1.8ex]
			& 	North &South & North &South \\ 
			\hline \\[-1.8ex] 
			20-24 & $\textbf{0.034}^{***}$ & $0.013$ & $\textbf{0.018}^{***}$ & $\textbf{0.016}^{*}$\\
			&  $(0.000)$ & $(0.143)$ & $(0.000)$& $(0.079)$ \\ 
			25-29 & $\textbf{0.017}^{**}$ & $\textbf{0.058}^{***}$& $\textbf{0.020}^{***}$&$\textbf{0.017}^{*}$ \\ 
			& $(0.022)$ & $(0.000)$ & $(0.000)$ & $(0.065)$ \\ 
			30-39 &  $\textbf{0.034}^{***}$ & $0.010$ &  $0.003$ &$\textbf{0.027}^{***}$  \\ 
			&$(0.000)$ &  $(0.154)$ & $(0.127)$ & $(0.000)$\\ 
			40-49 &  $\textbf{0.020}^{***}$ & $0.011$ & $\textbf{0.013}^{***}$ &$\textbf{0.016}^{***}$ \\ 
			&  $(0.000)$ & $(0.116)$ & $(0.000)$ & $(0.003)$\\
			\hline \hline \\[-1.8ex] 
			\multicolumn{5}{l}{\textit{Note}: The attained significance levels (ASL)  of the null}\\  \multicolumn{5}{l}{ hypothesis of equality between the  shares in the two}\\  \multicolumn{5}{l}{  periods computed using 1000 bootstraps are reported}\\  \multicolumn{5}{l}{ in parenthesis \citep[p.220]{efron1994introduction}.}\\  \multicolumn{5}{l}{North includes regions in the North and the Center.}\\  \multicolumn{5}{l}{ $^{*}$ASL$<$0.1; $^{**}$ASL$<$0.05; $^{***}$ASL$<$0.01.}\\
	\end{tabular} }
\end{table} 

A further understanding of the effect of the pandemic is given by the changes in the transition probabilities across labour market states, in particular by  comparing   the observed data with the counterfactual scenario of no pandemic shock, i.e.,  quarterly transition probabilities for categories of individuals across demographic groups against the forecasted quarterly transition probabilities during the pandemic quarters (quarter I of 2020- quarter IV of 2020).\footnote{The forecasted transition probabilities are computed using a combination of four forecasting models (ETS, TSLM, THETAF, and ARIMA) \citep{HyndmanAthanasopoulosforecasting2021} in the period 2013 quarter I- 2019 quarter IV.} 
In order to understand whether the increased female inactivity rate is due to discouraged unemployed workers who stopped looking for jobs, or to employed workers who left their jobs, we focus on the transitions from temporary and permanent employment and unemployment to the inactive state. 

\begin{figure}[!htbp]
	\caption{Transition probabilities of females from unemployment to the inactive state by age groups.}
	\label{fig:transProbUtoinactive}
	\centering
	\vspace{0.1cm}
	\caption*{\scriptsize{\textbf{Panel A: Age 25-29}}.}
	\vspace{0.1cm}
	\begin{subfigure}[t]{0.3\textwidth}
		\centering
		\includegraphics[width=\linewidth]{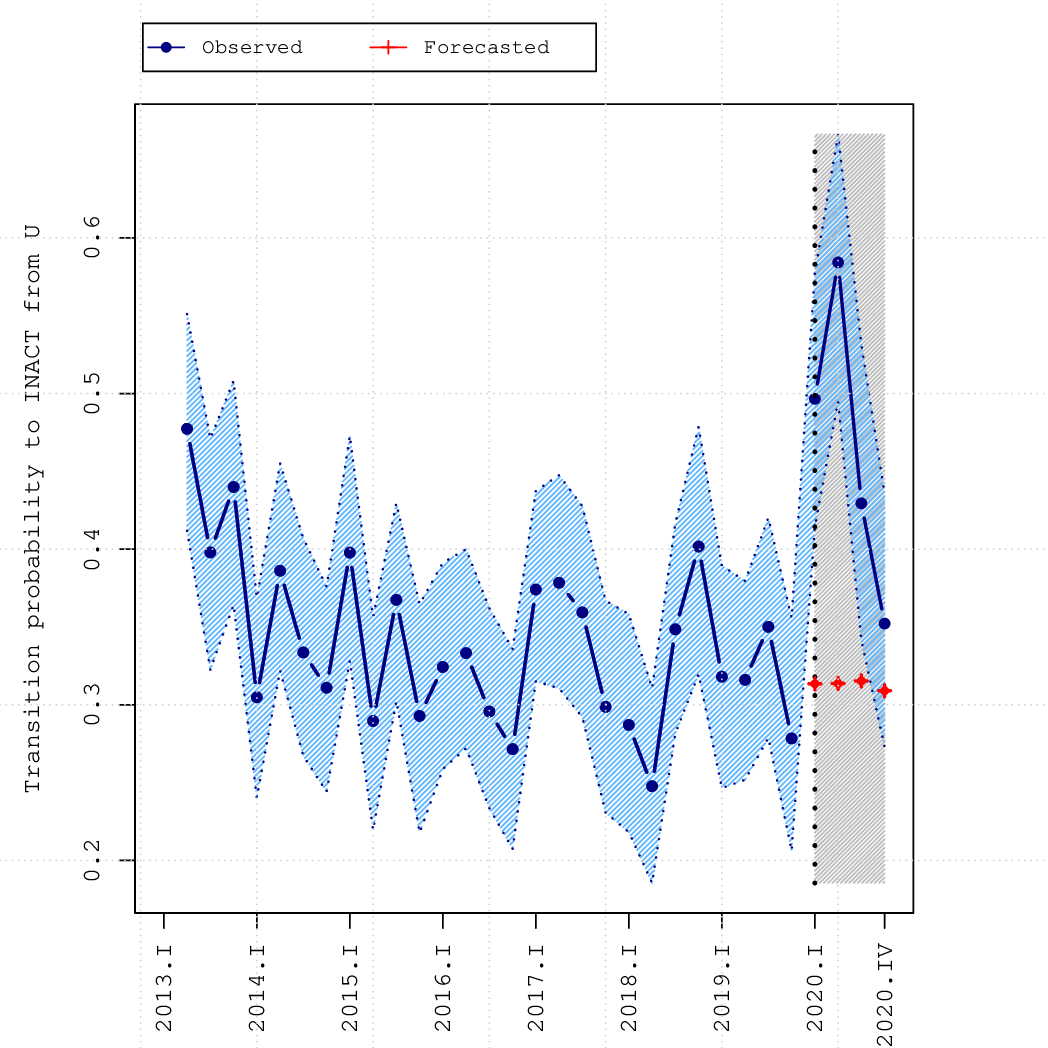}
		\caption{South.}
		\label{fig:transProbFromEDUtoSEApp}
		\vspace{0.2cm}
	\end{subfigure}
	\begin{subfigure}[t]{0.3\textwidth}
		\centering
		\includegraphics[width=\linewidth]{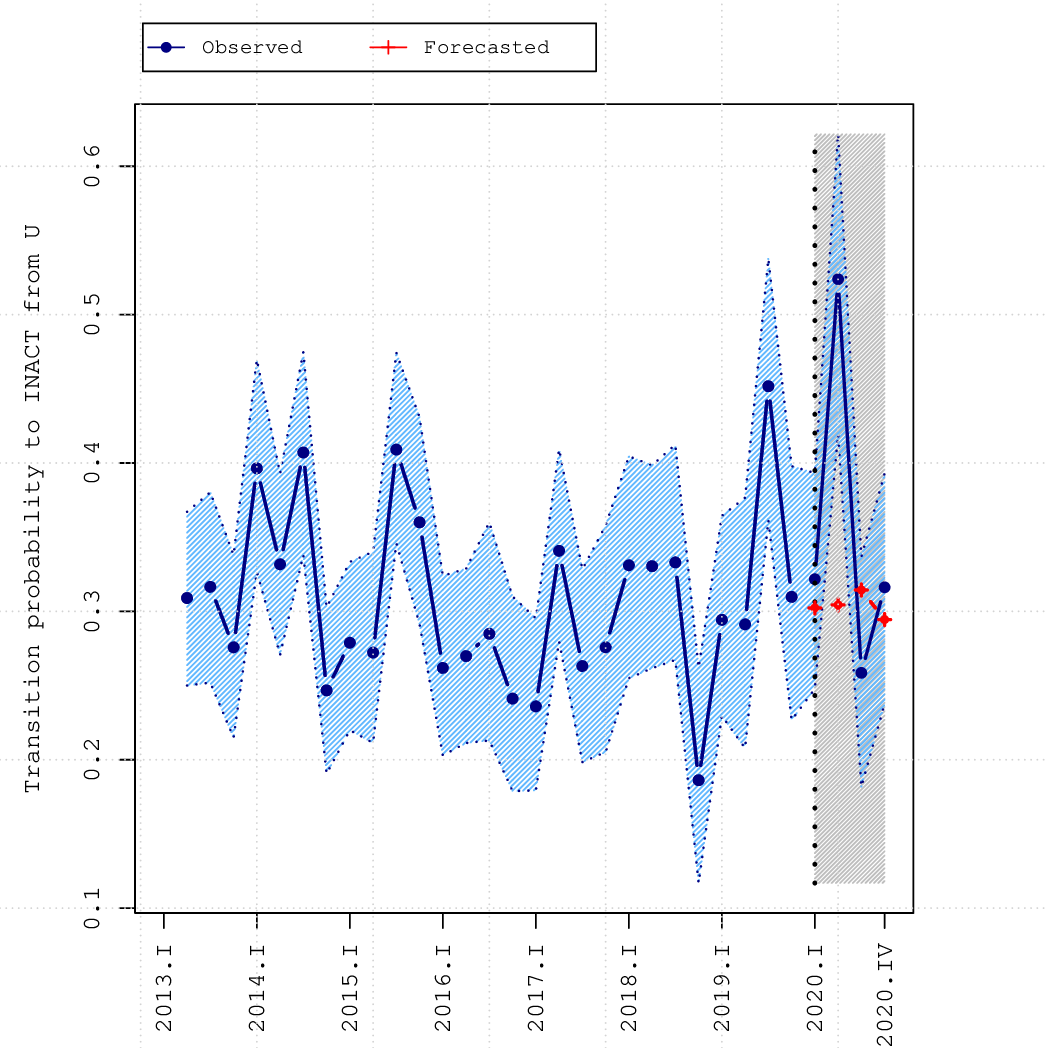}
		\caption{North.}
		\label{fig:transProbFromEDUtoTEApp_I}
	\end{subfigure}
	\vspace{0.1cm}
	\caption*{\scriptsize{\textbf{Panel B: Age 30-39}}.}
	\vspace{0.1cm}
	\begin{subfigure}[t]{0.3\textwidth}
		\centering
		\includegraphics[width=\linewidth]{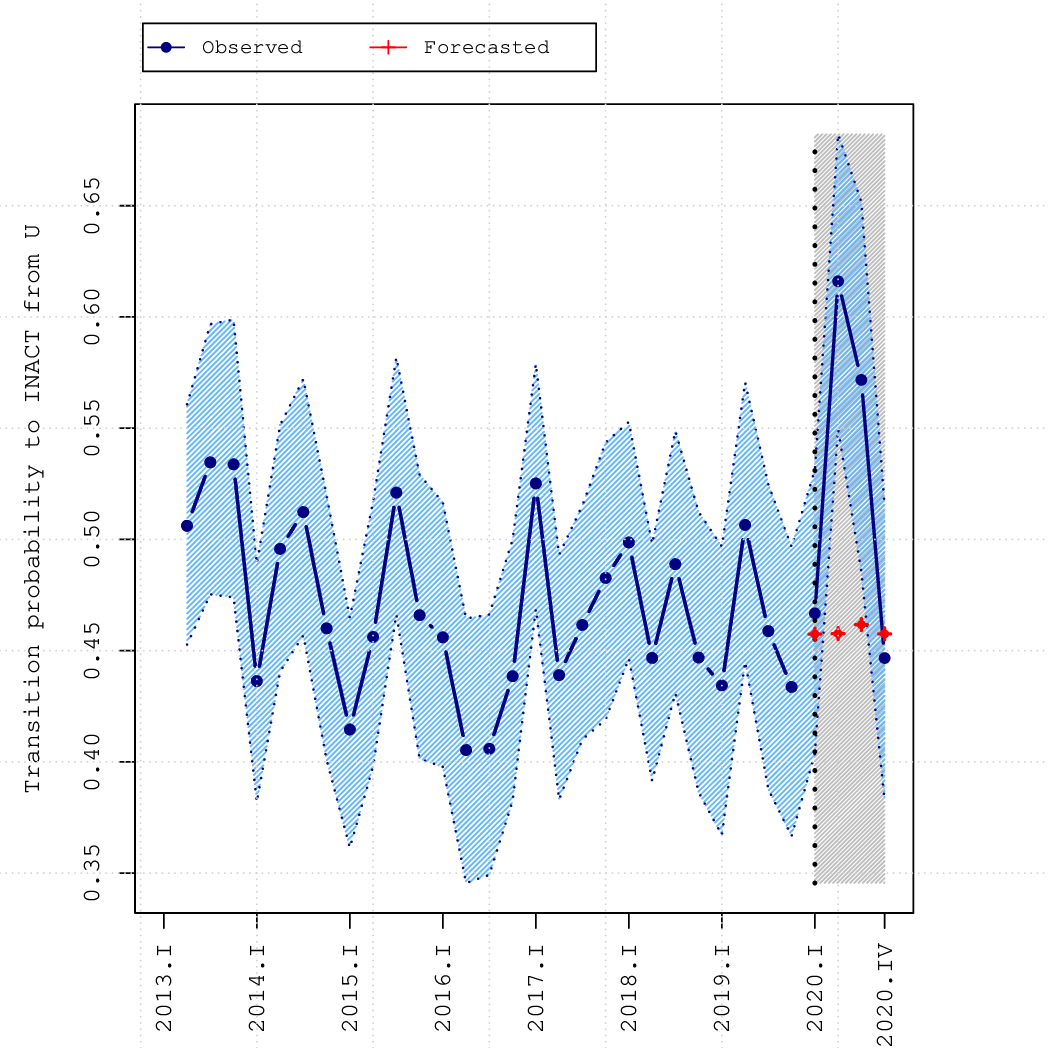}
		\caption{South.}
		\label{fig:transProbFromEDUtoSEApp_IIII}
	\end{subfigure}
	\begin{subfigure}[t]{0.3\textwidth}
		\centering
		\includegraphics[width=\linewidth]{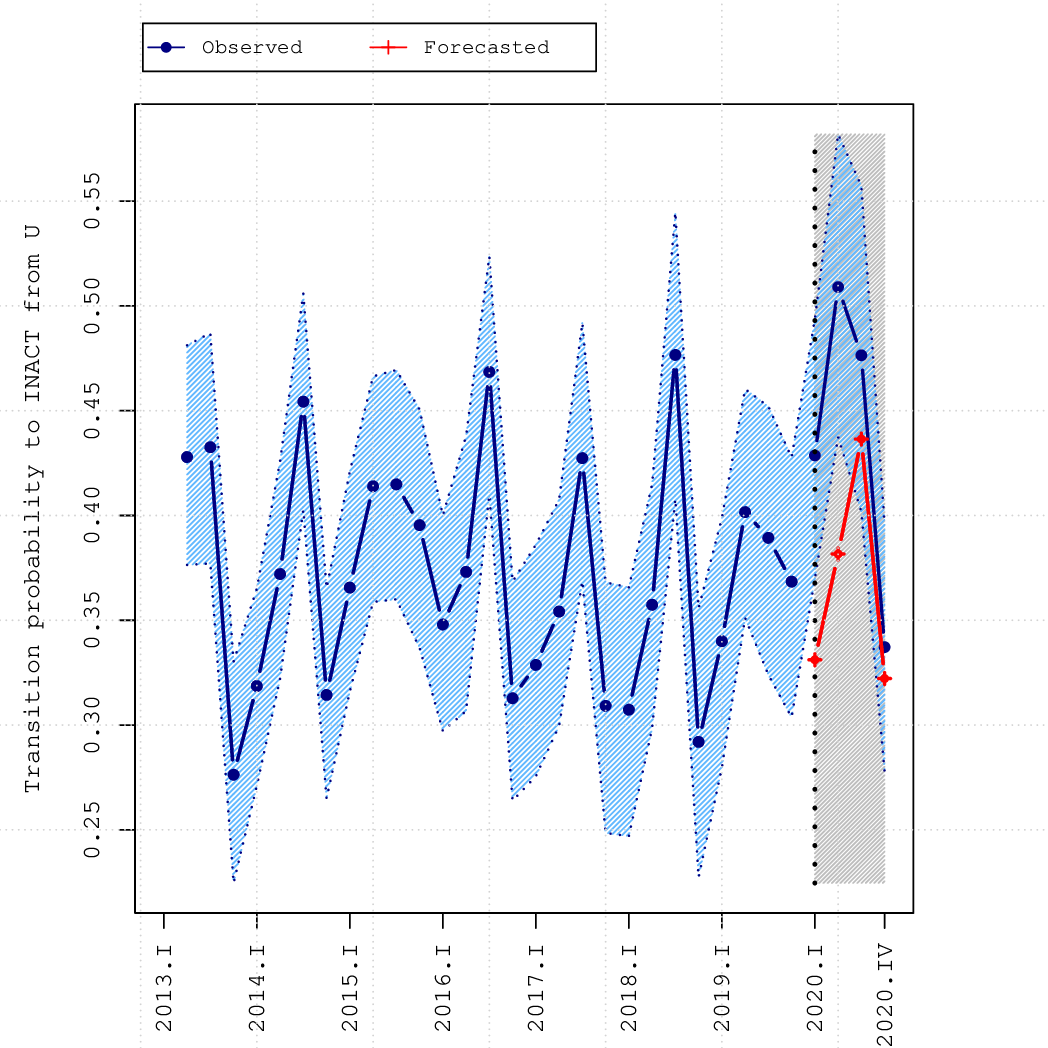}
		\caption{North.}
		\label{fig:transProbFromEDUtoTEApp_II}
	\end{subfigure}
	\vspace{0.1cm}
	\caption*{\scriptsize{\textbf{Panel C: Age 40-49}}.}
	\vspace{0.1cm}
	\begin{subfigure}[t]{0.3\textwidth}
		\centering
		\includegraphics[width=\linewidth]{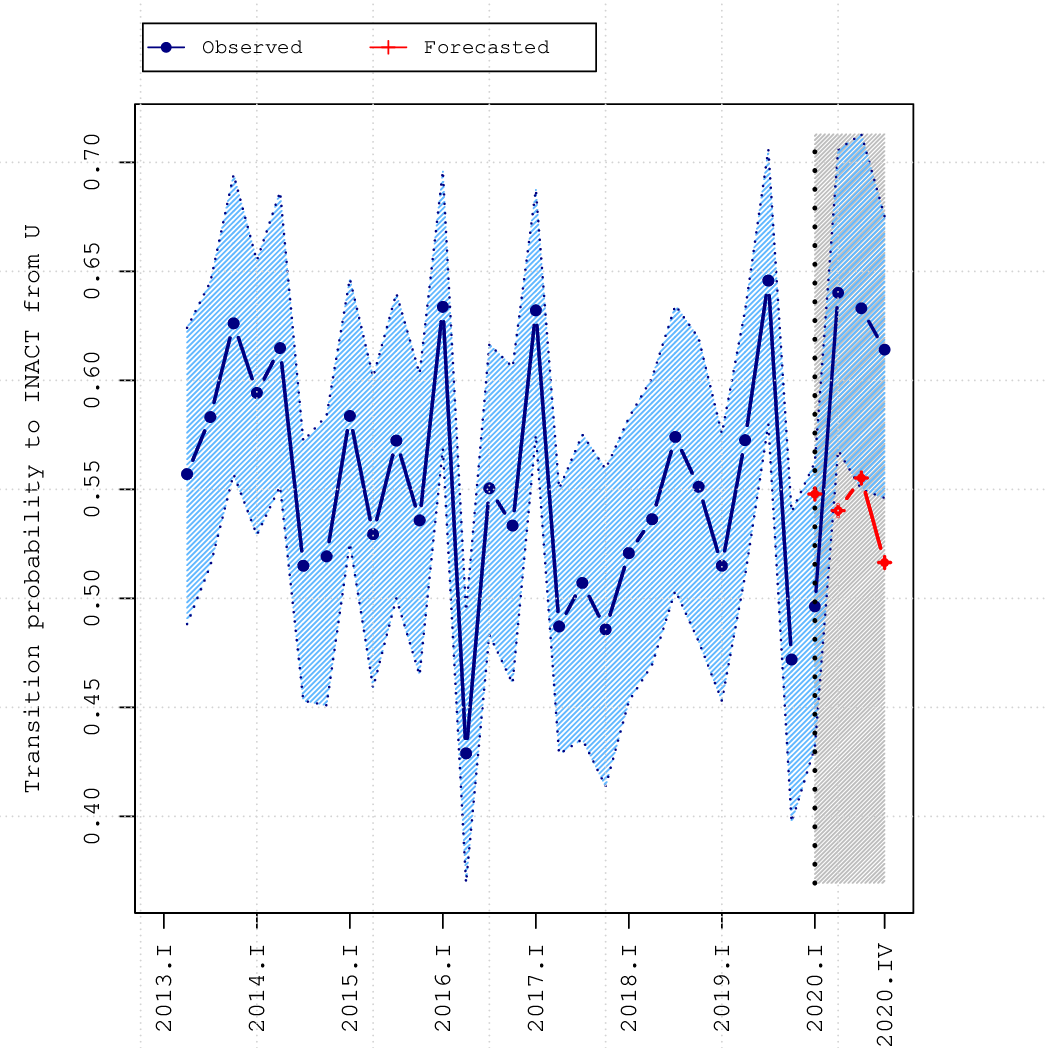}
		\caption{South.}
		\label{fig:transProbFromEDUtoSEApp_IIIII}
	\end{subfigure}
	\begin{subfigure}[t]{0.3\textwidth}
		\centering
		\includegraphics[width=\linewidth]{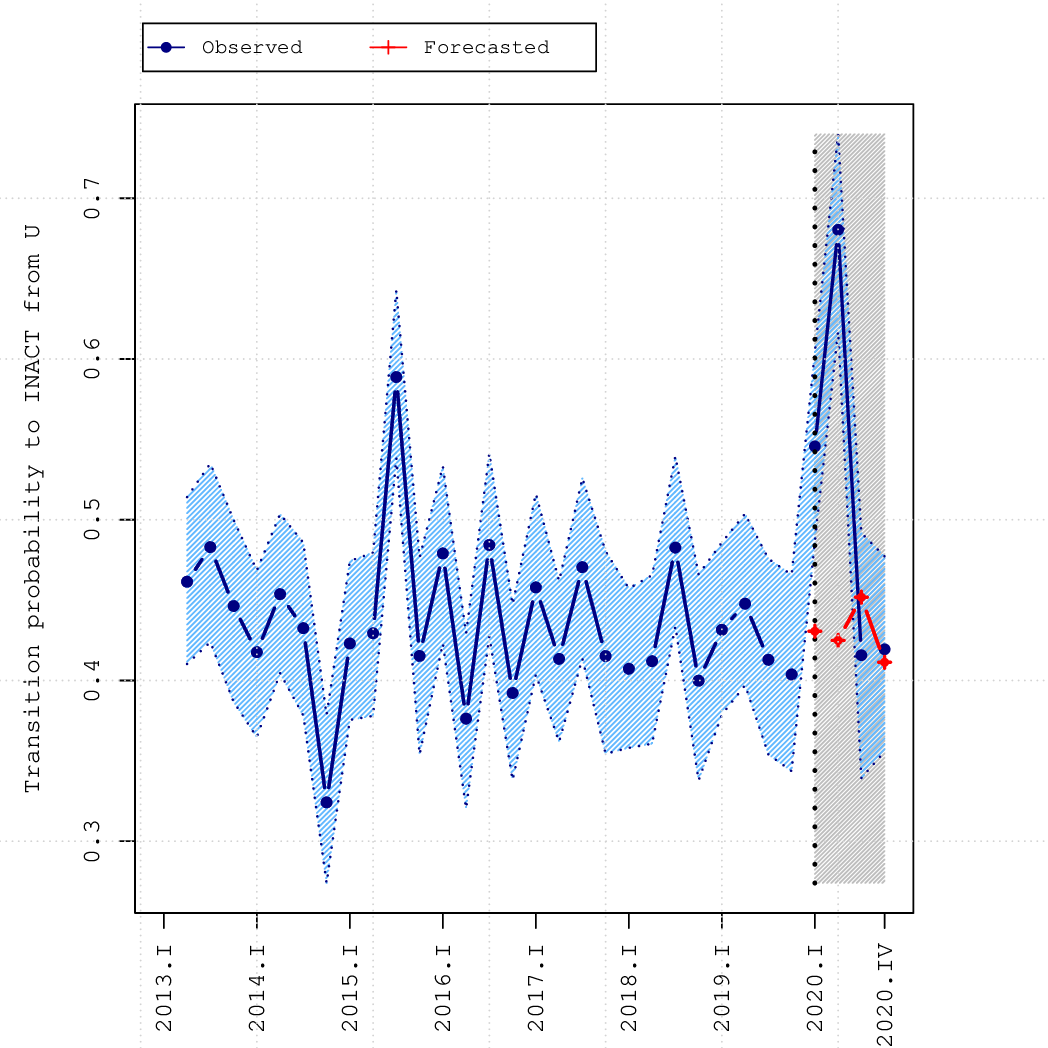}
		\caption{North.}
		\label{fig:transProbFromEDUtoTEApp_III}
	\end{subfigure}
	\vspace{0.2cm}
	\caption*{ \scriptsize{\textit{Note}: The forecasted transition probabilities are computed using a combination of four forecasting models (ETS, TSLM, THETAF, and ARIMA) \citep{HyndmanAthanasopoulosforecasting2021} in the period 2013 (quarter I)- 2019 (quarter IV). Confidence intervals at 90\% are computed using 1000 bootstraps and reported in parenthesis. The grey area identifies the COVID period. North includes regions in the North and the Center. \textit{Source}: LFS 3-month longitudinal data as provided by the Italian Institute of Statistics (ISTAT).}}
\end{figure}

As regards whether an increased number of workers became pessimistic about the probability to find a job and gave up on job search during the pandemic, Panel B of Figure \ref{fig:transProbUtoinactive} reports the transition probabilities from unemployment to the inactive  state for females in the 30-39 age cohort. Across both geographical locations these probabilities have significantly increased in quarter II of 2020, compared to the forecasted probabilities. However, the change appears to be temporary as the transition probabilities went back to the pre-pandemic values in quarter IV of 2020. Females in the  25-29 age cohort (Panel A of Figure \ref{fig:transProbUtoinactive}) show increased transition probabilities in both geographical locations, which persisted until quarter III of 2020. Females in the  40-49 age cohort (Panel C of Figure \ref{fig:transProbUtoinactive}) also appear to have  transited much more from unemployment to inactivity in both  locations, but while being a temporary change for women in the North, it persisted until quarter IV of 2020 among females in the South. Similar patterns are observed among 30-39 males in the South, while among both males and females in the North, we do not observe persistence (Figure \ref{appfig:tranProbUtoinactiveMales} in Appendix \ref{app:dynamicsduring}). To summarize, while at the outburst of the pandemic female unemployed workers across different locations and age categories got discouraged to some degree and left the labour market, the higher transition rates from unemployment to inactivity persisted mostly for females aged 40-49 in the South. 

\begin{figure}[!htbp]
	\caption{Transition probabilities of females from permanent employment to the inactive  state by age groups.}
	\label{fig:transProbPEtoinactiveFemales}
		\vspace{0.1cm}
	\caption*{\scriptsize{\textbf{Age 25-29}}.}
	\vspace{0.1cm}
	\centering
	\begin{subfigure}[t]{0.3\textwidth}
		\centering
		\includegraphics[width=\linewidth]{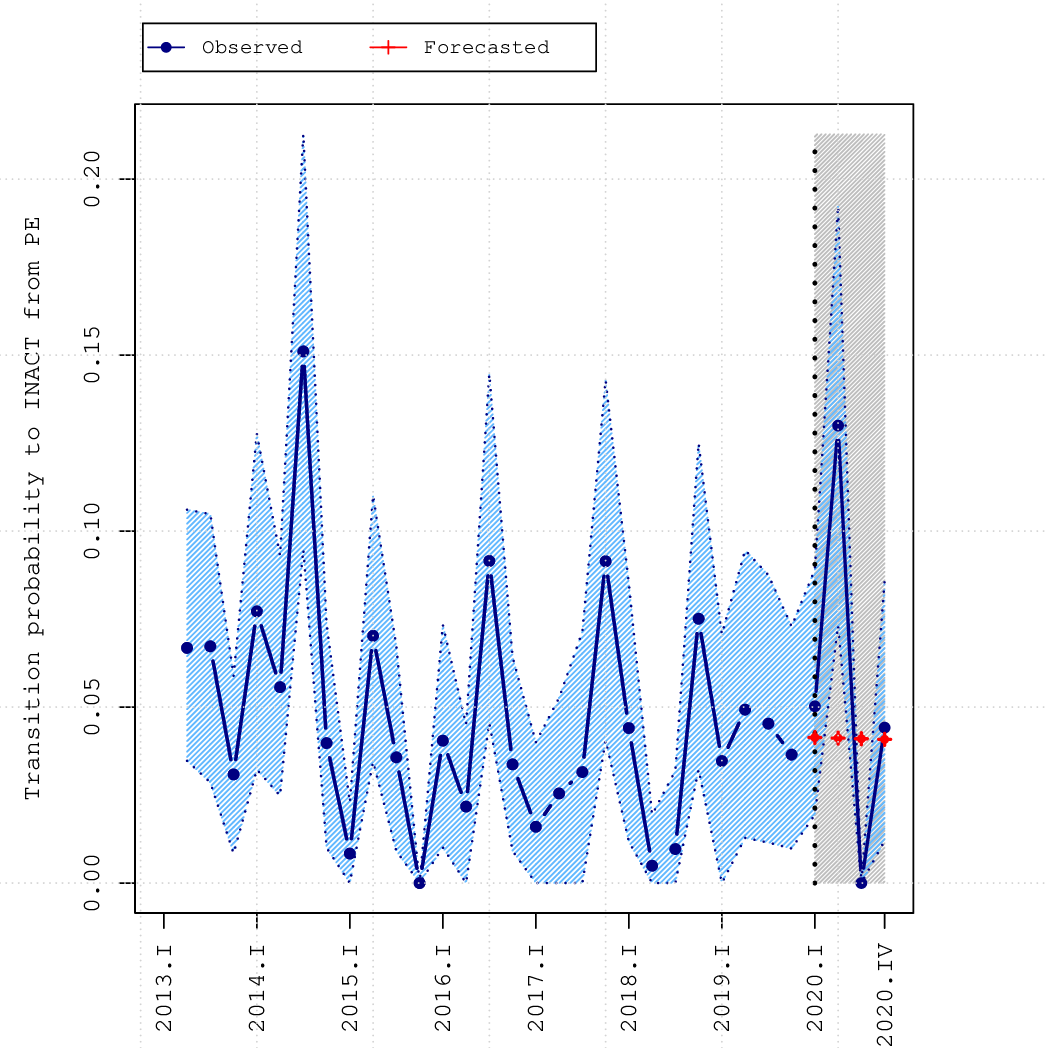}
		\caption{South.}
		\label{fig:transProbFromEDUtoSEApp_IIIIII}
		\vspace{0.2cm}
	\end{subfigure}
	\begin{subfigure}[t]{0.3\textwidth}
		\centering
		\includegraphics[width=\linewidth]{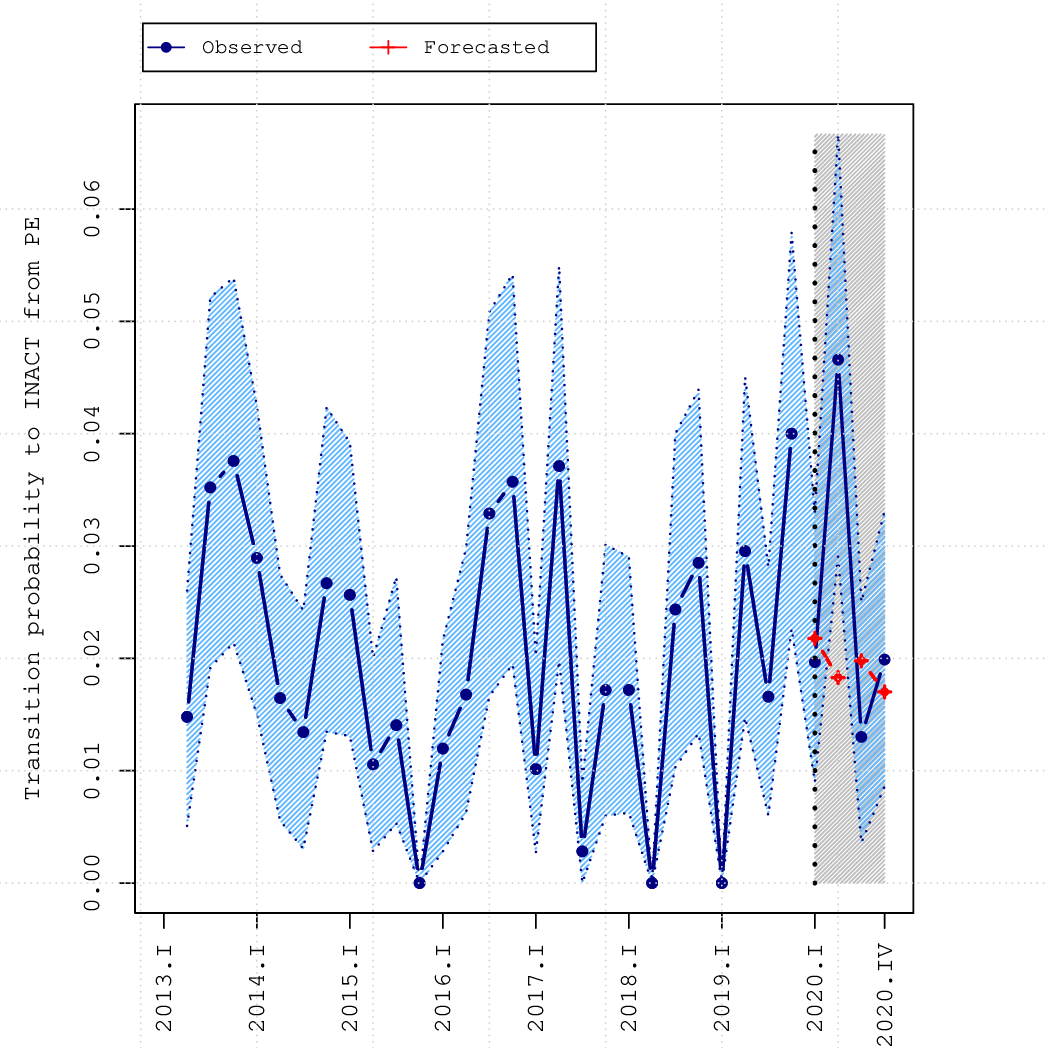}
		\caption{North.}
		\label{fig:transProbFromEDUtoTEApp_IIII}
	\end{subfigure}\vspace{0.1cm}
	\caption*{\scriptsize{\textbf{Age 30-39}}.}
	\vspace{0.1cm}
	\centering
	\begin{subfigure}[t]{0.3\textwidth}
		\centering
		\includegraphics[width=\linewidth]{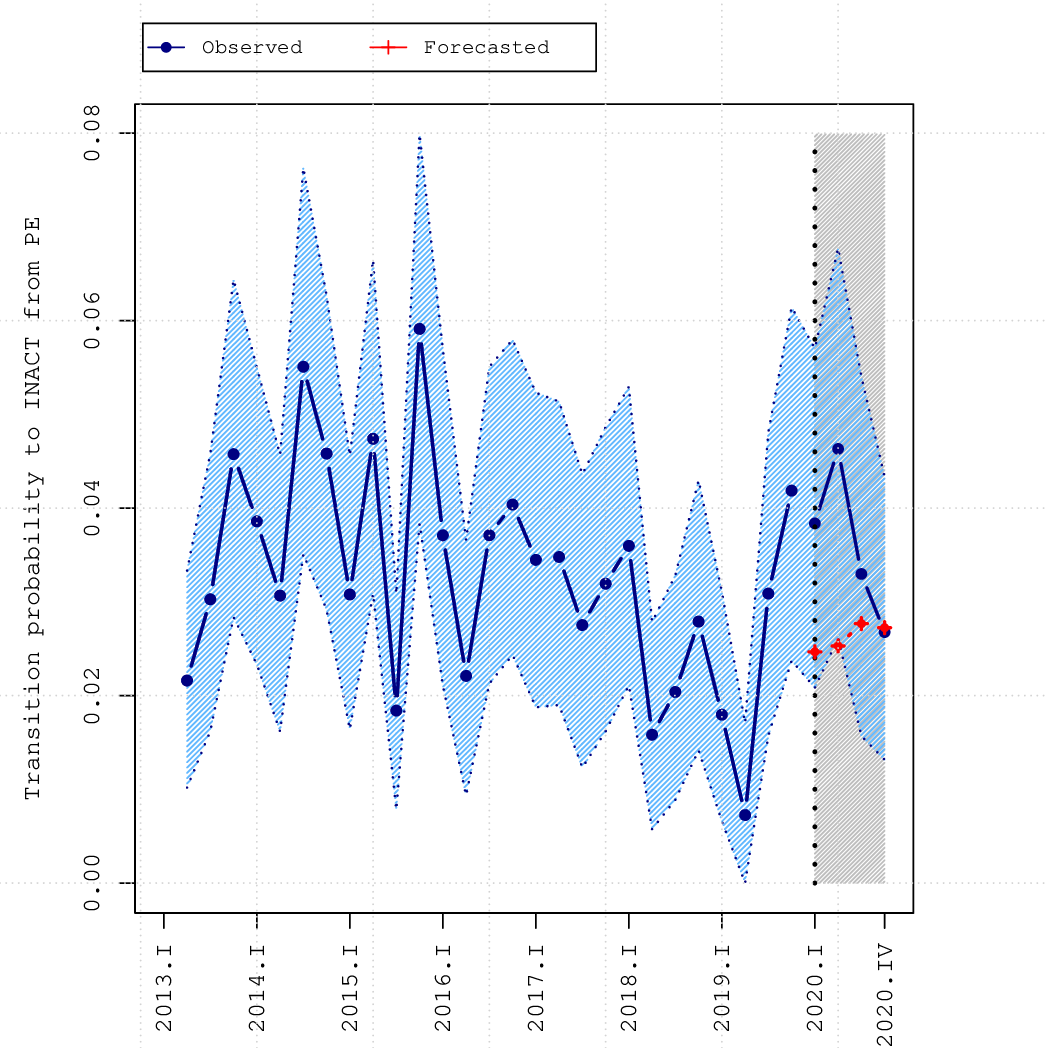}
		\caption{South.}
		\label{fig:transProbFromEDUtoSEApp_IIIIIII}
		\vspace{0.2cm}
	\end{subfigure}
	\begin{subfigure}[t]{0.3\textwidth}
		\centering
		\includegraphics[width=\linewidth]{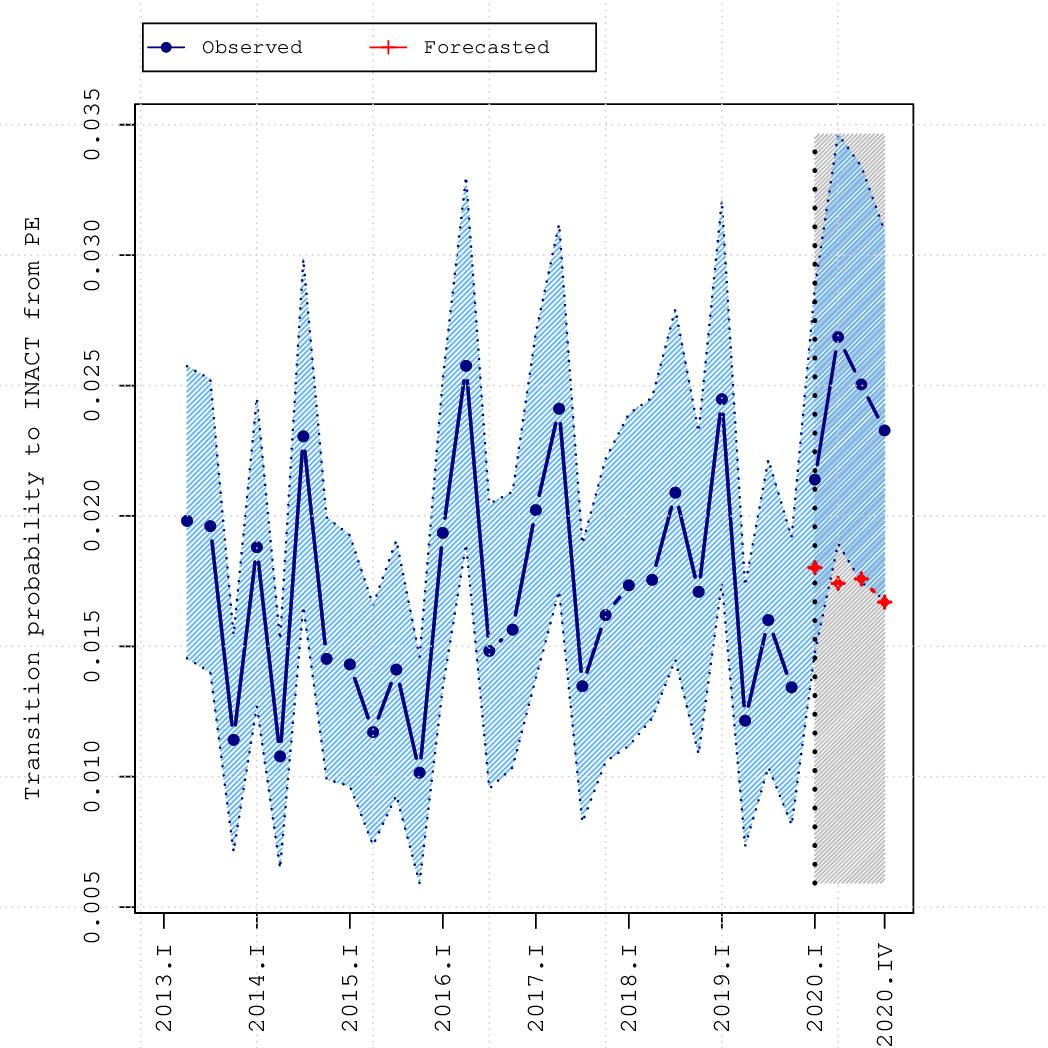}
		\caption{North.}
		\label{fig:transProbFromEDUtoTEApp_IIIII}
	\end{subfigure}
	\vspace{0.1cm}
	\caption*{\scriptsize{\textbf{Age 40-49}}.}
	\vspace{0.1cm}
	\centering
	\begin{subfigure}[t]{0.3\textwidth}
		\centering
		\includegraphics[width=\linewidth]{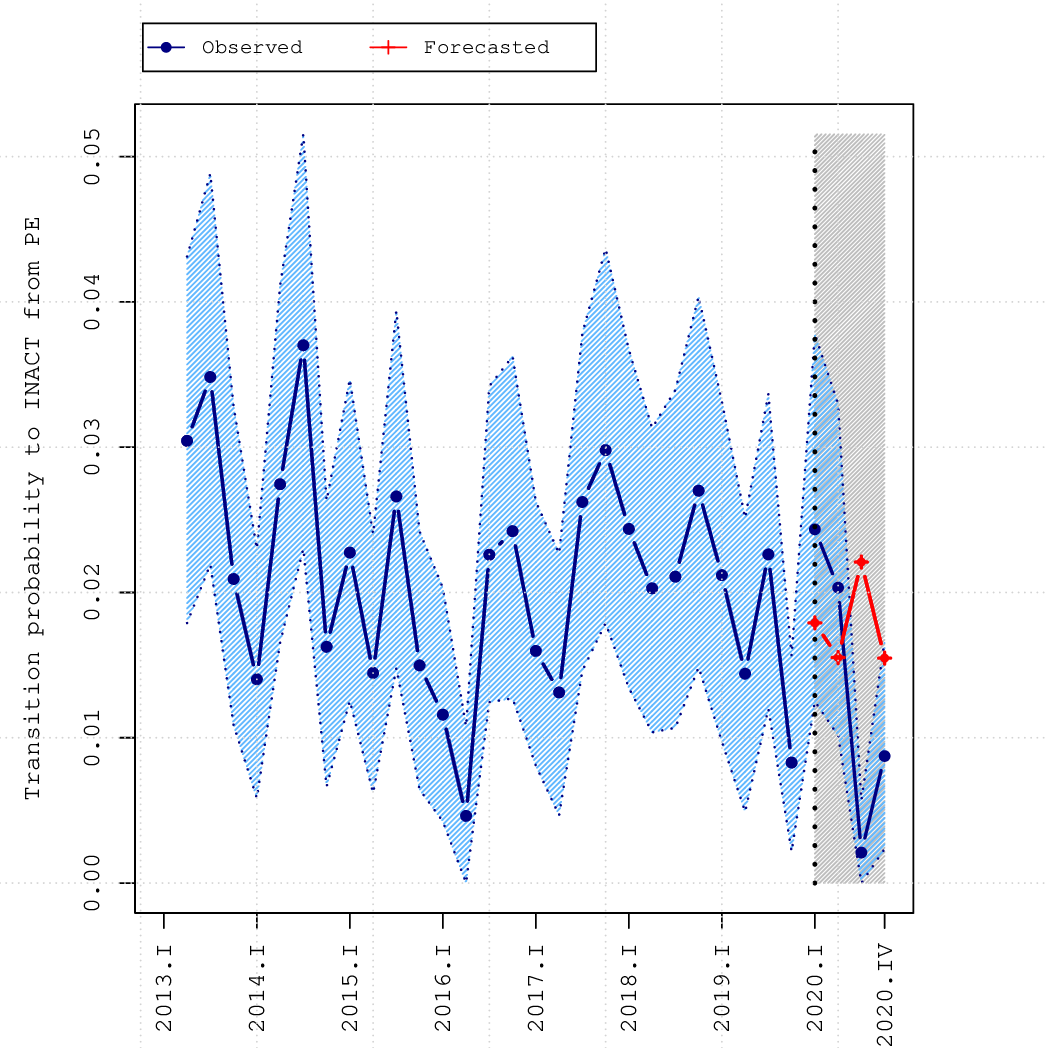}
		\caption{South.}
		\label{fig:transProbFromEDUtoSEApp_IIIIIIII}
		\vspace{0.2cm}
	\end{subfigure}
	\begin{subfigure}[t]{0.3\textwidth}
		\centering
		\includegraphics[width=\linewidth]{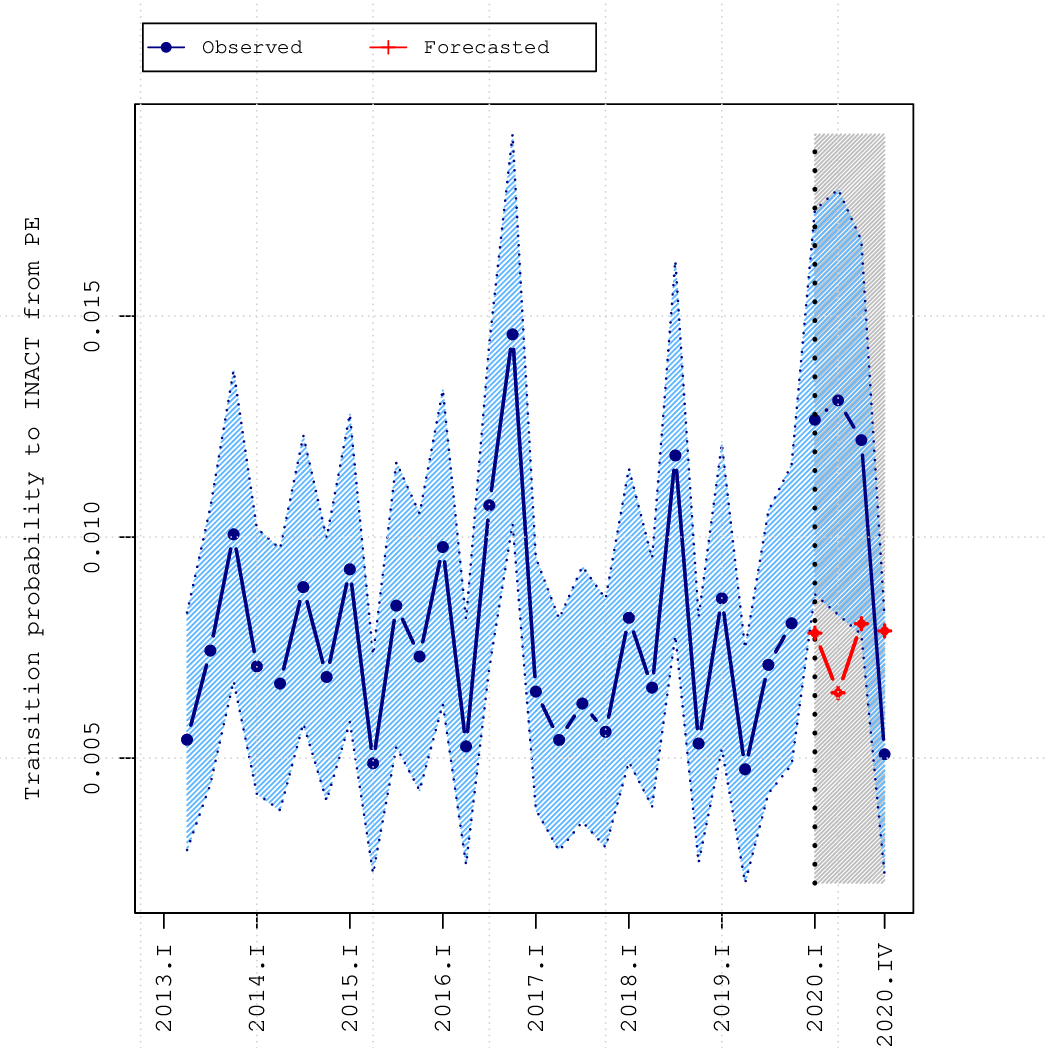}
		\caption{North.}
		\label{fig:transProbFromEDUtoTEApp_IIIIII}
	\end{subfigure}
	\vspace{0.2cm}
	\caption*{ \scriptsize{\textit{Note}: The forecasted transition probabilities are computed using a combination of four forecasting models (ETS, TSLM, THETAF, and ARIMA) \citep{HyndmanAthanasopoulosforecasting2021} in the period 2013 (quarter I)- 2019 (quarter IV). Confidence intervals at 90\% are computed using 1000 bootstraps and reported in parenthesis. The grey area identifies the COVID period. North includes regions in the North and the Center. \textit{Source}: LFS 3-month longitudinal data as provided by the Italian Institute of Statistics (ISTAT).}}
\end{figure}

As regards the effect of the pandemic on the decision to leave employment, the pandemic significantly increased the transition probabilities from  permanent employment to the inactive  state for females in the 30-39 age category living in the North (Figure \ref{fig:transProbPEtoinactiveFemales}). In quarter III of 2020 the probability of females aged 30-39 to transit from permanent employment to the inactive  state jumped to 2.5\% compared to a forecasted probability of 1.7\%. Similar patterns are found for the probability to transit from temporary employment to the inactive state: it jumped to 25\% compared to the forecasted 10\% (Table \ref{appfig:transProbTEtoinactiveFemales} in Appendix \ref{app:dynamicsduring}).

On the contrary, the transition probabilities from temporary and permanent employment for women in the 30-39 age group living in the South do not show any significant change. Moreover, the transition probabilities from permanent employment to inactivity for women in the 40-49 age group living in the North increased just in the first three quarters of 2022, while no significant  changes happened for women living in the South. Finally, no changes are observed  in the transition probabilities of males across different age categories (Figures \ref{appfig:transProbTEtoinactiveMales}-\ref{appfig:transProbPEtoinactiveMales} in Appendix \ref{app:dynamicsduring}).

To summarize, during the outburst of the pandemic there was a significant outflow of women in the age 30-39 in the North from permanent and temporary employment to inactivity. On the contrary, there is no significant evidence of the same phenomenon among women in other age categories, and among males and females in the 30-39 age group in the South. 

\subsection{Female labour market participation and household composition}

A possible explanation of the significant and negative impact of the pandemic on  female labour force participation is the role played by  childcare responsibilities.
Across all Italian regions, more than 60\% of women have children when they are between 30 and 40 years old (Table \ref{apptab:birthbyage} in the Appendix). Unfortunately, the Labour Force Survey (LFS) does not provide information about the number and age of children, but only about the household size. Therefore, we use  the European Labour Force Survey data for Italy for 2019 to compute the shares of females in the age cohorts 30-39 and 40-49 with at least one child below the age of 11, i.e., an age below which children need the presence of their parents, by employment status and geographical location (Table \ref{sharechangesELFS2019}).\footnote{In Appendix \ref{app:householdSize} we report similar statistics for 2020.}
  
\begin{table}[!htbp]
	\centering 
  	\caption{Percentage of females with at least one child below the age of 11 by geographical area and employment status in 2019.} 
  	\label{sharechangesELFS2019} 
  	\scriptsize
  	\begin{tabular}{@{\extracolsep{0pt}} lcccc} 
  	     \hline \hline \\[-1.8ex] 
  		&\multicolumn{2}{c}{Age 30-39}&\multicolumn{2}{c}{Age 40-49}\\	
  		\hline \\[-1.8ex]
  		&North&South&North&South\\
  		\hline \\[-1.8ex]
  		Employed&64.8&33.1&73.5&43.2\\\\[-1.8ex]
  \hspace{0.5cm}Permanent  &47.5&19.1&55.0&28.1\\
  \hspace{0.5cm}Temporary&8.7&7.1&6.4&5.9\\
  \hspace{0.5cm}Self-employed&8.6&6.9&12.1&9.2\\\\[-1.8ex]
    Unemployed& 6.2&10.2&5.4&8.4\\
\textbf{Inactive}&	28.2&55.7&20.2&48.0\\
  \hline \\[-1.8ex]
Total (in 000s)& 1270&736&1249&559\\
  	\hline \hline
  	\multicolumn{5}{l}{\tiny{\textit{Note}: North includes regions in the North and the Center.}}\\\multicolumn{5}{l}{\tiny{\textit{Source}: ELFS data.}}
  	\end{tabular}	
  \end{table}

In the North,  47.5\% of females in the 30-39 age cohort with at least one young child is employed on a permanent contract, compared to 19.1\% in the South. While the share on temporary employment and self-employment is comparable, and the share in unemployment is slightly higher in the South (10.2\% against 6.2\%), the striking difference is in the share of inactive females, which is much higher in the South, i.e., 55.7\% versus 28.2\%. The 40-49 age cohort shows very similar patterns.

Table \ref{shareHH2ELFS2019} reports the shares of females with at least one child below the age of 11 by household size, geographical area and age cohort. Among females in the 30-39 age cohort, 81.2\% live in a household with more than two components in the North, and 71.5\% in the South. Hence, household size (the only information available in LFS on individuals' family) appears to be a good proxy for the presence of small children in the household.  

\begin{table}[!htbp]
	\centering 
	\caption{Percentage of females with at least one child below the age of 11 by geographical area and household size.}  
	\label{shareHH2ELFS2019} 
	\scriptsize
	\begin{tabular}{@{\extracolsep{5pt}} ccccc} 
		\hline \hline \\[-1.8ex] 
		Females with at least one child below the age of 11	&\multicolumn{2}{c}{Age 30-39}&\multicolumn{2}{c}{Age 40-49}\\	
		\hline \\[-1.8ex]
		Household size & North&South&North&South\\
		\hline \\[-1.8ex]
		$>$ 2 components&81.2& 71.5&55.2&44.0\\
		$\leq$ 2 components& 6.6&9.3&6.0&4.8\\
		\hline \hline
		\multicolumn{5}{l}{\tiny{\textit{Note}: North includes regions in the North and the Center. \textit{Source}: ELFS data.}}
	\end{tabular}	
\end{table}	

\begin{figure}[!htbp]
	\caption{Transition probabilities of females aged 30-39 from permanent employment to the inactive state in the North and South of Italy by household size.}
	\label{fig:transProbHHsize}
	\centering
		\begin{subfigure}[t]{0.3\textwidth}
		\centering
		\includegraphics[width=\linewidth]{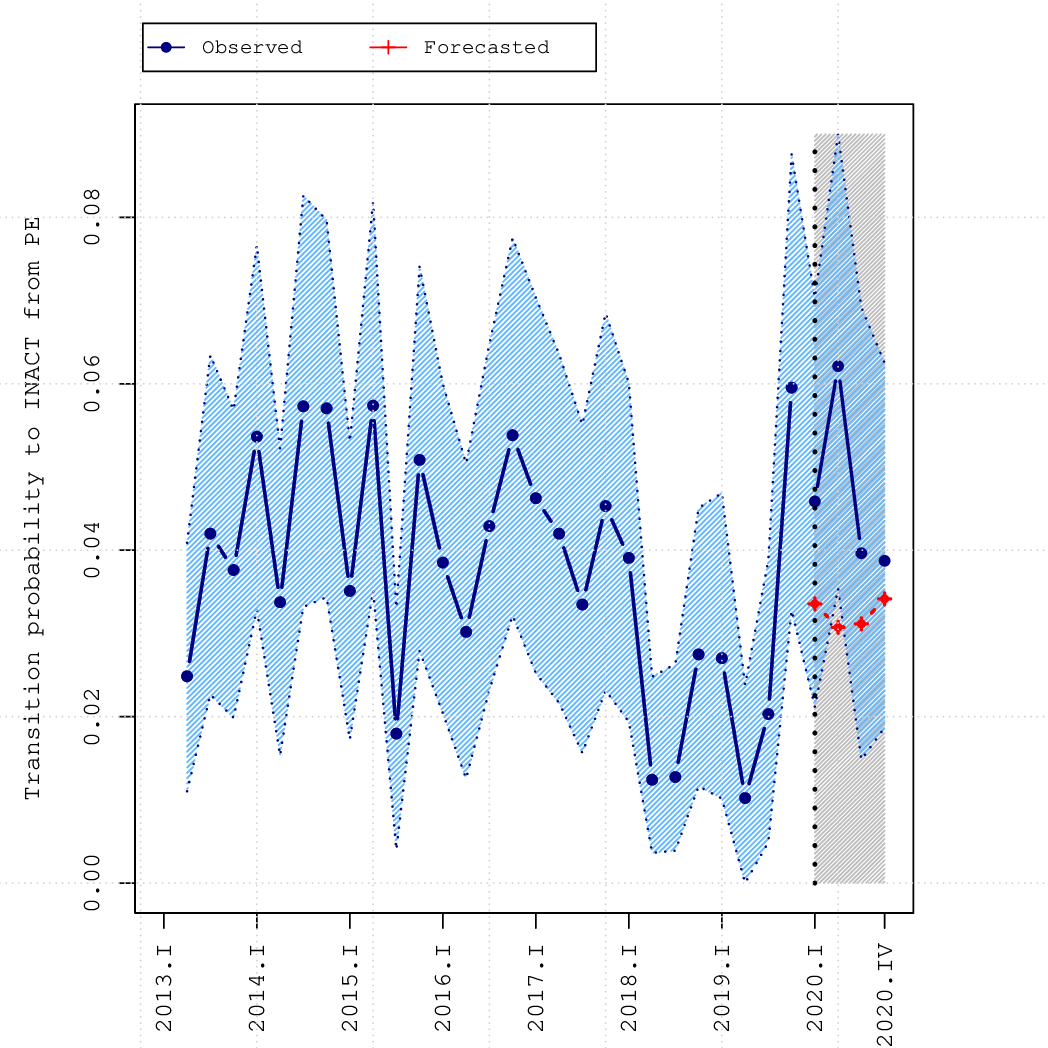}
		\caption{South - Household size $>$ 2.}
		\label{fig:transProbFromEDUtoPE_II}
	\end{subfigure}
	\begin{subfigure}[t]{0.3\textwidth}
		\centering
		\includegraphics[width=\linewidth]{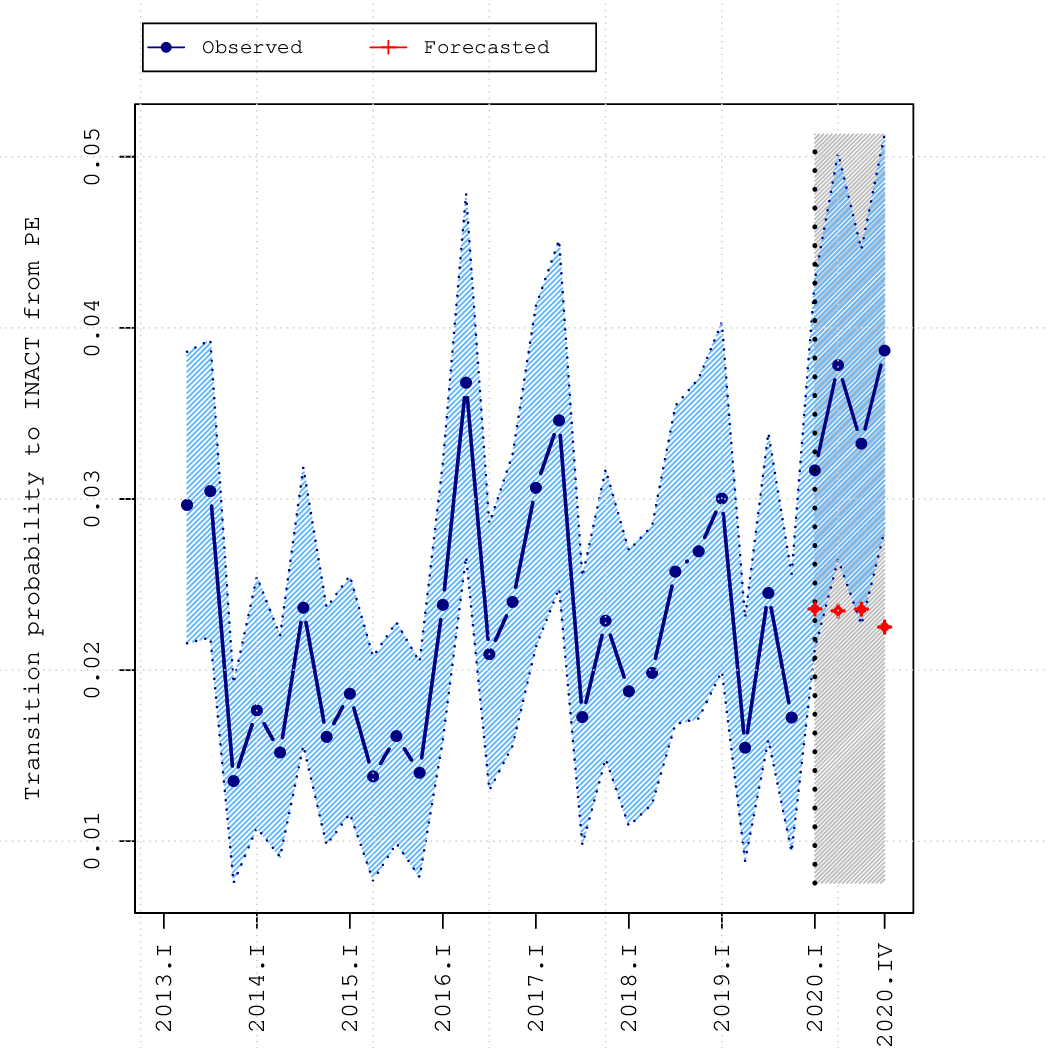}
		\caption{North - Household size $>$ 2.}
		\label{fig:transProbFromEDUtoSE_I}
		\vspace{0.1cm}
	\end{subfigure}\\
	\begin{subfigure}[t]{0.3\textwidth}
	\centering
	\includegraphics[width=\linewidth]{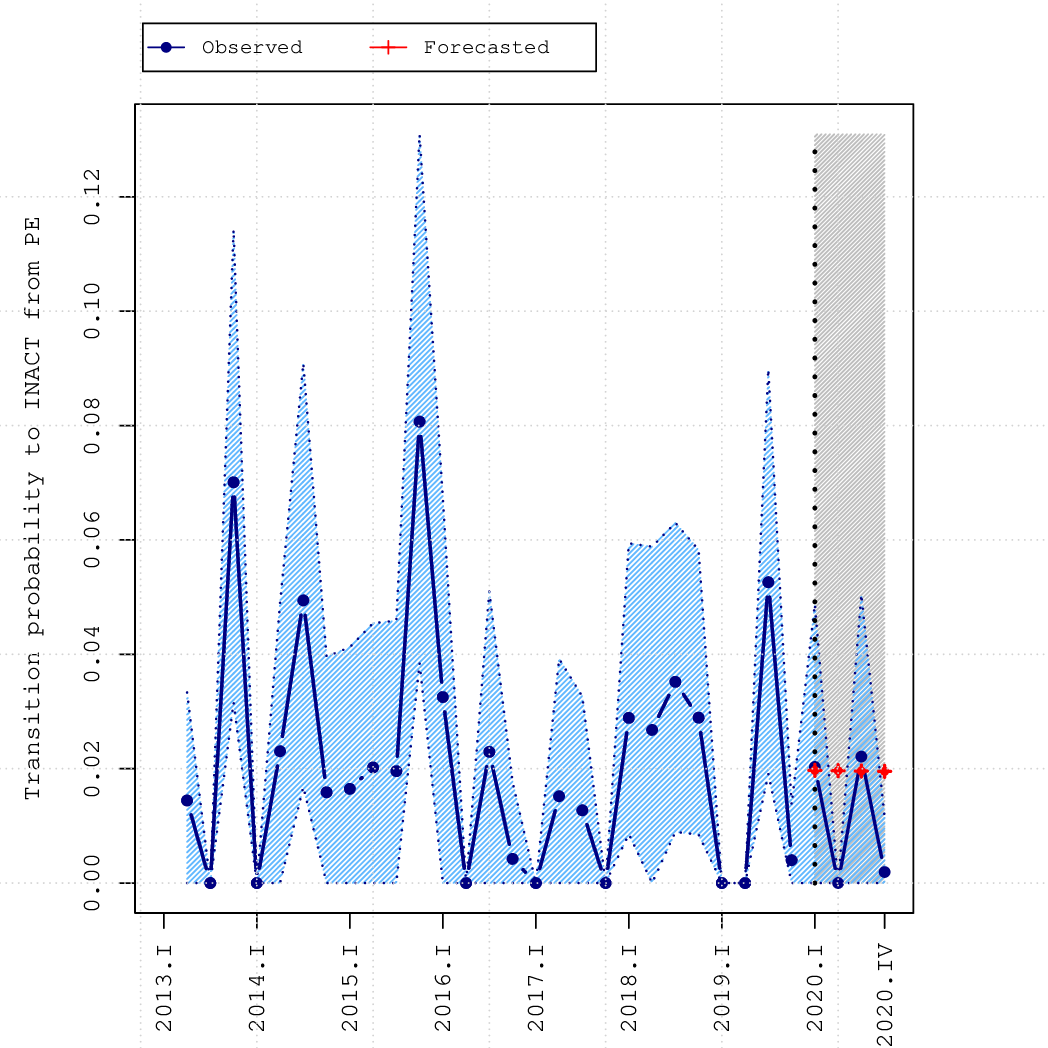}
	\caption{South - Household size $\leq$ 2.}
	\label{fig:transProbFromEDUtoPE_III}
\end{subfigure}
	\begin{subfigure}[t]{0.3\textwidth}
		\centering
		\includegraphics[width=\linewidth]{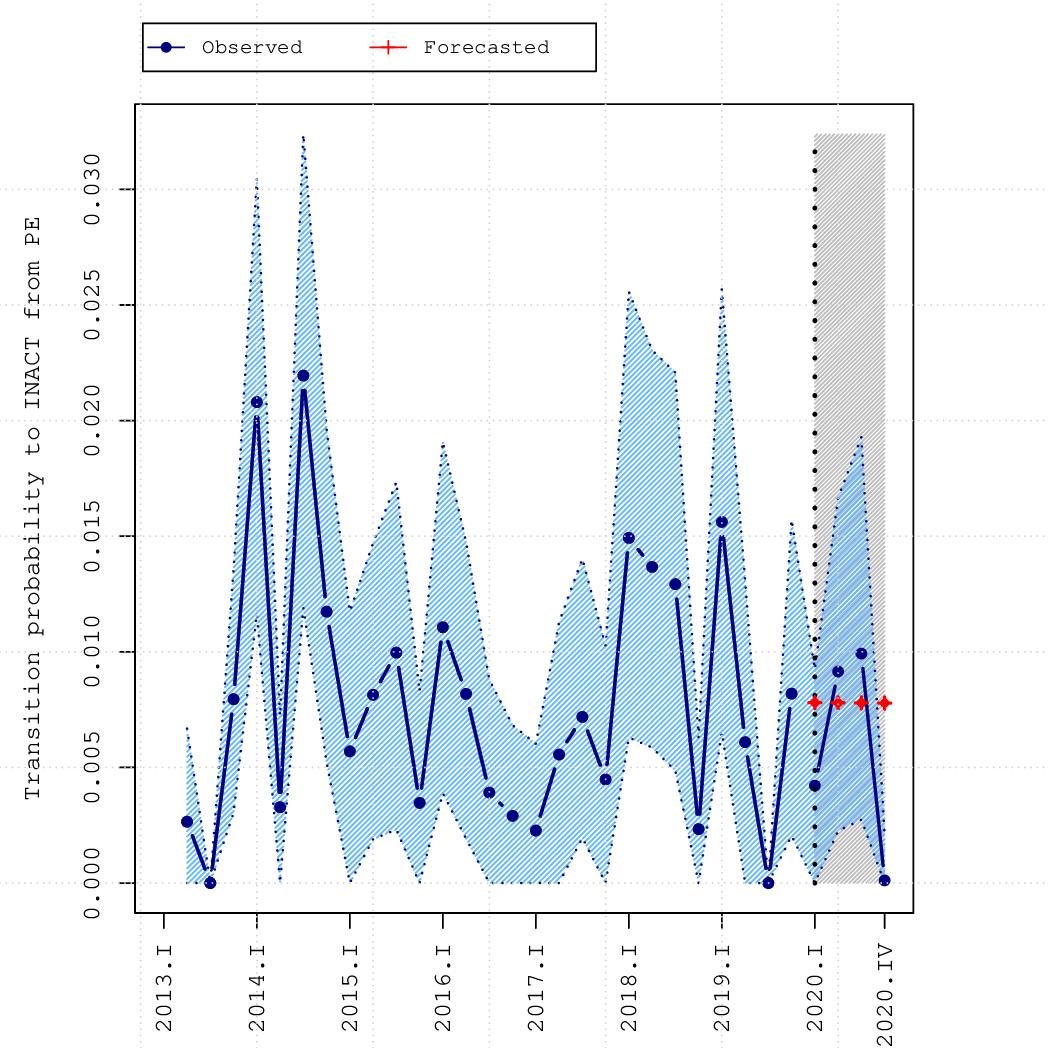}
		\caption{North - Household size $\leq$ 2.}
		\label{fig:transProbFromEDUtoPE_IIII}
	\end{subfigure}\\
	\vspace{0.2cm}
\caption*{\scriptsize{\textit{Note}: Confidence intervals at 90\% are computed using 1000 bootstraps. The grey area identifies the COVID period. North includes regions in the North and the Center. \textit{Source}: LFS 3-month longitudinal data as provided by the Italian Institute of Statistics (ISTAT).}}
\end{figure}

Using the household size of more than 2 members as a cutoff to split the sample of females, the estimated transition probabilities from permanent employment to the inactive state show an even clearer geographical pattern (Figure \ref{fig:transProbHHsize}).
In a household with more than two individuals the transition probabilities from permanent employment to the inactive state have significantly and persistently increased for females in the 30-39 age cohort living in the North. On the contrary, no statistically significant effect is present for females living in households with less than two components neither in the North or the South, neither from permanent nor from temporary employment. No effect is observed also among females living in the South in a household with more than two components. The same patterns are found when looking at the transitions from temporary employment (Figure \ref{appfig:transProbHHsizefromTE} in Appendix \ref{app:dynamicsduring}).

Females with young children living in the North of Italy have therefore been charged with the highest toll. In light of the pre-pandemic evidence of weaker attachment of females to the labour market compared to males, and of high female state dependency in the inactivity state \citep{duhautois2018state}, an additional worry is the persistence of the increased outflow toward inactivity until the end of 2020.

As explained by \cite{honore2000panel}, this higher persistence in the inactive state might be due two different factors. First, it may be ascribable to lagged decisions (\textit{true state dependency}); for instance, the choice of women to remain in the inactive state is linked to their fertility choice, i.e., the choice to leave the labour market to have children and remain inactive to take care of them. Second, it could be due to a reaction to the shock, i.e., individual preferences have changed as a consequence of the pandemic (\textit{spurious state dependency}). For instance, a woman might have decided to remain inactive due to the worsening of the labour market conditions.
Distinguishing between these two types of state dependency is extremely important for policy reasons. However, the identification would require long longitudinal data on individual histories and information on individual preferences, which unfortunately are not available. In the next section, we move a first step toward disentangling the two explanations.

\subsection{The determinants of labour force participation}\label{sec:logit}

The non-parametric analysis based on transition probabilities cannot easily handle the strong heterogeneity  between the North and South of Italy along several dimensions. This is even more severe in our case due to the (relatively small) sample size.
To circumvent this issue, we use a parametric approach based on a \textit{logit} model, which allows to control for several individual and job characteristics, such as gender, age, sector, education, household size, etc.

We use the sample  of females in the 30-39 age cohort and repeat the same estimation on other subsamples including different age groups and gender. The estimation of a logit model using design-based longitudinal weights may create severe numerical issues \citep{train2009discrete}. Hence, we run 1000 bootstraps using the longitudinal sample weights to estimate the model's coefficients and their 90\% and 95\% confidence intervals.

Table \ref{tab:logit3039femalepart} displays the estimated odds-ratios of the probability to remain active on the labour market, conditional on being active the quarter before, given a set of explanatory variables for the subsample of females in the 30-39 age cohort.\footnote{The odds-ratio represents the ratio between the probability that the event will occur with respect to the probability the event will not occur, conditioned to a given explanatory variable; hence, an odds-ratio grater than one implies an increased occurrence of the event, while an odds-ratio lower than one implies a decreased occurrence of the event with respect to a given explanatory variable.} The reference category includes all individuals with a tertiary level of education living in the South in a household with less than two people working in the agriculture sector.\footnote{The full regression with all the variables included is reported in Table \ref{apptab:logit3039femalepart} in Appendix \ref{app:logit}.}

\begin{table}[!htbp]
	\centering
		\caption{Odds-ratios of being active next quarter for an individual currently active in the labour  market (females age 30-39).} 
	\label{tab:logit3039femalepart}
	\begin{scriptsize}
	\begin{tabular}{lccccc}
		\hline \hline
			\\[-1.8ex]
	 \textbf{Females age 30-39}	& C.I. & C.I.  & Mean & C.I.  & C.I.  \\ 
			&  2.5\% &  5\% &   bootstrap &  95\% &  97.5\% \\ 	\\[-1.8ex]
		\hline 
		\\
		\textbf{	COVID $\times$ North  $\times$ } &  &  &  &  &  \\ 
	\textbf{	$\times$ Household members$>$2} & 0.445 & 0.470 & \textbf{0.690} & 0.964 & 1.035 \\ 
		\hline \\[-1.8ex]
		Industry & 2.379 & 2.450 & \textbf{2.984} & 3.542 & 3.676 \\ 
		Constructions & 1.915 & 2.015 & \textbf{2.886} & 3.991 & 4.294 \\ 
		Commerce & 2.667 & 2.753 & \textbf{3.303} & 3.923 & 4.036 \\ 
		Hotels and restaurants & 1.130 & 1.170 & \textbf{1.400} & 1.672 & 1.701 \\ 
		Transport & 2.382 & 2.486 & \textbf{3.377} & 4.555 & 4.792 \\ 
		Communications & 2.210 & 2.319 & \textbf{3.226} & 4.394 & 4.700 \\ 
		Finance & 3.385 & 3.619 &\textbf{ 4.917} & 6.608 & 7.105 \\ 
		Real estate & 2.450 & 2.539 & \textbf{3.053} & 3.631 & 3.749 \\ 
		Public administration & 3.276 & 3.492 & \textbf{5.111} & 7.299 & 7.907 \\ 
		Education and health & 1.709 & 1.760 & \textbf{2.077} & 2.457 & 2.537 \\ 
		Other sectors & 1.574 & 1.624 & \textbf{1.958} & 2.320 & 2.388 \\ 
		
		Primary education & 0.551 & 0.564 & \textbf{0.629} & 0.694 & 0.705 \\ 
		Secondary education & 0.772 & 0.785 & \textbf{0.862} & 0.943 & 0.959 \\ 

		North  & 1.553 & 1.595 & \textbf{1.836 }& 2.094 & 2.139 \\ 

		Household members$>$2 & 0.592 & 0.610 & \textbf{0.705} & 0.804 & 0.819 \\ 
		North x Household members$>$2 & 0.833 & 0.856 & 1.009 & 1.181 & 1.222 \\ 
		\\[-1.8ex] 
		COVID & 0.475 & 0.509 & 0.886 & 1.436 & 1.574 \\ 
		COVID x Industry & 0.358 & 0.393 & 0.674 & 1.014 & 1.098 \\ 
		COVID x Constructions & 0.175 & 0.204 & 0.548 & 1.194 & 1.412 \\ 
		COVID x Commerce & 0.295 & 0.344 & \textbf{0.572} & 0.861 & 0.936 \\ 
		COVID x Hotels and restaurants & 0.218 & 0.247 & \textbf{0.411} & 0.619 & 0.665 \\ 
		COVID x Transport & 0.161 & 0.179 & \textbf{0.362} & 0.624 & 0.700 \\ 
		COVID x Communications & 0.307 & 0.358 & 0.800 & 1.488 & 1.745 \\ 
		COVID x Finance & 0.381 & 0.442 & 1.076 & 2.068 & 2.614 \\ 
		COVID x Real estate & 0.514 & 0.560 & 0.974 & 1.458 & 1.565 \\ 
		COVID x Public administration & 0.594 & 0.668 & 2.422 & 5.852 & 6.747 \\ 
		COVID x Education and health & 0.372 & 0.413 & 0.678 & 1.010 & 1.091 \\ 
		COVID x Others & 0.425 & 0.458 & 0.777 & 1.168 & 1.265 \\ 
		COVID x Primary education & 0.853 & 0.891 & 1.162 & 1.461 & 1.543 \\ 
		COVID x Secondary education & 0.700 & 0.725 & 0.878 & 1.045 & 1.073 \\ 
		COVID x North  & 0.688 & 0.730 & 1.014 & 1.356 & 1.420 \\ 
		COVID x Household members$>$2 & 0.939 & 0.993 & 1.379 & 1.811 & 1.912 \\
		\hline
		Observations&\multicolumn{5}{c}{57264}\\
		\hline 
		\hline
		\multicolumn{6}{l}{\tiny{Note: year and quarter fixed effects and country of origin of the individual are included.}} \\
		\multicolumn{6}{l}{\tiny{North includes regions in the North and the Center.  \textit{Source}: LFS 3-month longitudinal data as provided }}\\ \multicolumn{6}{l}{\tiny{by the Italian Institute of Statistics (ISTAT).}}
	\end{tabular}
	\end{scriptsize}
\end{table}

Independently on the pandemic, the probability to remain active is on average higher for workers in all sectors, compared to agriculture, which is our baseline. However, the coefficients vary widely across sectors, ranging from 1.4 for hotels and restaurant, to 5.1 for public administration. Not surprisingly, primary and secondary educated individuals show a lower likelihood to persist in the active state, compared to tertiary educated individuals. The geographical location is important: living in the North increases the likelihood to remain active, compared to living in the South. Appendix \ref{app:logit} shows that the same applies for 30-39 years old males and for both females and males in the 40-49 age group. These results point to a higher efficiency of the labour market in the North, likely due to its structural and institutional settings.

Finally, living in a household with more than two people significantly decreases the probability to remain active on the labour market. 
As regards the effect of the pandemic, we do not find asymmetric effects by education and household size. The shock appears to be symmetric even across the North and the South, i.e., COVID did not change the differential geographical efficiency of the labour market (the same applies for males and for both females and males in the age group 40-49, see Appendix \ref{app:logit}). The observed differential geographical changes in the labour force participation cannot be therefore attributed to changes in structural and institutional settings.

The only factor that plays a role is the sector of work: the odds-ratio is significantly below one in sectors such as hotel and restaurants, commerce and transportation, while it is not statistically significant across all other sectors.

The estimate of a lower-than-one coefficient for the interaction among the COVID period dummy, the North dummy, and the household-size-bigger-than-2 dummy suggests a significant reduction in the probability to be active on the labour market for females in the 30-39 age cohort, living in the North of Italy and in a household with more than two people. On the contrary, among older women we do not observe any significant change (Tables \ref{apptab:logit4049femalepart}-\ref{apptab:logit4049malepart} in Appendix \ref{app:logit}). 
Among males in the 30-39 age category, the same coefficient is marginally significant, suggesting that they have been struggling too during the pandemic (Table \ref{apptab:logit3039malepart} in Appendix \ref{app:logit}). 
Among females in the 30-39 age cohort, the probability to persist in inactivity is not affected by the pandemic. On the contrary, for males in the 30-39 age cohort with primary and secondary education, the estimates point to a return to the labour force over the 2020 (Table \ref{tab:logit3039nonpart}).\footnote{The full estimates are reported in Tables \ref{apptab:logit3039femalenonpart} and \ref{apptab:logit3039malenonpart} in Appendix \ref{app:logit}.} 

To summarize, although males in the 30-39 age cohort were also hurt by the pandemic in a similar manner as females, they went back to the labour market towards the end of 2020, while females of the same age cohort did not. 
	
\begin{table}[!htbp]
		\centering 
		\caption{Odds-ratios of being \textbf{inactive} next quarter for an individual currently inactive in the labour market.} 
		\label{tab:logit3039nonpart} 
		\begin{scriptsize}
		\begin{tabular}{lccccc}
		\hline \hline 
		\\[-1.8ex]
		& C.I. & C.I.  & Mean & C.I.  & C.I.  \\ 
		&  2.5\% &  5\% &   bootstrap &  95\% &  97.5\% \\ 
		\\[-1.8ex]
		\hline
		\\[-1.8ex]
		\multicolumn{5}{l}{\textbf{Females 30-39}}\\
		\\[-1.8ex]
		\hline
		\\[-1.8ex]
		COVID x EU citizen & 0.647 & 0.669 & 0.840 & 1.045 & 1.091 \\ 
		COVID x No EU citizen & 0.949 & 0.981 & 1.189 & 1.425 & 1.469 \\ 
		COVID x Primary education & 0.833 & 0.858 & 1.021 & 1.195 & 1.226 \\ 
		COVID x Secondary education & 0.767 & 0.789 & 0.914 & 1.055 & 1.089 \\ 
		COVID x North  & 0.833 & 0.866 & 1.111 & 1.406 & 1.470 \\ 
		COVID x Household members$>$2 & 0.854 & 0.892 & 1.098 & 1.331 & 1.366  \\ 
		\\[-1.8ex]
		\hline
		
		Observations&\multicolumn{5}{c}{44428}\\
		\hline
		\\[-1.8ex]
		\multicolumn{5}{l}{\textbf{Males 30-39}}\\
		\\[-1.8ex]
		\hline
		\\[-1.8ex]
		COVID x EU citizen & 0.911 & 0.971 & 1.526 & 2.191 & 2.343 \\ 
		COVID x No EU citizen & 0.671 & 0.705 & 0.917 & 1.161 & 1.215 \\ 
		COVID x Primary education & 0.502 & 0.527 & \textbf{0.677} & 0.842 & 0.883 \\ 
		COVID x Secondary education & 0.510 & 0.532 & \textbf{0.679} & 0.848 & 0.887 \\ 
		COVID x North & 0.664 & 0.696 & 0.917 & 1.177 & 1.239 \\ 
		COVID x Household members$>$2 & 0.967 & 1.021 & 1.278 & 1.568 & 1.641 \\ 
		\hline
		Observations&\multicolumn{5}{c}{20358}\\
		\hline
		\hline
		\multicolumn{6}{l}{\tiny{Note: year and quarter fixed effects and  individual characteristics are included.}} \\
	\multicolumn{6}{l}{\tiny{North includes regions in the North and the Center. \textit{Source}: LFS 3-month longitudinal}}\\ 
		\multicolumn{6}{l}{\tiny{ data as provided by the Italian Institute of Statistics (ISTAT).} }\\
	\end{tabular}
\end{scriptsize}
\end{table}

\section{Discussion and concluding remarks \label{sec:concludingRemarks}}
	
On top of an increased number of workers who exited the labour force across gender, age groups and geographical location at the outburst of the pandemic, there exists a substantial outflow of females in the North of Italy in their 30s with small children  leaving employment, either permanent or temporary, and becoming inactive (at least until the end of 2020). To appreciate the severity of the phenomenon, they correspond to approximately 40.000 women in the  30-39 age cohort moving from employment to inactivity in a quarter. But, surprisingly, the same significant outflows of women are not present in the South.

\begin{table}[!htbp] \centering 
	\caption{Parents' employment and individual beliefs of Italian individuals aged 25-39 by gender, geographical location and labour market participation.} 
	\label{tab:attitudeWork} 
	\begin{scriptsize}
		\begin{tabular}{lccccccccHH}
			\hline \hline
			\\[-1.8ex]
			& \multicolumn{2}{c}{Active males} & \multicolumn{2}{c}{Non-active males} & \multicolumn{2}{c}{Active females} & \multicolumn{2}{c}{Non-active females}\\
			\\[-1ex]
			\\[-1.8ex]
			\hline
			\\[-1.8ex]
			&North&South	&North&South	&North&South&North&South&North&South\\
			\\
			\multicolumn{6}{l}{Percentage of individuals aged 25-39 who responded positively.} \\
			\hline
			\\[-1.8ex]
			(1) At age 14, father employed&98.33&95.06&87.50&100.00&\textbf{98.26}&\textbf{98.18}&\textbf{100.00}&\textbf{89.13}&93.33&84.44\\
			&(1.22)&(2.49)&(12.24)&(0)&(1.22)&(1.84)&(0)&(4.62)&(2.4)&(5.6)\\
			\hline \\[-1.8ex]
			Observations&120&81&8&5&115&55&22&46&105&45\\
			\\
			\hline
			\\[-1.8ex]
			(2) At age 14, mother employed&59.68&32.50&88.89&16.67&\textbf{65.49}&\textbf{38.71}&\textbf{55.00}&\textbf{27.67}&60.95&26.53\\
			&(4.43)&(5.23)&(11.02)&(16.16)&(4.57)&(6.31)&(11.03)&(6.64)&(3.60)&(3.69)\\
			\hline \\[-1.8ex]
			Observations&124&80&9&6&113&62&20&47&105&45\\
			\\
			\\[-1.8ex]
			\\[-1.8ex]
				\multicolumn{6}{l}{Percentage of individuals aged 25-39 who responded they agree with the statement.} \\
			\hline
			\\[-1.8ex]
			(3) Job needed to develop talent&76.34&88.37&25.00& 85.71&\textbf{75.41} &\textbf{89.06}&58.33 &79.59&75.24&84.44\\
			&(3.62)&(3.47)&(13.01)&(14.20)&(3.88)&(3.84)&(9.82)&(5.94)&(0.042)&(0.053)\\
			(4) Lazy if not working& 66.41 & 68.60 &33.33&57.14& \textbf{63.11} & \textbf{71.88 }& 54.17 & 51.02&59.05&75.56 \\
			&(4.20)&(5.07)&(13.95)&(19.84)&(4.31)&(5.84)&(10.37)&(7.17)&(0.048)&(0.066)\\
			(5)	Child suffers with working mother & \textbf{25.19}& \textbf{48.83} &16.67&0& \textbf{30.33} & \textbf{32.81}& \textbf{29.17}& \textbf{55.10}&29.50&31.11 \\
			& (3.79)&(5.59)&(10.68)&(0)&(4.30)&(5.91)&(9.55)&(7.11)&(0.046)&(0.070)\\
			\hline
			Observations&131&86&12&7&122&64&24&49&105&45\\
			\hline
			\hline
			\multicolumn{11}{l}{\tiny{\textit{Note}: The specific questions asked are: (1)  "When you were 14, was your mother employed, self-employed or not?"}}\\ \multicolumn{11}{l}{\tiny{(2)  "When you were 14, was your father employed, self-employed or not?" The following questions are:}}\\ \multicolumn{11}{l}{\tiny{  "Do you agree or disagree with the following statements?" (3) "To fully develop your talents, you need to have a job"}}\\\multicolumn{11}{l}{\tiny{   (4) "People who don’t work turn
					lazy" (5) "When a mother works for pay, the children
					suffer". Standard errors}}\\\multicolumn{11}{l}{\tiny{   are computed using bootstrap (1000 draws). North includes regions in the North and the Center.}}\\\multicolumn{11}{l}{\tiny{  \textit{Source}: European Value Survey, 2017.}}
		\end{tabular}
	\end{scriptsize}
\end{table}	

The reasons behind this asymmetric pattern could be several and are hard to be precisely pinned down, due to the lack of appropriate data at individual level. One possible explanation is that females in the North were less afraid to pause their career knowing that the labour market is more efficient, and they will have the chance to re-enter later, compared to women in the South. Testing this hypothesis, which carries relevant policy implications, would require more recent data, currently not available.

Instead, inspired by the socio-cultural literature, we argue that part of the explanation lays in the presence of heterogeneous cultural factors, in particular related to the attitude towards work, across Italian regions \citep{McGinn2019, Stevens1980, fernandez2011does}, and in the self-selection into the labour market of females in the South.

Table \ref{tab:attitudeWork} reports the answer to specific questions which could pin down the different attitude towards work in the North and South of Italy of individuals between the age of 25 and 39 included in the most recent wave of the European Value Survey of 2017. 
In particular, the employment status of the father when the individual was 14 years old does not play any role in the labour force participation decision of males and females, neither in the North nor in the South (Question (1) in Table \ref{tab:attitudeWork}).\footnote{The statistics for males non active both in the North and in the South are not to be considered as based on too few observations.} On the contrary, individuals with a non-working mother are more likely to be inactive (Question (2) in Table \ref{tab:attitudeWork}), which is further evidence of the strong relationship between maternal labour market participation and adult daughters’ activity in the labour market \citep{Stevens1980}.
Not surprisingly, the percentage of females (and males) with a working mother is higher in the North, due to the historically higher female labour force participation.

As regards to the different attitude towards work, a larger percentage of individuals in the South considers work important to develop individual talents, compared to the North (Question (3) in Table \ref{tab:attitudeWork}). Specifically, the large difference between active females in North and in the South suggests a stronger attachment of the latter to the labour market (89\% versus 75\%).
Moreover, among active females, not working represents a stigma much more in the South, compared to the North (72\% versus 63\%), suggesting larger costs for women in the South to leave the labour market. On the contrary, there is no difference among non-active females in the North and South.
Finally, while 55\% of non-active females in the South (and 49\% of active males in the South), are of the opinion that children suffer when the mother works, only 30\% of active females share this belief (Question (5) in Table \ref{tab:attitudeWork}). 

In conclusion, the intersection of socio-cultural and economic factors in the North created an environment more favourable to reconcile career and family. On the contrary, active females aged 30-39 in the South were proportionally fewer but much more motivated. 
Thus, when the COVID-19 pandemic hit, the latter appeared to be more resilient, while the former more vulnerable. In addition to this short-term cost, as the future female labour force participation appears to be strongly dependent on the current female labour force participation, the COVID-19 pandemic could impose a further long-term toll caused by a persistent female labour force outflow.

\clearpage
	
\bibliographystyle{chicago}
\bibliography{references}

\begin{thebibliography}{}

\bibitem[\protect\citeauthoryear{Adams-Prassl, Boneva, Golin, and
  Rauh}{Adams-Prassl et~al.}{2020}]{adams2020inequality}
Adams-Prassl, A., T.~Boneva, M.~Golin, and C.~Rauh (2020).
\newblock Inequality in the impact of the coronavirus shock: Evidence from real
  time surveys.
\newblock {\em Journal of Public Economics\/}~{\em 189}, 104245.

\bibitem[\protect\citeauthoryear{Agovino, Garofalo, and Cerciello}{Agovino
  et~al.}{2019}]{agovino2019local}
Agovino, M., A.~Garofalo, and M.~Cerciello (2019).
\newblock Do local institutions affect labour market participation? {T}he
  {I}talian case.
\newblock {\em The BE Journal of Economic Analysis \& Policy\/}~{\em 19\/}(2).

\bibitem[\protect\citeauthoryear{Albanesi and Kim}{Albanesi and
  Kim}{2021}]{albanesi2021gendered}
Albanesi, S. and J.~Kim (2021).
\newblock The gendered impact of the covid-19 recession on the {US} labor
  market.
\newblock Technical report, National Bureau of Economic Research.

\bibitem[\protect\citeauthoryear{Albanesi and Olivetti}{Albanesi and
  Olivetti}{2016}]{albanesi2016gender}
Albanesi, S. and C.~Olivetti (2016).
\newblock Gender roles and medical progress.
\newblock {\em Journal of Political Economy\/}~{\em 124\/}(3), 650--695.

\bibitem[\protect\citeauthoryear{Alon, Coskun, Doepke, Koll, and Tertilt}{Alon
  et~al.}{2021}]{alon2021mancession}
Alon, T., S.~Coskun, M.~Doepke, D.~Koll, and M.~Tertilt (2021).
\newblock From mancession to shecession: Women's employment in regular and
  pandemic recessions.
\newblock Technical report, National Bureau of Economic Research.

\bibitem[\protect\citeauthoryear{Andresen and Havnes}{Andresen and
  Havnes}{2019}]{andresen2019child}
Andresen, M.~E. and T.~Havnes (2019).
\newblock Child care, parental labor supply and tax revenue.
\newblock {\em Labour Economics\/}~{\em 61}, 101762.

\bibitem[\protect\citeauthoryear{Barua}{Barua}{2014}]{BARUA2014129}
Barua, R. (2014).
\newblock Intertemporal substitution in maternal labor supply: Evidence using
  state school entrance age laws.
\newblock {\em Labour Economics\/}~{\em 31}, 129--140.

\bibitem[\protect\citeauthoryear{Baussola and Mussida}{Baussola and
  Mussida}{2014}]{baussola2014disadvantaged}
Baussola, M. and C.~Mussida (2014).
\newblock Disadvantaged workers in the {I}talian labour market: gender and
  regional gaps.
\newblock In {\em Disadvantaged Workers}, pp.\  231--256. Springer.

\bibitem[\protect\citeauthoryear{Bertola and Garibaldi}{Bertola and
  Garibaldi}{2003}]{bertola2003structure}
Bertola, G. and P.~Garibaldi (2003).
\newblock {\em The structure and history of Italian unemployment: presented at
  CESifo Conference on Unemployment in Europe, December 2002}.
\newblock CES.

\bibitem[\protect\citeauthoryear{Bettendorf, Jongen, and Muller}{Bettendorf
  et~al.}{2015}]{BETTENDORF2015112}
Bettendorf, L.~J., E.~L. Jongen, and P.~Muller (2015).
\newblock Childcare subsidies and labour supply. evidence from a large {D}utch
  reform.
\newblock {\em Labour Economics\/}~{\em 36}, 112--123.

\bibitem[\protect\citeauthoryear{Bettio and Pastore}{Bettio and
  Pastore}{2017}]{SOAS2021gender}
Bettio, F. and F.~Pastore (2017).
\newblock Overview of female employment issues in {I}taly.
\newblock Technical report.

\bibitem[\protect\citeauthoryear{Blau and Kahn}{Blau and Kahn}{2013}]{Blau2013}
Blau, F.~D. and L.~M. Kahn (2013).
\newblock Female labor supply: Why is the {U}nited {S}tates falling behind?
\newblock {\em American Economic Review\/}~{\em 103\/}(3), 251--56.

\bibitem[\protect\citeauthoryear{Bluedorn, Caselli, Hansen, Shibata, and
  Tavares}{Bluedorn et~al.}{2021}]{bluedorn2021gender}
Bluedorn, J., F.~Caselli, N.-J. Hansen, I.~Shibata, and M.~M. Tavares (2021).
\newblock Gender and employment in the covid-19 recession: Evidence on
  “she-cessions”.
\newblock Technical report, IMF Working Paper 2021/95.

\bibitem[\protect\citeauthoryear{Casarico and Lattanzio}{Casarico and
  Lattanzio}{2020}]{casarico2020heterogeneous}
Casarico, A. and S.~Lattanzio (2020).
\newblock The heterogeneous effects of covid-19 on labor market flows: Evidence
  from administrative data.
\newblock {\em Covid Economics\/}~{\em 52}, 152--174.

\bibitem[\protect\citeauthoryear{Cascio, Haider, and Nielsen}{Cascio
  et~al.}{2015}]{cascio2015effectiveness}
Cascio, E.~U., S.~J. Haider, and H.~S. Nielsen (2015).
\newblock The effectiveness of policies that promote labor force participation
  of women with children: A collection of national studies.
\newblock {\em Labour Economics\/}~{\em 36}, 64--71.

\bibitem[\protect\citeauthoryear{Caselli, Grigoli, Sandri, and
  Spilimbergo}{Caselli et~al.}{2021}]{caselli2021mobility}
Caselli, F., F.~Grigoli, D.~Sandri, and A.~Spilimbergo (2021).
\newblock Mobility under the covid-19 pandemic: Asymmetric effects across
  gender and age.
\newblock {\em IMF Economic Review\/}, 1--34.

\bibitem[\protect\citeauthoryear{Chetty, Friedman, Hendren, and Stepner}{Chetty
  et~al.}{2020}]{chetty2020economic}
Chetty, R., J.~Friedman, N.~Hendren, and M.~Stepner (2020).
\newblock The economic impacts of covid-19: Evidence from a new public database
  built from private sector data.
\newblock {\em Opportunity Insights\/}.

\bibitem[\protect\citeauthoryear{Compton and Pollak}{Compton and
  Pollak}{2014}]{COMPTON201472}
Compton, J. and R.~A. Pollak (2014).
\newblock Family proximity, childcare, and women’s labor force attachment.
\newblock {\em Journal of Urban Economics\/}~{\em 79}, 72--90.

\bibitem[\protect\citeauthoryear{Da~Rocha and Fuster}{Da~Rocha and
  Fuster}{2006}]{da2006fertility}
Da~Rocha, J.~M. and L.~Fuster (2006).
\newblock Why are fertility rates and female employment ratios positively
  correlated across oecd countries?
\newblock {\em International Economic Review\/}~{\em 47\/}(4), 1187--1222.

\bibitem[\protect\citeauthoryear{Dang and Nguyen}{Dang and
  Nguyen}{2021}]{dang2021gender}
Dang, H.-A.~H. and C.~V. Nguyen (2021).
\newblock Gender inequality during the covid-19 pandemic: Income, expenditure,
  savings, and job loss.
\newblock {\em World Development\/}~{\em 140}, 105296.

\bibitem[\protect\citeauthoryear{Del~Boca}{Del~Boca}{2002}]{del2002effect}
Del~Boca, D. (2002).
\newblock The effect of child care and part time opportunities on participation
  and fertility decisions in {I}taly.
\newblock {\em Journal of Population Economics\/}~{\em 15\/}(3), 549--573.

\bibitem[\protect\citeauthoryear{Del~Boca, Oggero, Profeta, and Rossi}{Del~Boca
  et~al.}{2020}]{del2020women}
Del~Boca, D., N.~Oggero, P.~Profeta, and M.~Rossi (2020).
\newblock Women’s and men’s work, housework and childcare, before and
  during covid-19.
\newblock {\em Review of Economics of the Household\/}~{\em 18\/}(4),
  1001--1017.

\bibitem[\protect\citeauthoryear{Doepke, Hannusch, Kindermann, and
  Tertilt}{Doepke et~al.}{2022}]{doepke2022new}
Doepke, M., A.~Hannusch, F.~Kindermann, and M.~Tertilt (2022).
\newblock A new era in the economics of fertility.
\newblock {\em VoxEU. org\/}~{\em 11}.

\bibitem[\protect\citeauthoryear{Duhautois, Erhel, and
  Guergoat-Larivi{\`e}re}{Duhautois et~al.}{2018}]{duhautois2018state}
Duhautois, R., C.~Erhel, and M.~Guergoat-Larivi{\`e}re (2018).
\newblock State dependence and labor market transitions in the {E}uropean
  {U}nion.
\newblock {\em Annals of Economics and Statistics/Annales d'{\'E}conomie et de
  Statistique\/}~(131), 59--82.

\bibitem[\protect\citeauthoryear{Efron and Tibshirani}{Efron and
  Tibshirani}{1994}]{efron1994introduction}
Efron, B. and R.~J. Tibshirani (1994).
\newblock {\em An introduction to the bootstrap}.
\newblock CRC press.

\bibitem[\protect\citeauthoryear{Fabrizio, Gomes, and Tavares}{Fabrizio
  et~al.}{2021}]{fabrizio2021covid}
Fabrizio, M.~S., D.~B. Gomes, and M.~M.~M. Tavares (2021).
\newblock {\em COVID-19 She-Cession: The Employment Penalty of Taking Care of
  Young Children}.
\newblock International Monetary Fund.

\bibitem[\protect\citeauthoryear{Fern{\'a}ndez}{Fern{\'a}ndez}{2011}]{fernandez2011does}
Fern{\'a}ndez, R. (2011).
\newblock Does culture matter?
\newblock {\em Handbook of social economics\/}~{\em 1}, 481--510.

\bibitem[\protect\citeauthoryear{Fernández, Fogli, and Olivetti}{Fernández
  et~al.}{2004}]{Fernandez2004}
Fernández, R., A.~Fogli, and C.~Olivetti (2004).
\newblock Mothers and sons: Preference formation and female labor force
  dynamics.
\newblock {\em The Quarterly Journal of Economics\/}~{\em 119\/}(4),
  1249--1299.

\bibitem[\protect\citeauthoryear{Fiaschi and Tealdi}{Fiaschi and
  Tealdi}{2022}]{fiaschi2022young}
Fiaschi, D. and C.~Tealdi (2022).
\newblock Young people between education and the labour market during the
  covid-19 pandemic in {I}taly.
\newblock {\em International Journal of Manpower\/}~{\em 43\/}(7).

\bibitem[\protect\citeauthoryear{Givord and Marbot}{Givord and
  Marbot}{2015}]{GIVORD201599}
Givord, P. and C.~Marbot (2015).
\newblock Does the cost of child care affect female labor market participation?
  {A}n evaluation of a {F}rench reform of childcare subsidies.
\newblock {\em Labour Economics\/}~{\em 36}, 99--111.

\bibitem[\protect\citeauthoryear{Goldin}{Goldin}{2021}]{goldin2021career}
Goldin, C. (2021).
\newblock {\em Career and Family: Women’s Century-Long Journey toward
  Equity}.
\newblock Princeton University Press.

\bibitem[\protect\citeauthoryear{Goldin and Katz}{Goldin and
  Katz}{2002}]{goldin2002power}
Goldin, C. and L.~F. Katz (2002).
\newblock The power of the pill: Oral contraceptives and women’s career and
  marriage decisions.
\newblock {\em Journal of Political Economy\/}~{\em 110\/}(4), 730--770.

\bibitem[\protect\citeauthoryear{Greenwood, Seshadri, and Yorukoglu}{Greenwood
  et~al.}{2005}]{greenwood2005engines}
Greenwood, J., A.~Seshadri, and M.~Yorukoglu (2005).
\newblock Engines of liberation.
\newblock {\em The Review of Economic Studies\/}~{\em 72\/}(1), 109--133.

\bibitem[\protect\citeauthoryear{Honor{\'e} and Kyriazidou}{Honor{\'e} and
  Kyriazidou}{2000}]{honore2000panel}
Honor{\'e}, B.~E. and E.~Kyriazidou (2000).
\newblock Panel data discrete choice models with lagged dependent variables.
\newblock {\em Econometrica\/}~{\em 68\/}(4), 839--874.

\bibitem[\protect\citeauthoryear{Huebener, Pape, and Spiess}{Huebener
  et~al.}{2020}]{HUEBENER2020510}
Huebener, M., A.~Pape, and C.~K. Spiess (2020).
\newblock Parental labour supply responses to the abolition of day care fees.
\newblock {\em Journal of Economic Behavior \& Organization\/}~{\em 180},
  510--543.

\bibitem[\protect\citeauthoryear{Hupkau and Petrongolo}{Hupkau and
  Petrongolo}{2020}]{hupkau2020work}
Hupkau, C. and B.~Petrongolo (2020).
\newblock Work, care and gender during the covid-19 crisis.
\newblock {\em Fiscal Studies\/}~{\em 41\/}(3), 623--651.

\bibitem[\protect\citeauthoryear{ISTAT}{ISTAT}{2019}]{ISTAT}
ISTAT (2019).
\newblock Italian institute of statistics.
\newblock Technical report, Italian Institute of Statistics.

\bibitem[\protect\citeauthoryear{Jones, Manuelli, and McGrattan}{Jones
  et~al.}{2015}]{jones2015married}
Jones, L.~E., R.~E. Manuelli, and E.~R. McGrattan (2015).
\newblock Why are married women working so much?
\newblock {\em Journal of Demographic Economics\/}~{\em 81\/}(1), 75--114.

\bibitem[\protect\citeauthoryear{Lafond, Goldin, Koutroumpis, and
  Winkler}{Lafond et~al.}{2021}]{goldin2020productivity}
Lafond, F., I.~Goldin, P.~Koutroumpis, and J.~Winkler (2021).
\newblock Why is productivity slowing down?
\newblock Technical report.

\bibitem[\protect\citeauthoryear{McGinn, Castro, and Lingo}{McGinn
  et~al.}{2019}]{McGinn2019}
McGinn, K.~L., M.~R. Castro, and E.~L. Lingo (2019).
\newblock Learning from mum: Cross-national evidence linking maternal
  employment and adult children’s outcomes.
\newblock {\em Work, Employment and Society\/}~{\em 33\/}(3), 374--400.

\bibitem[\protect\citeauthoryear{Myong, Park, and Yi}{Myong
  et~al.}{2021}]{myong2021social}
Myong, S., J.~Park, and J.~Yi (2021).
\newblock Social norms and fertility.
\newblock {\em Journal of the European Economic Association\/}~{\em 19\/}(5),
  2429--2466.

\bibitem[\protect\citeauthoryear{OECD}{OECD}{2018}]{OECD2018}
OECD (2018).
\newblock How does {I}taly compare?
\newblock Technical report.

\bibitem[\protect\citeauthoryear{OECD}{OECD}{2019}]{OECDItaly}
OECD (2019).
\newblock {\em Recent trends in the Italian Labour Market}.

\bibitem[\protect\citeauthoryear{Olivetti}{Olivetti}{2006}]{olivetti2006changes}
Olivetti, C. (2006).
\newblock Changes in women's hours of market work: The role of returns to
  experience.
\newblock {\em Review of Economic Dynamics\/}~{\em 9\/}(4), 557--587.

\bibitem[\protect\citeauthoryear{Olivetti and Petrongolo}{Olivetti and
  Petrongolo}{2017}]{olivetti2017economic}
Olivetti, C. and B.~Petrongolo (2017).
\newblock The economic consequences of family policies: lessons from a century
  of legislation in high-income countries.
\newblock {\em Journal of Economic Perspectives\/}~{\em 31\/}(1), 205--30.

\bibitem[\protect\citeauthoryear{Pacelli, Pasqua, and Villosio}{Pacelli
  et~al.}{2013}]{pacelli2013labor}
Pacelli, L., S.~Pasqua, and C.~Villosio (2013).
\newblock Labor market penalties for mothers in {I}taly.
\newblock {\em Journal of Labor Research\/}~{\em 34\/}(4), 408--432.

\bibitem[\protect\citeauthoryear{Panagiotelis, Athanasopoulos, Gamakumara, and
  Hyndman}{Panagiotelis et~al.}{2021}]{HyndmanAthanasopoulosforecasting2021}
Panagiotelis, A., G.~Athanasopoulos, P.~Gamakumara, and R.~J. Hyndman (2021).
\newblock Forecast reconciliation: A geometric view with new insights on bias
  correction.
\newblock {\em International Journal of Forecasting\/}~{\em 37\/}(1), 343--359.

\bibitem[\protect\citeauthoryear{Picchio, Pigini, Staffolani, and
  Verashchagina}{Picchio et~al.}{2021}]{picchio2021if}
Picchio, M., C.~Pigini, S.~Staffolani, and A.~Verashchagina (2021).
\newblock If not now, when? {T}he timing of childbirth and labor market
  outcomes.
\newblock {\em Journal of Applied Econometrics\/}~{\em 36\/}(6), 663--685.

\bibitem[\protect\citeauthoryear{Schiattarella and Piacentini}{Schiattarella
  and Piacentini}{2018}]{schiattarella2018old}
Schiattarella, R. and P.~Piacentini (2018).
\newblock Old and new dualisms in the {I}talian labour market.
\newblock In {\em The Italian Economy at the Dawn of the 21st Century}, pp.\
  81--100. Routledge.

\bibitem[\protect\citeauthoryear{Shibata}{Shibata}{2020}]{shibata2020distributional}
Shibata, I. (2020).
\newblock The distributional impact of recessions: the global financial crisis
  and the pandemic recession.
\newblock {\em IMF Working Paper\/}.

\bibitem[\protect\citeauthoryear{Stevens and Boyd}{Stevens and
  Boyd}{1980}]{Stevens1980}
Stevens, G. and M.~Boyd (1980).
\newblock The importance of mother: Labor force participation and
  intergenerational mobility of women.
\newblock {\em Social Forces\/}~{\em 59\/}(1), 186--199.

\bibitem[\protect\citeauthoryear{Train}{Train}{2009}]{train2009discrete}
Train, K.~E. (2009).
\newblock {\em Discrete choice methods with simulation}.
\newblock Cambridge University Press.

\bibitem[\protect\citeauthoryear{Zamarro, Perez-Arce, and Prados}{Zamarro
  et~al.}{2020}]{zamarro2020gender}
Zamarro, G., F.~Perez-Arce, and M.~J. Prados (2020).
\newblock Gender differences in the impact of covid-19.
\newblock Technical report, Working Paper. Switzerland: Frontiers in Public
  Health.

\end{thebibliography}

\clearpage

\appendix

\begin{Large}
	\textbf{Appendix}
\end{Large}

\section{COVID-19 in Italy}\label{appsec:italianPolicies}

The first cases of COVID-19 in Italy were registered on January 31, 2020, but the virus began to spread exponentially in the second half of February. At the beginning, the virus circulated predominantly in Northern regions but by the beginning of March, it had reached all regions. The first measures limiting the mobility were implemented on February 23, and applied to a restricted number of municipalities in the Lombardy region; soon after, on  March 10, the whole country went into a full lockdown. On  March 11 the government prohibited nearly all commercial activity except for supermarkets and pharmacies and, on March 21, it restricted the movement of people and closed all non-essential businesses and industries. Sectors identified as essential, which could continue operating, include mainly agriculture, some manufacturing, energy and water supply, transports and logistics, ICT, banking and insurance, professional and scientific activities, public administration, education, health care and some service activities. Non-essential sectors which were completely shut include most manufacturing, wholesale and retail trade, hotels, restaurants and bars, entertainment and sport activities \cite{casarico2020heterogeneous}.
The first national lockdown was relaxed on May 3 and, finally, ended on June 14. On October 5, new restrictions where imposed at national level, which were further strengthened on November 6, but allowing for regional differences based on the local pandemic spreading. However, almost all regions were locked in a strict regime until the end of December 2020.  

On March 17 the Italian government introduced  two new labour market policies to protect workers: (i) a COVID-19 furlough scheme  and (ii) a ban on layoffs. The former was implemented for an initial duration of 9 weeks, and it applied retroactively starting from February 23. It represents an extension of the regular furlough scheme to all firms, independently on size. This measure aimed at preserving employment and allowed firms to cut labour costs during the lockdown period, by reducing hours of work thanks to a wage subsidy granted by the government. Firms using the COVID-19 furlough scheme could renew temporary contracts, waiving to the norms of the standard regulation. Upon completion of the furlough period, firms were allowed to dismiss employees for redundancy. The ban on layoffs prevented firm to fire workers for 60 days, starting from  March 17; this ban could be applied retroactively to pending, but already validated, layoffs from February 23. Two later decrees extended the validity of these measures, which were in place until the end of 2021.

\clearpage

\section{Labour market dynamics  pre-COVID-19}\label{appsec:dynamicsprecovid}

\subsection{Shares of individuals by age, gender and geographical location} \label{appsec:dynamicsprecovidIN}

\begin{figure}[!htbp]
	\caption{Shares of inactive individuals (INACT) in the North and South of Italy.}
	\label{appfig:shares1519text}
	\centering
	\begin{subfigure}[t]{0.24\textwidth}
		\centering
		\includegraphics[width=\linewidth]{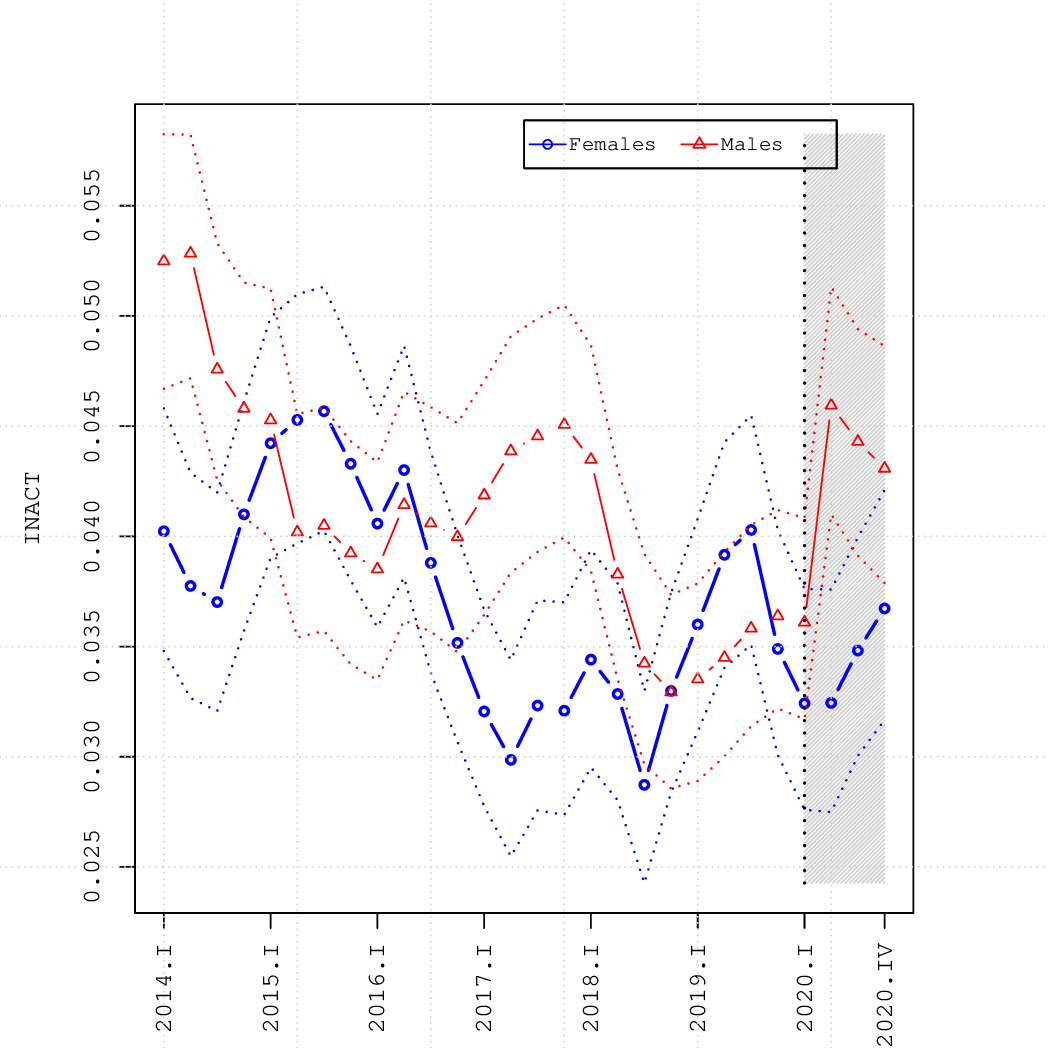}
		\caption{North - Age 15-19.}
		\label{fig:transProbFromEDUtoPE_IIIII}
	\end{subfigure}
	\begin{subfigure}[t]{0.24\textwidth}
		\centering
		\includegraphics[width=\linewidth]{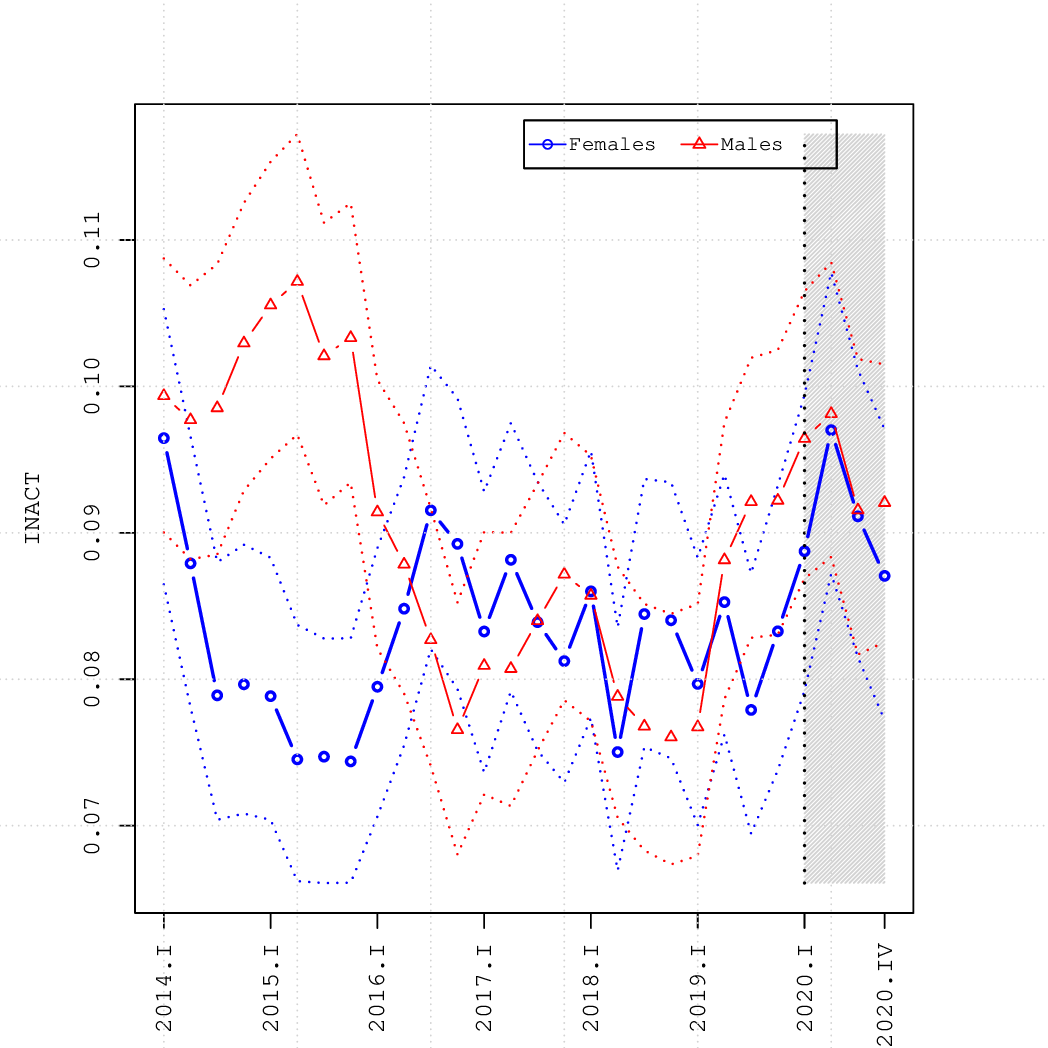}
		\caption{South - Age 15-19.}
		\label{fig:transProbFromEDUtoU_II}
	\end{subfigure}
	\begin{subfigure}[t]{0.24\textwidth}
		\centering
		\includegraphics[width=\linewidth]{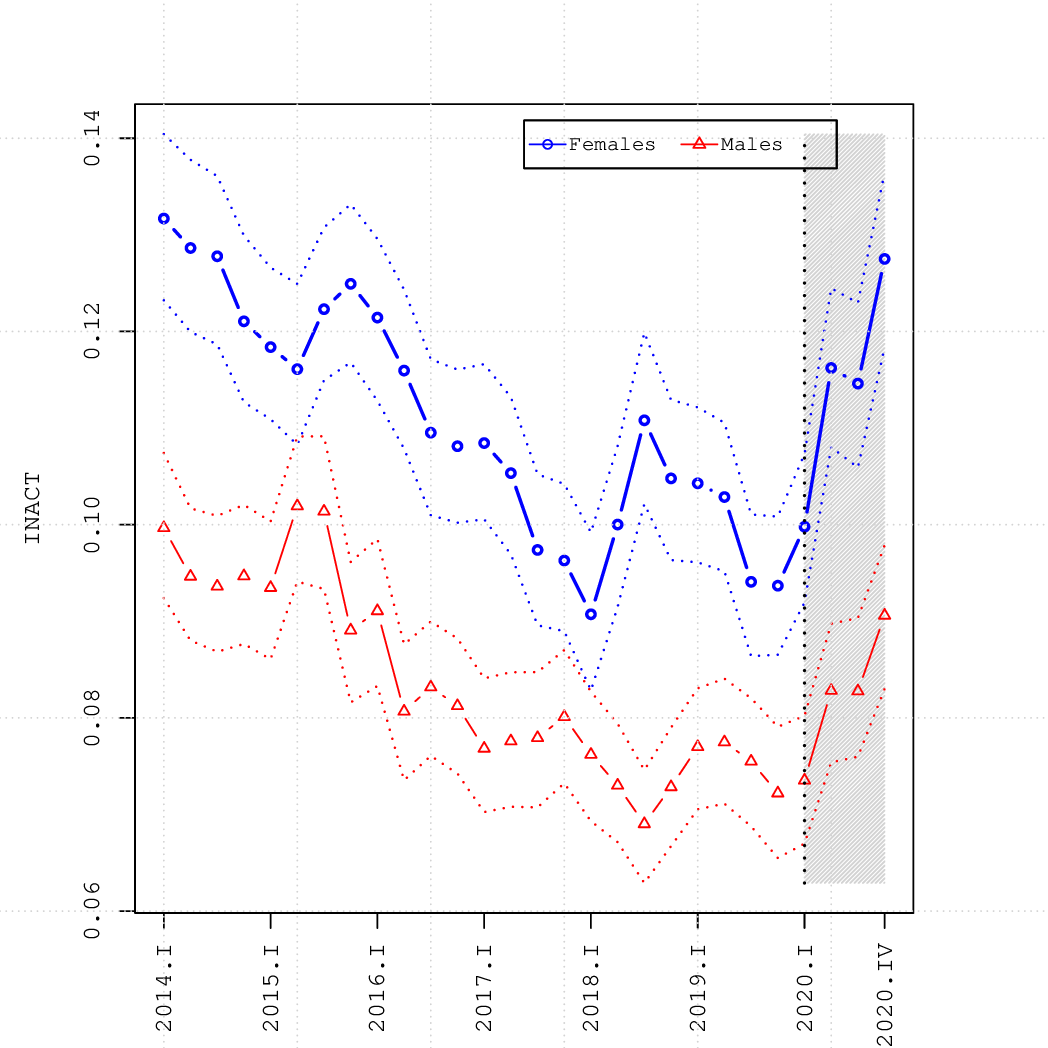}
		\caption{North - Age 20-24.}
		\label{fig:transProbFromEDUtoPE_IIIIII}
	\end{subfigure}
	\begin{subfigure}[t]{0.24\textwidth}
		\centering
		\includegraphics[width=\linewidth]{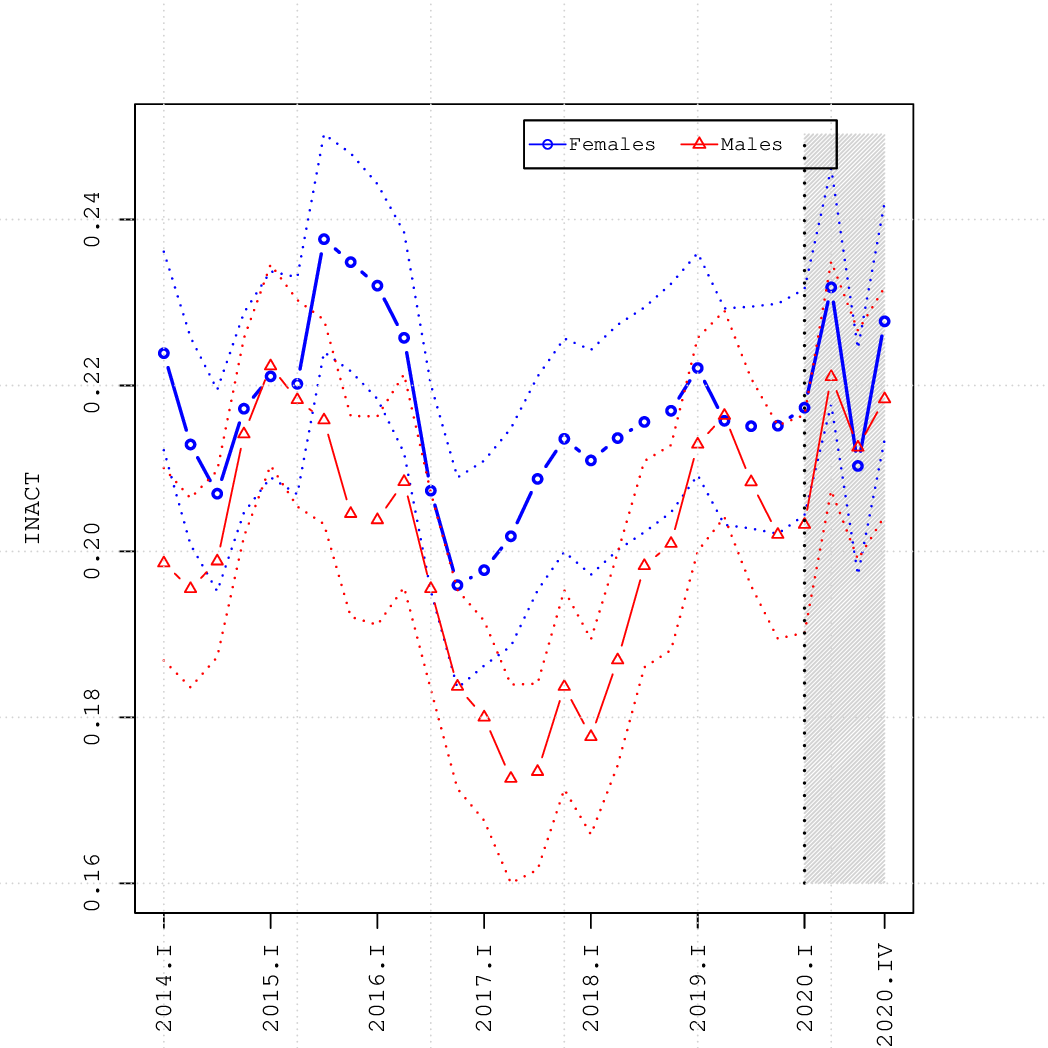}
		\caption{South - Age 20-24.}
		\label{fig:transProbFromEDUtoU_III}
	\end{subfigure}
	\begin{subfigure}[t]{0.24\textwidth}
		\centering
		\includegraphics[width=\linewidth]{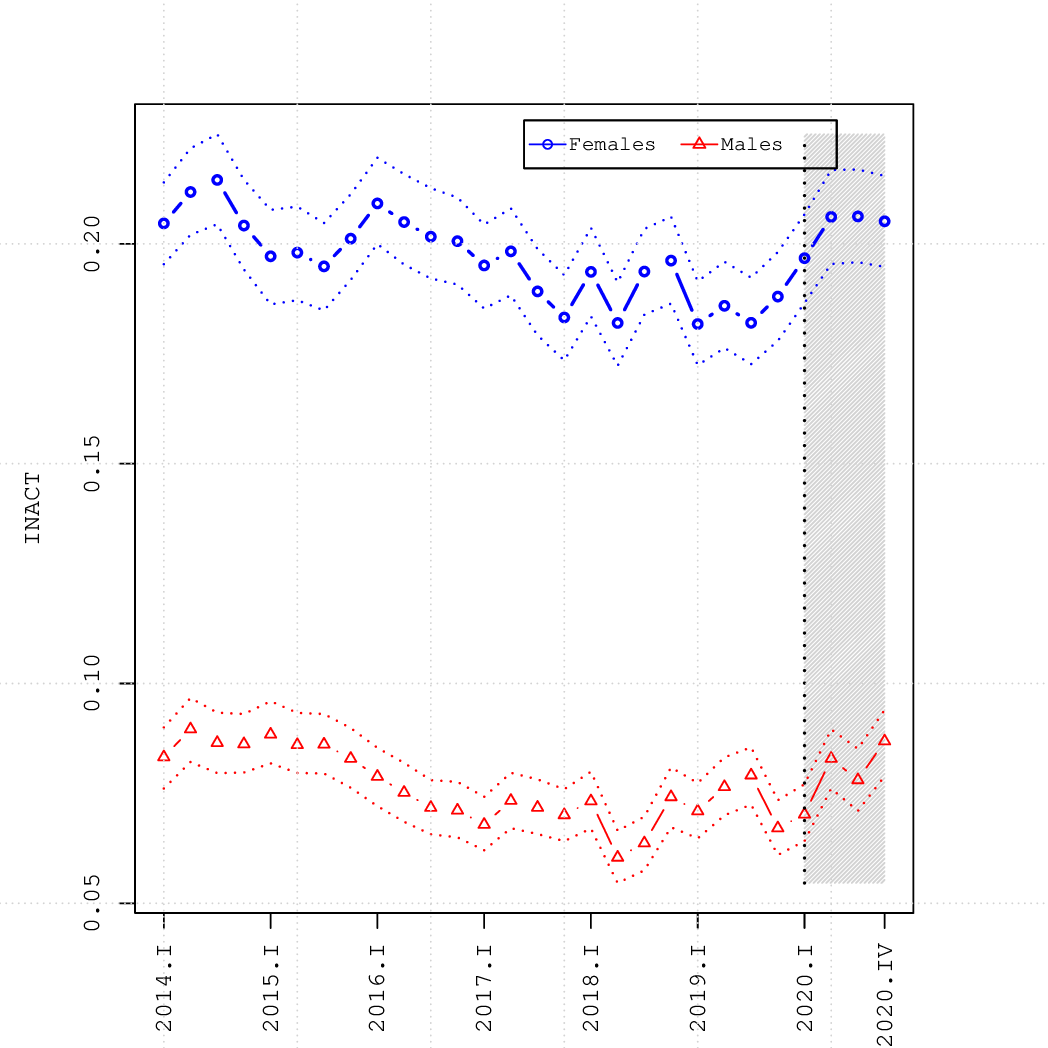}
		\caption{North - Age 25-29.}
		\label{fig:transProbFromEDUtoPE_IIIIIII}
	\end{subfigure}
	\begin{subfigure}[t]{0.24\textwidth}
		\centering
		\includegraphics[width=\linewidth]{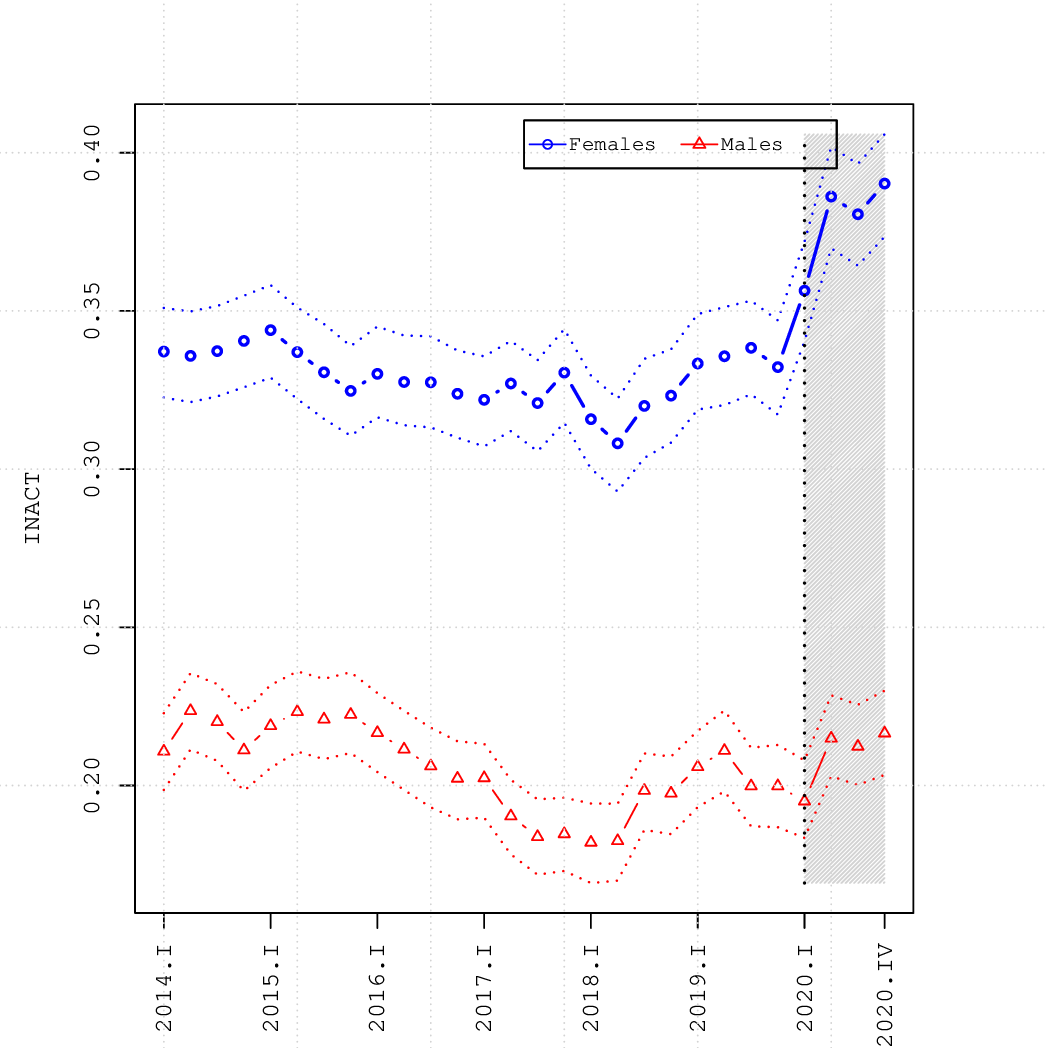}
		\caption{South - Age 25-29.}
		\label{fig:transProbFromEDUtoU_IIIII}
	\end{subfigure}
	\begin{subfigure}[t]{0.24\textwidth}
		\centering
		\includegraphics[width=\linewidth]{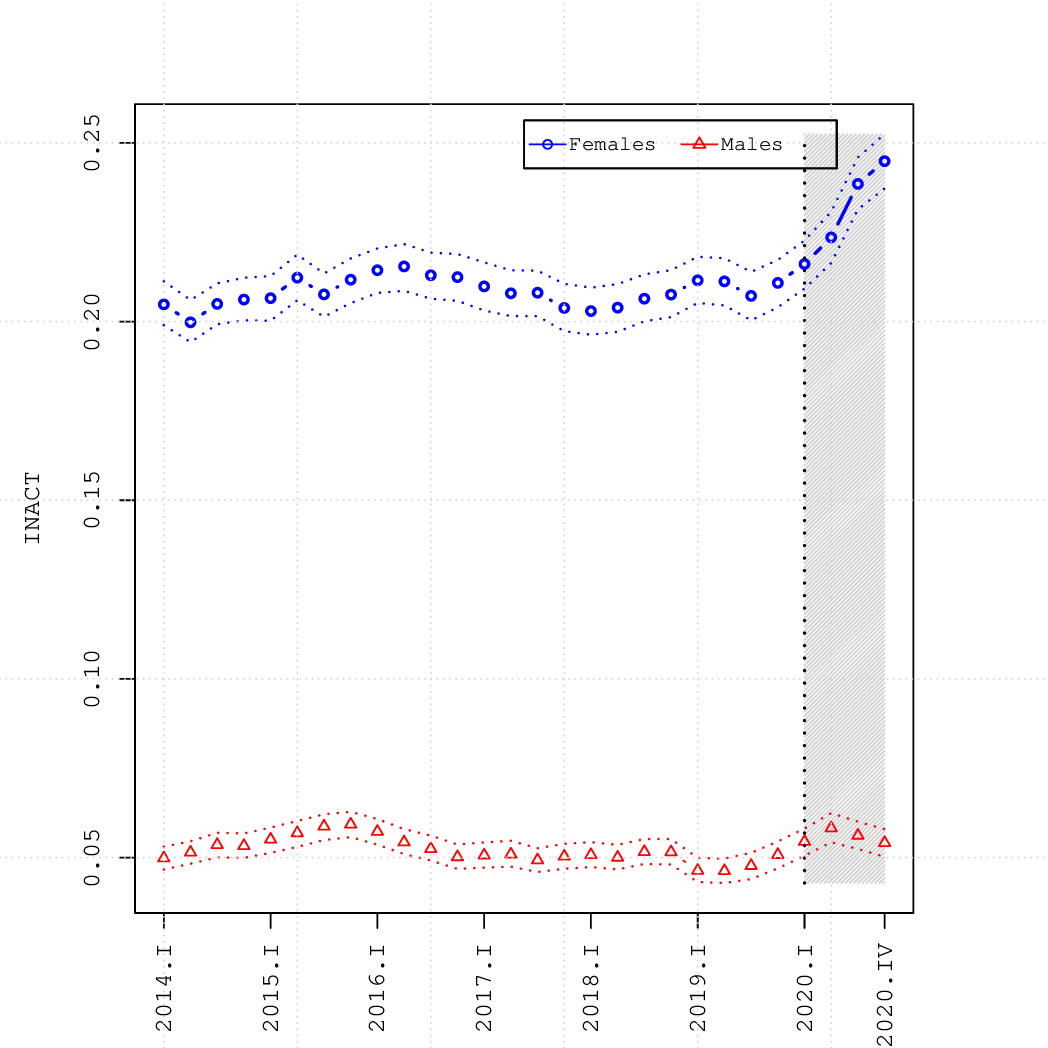}
		\caption{North - Age 30-39.}
		\label{fig:transProbFromEDUtoPE_IIIIIIII}
	\end{subfigure}
	\begin{subfigure}[t]{0.24\textwidth}
		\centering
		\includegraphics[width=\linewidth]{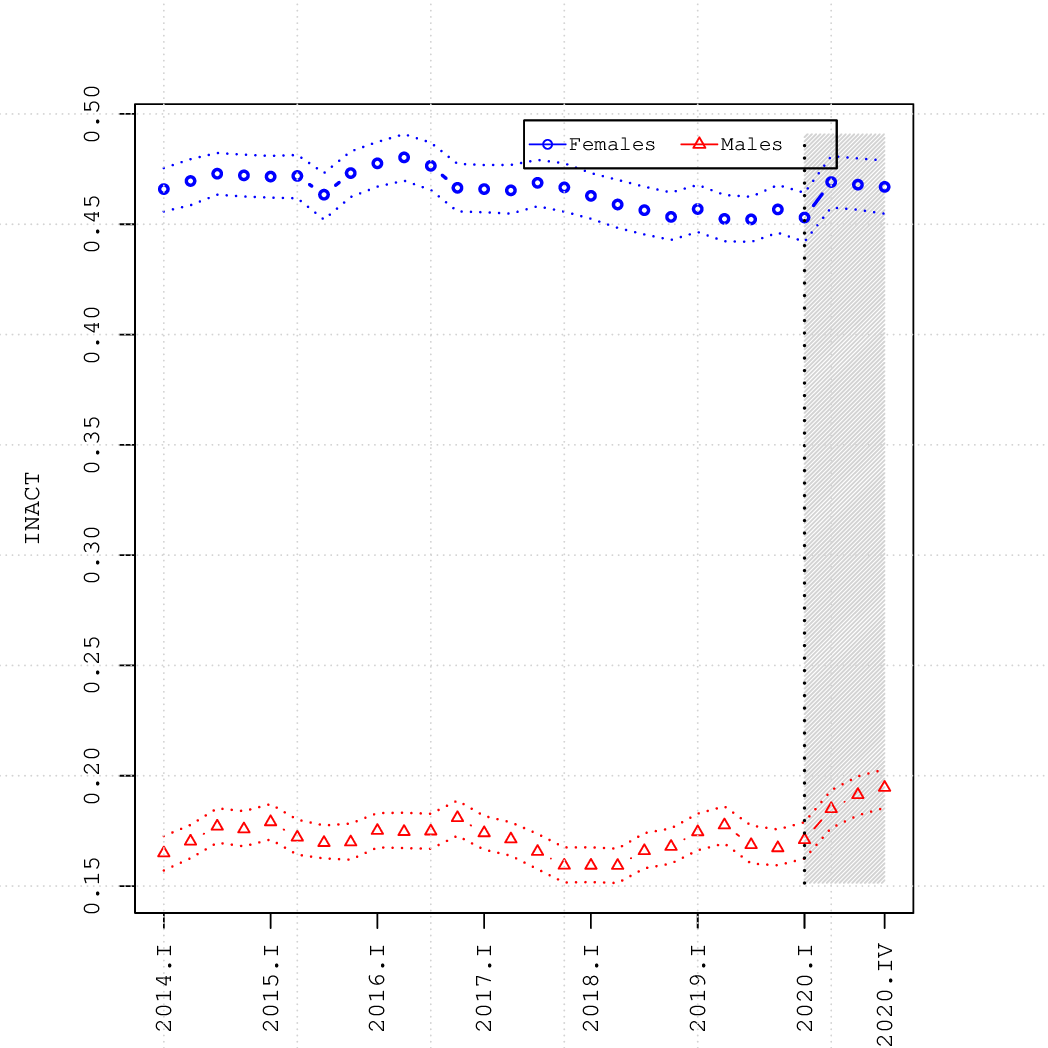}
		\caption{South - Age 30-39.}
		\label{fig:transProbFromEDUtoU_IIII}
	\end{subfigure}
	\begin{subfigure}[t]{0.24\textwidth}
		\centering
		\includegraphics[width=\linewidth]{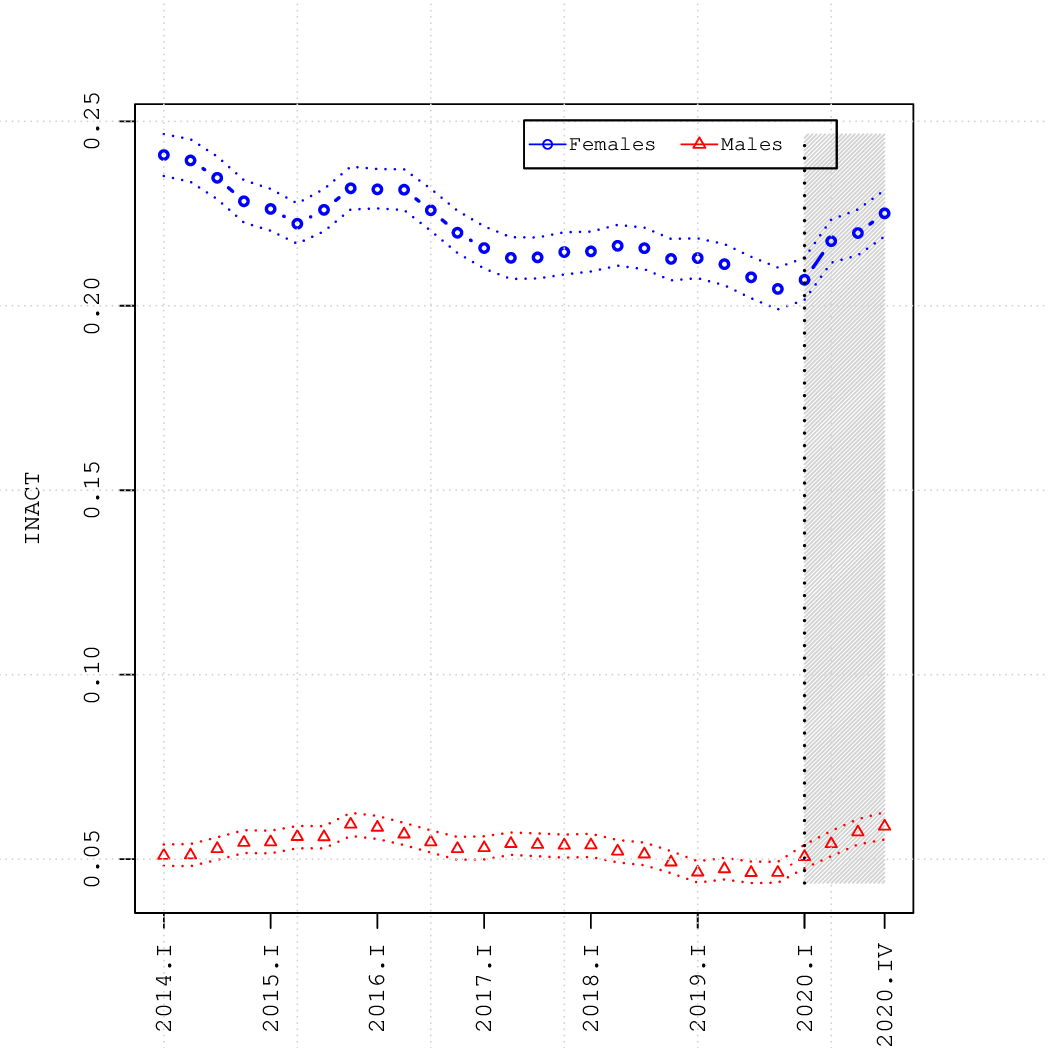}
		\caption{North - Age 40-49.}
		\label{fig:transProbFromEDUtoPE_IIIIIIIII}
	\end{subfigure}
	\begin{subfigure}[t]{0.24\textwidth}
		\centering
		\includegraphics[width=\linewidth]{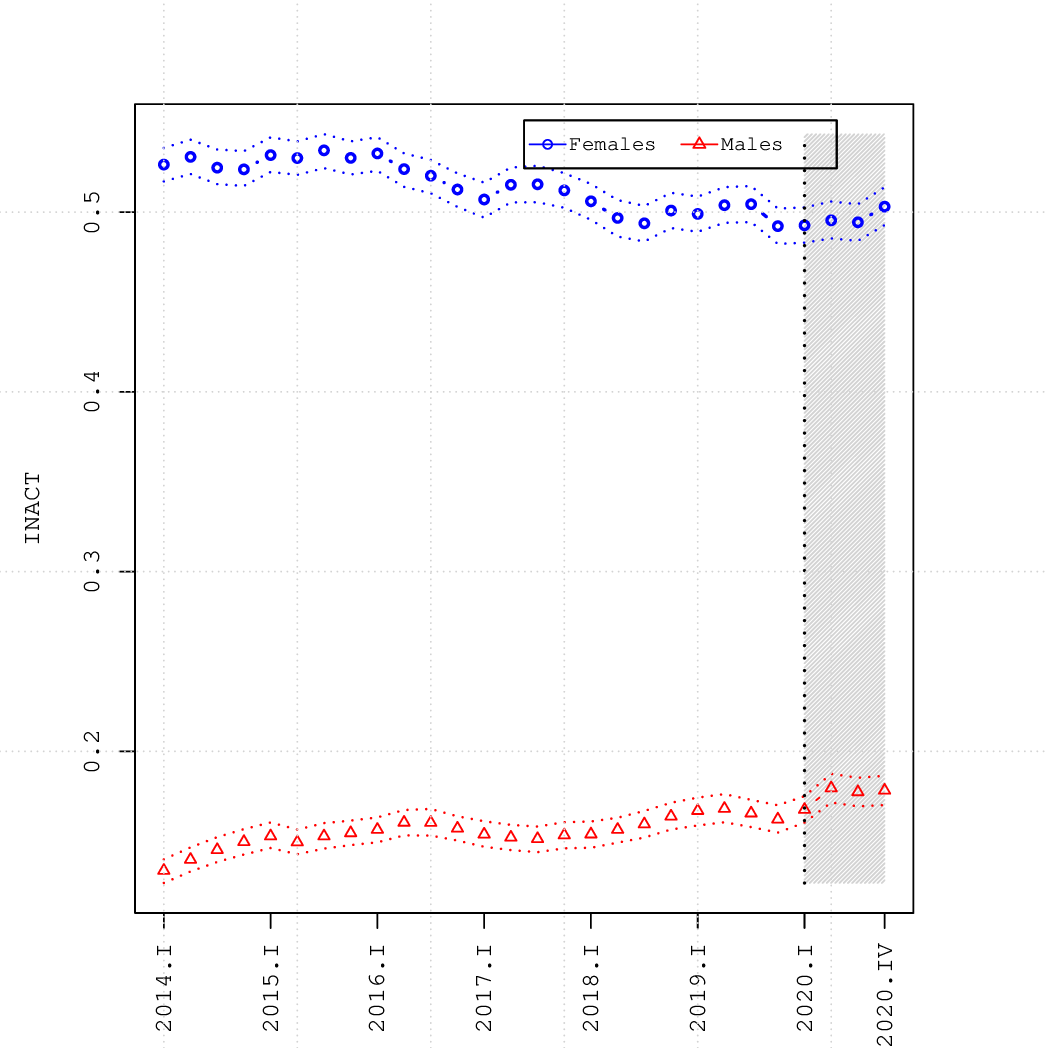}
		\caption{South - Age 40-49.}
		\label{fig:transProbFromEDUtoU_IIIIII}
	\end{subfigure}
	\vspace{0.2cm}
	\caption*{\scriptsize{\textit{Note}: Confidence intervals at 90\% are computed using 1000 bootstraps. The gray area identifies the COVID period. North includes regions in the North and the Center.	 \textit{Source}: LFS 3-month longitudinal data as provided by the Italian Institute of Statistics (ISTAT).}}
\end{figure}

\clearpage



\begin{figure}[!htbp]
	\caption{Shares of individuals aged 25-29 in the temporary employment, unemployment, inactive, and education states in the North and South of Italy.}
	\label{fig:shares2529}
	\caption*{\scriptsize{\textbf{North}}.}
	\centering
	\begin{subfigure}[t]{0.24\textwidth}
		\centering
		\includegraphics[width=\linewidth]{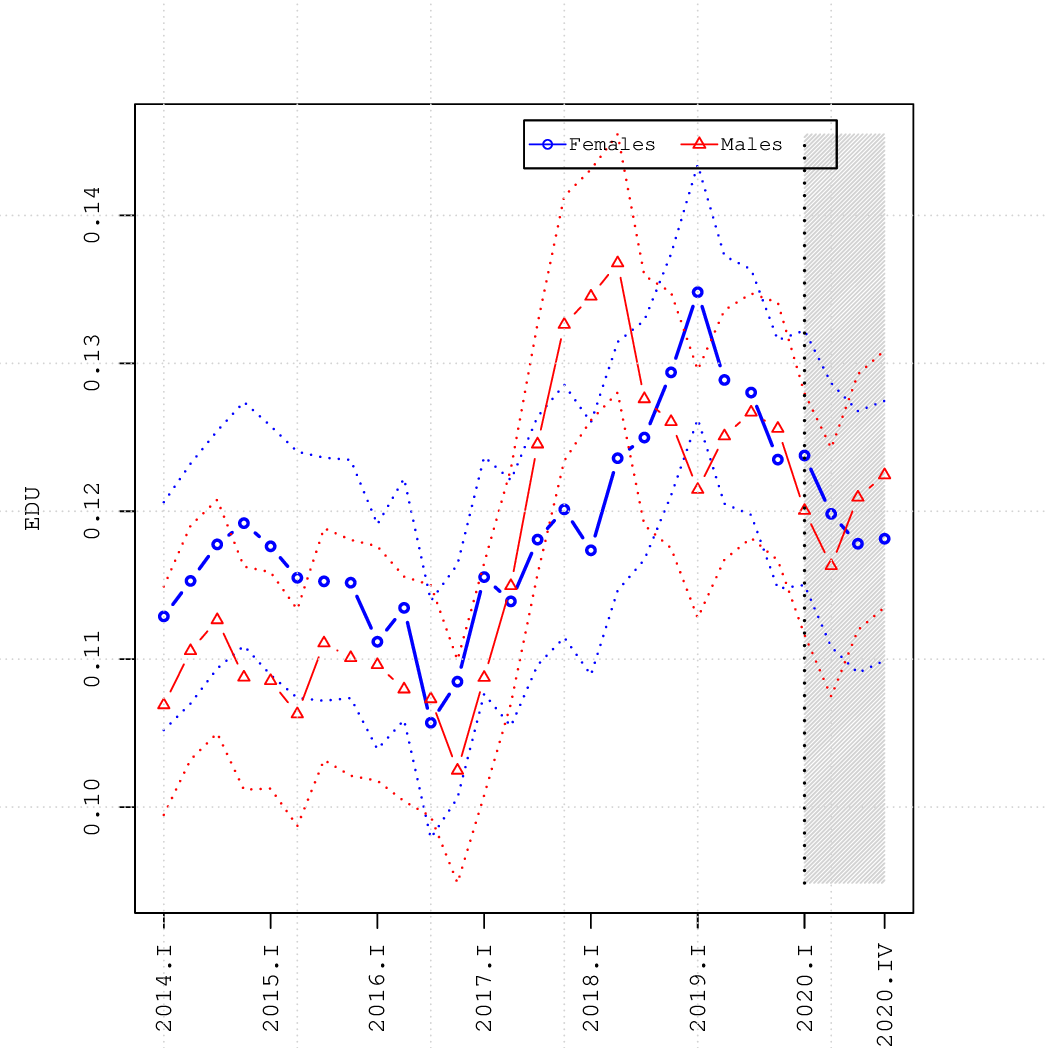}
		\caption{EDU.}
		\label{fig:transProbFromEDUtoSE_II}
		\vspace{0.2cm}
	\end{subfigure}
	\begin{subfigure}[t]{0.24\textwidth}
		\centering
		\includegraphics[width=\linewidth]{actualAverageMass_North2529INACT.eps}
		\caption{INACT.}
		\label{fig:transProbFromEDUtoPE_IIIIIIIIII}
	\end{subfigure}
	\begin{subfigure}[t]{0.24\textwidth}
		\centering
		\includegraphics[width=\linewidth]{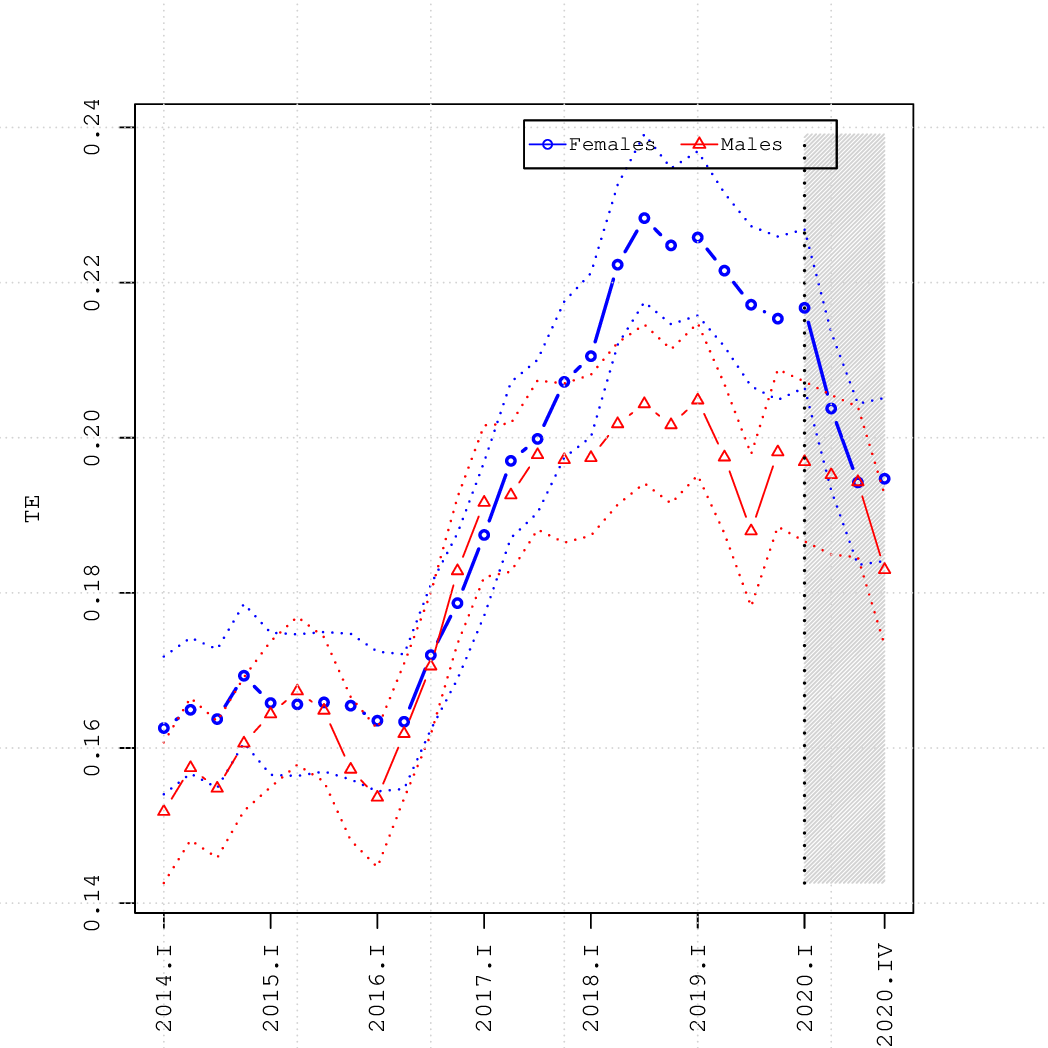}
		\caption{TE.}
		\label{fig:transProbFromEDUtoPE_IIIIIIIIIII}
	\end{subfigure}
	\begin{subfigure}[t]{0.24\textwidth}
		\centering
		\includegraphics[width=\linewidth]{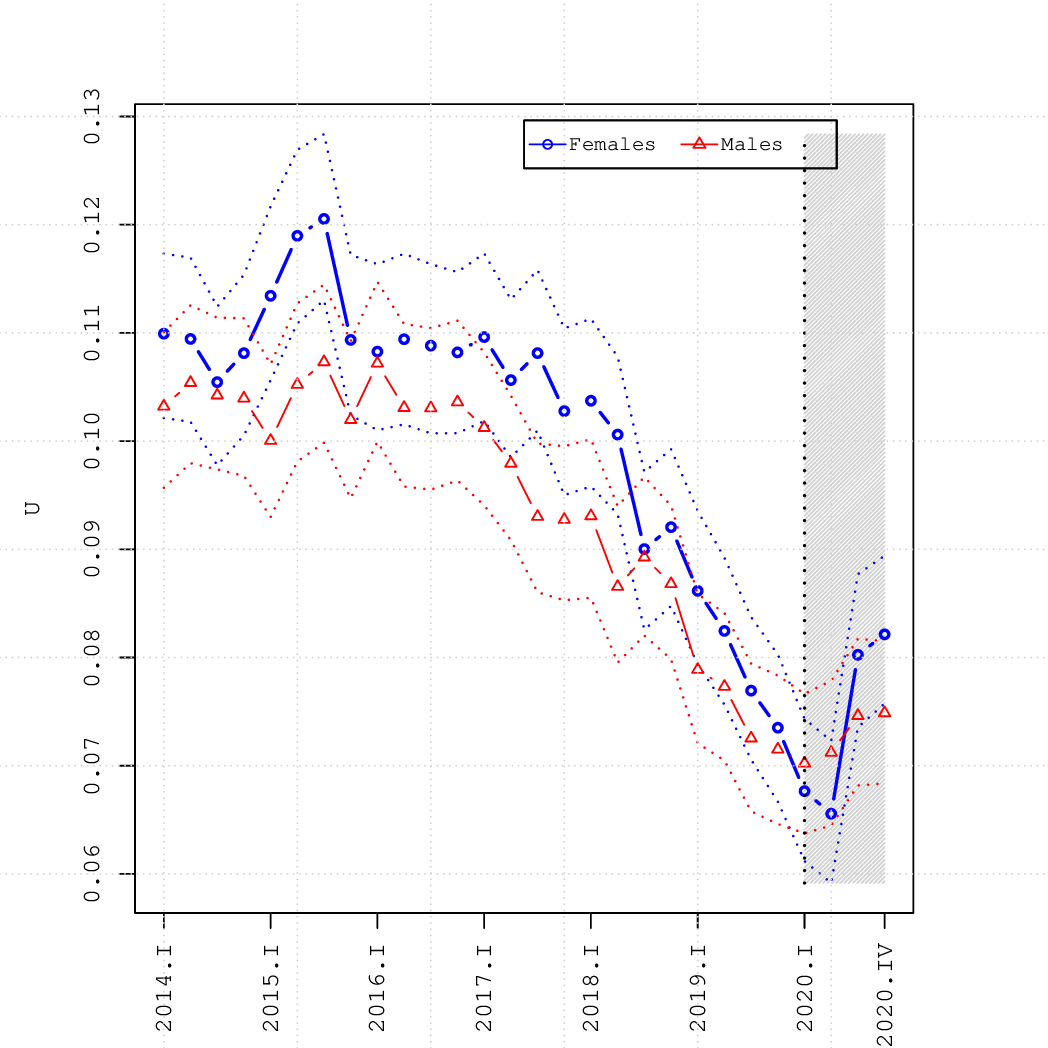}
		\caption{U.}
		\label{fig:transProbFromEDUtoPE_IIIIIIIIIIII}
	\end{subfigure}
	\caption*{\scriptsize{\textbf{South}}.}
	\begin{subfigure}[t]{0.24\textwidth}
		\centering
		\includegraphics[width=\linewidth]{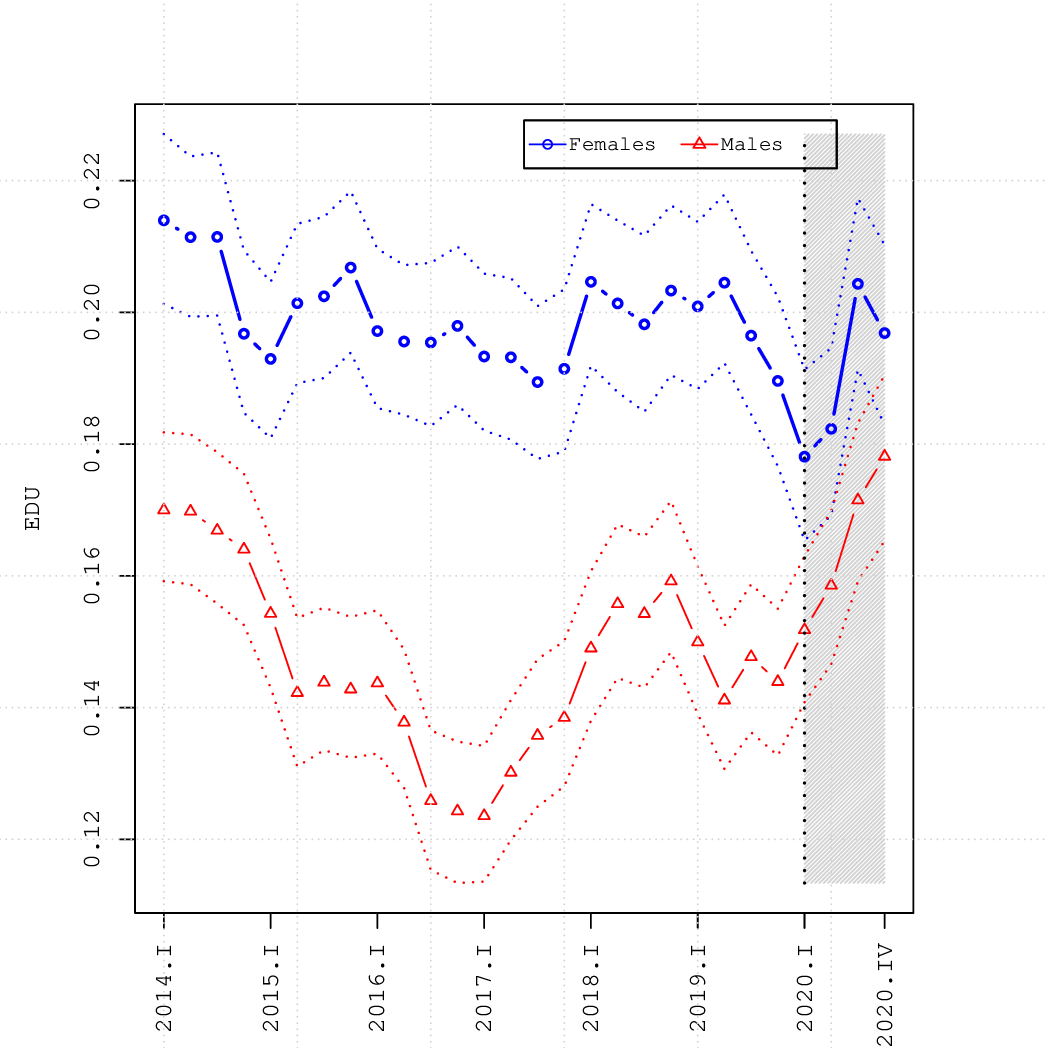}
		\caption{EDU.}
		\label{fig:transProbFromEDUtoTE_I}
	\end{subfigure}
	\begin{subfigure}[t]{0.24\textwidth}
		\centering
		\includegraphics[width=\linewidth]{actualAverageMass_South2529INACT.eps}
		\caption{INACT.}
		\label{fig:transProbFromEDUtoU_IIIIIII}
	\end{subfigure}
	\begin{subfigure}[t]{0.24\textwidth}
		\centering
		\includegraphics[width=\linewidth]{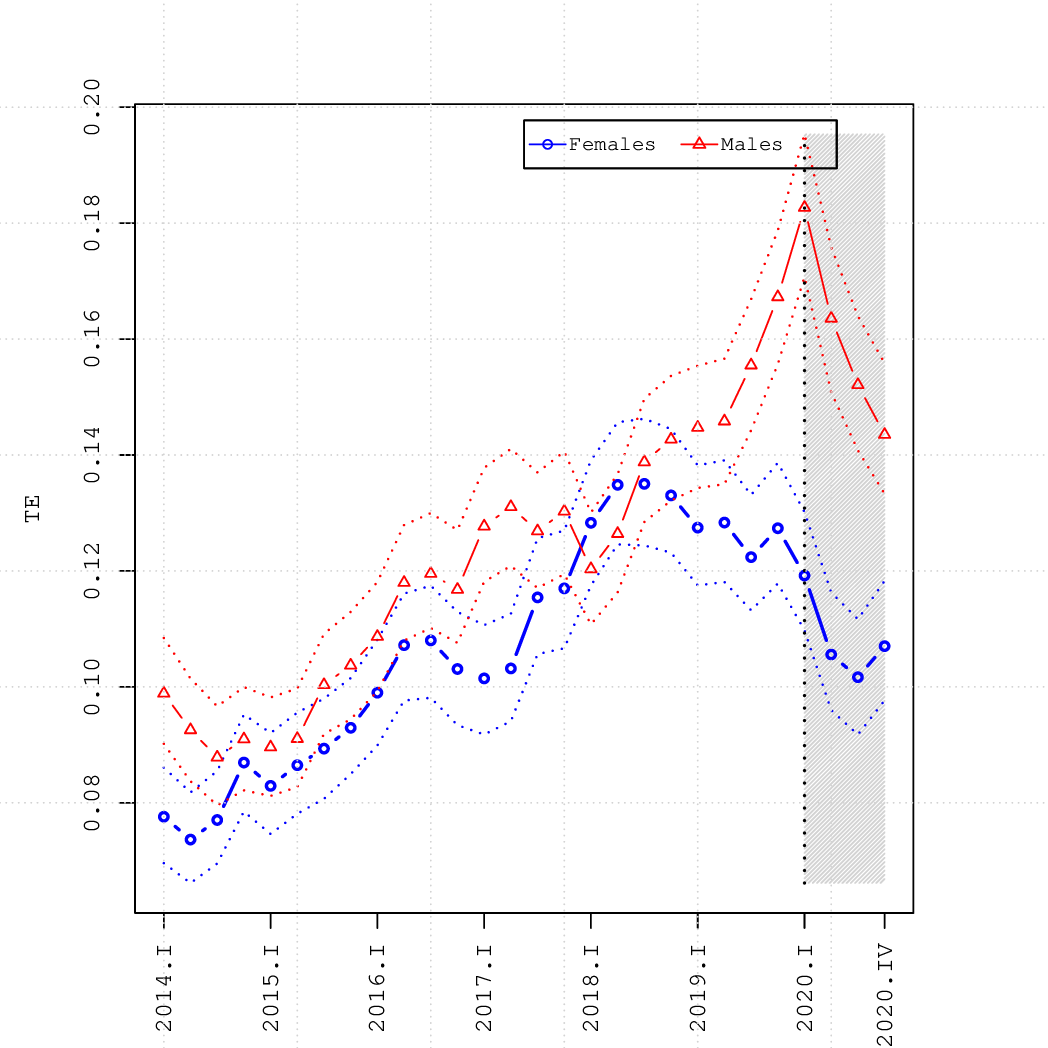}
		\caption{TE.}
		\label{fig:transProbFromEDUtoU_IIIIIIII}
	\end{subfigure}
	\begin{subfigure}[t]{0.24\textwidth}
		\centering
		\includegraphics[width=\linewidth]{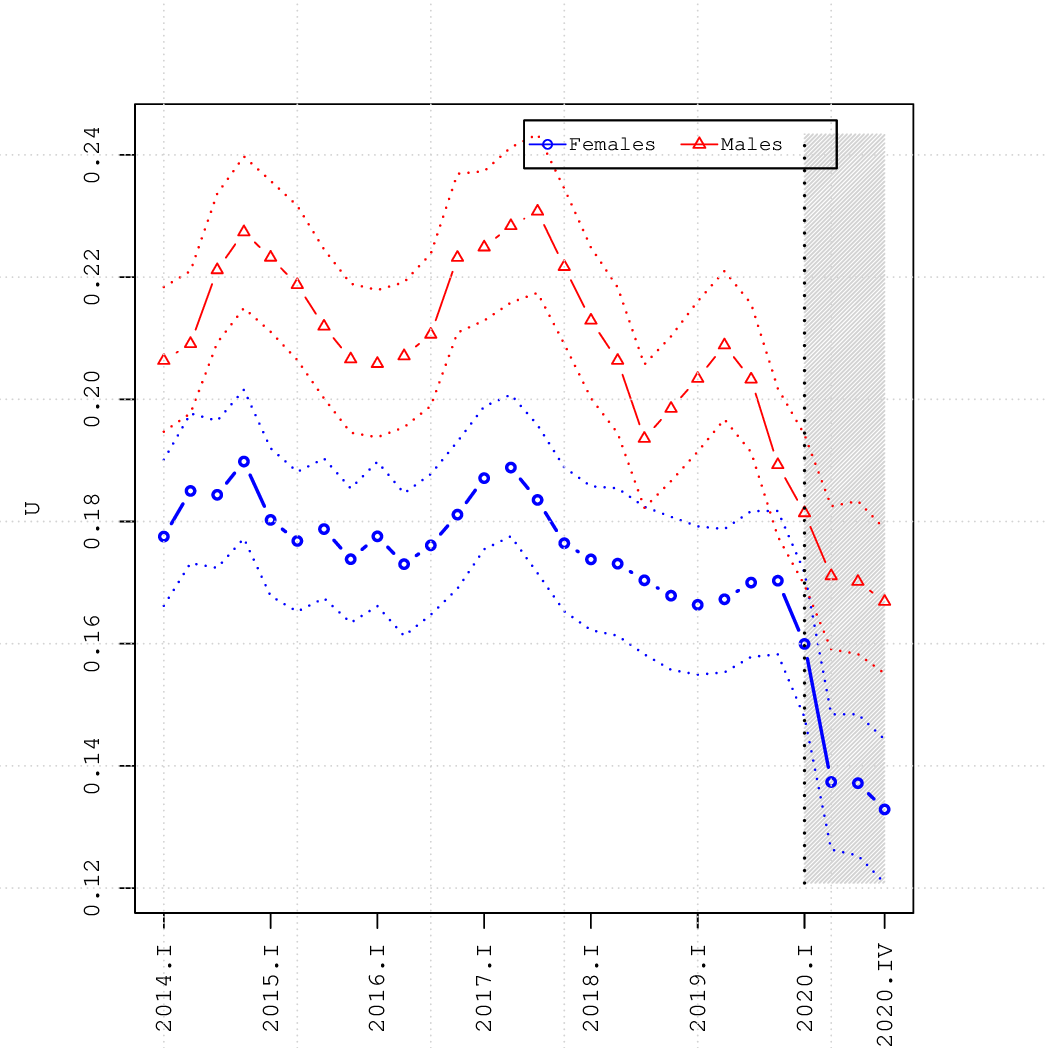}
		\caption{U.}
		\label{fig:transProbFromEDUtoU_IIIIIIIII}
	\end{subfigure}
	\vspace{0.2cm}
	\caption*{\scriptsize{\textit{Note}: Confidence intervals at 90\% are computed using 1000 bootstraps. The gray area identifies the COVID period. North includes regions in the North and the Center. \textit{Source}: LFS 3-month longitudinal data as provided by the Italian Institute of Statistics (ISTAT).}}
\end{figure}


\begin{figure}[!htbp]
	\caption{Shares of individuals aged 30-39  in the temporary employment, permanent employment, inactive, and unemployment states in the North and South of Italy.}
	\label{fig:shares3039}
	\caption*{\scriptsize{\textbf{North}}.}
	\centering
	\begin{subfigure}[t]{0.24\textwidth}
		\centering
		\includegraphics[width=\linewidth]{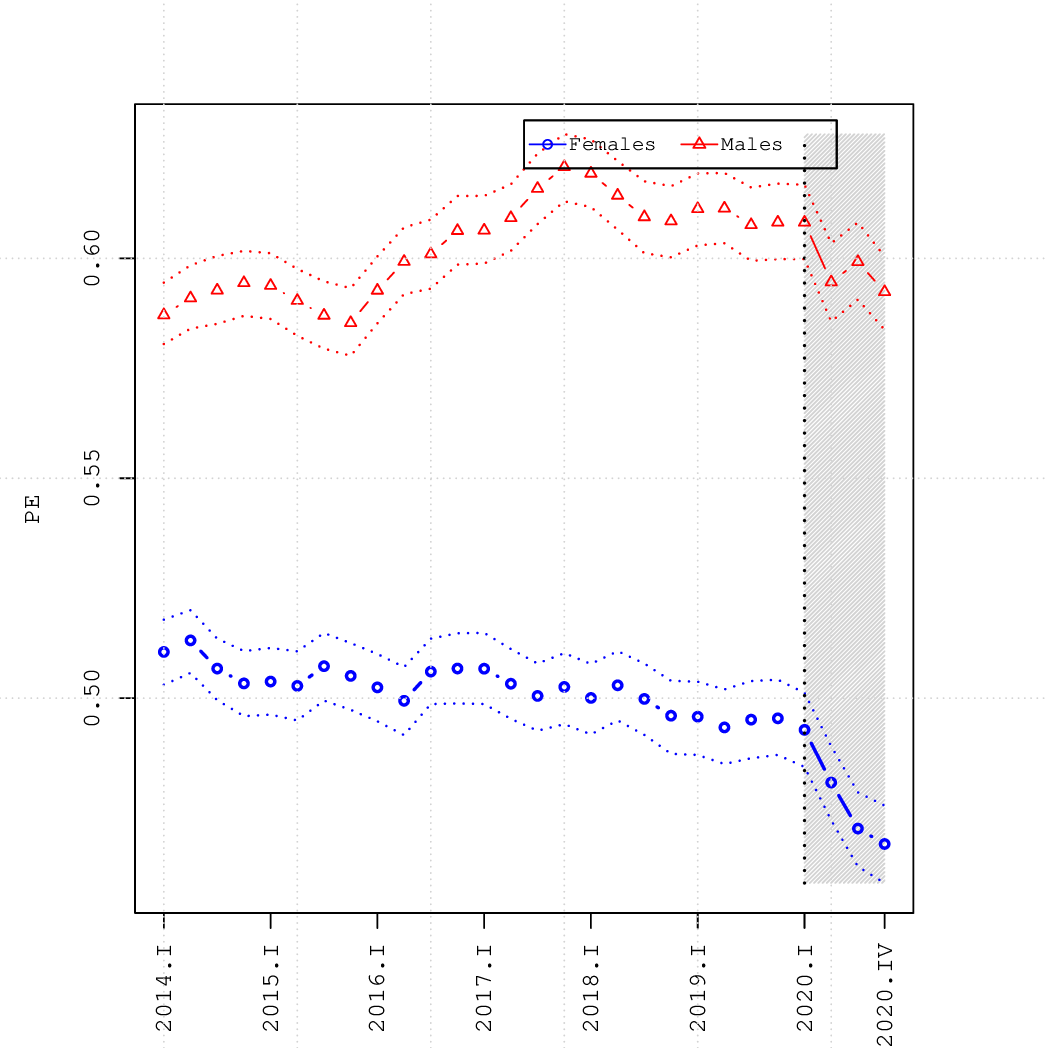}
		\caption{PE.}
		\label{fig:transProbFromEDUtoSE_III}
		\vspace{0.2cm}
	\end{subfigure}
	\begin{subfigure}[t]{0.24\textwidth}
		\centering
		\includegraphics[width=\linewidth]{actualAverageMass_North3039INACT.eps}
		\caption{INACT.}
		\label{fig:transProbFromEDUtoPE_IIIIIIIIIIIII}
	\end{subfigure}
	\begin{subfigure}[t]{0.24\textwidth}
		\centering
		\includegraphics[width=\linewidth]{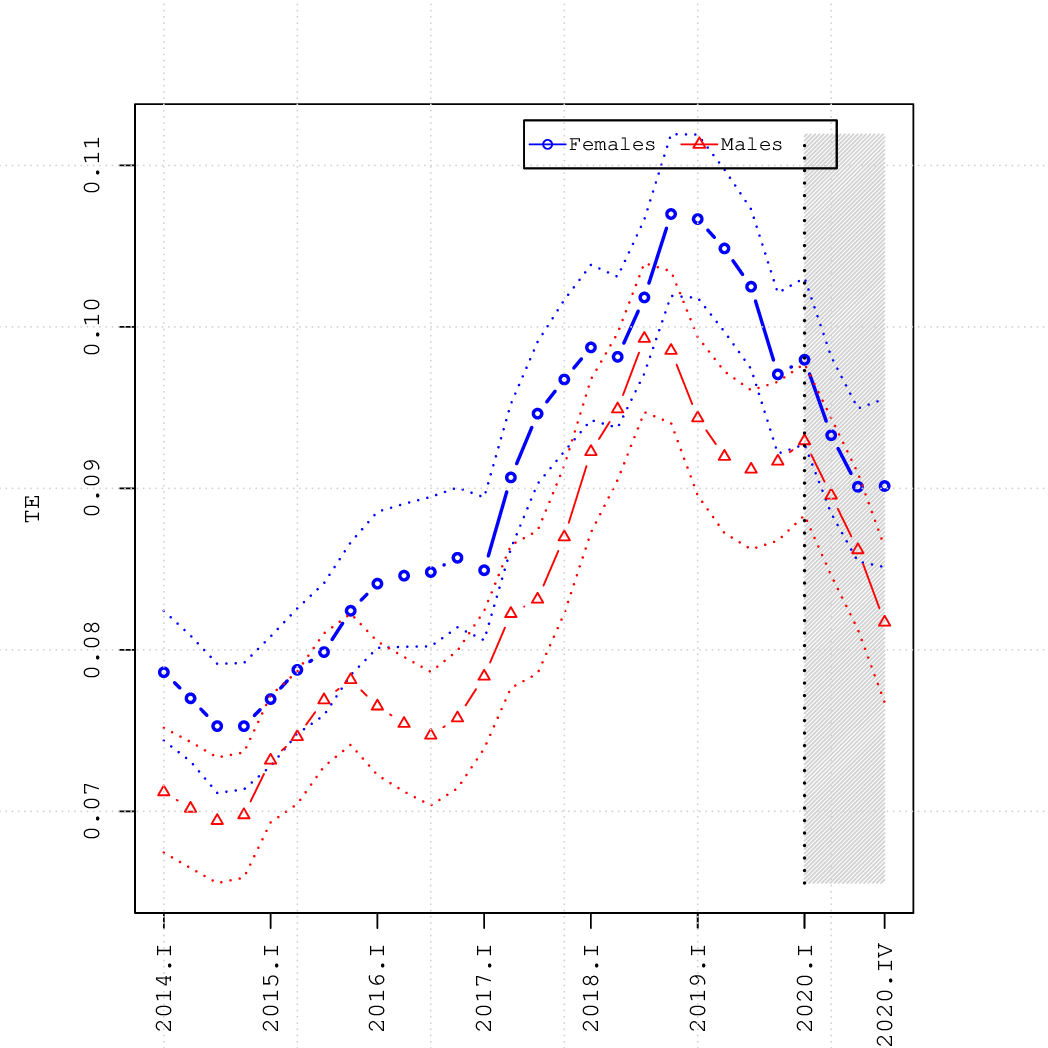}
		\caption{TE.}
		\label{fig:transProbFromEDUtoPE_IIIIIIIIIIIIII}
	\end{subfigure}
	\begin{subfigure}[t]{0.24\textwidth}
		\centering
		\includegraphics[width=\linewidth]{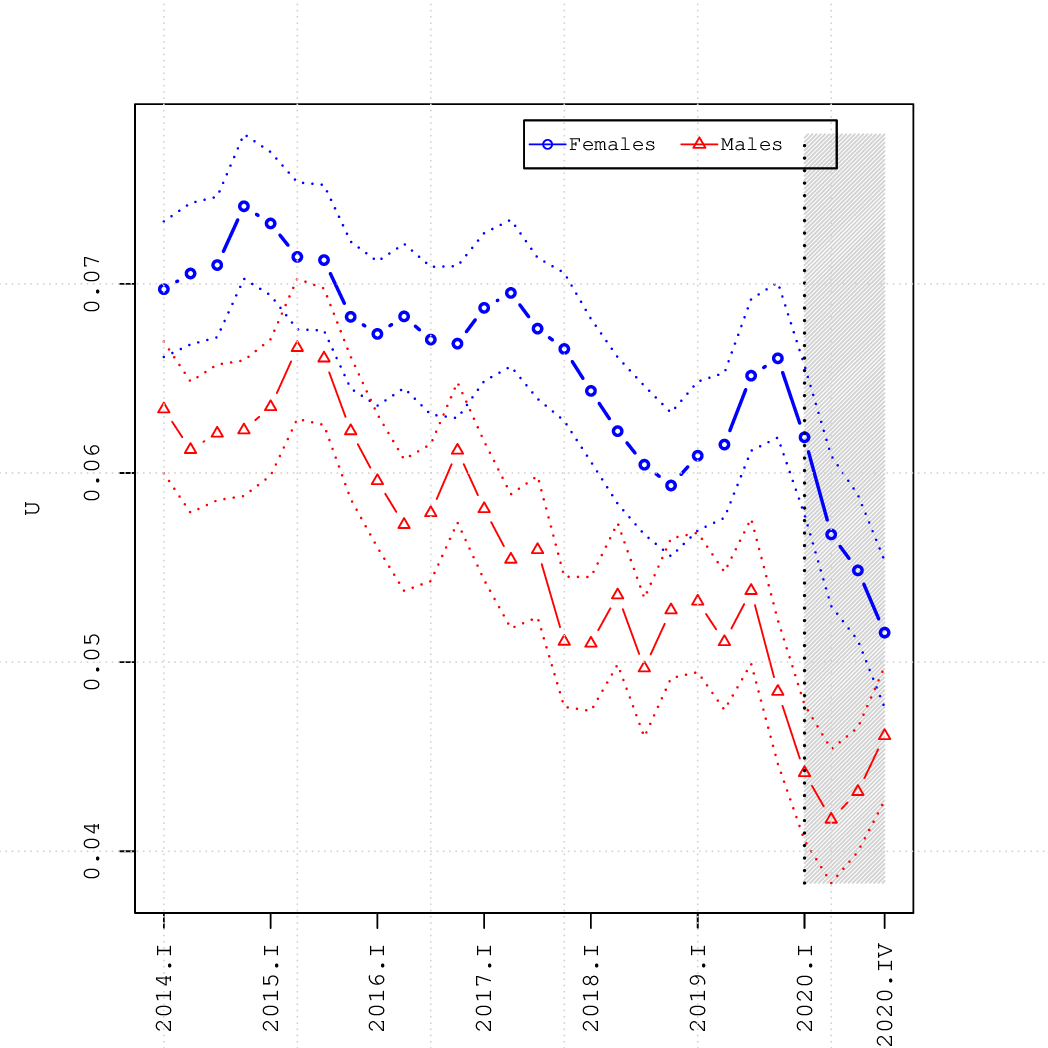}
		\caption{U.}
		\label{fig:transProbFromEDUtoPE_IIIIIIIIIIIIIII}
	\end{subfigure}
	\caption*{\scriptsize{\textbf{South}}.}
	\begin{subfigure}[t]{0.24\textwidth}
		\centering
		\includegraphics[width=\linewidth]{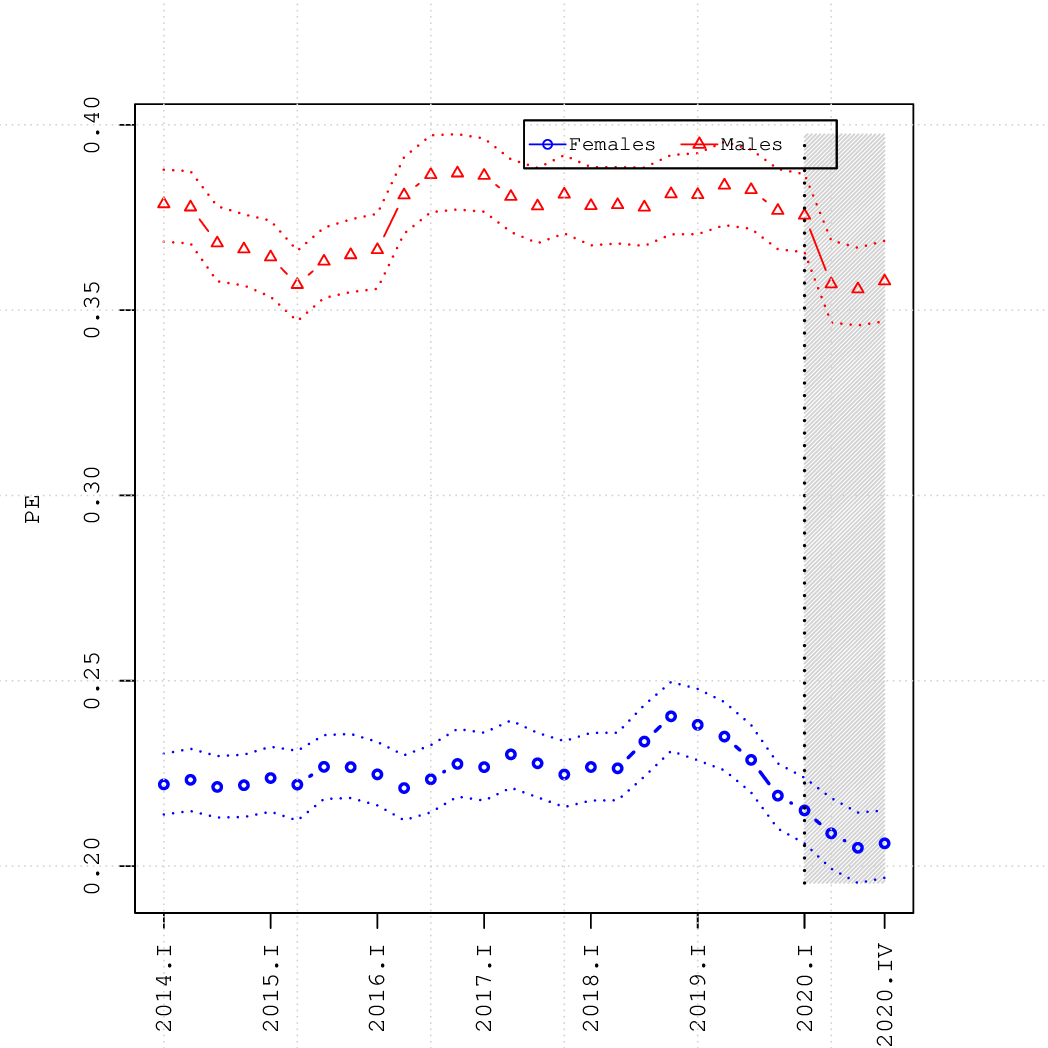}
		\caption{PE.}
		\label{fig:transProbFromEDUtoTE_II}
	\end{subfigure}
	\begin{subfigure}[t]{0.24\textwidth}
		\centering
		\includegraphics[width=\linewidth]{actualAverageMass_South3039INACT.eps}
		\caption{INACT.}
		\label{fig:transProbFromEDUtoU_IIIIIIIIII}
	\end{subfigure}
	\begin{subfigure}[t]{0.24\textwidth}
		\centering
		\includegraphics[width=\linewidth]{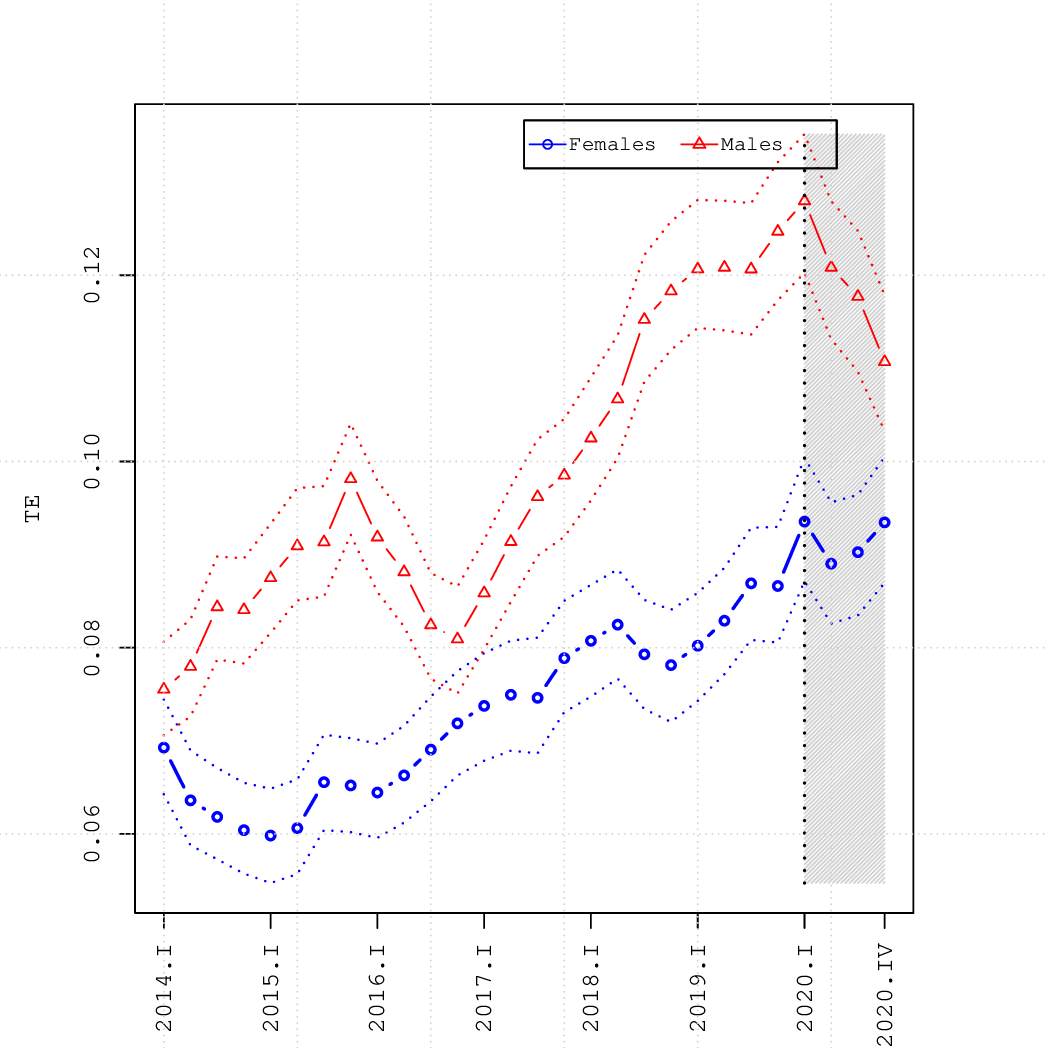}
		\caption{TE.}
		\label{fig:transProbFromEDUtoU_IIIIIIIIIII}
	\end{subfigure}
	\begin{subfigure}[t]{0.24\textwidth}
		\centering
		\includegraphics[width=\linewidth]{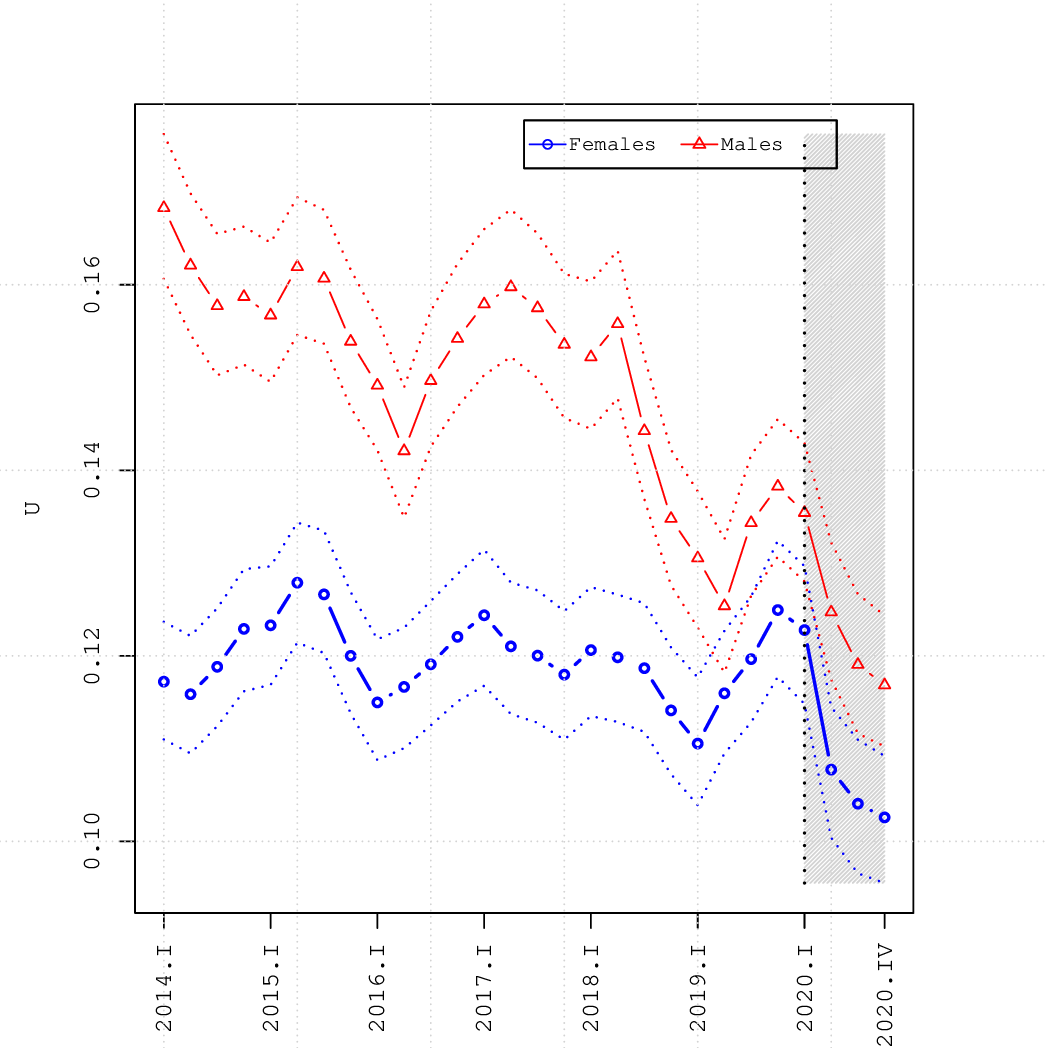}
		\caption{U.}
		\label{fig:transProbFromEDUtoU_IIIIIIIIIIII}
	\end{subfigure}
	\vspace{0.2cm}
	\caption*{\scriptsize{\textit{Note}: Confidence intervals at 90\% are computed using 1000 bootstraps. The gray area identifies the COVID period. North includes regions in the North and the Center. \textit{Source}: LFS 3-month longitudinal data as provided by the Italian Institute of Statistics (ISTAT).}}
\end{figure}


\begin{figure}[!htbp]
	\caption{Shares of individuals aged 40-49  in the temporary employment, permanent employment, inactive, and unemployment states in the North and South of Italy.}
	\label{fig:shares4049}
	\caption*{\scriptsize{\textbf{North}}.}
	\centering
	\begin{subfigure}[t]{0.24\textwidth}
		\centering
		\includegraphics[width=\linewidth]{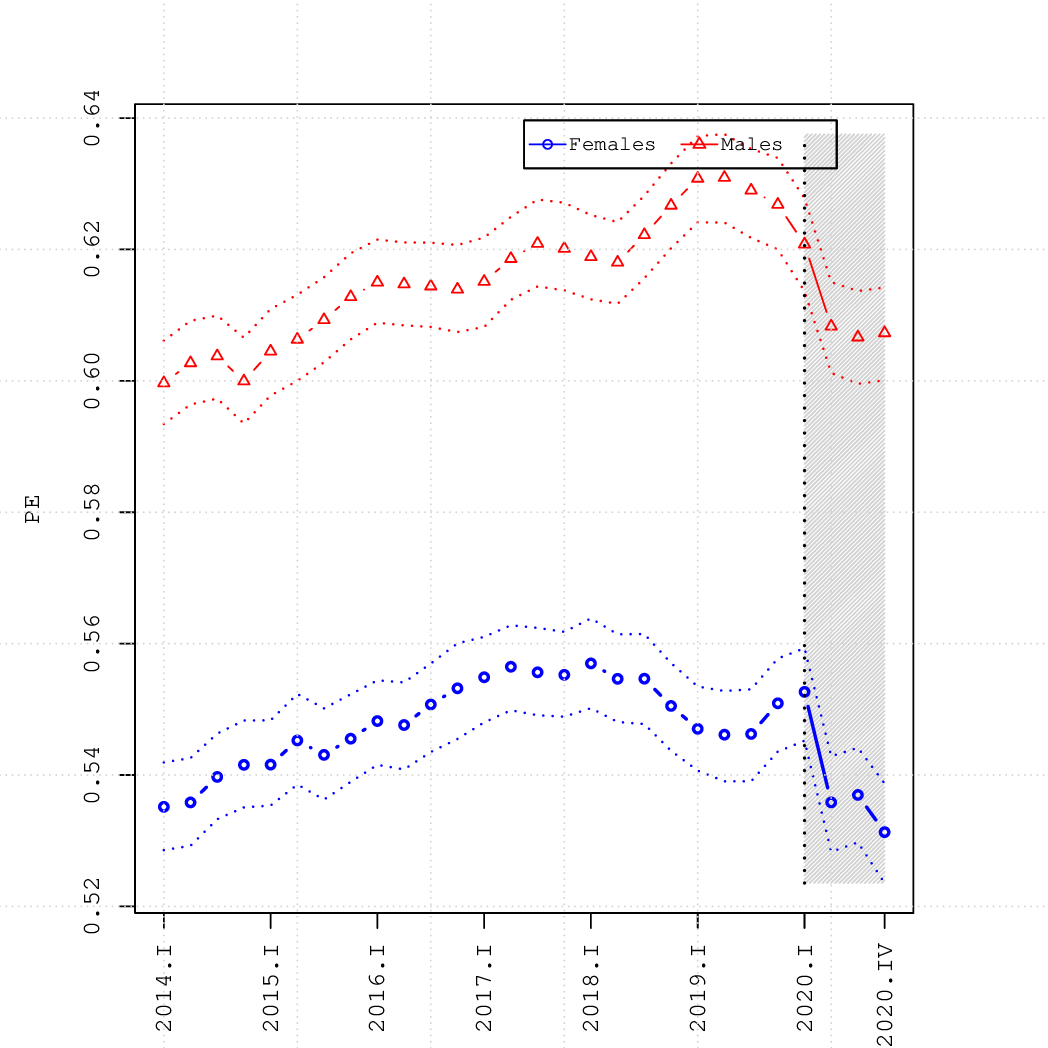}
		\caption{PE.}
		\label{fig:transProbFromEDUtoSE_IIII}
		\vspace{0.2cm}
	\end{subfigure}
	\begin{subfigure}[t]{0.24\textwidth}
		\centering
		\includegraphics[width=\linewidth]{actualAverageMass_North4049INACT.eps}
		\caption{INACT.}
		\label{fig:transProbFromEDUtoPE_IIIIIIIIIIIIIIII}
	\end{subfigure}
	\begin{subfigure}[t]{0.24\textwidth}
		\centering
		\includegraphics[width=\linewidth]{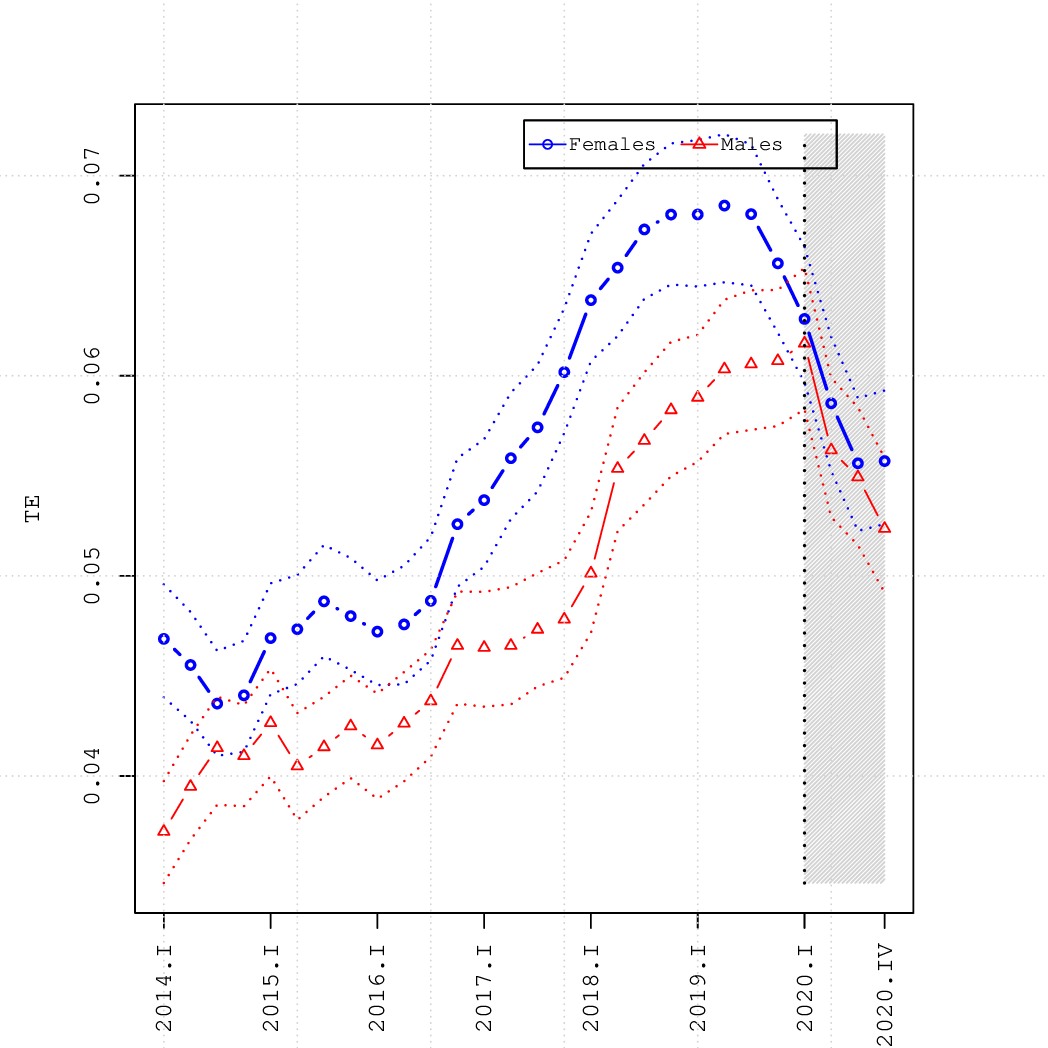}
		\caption{TE.}
		\label{fig:transProbFromEDUtoPE_IIIIIIIIIIIIIIIII}
	\end{subfigure}
	\begin{subfigure}[t]{0.24\textwidth}
		\centering
		\includegraphics[width=\linewidth]{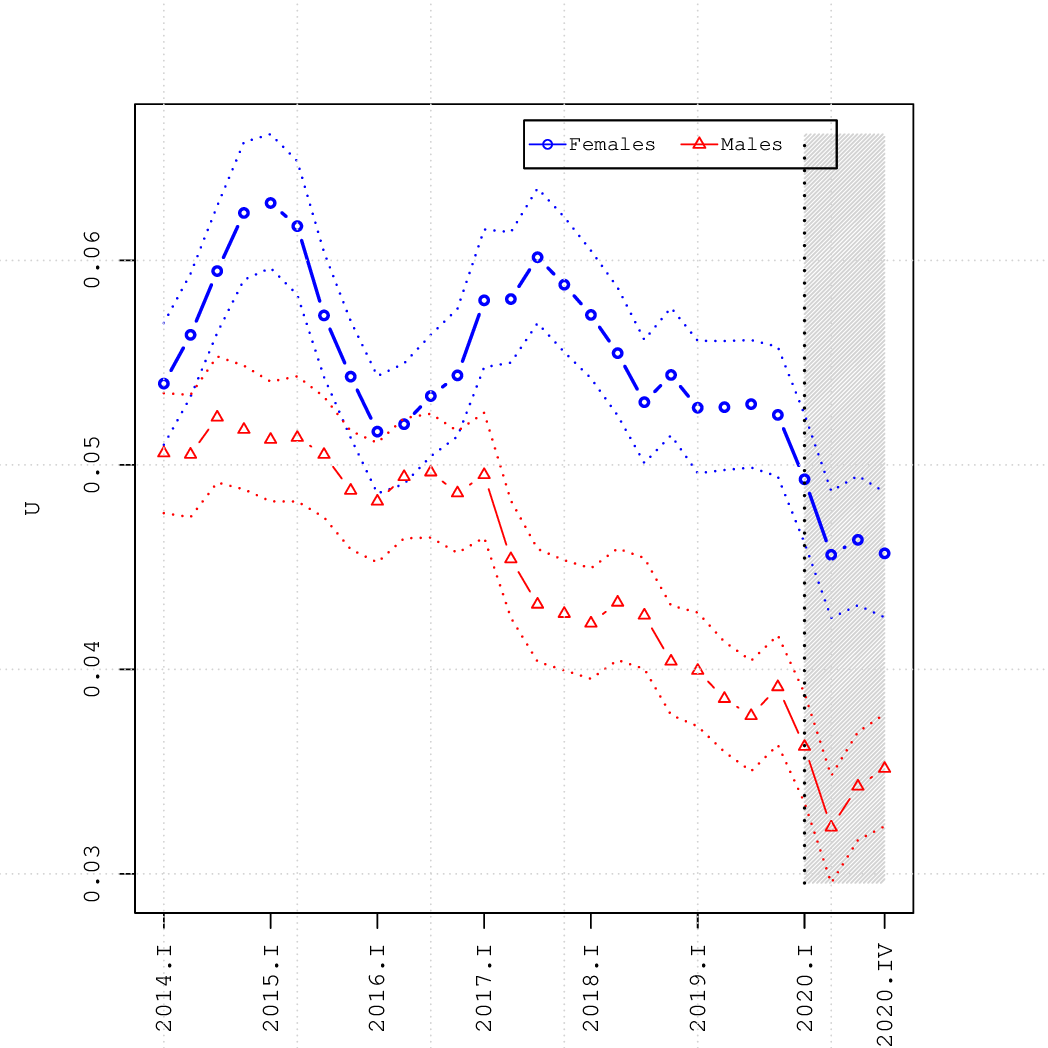}
		\caption{U.}
		\label{fig:transProbFromEDUtoPE_IIIIIIIIIIIIIIIIII}
	\end{subfigure}\\
	\caption*{\scriptsize{\textbf{South}}.}
	\begin{subfigure}[t]{0.24\textwidth}
		\centering
		\includegraphics[width=\linewidth]{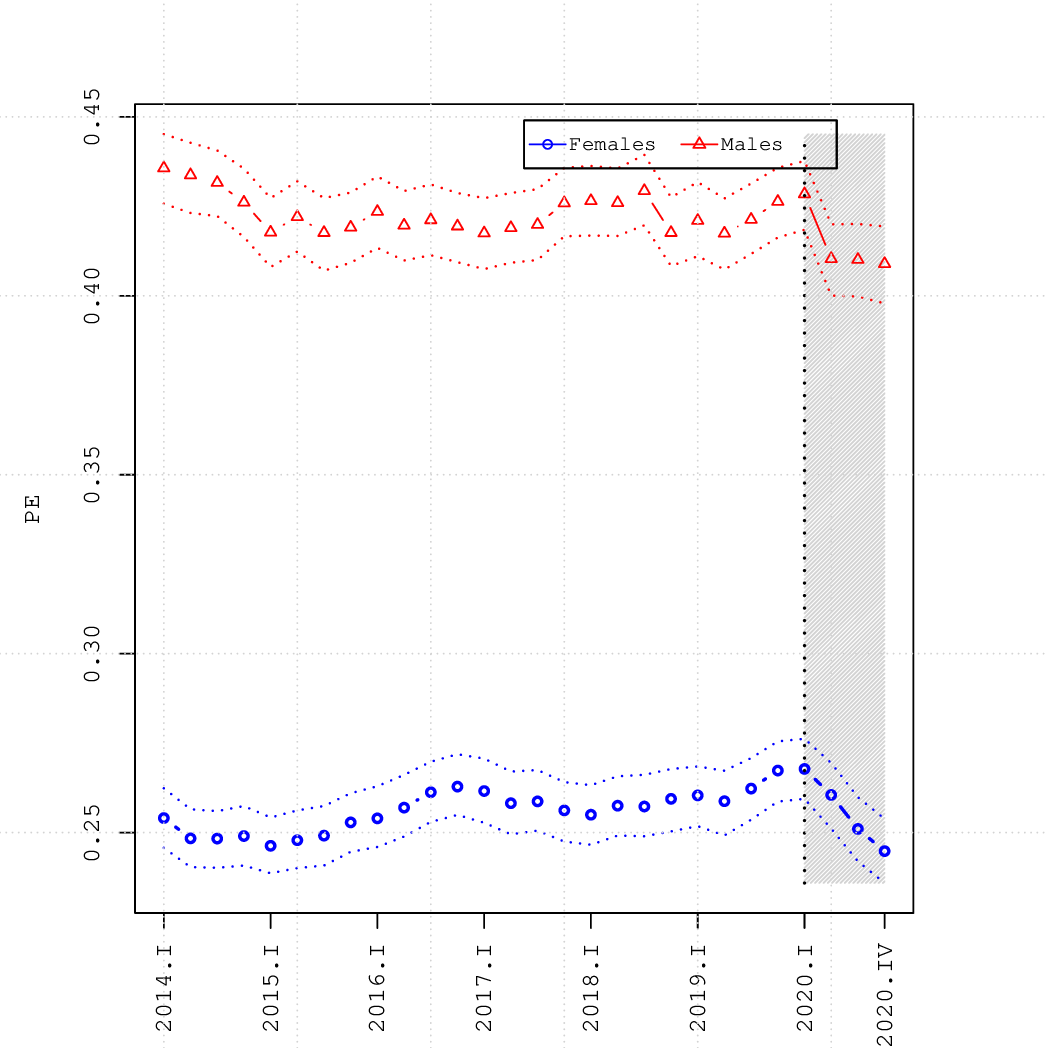}
		\caption{PE.}
		\label{fig:transProbFromEDUtoTE_III}
	\end{subfigure}
	\begin{subfigure}[t]{0.24\textwidth}
		\centering
		\includegraphics[width=\linewidth]{actualAverageMass_South4049INACT.eps}
		\caption{INACT.}
		\label{fig:transProbFromEDUtoU_IIIIIIIIIIIII}
	\end{subfigure}
	\begin{subfigure}[t]{0.24\textwidth}
		\centering
		\includegraphics[width=\linewidth]{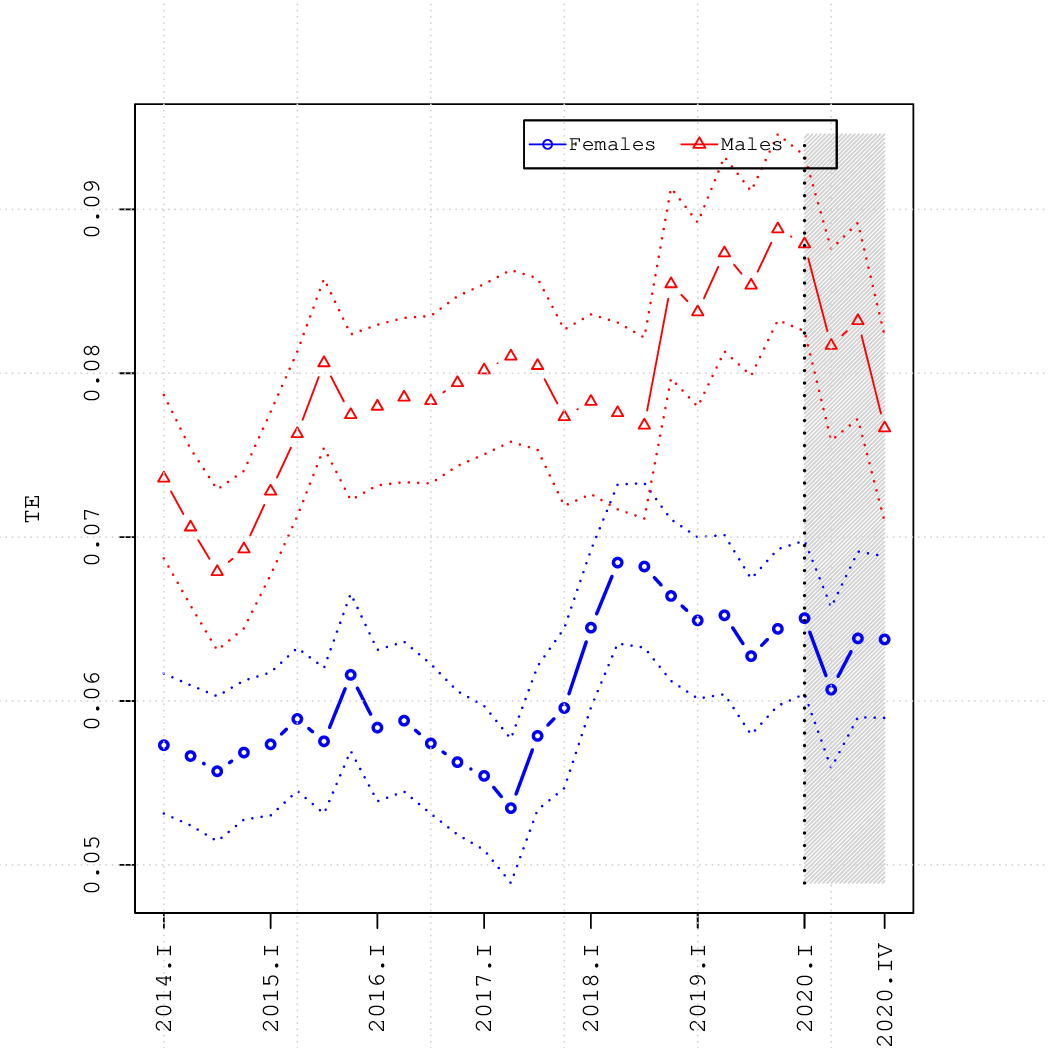}
		\caption{TE.}
		\label{fig:transProbFromEDUtoU_IIIIIIIIIIIIII}
	\end{subfigure}
	\begin{subfigure}[t]{0.24\textwidth}
		\centering
		\includegraphics[width=\linewidth]{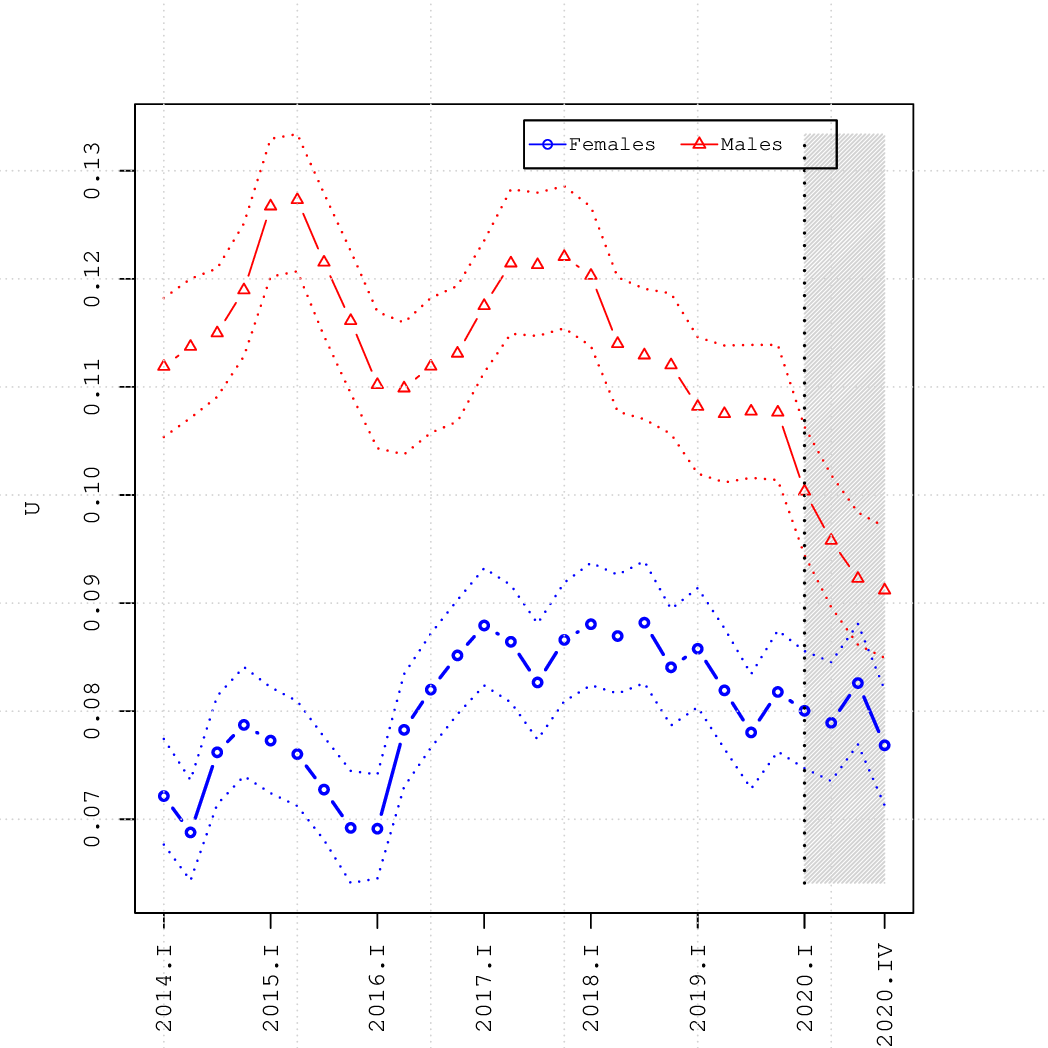}
		\caption{U.}
		\label{fig:transProbFromEDUtoU_IIIIIIIIIIIIIII}
	\end{subfigure}
	\vspace{0.2cm}
	\caption*{\scriptsize{\textit{Note}: Confidence intervals at 90\% are computed using 1000 bootstraps. The gray area identifies the COVID period. North includes regions in the North and the Center. \textit{Source}: LFS 3-month longitudinal data as provided by the Italian Institute of Statistics (ISTAT).}}
\end{figure}

\clearpage

\subsection{Transition probabilities by age, gender and geographical location}


\begin{figure}[!htbp]
	\caption{Annual transition probabilities of individuals aged 25-29 from education to temporary employment, unemployment, inactive  states in the North and South of Italy.}
	\label{fig:transprob2529females}
	\caption*{\scriptsize{\textbf{Females}}.}
	\centering
	\begin{subfigure}[t]{0.24\textwidth}
		\centering
		\includegraphics[width=\linewidth]{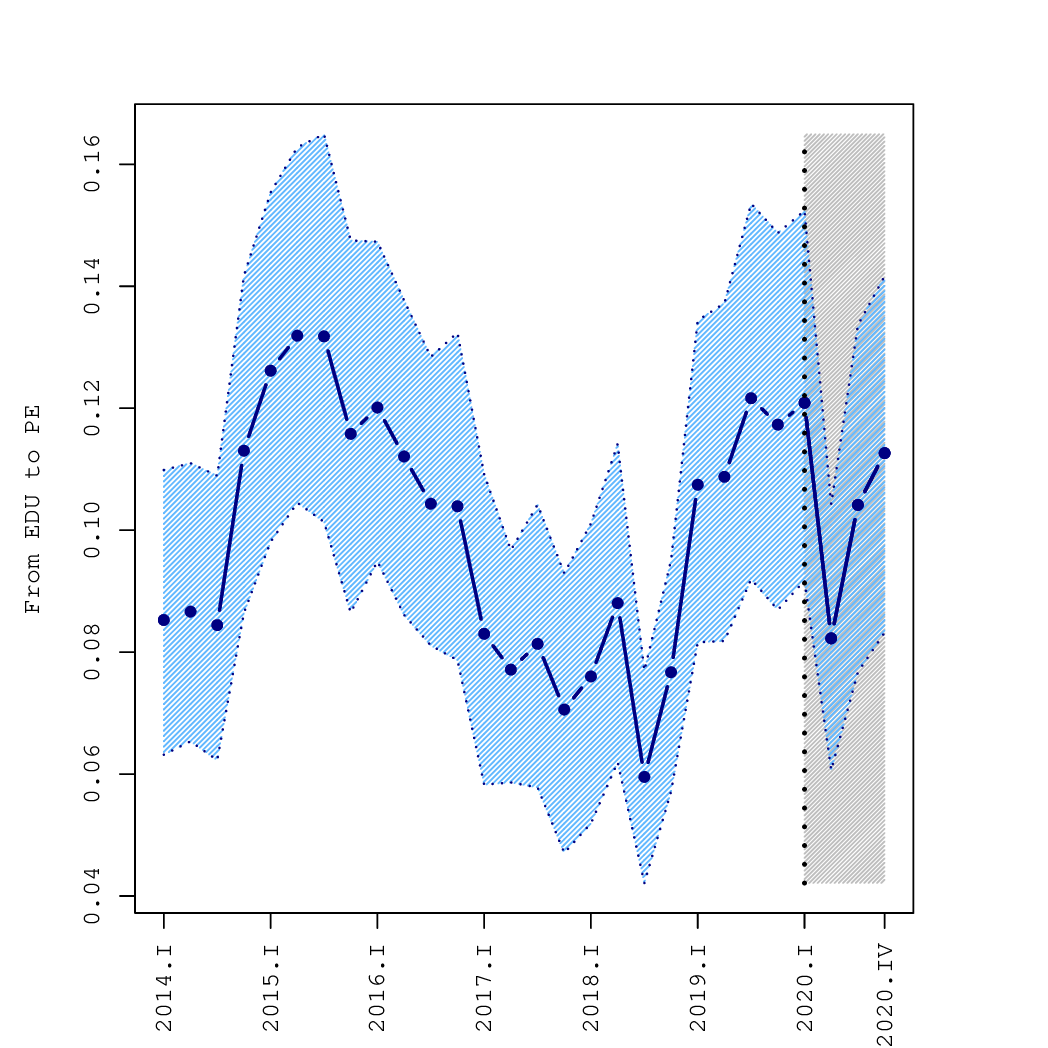}
		\caption{PE - North.}
		\label{fig:transProbFromEDUtoSE_IIIII}
		\vspace{0.2cm}
	\end{subfigure}
	\begin{subfigure}[t]{0.24\textwidth}
		\centering
		\includegraphics[width=\linewidth]{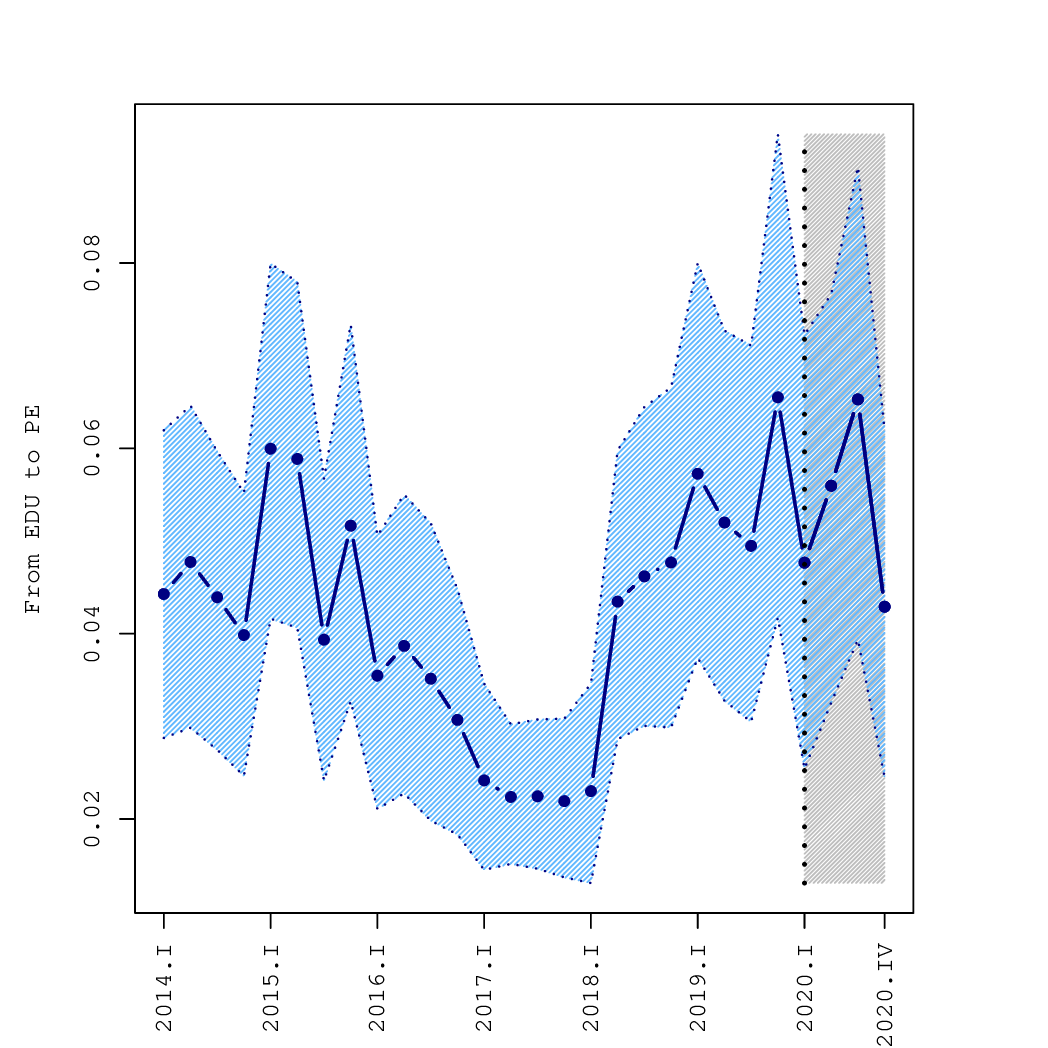}
		\caption{PE - South.}
		\label{fig:transProbFromEDUtoTE_IIII}
	\end{subfigure}
	\begin{subfigure}[t]{0.24\textwidth}
		\centering
		\includegraphics[width=\linewidth]{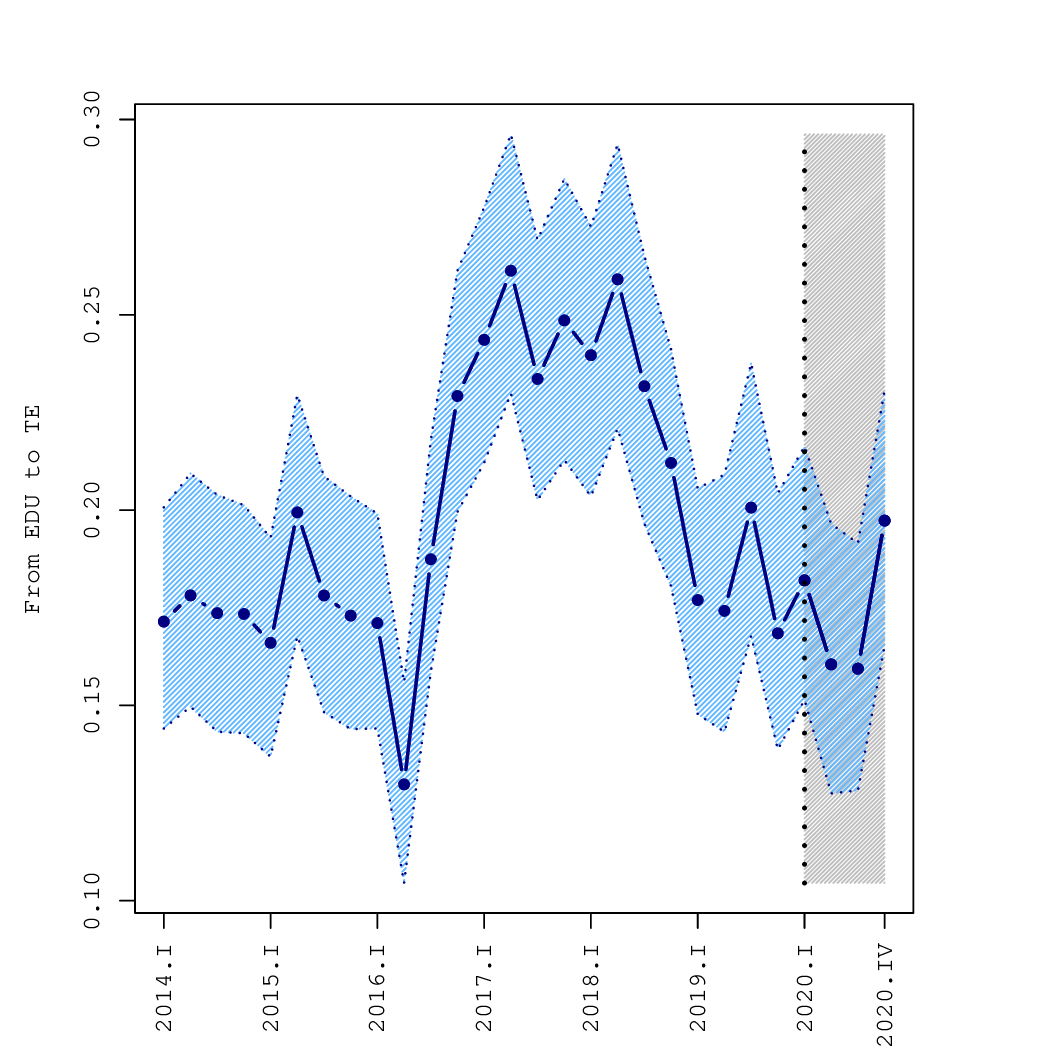}
		\caption{TE - North.}
		\label{fig:transProbFromEDUtoPE_IIIIIIIIIIIIIIIIIII}
	\end{subfigure}
	\begin{subfigure}[t]{0.24\textwidth}
		\centering
		\includegraphics[width=\linewidth]{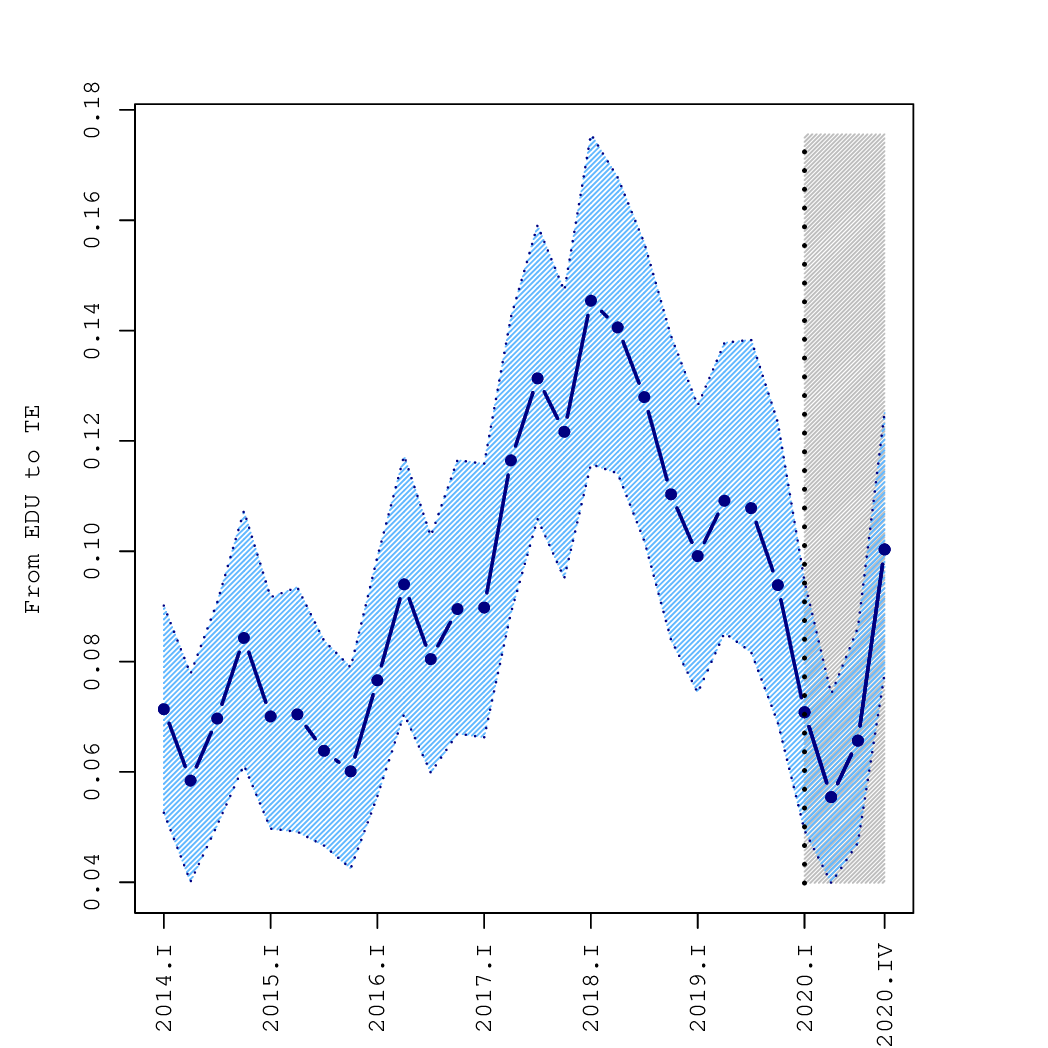}
		\caption{TE - South.}
		\label{fig:transProbFromEDUtoU_IIIIIIIIIIIIIIII}
	\end{subfigure}
	\caption*{\scriptsize{\textbf{Males}}.}
	\centering
	\begin{subfigure}[t]{0.24\textwidth}
		\centering
		\includegraphics[width=\linewidth]{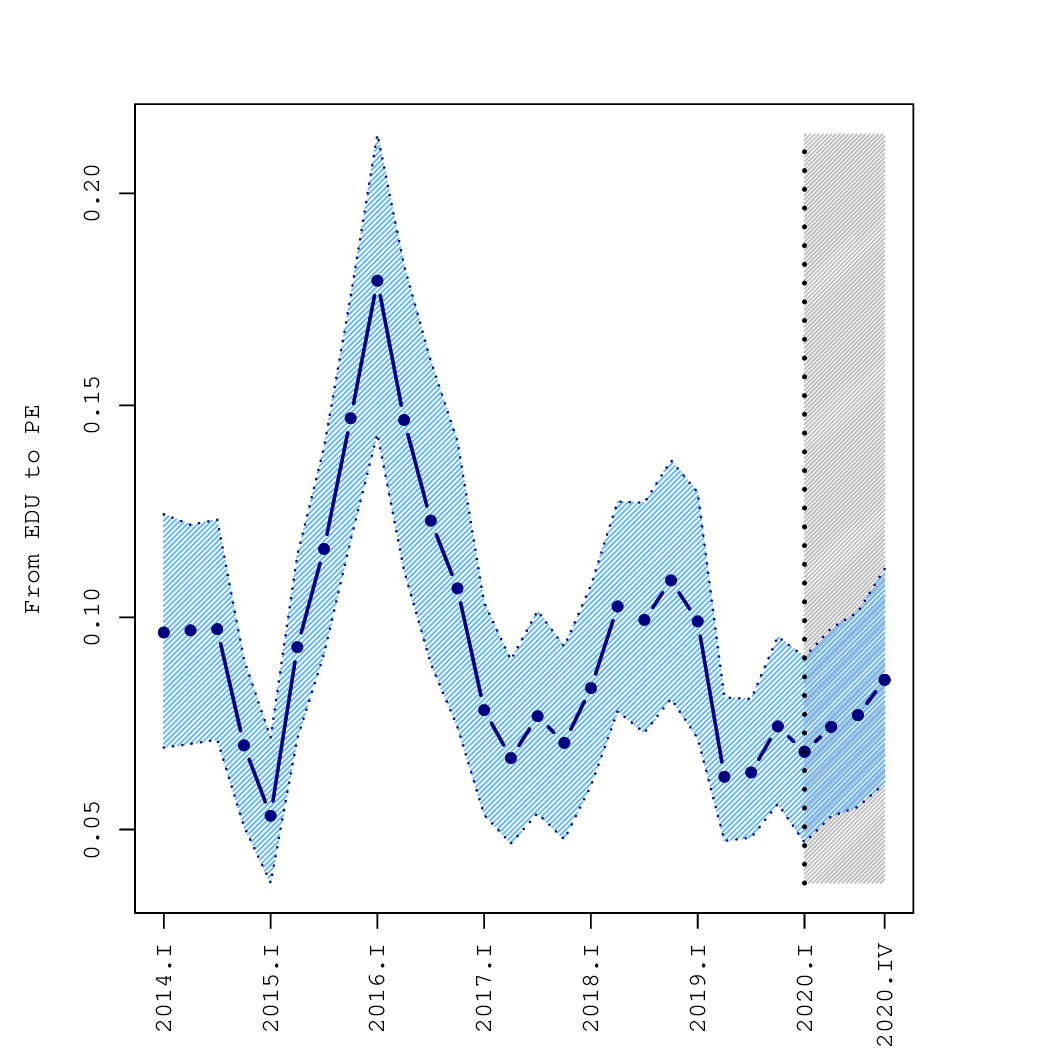}
		\caption{PE - North.}
		\label{fig:transProbFromEDUtoSE_IIIIII}
		\vspace{0.2cm}
	\end{subfigure}
	\begin{subfigure}[t]{0.24\textwidth}
		\centering
		\includegraphics[width=\linewidth]{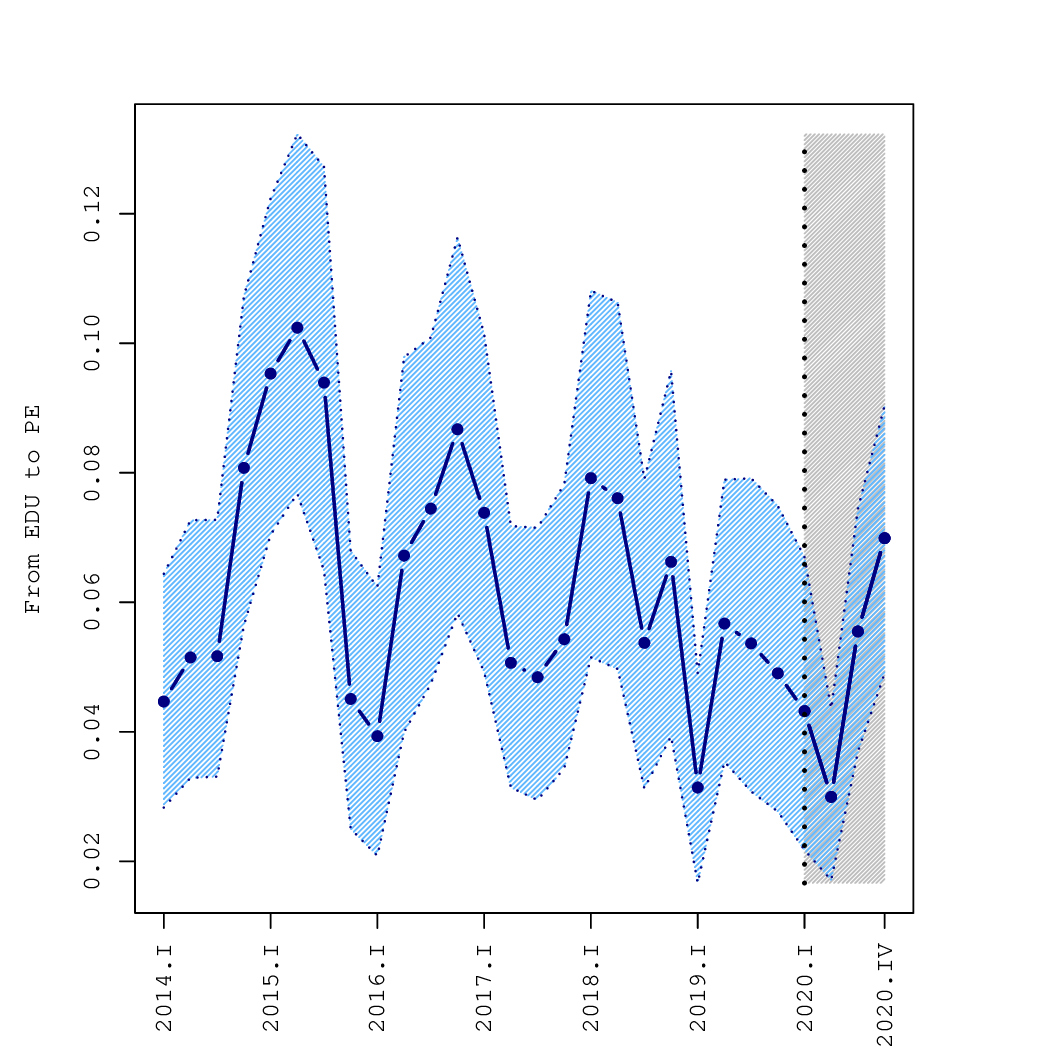}
		\caption{PE - South.}
		\label{fig:transProbFromEDUtoTE_IIIIII}
	\end{subfigure}
	\begin{subfigure}[t]{0.24\textwidth}
		\centering
		\includegraphics[width=\linewidth]{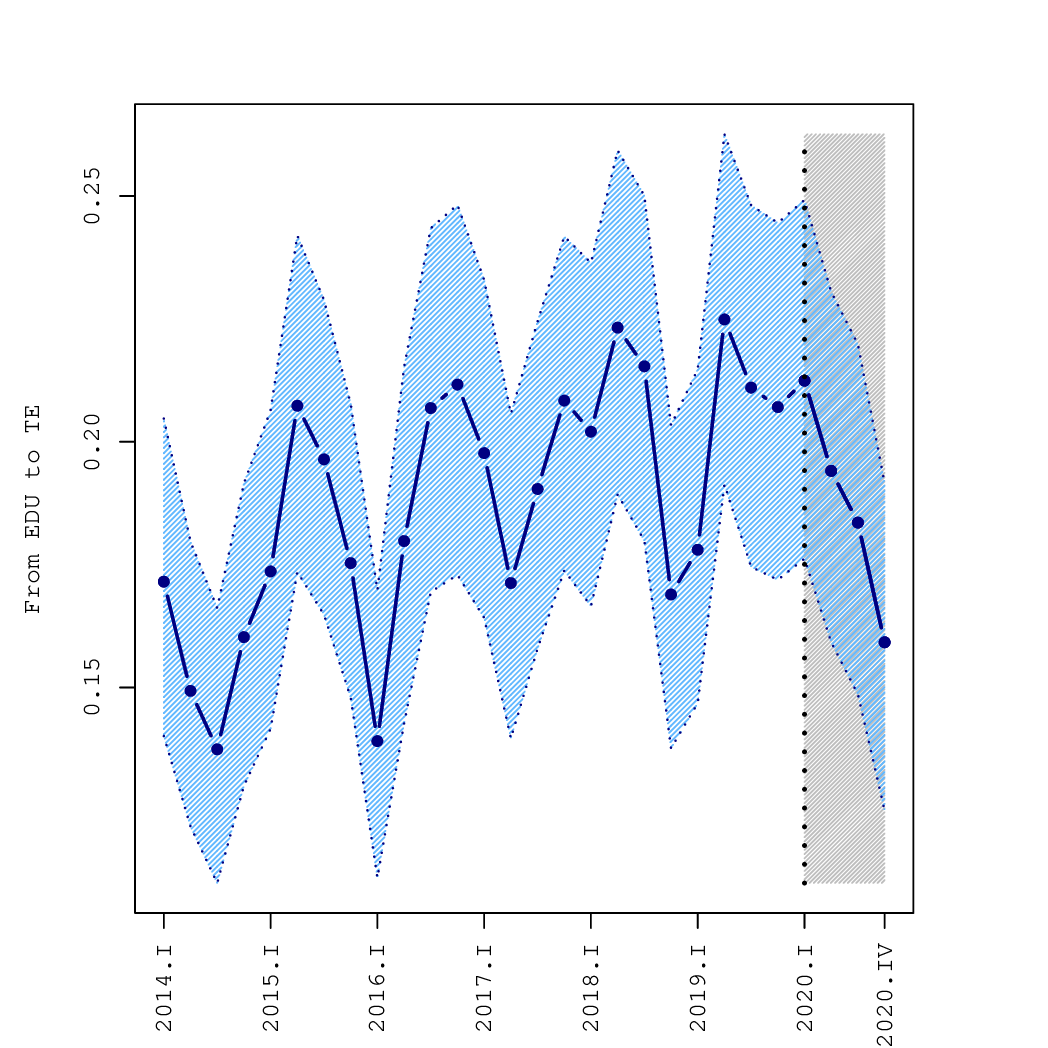}
		\caption{TE - North.}
		\label{fig:transProbFromEDUtoPE_IIIIIIIIIIIIIIIIIIII}
	\end{subfigure}
	\begin{subfigure}[t]{0.24\textwidth}
		\centering
		\includegraphics[width=\linewidth]{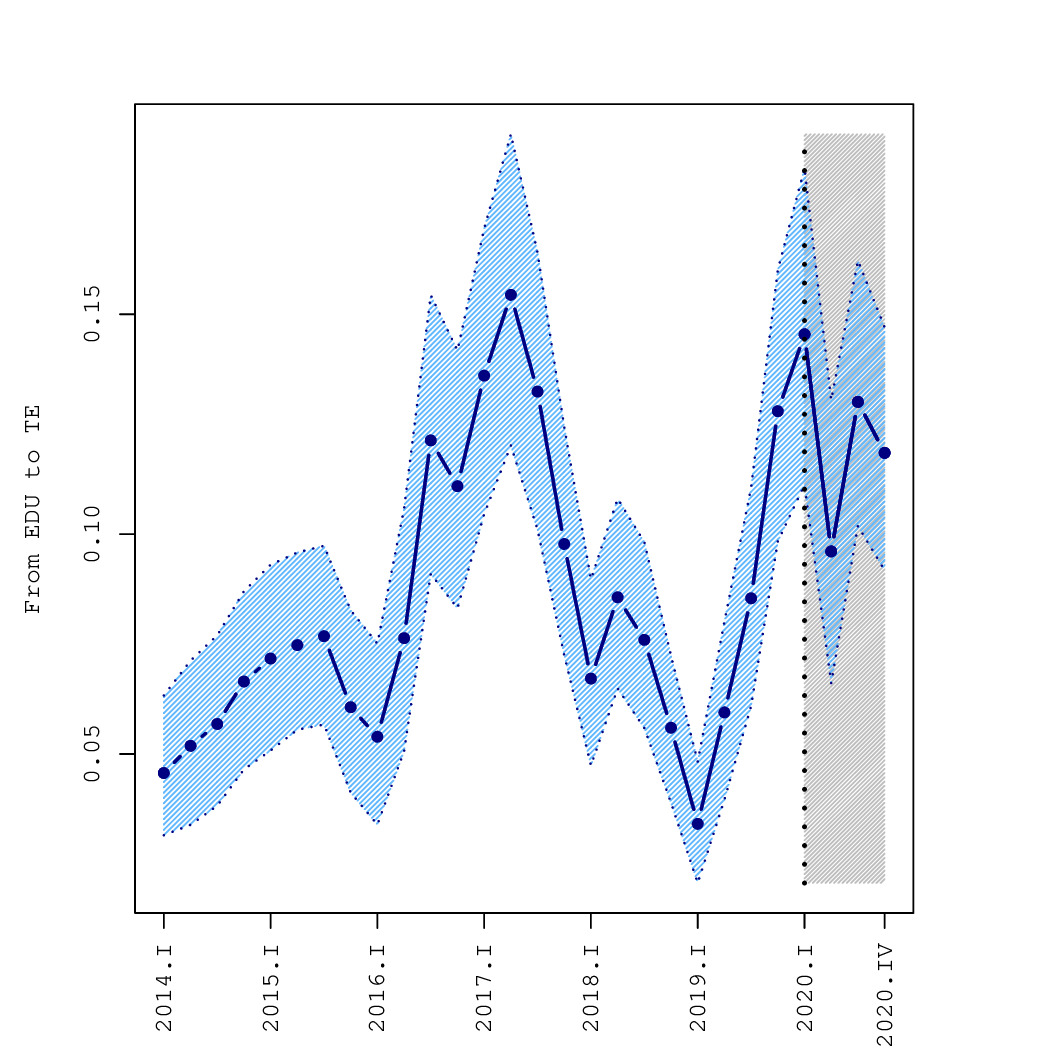}
		\caption{TE - South.}
		\label{fig:transProbFromEDUtoU_IIIIIIIIIIIIIIIII}
	\end{subfigure}
	\begin{subfigure}[t]{0.24\textwidth}
		\centering
		\includegraphics[width=\linewidth]{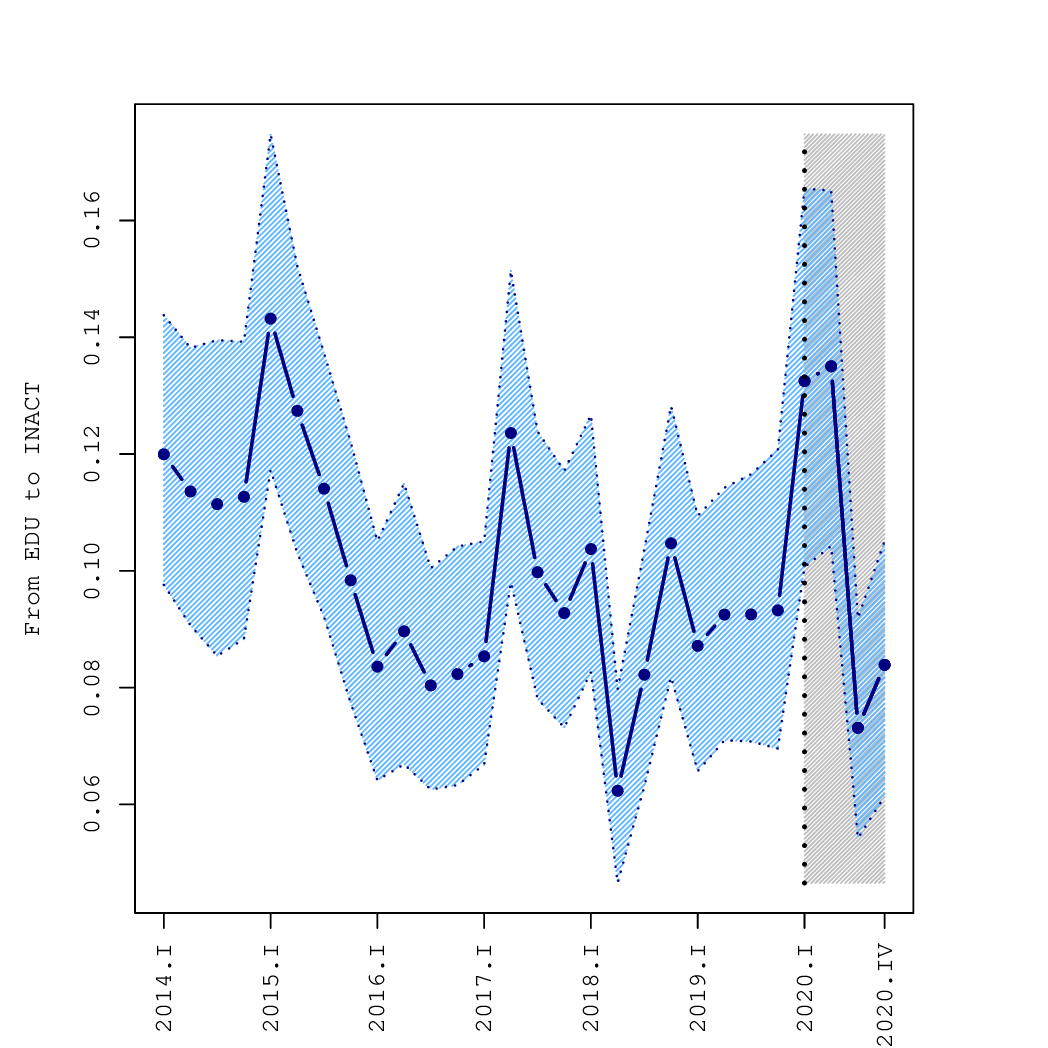}
		\caption{INACT - North.}
		\label{fig:transProbFromEDUtoPE_IIIIIIIIIIIIIIIIIIIII}
	\end{subfigure}
	\begin{subfigure}[t]{0.24\textwidth}
		\centering
		\includegraphics[width=\linewidth]{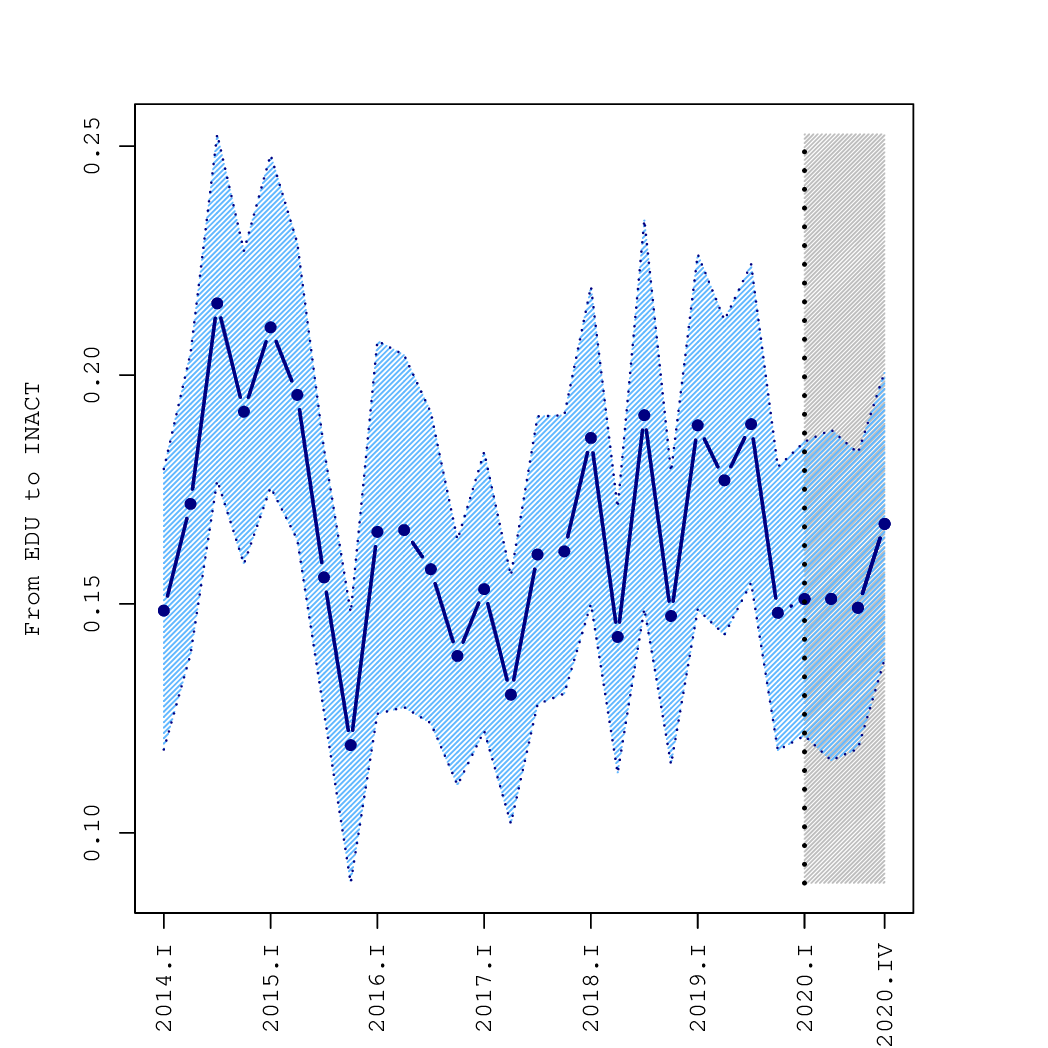}
		\caption{INACT - South.}
		\label{fig:transProbFromEDUtoU_IIIIIIIIIIIIIIIIII}
	\end{subfigure}
	\begin{subfigure}[t]{0.24\textwidth}
		\centering
		\includegraphics[width=\linewidth]{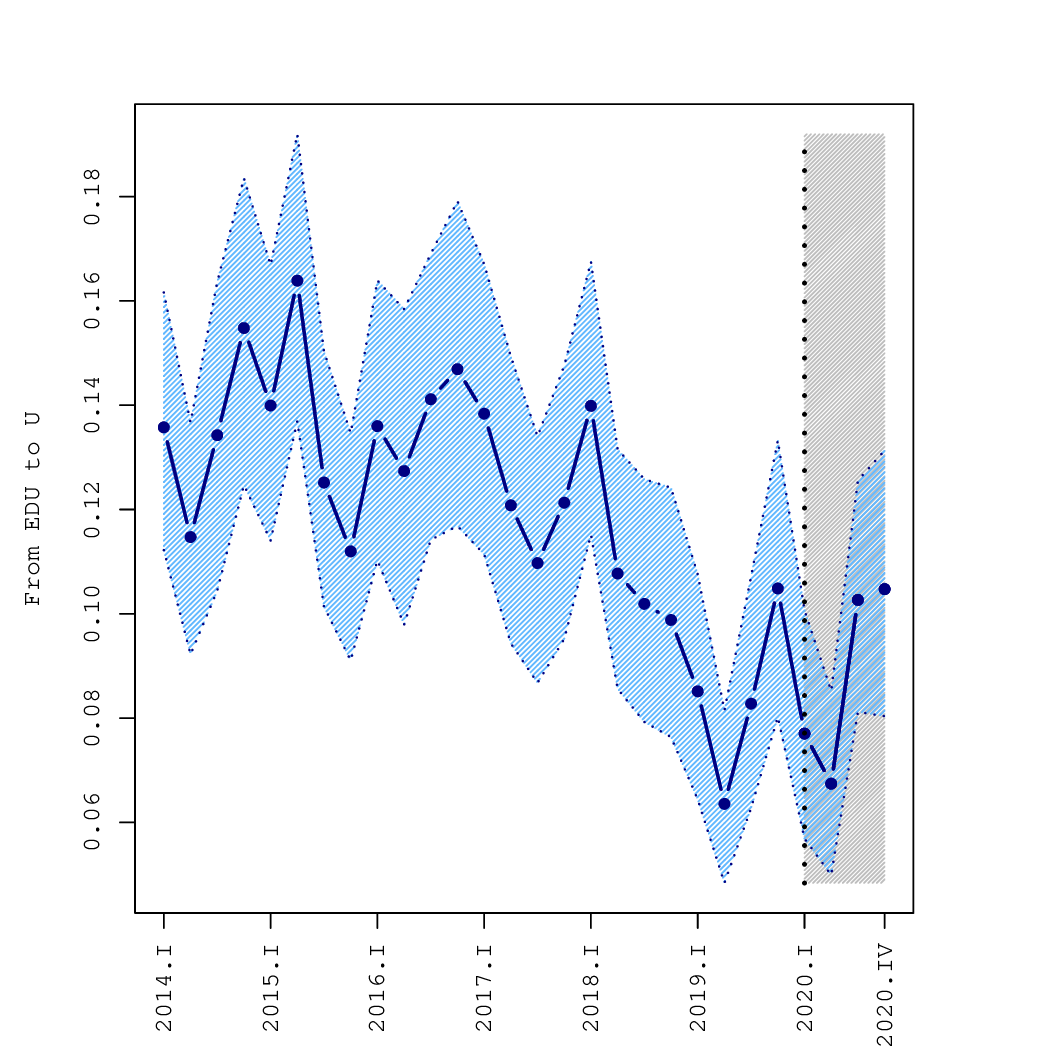}
		\caption{U - North.}
		\label{fig:transProbFromEDUtoPE_IIIIIIIIIIIIIIIIIIIIII}
	\end{subfigure}
	\begin{subfigure}[t]{0.24\textwidth}
		\centering
		\includegraphics[width=\linewidth]{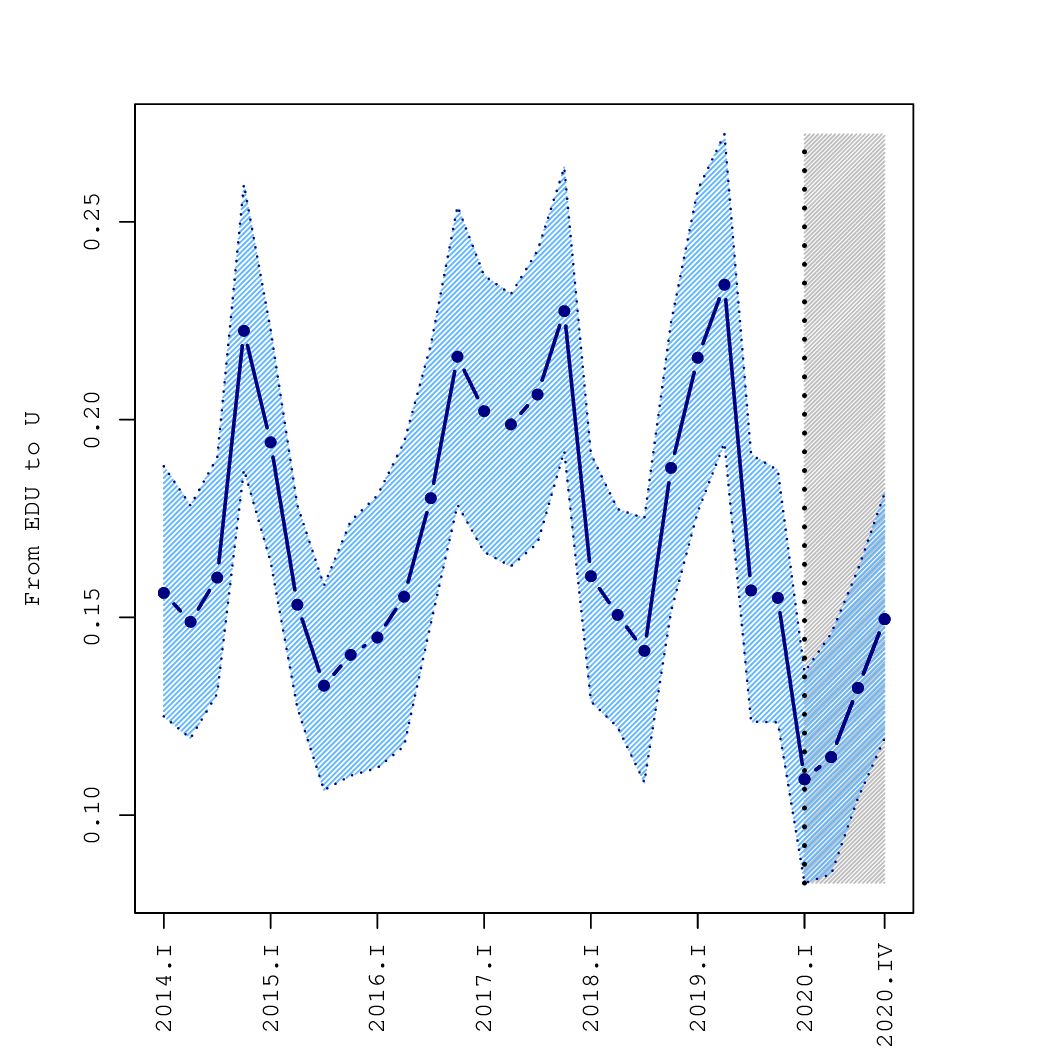}
		\caption{U  - South.}
		\label{fig:transProbFromEDUtoU_IIIIIIIIIIIIIIIIIII}
	\end{subfigure}
	\vspace{0.2cm}
	\caption*{\scriptsize{\textit{Note}: Confidence intervals at 90\% are computed using 1000 bootstraps. The gray area identifies the COVID period. North includes regions in the North and the Center. \textit{Source}: LFS 3-month longitudinal data as provided by the Italian Institute of Statistics (ISTAT).}}
\end{figure}


\clearpage

\begin{figure}[!htbp]
	\caption{Annual transition probabilities of females aged 30-39 from  temporary employment, unemployment, inactive, and permanent employment to the inactive  state in the North and South of Italy.}
	\label{fig:transprob3039females}
	\vspace{0.1cm}
	\caption*{\scriptsize{\textbf{North}}.}
	\centering
	\begin{subfigure}[t]{0.24\textwidth}
		\centering
		\includegraphics[width=\linewidth]{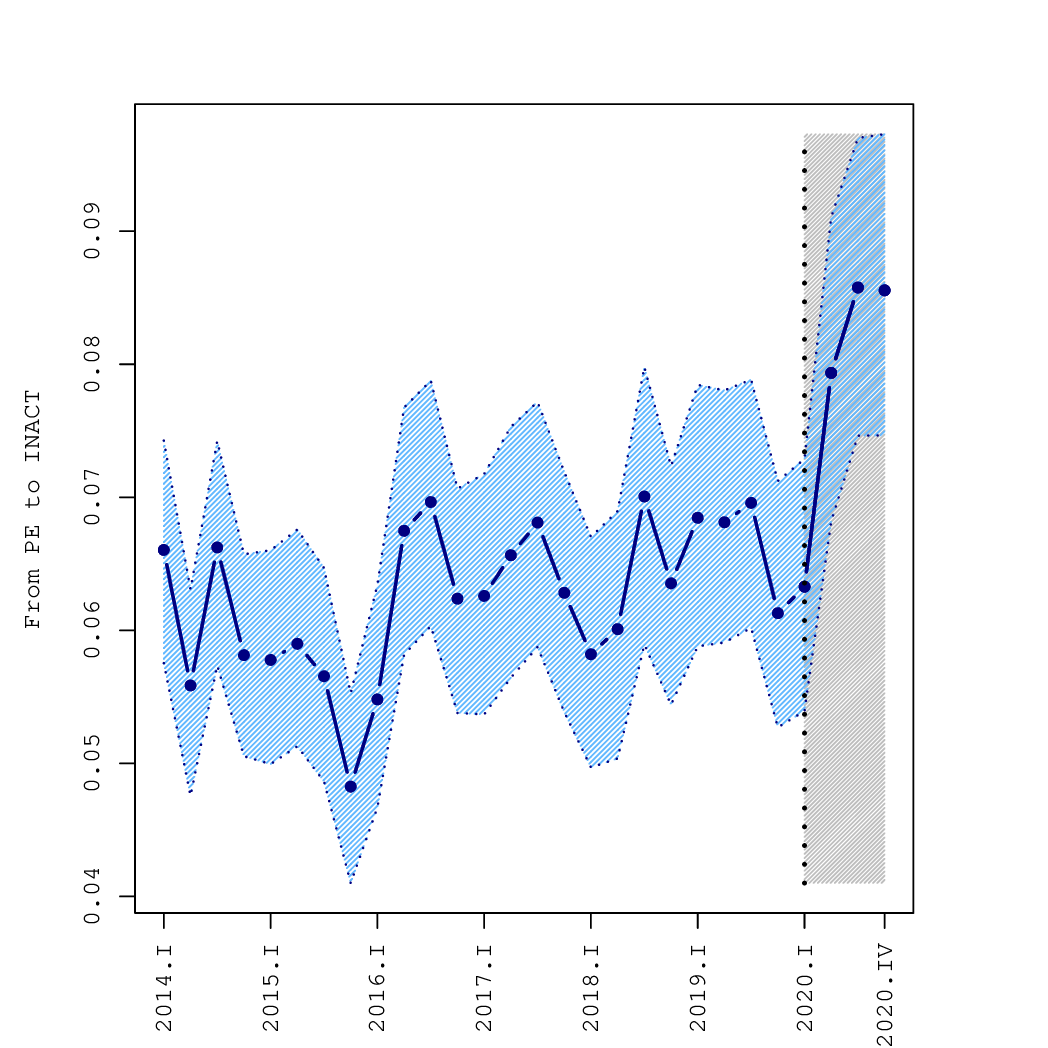}
		\caption{PE.}
		\label{fig:transProbFromEDUtoSE_IIIIIII}
		\vspace{0.2cm}
	\end{subfigure}
	\begin{subfigure}[t]{0.24\textwidth}
		\centering
		\includegraphics[width=\linewidth]{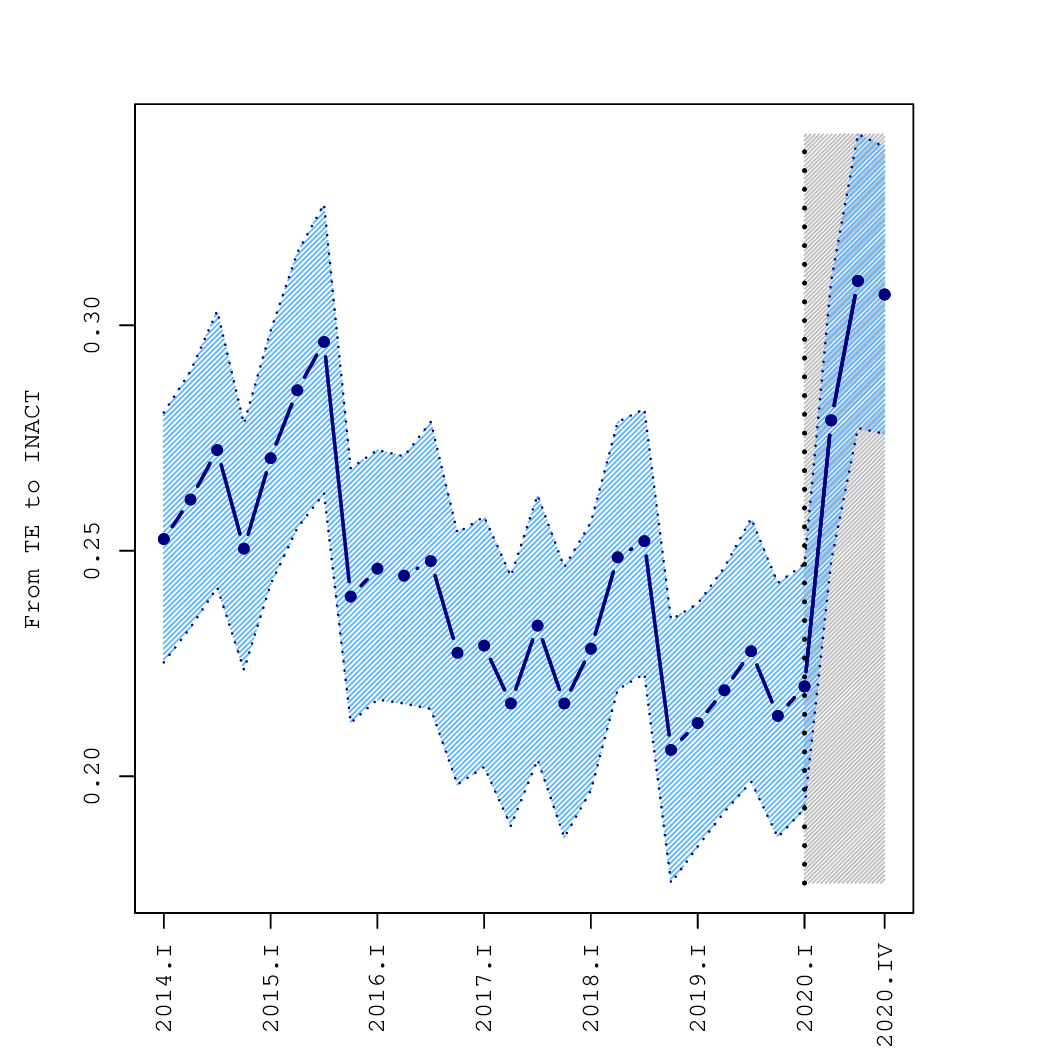}
		\caption{TE.}
		\label{fig:transProbFromEDUtoPE_IIIIIIIIIIIIIIIIIIIIIII}
	\end{subfigure}
	\begin{subfigure}[t]{0.24\textwidth}
		\centering
		\includegraphics[width=\linewidth]{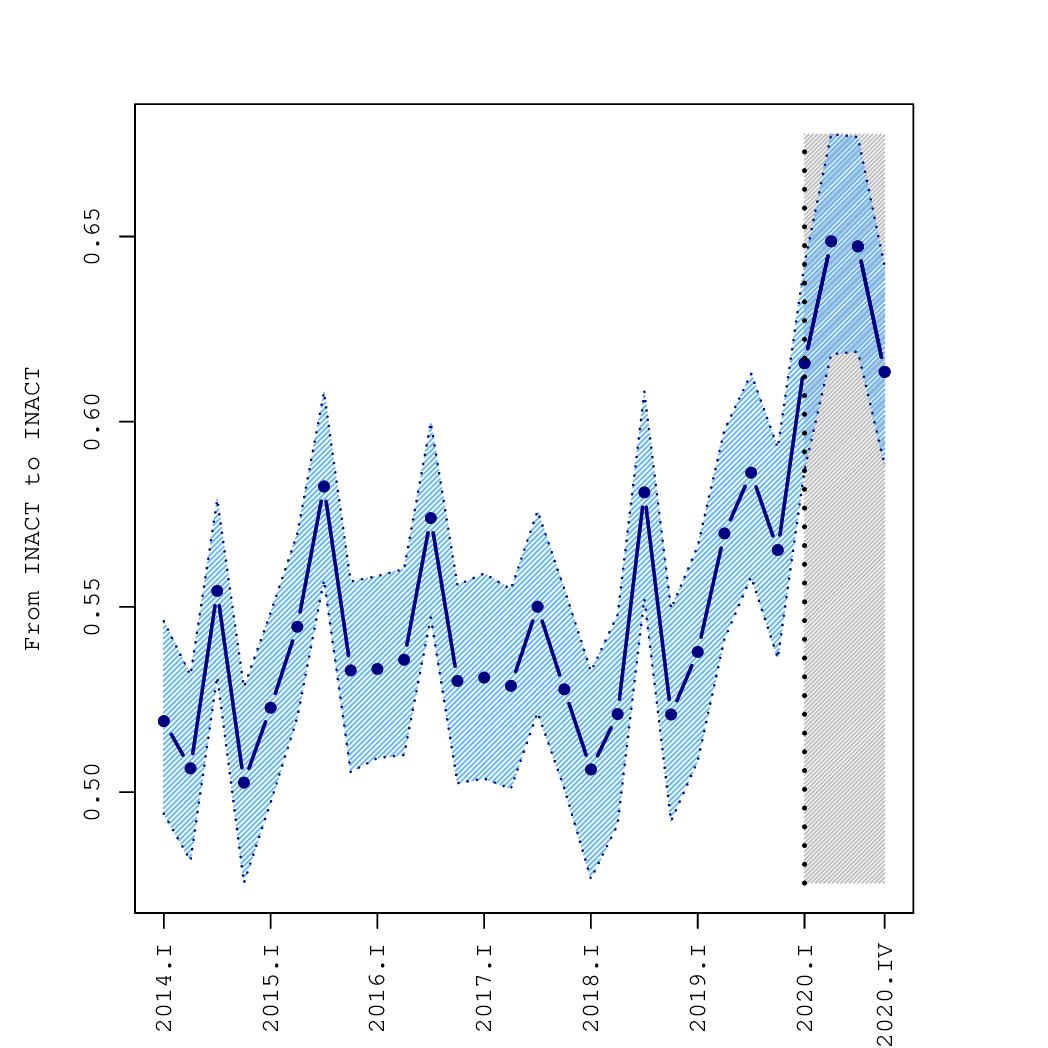}
		\caption{INACT.}
		\label{fig:transProbFromEDUtoPE_IIIIIIIIIIIIIIIIIIIIIIII}
	\end{subfigure}
	\begin{subfigure}[t]{0.24\textwidth}
		\centering
		\includegraphics[width=\linewidth]{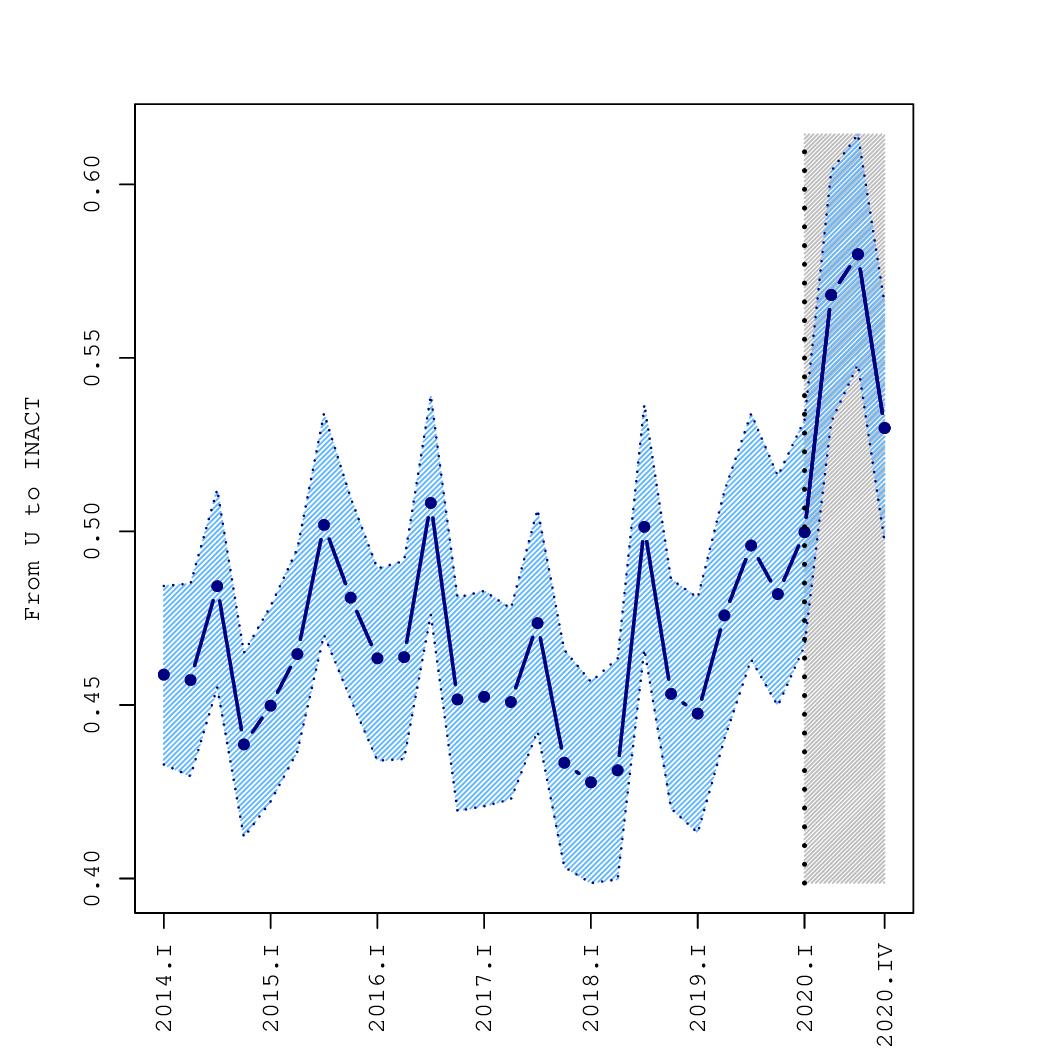}
		\caption{U.}
		\label{fig:transProbFromEDUtoPE_IIIIIIIIIIIIIIIIIIIIIIIII}
	\end{subfigure}
	\vspace{0.1cm}
	\caption*{\scriptsize{\textbf{South}}.}
	\begin{subfigure}[t]{0.24\textwidth}
		\centering
		\includegraphics[width=\linewidth]{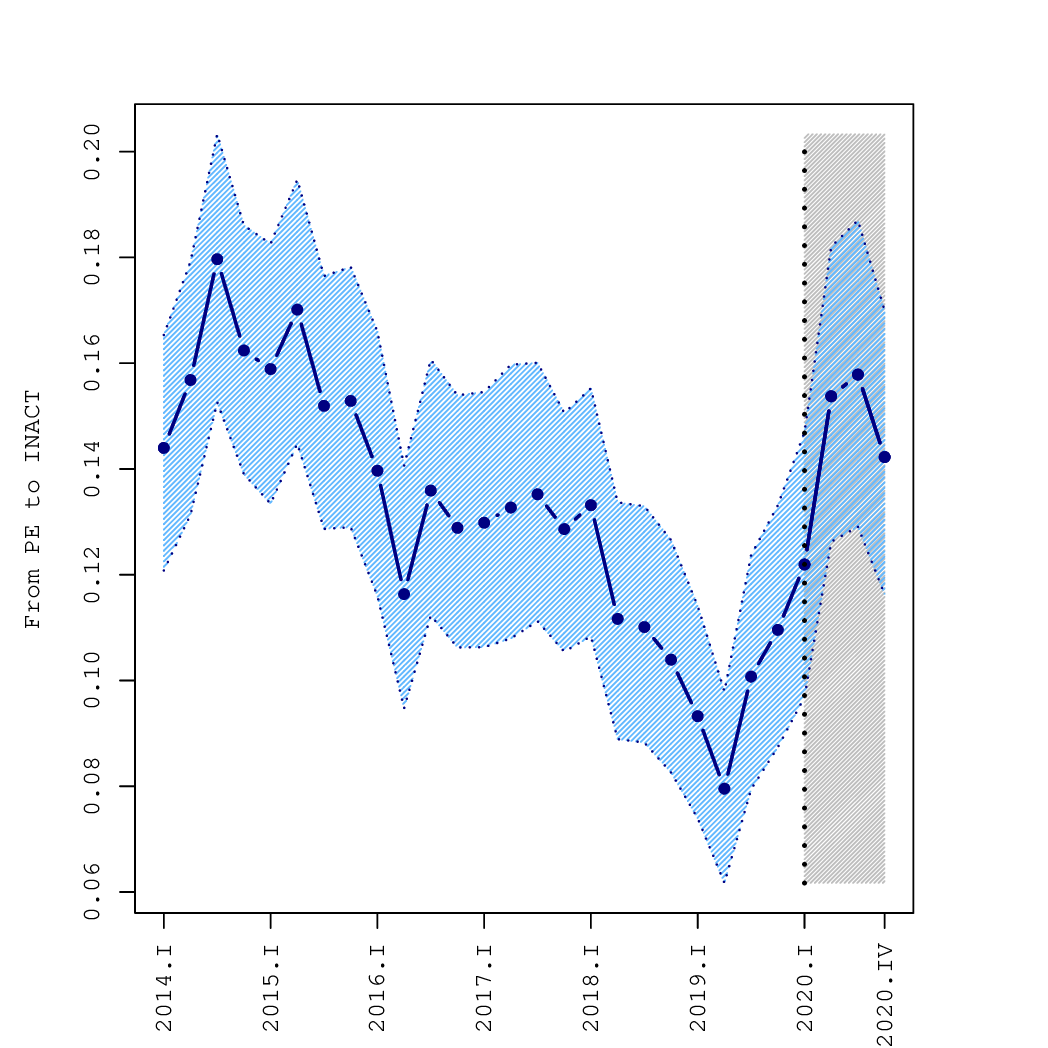}
		\caption{PE.}
		\label{fig:transProbFromEDUtoTE_IIIII}
	\end{subfigure}
	\begin{subfigure}[t]{0.24\textwidth}
		\centering
		\includegraphics[width=\linewidth]{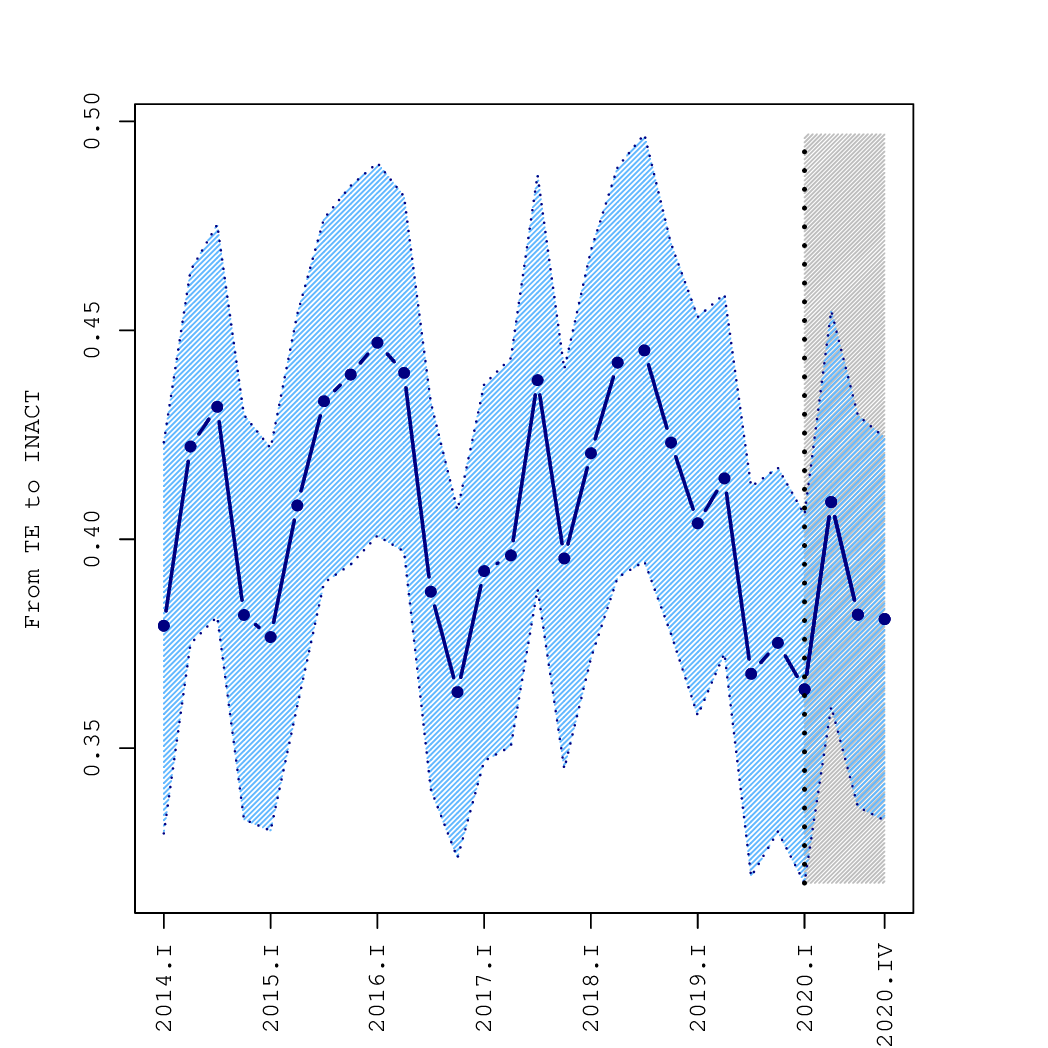}
		\caption{TE.}
		\label{fig:transProbFromEDUtoU_IIIIIIIIIIIIIIIIIIII}
	\end{subfigure}
	\begin{subfigure}[t]{0.24\textwidth}
		\centering
		\includegraphics[width=\linewidth]{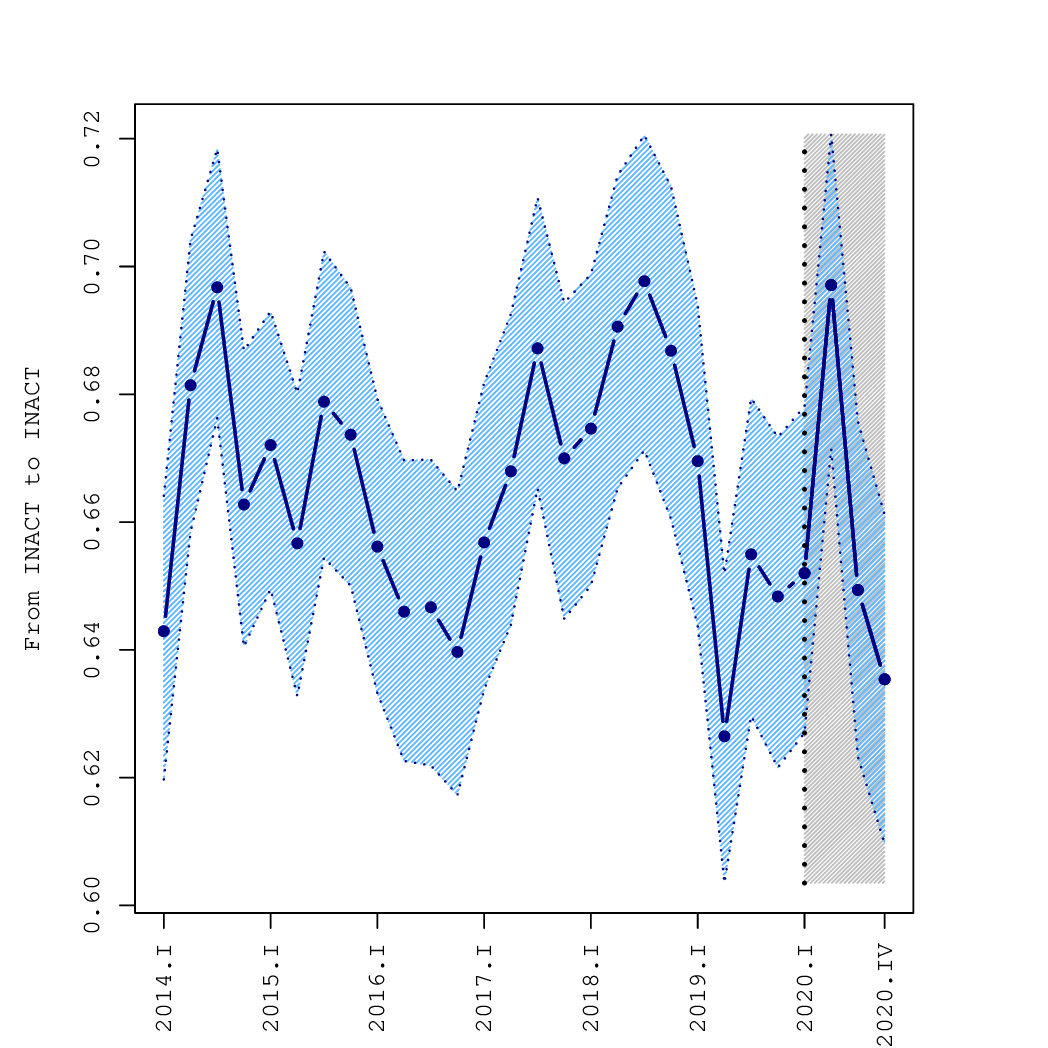}
		\caption{INACT.}
		\label{fig:transProbFromEDUtoU_IIIIIIIIIIIIIIIIIIIII}
	\end{subfigure}
	\begin{subfigure}[t]{0.24\textwidth}
		\centering
		\includegraphics[width=\linewidth]{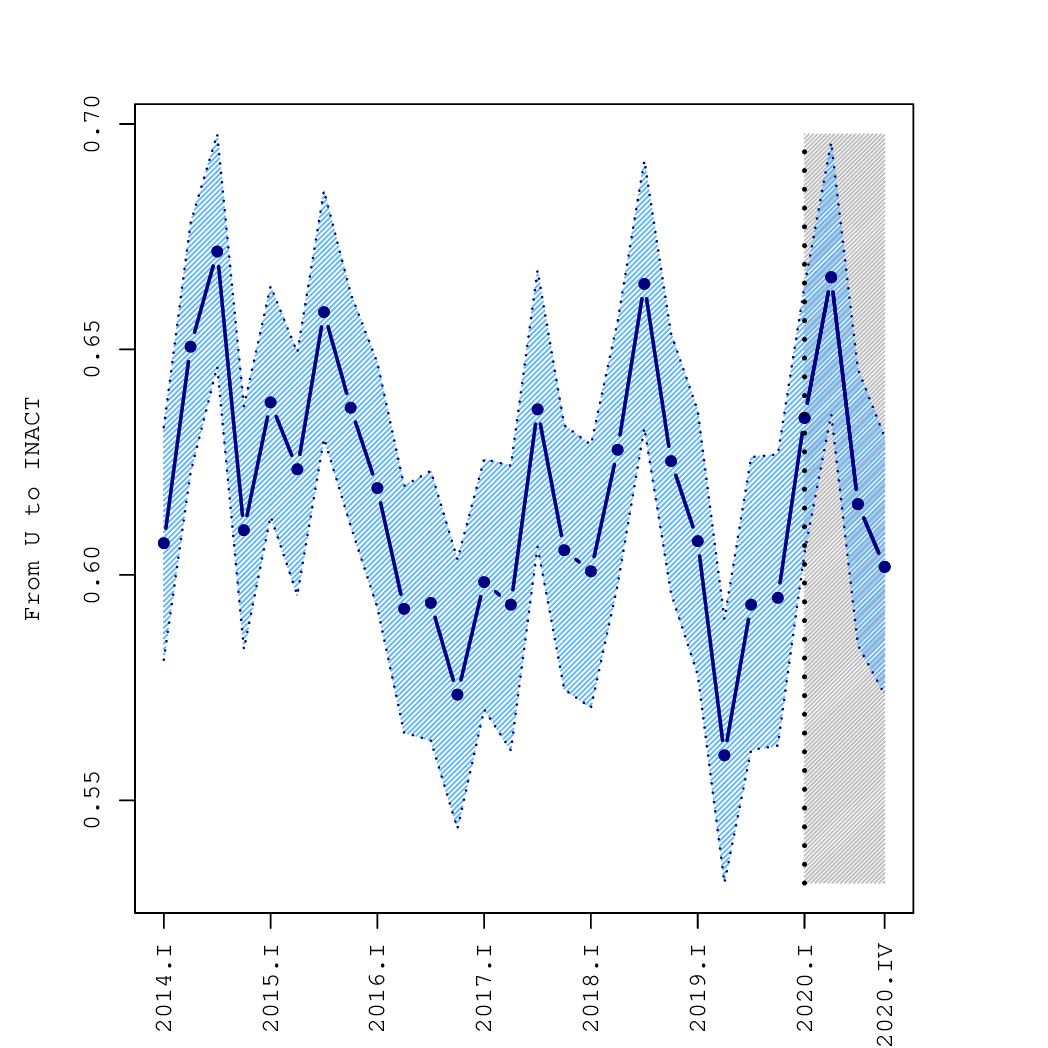}
		\caption{U.}
		\label{fig:transProbFromEDUtoU_IIIIIIIIIIIIIIIIIIIIII}
	\end{subfigure}
	\vspace{0.2cm}
	\caption*{\scriptsize{\textit{Note}: Confidence intervals at 90\% are computed using 1000 bootstraps. The gray area identifies the COVID period. North includes regions in the North and the Center. \textit{Source}: LFS 3-month longitudinal data as provided by the Italian Institute of Statistics (ISTAT).}}
\end{figure}


\begin{figure}[!htbp]
	\caption{Annual transition probabilities of males aged 30-39 from  temporary employment, unemployment, inactive, and permanent employment to the inactive  state in the North and South of Italy.}
	\label{fig:transprob3039males}
	\vspace{0.1cm}
	\caption*{\scriptsize{\textbf{North}}.}
	\centering
	\begin{subfigure}[t]{0.24\textwidth}
		\centering
		\includegraphics[width=\linewidth]{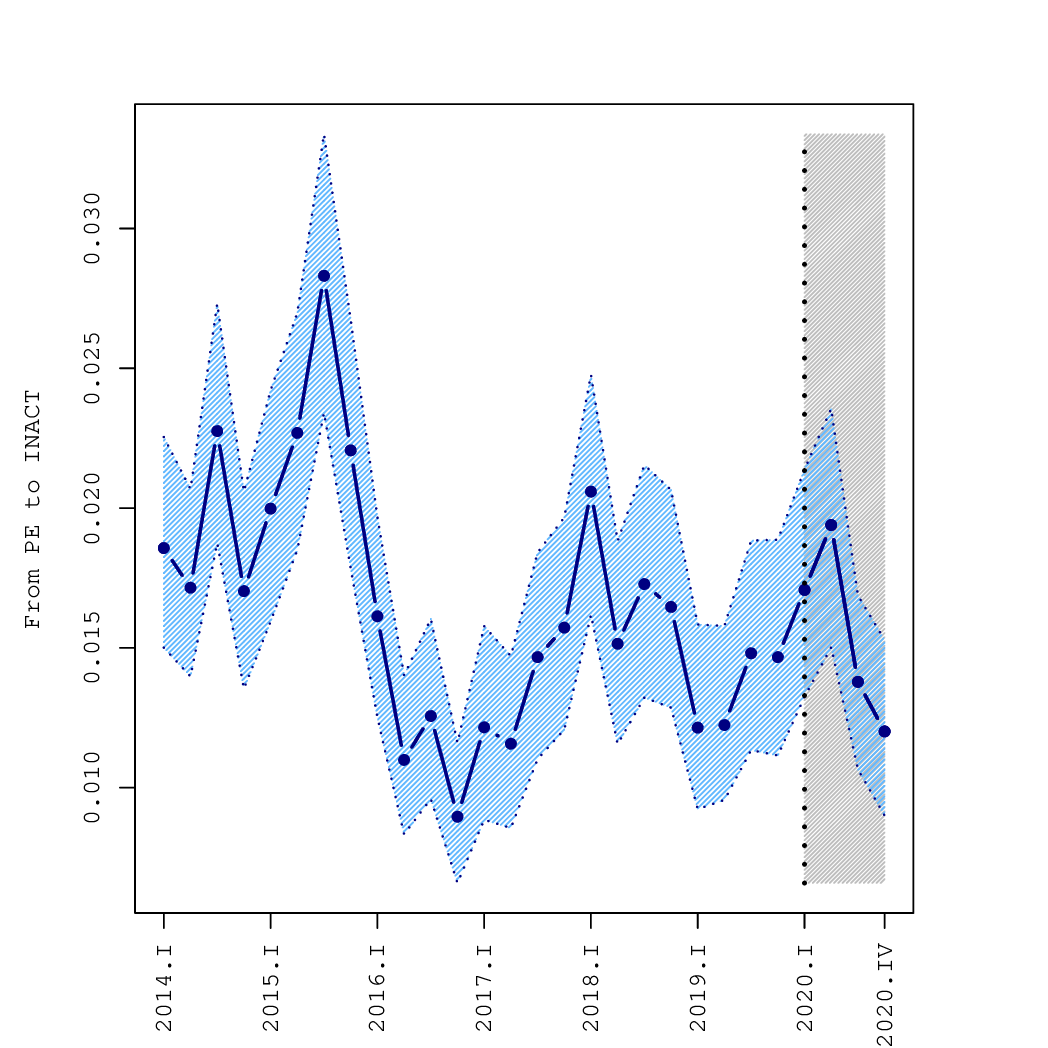}
		\caption{PE.}
		\label{fig:transProbFromEDUtoSE_IIIIIIII}
		\vspace{0.2cm}
	\end{subfigure}
	\begin{subfigure}[t]{0.24\textwidth}
		\centering
		\includegraphics[width=\linewidth]{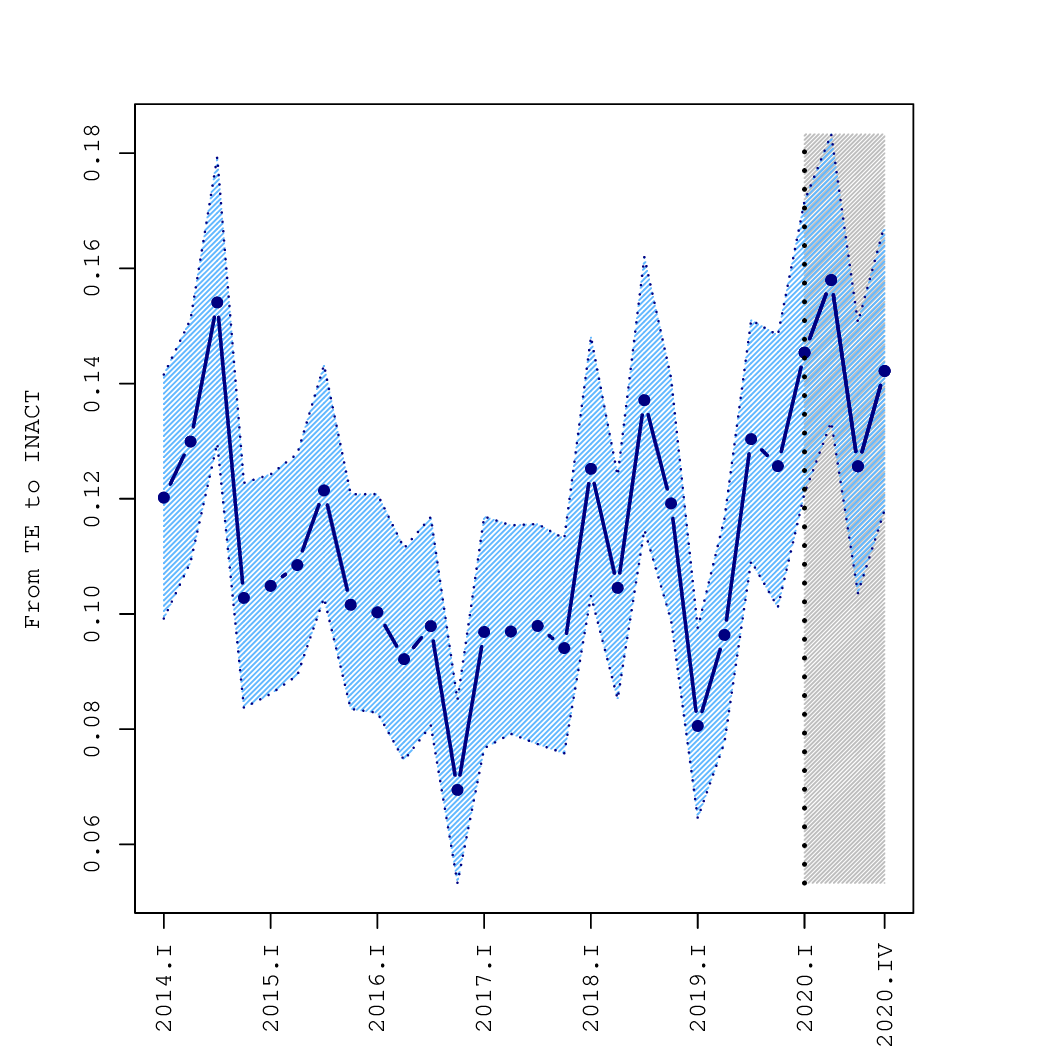}
		\caption{TE.}
		\label{fig:transProbFromEDUtoPE_IIIIIIIIIIIIIIIIIIIIIIIIII}
	\end{subfigure}
	\begin{subfigure}[t]{0.24\textwidth}
		\centering
		\includegraphics[width=\linewidth]{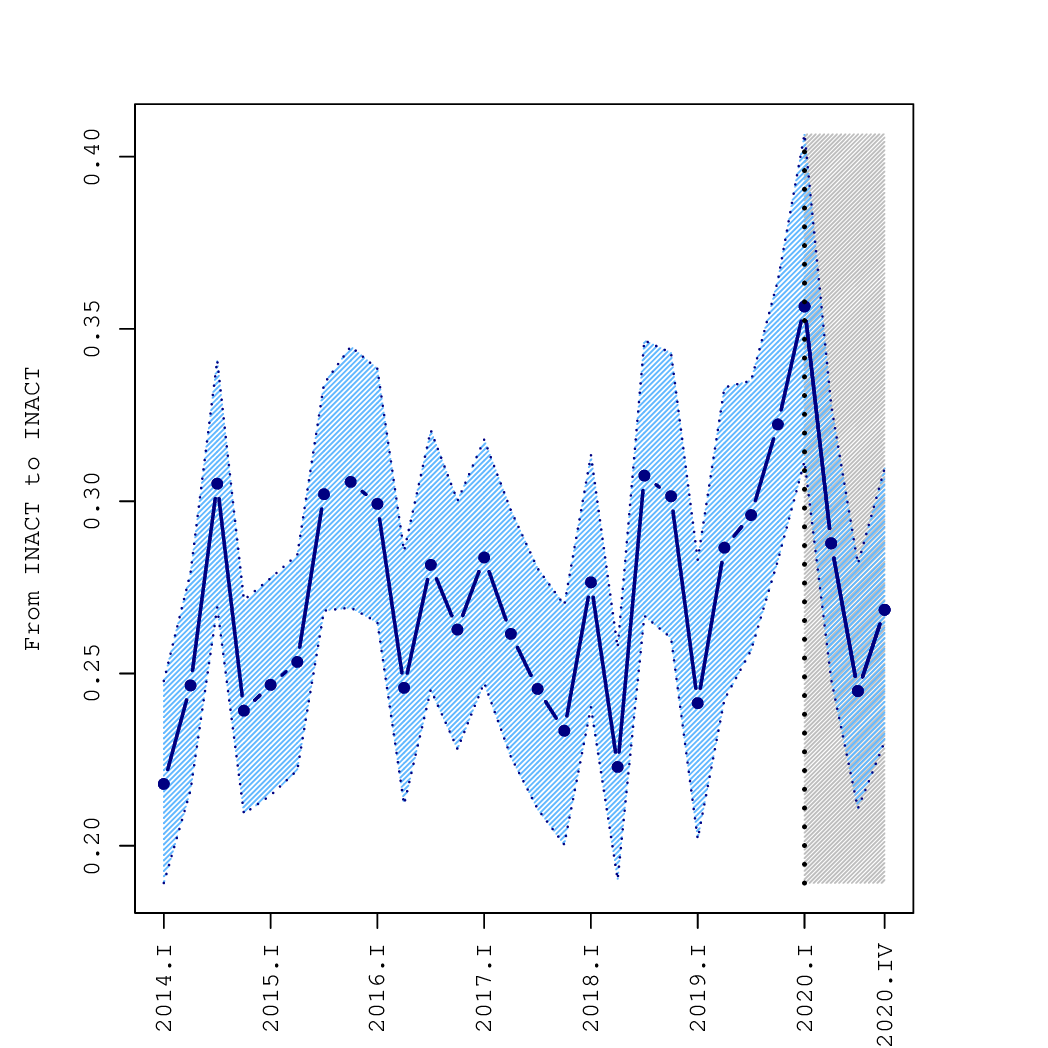}
		\caption{INACT.}
		\label{fig:transProbFromEDUtoPE_IIIIIIIIIIIIIIIIIIIIIIIIIII}
	\end{subfigure}
	\begin{subfigure}[t]{0.24\textwidth}
		\centering
		\includegraphics[width=\linewidth]{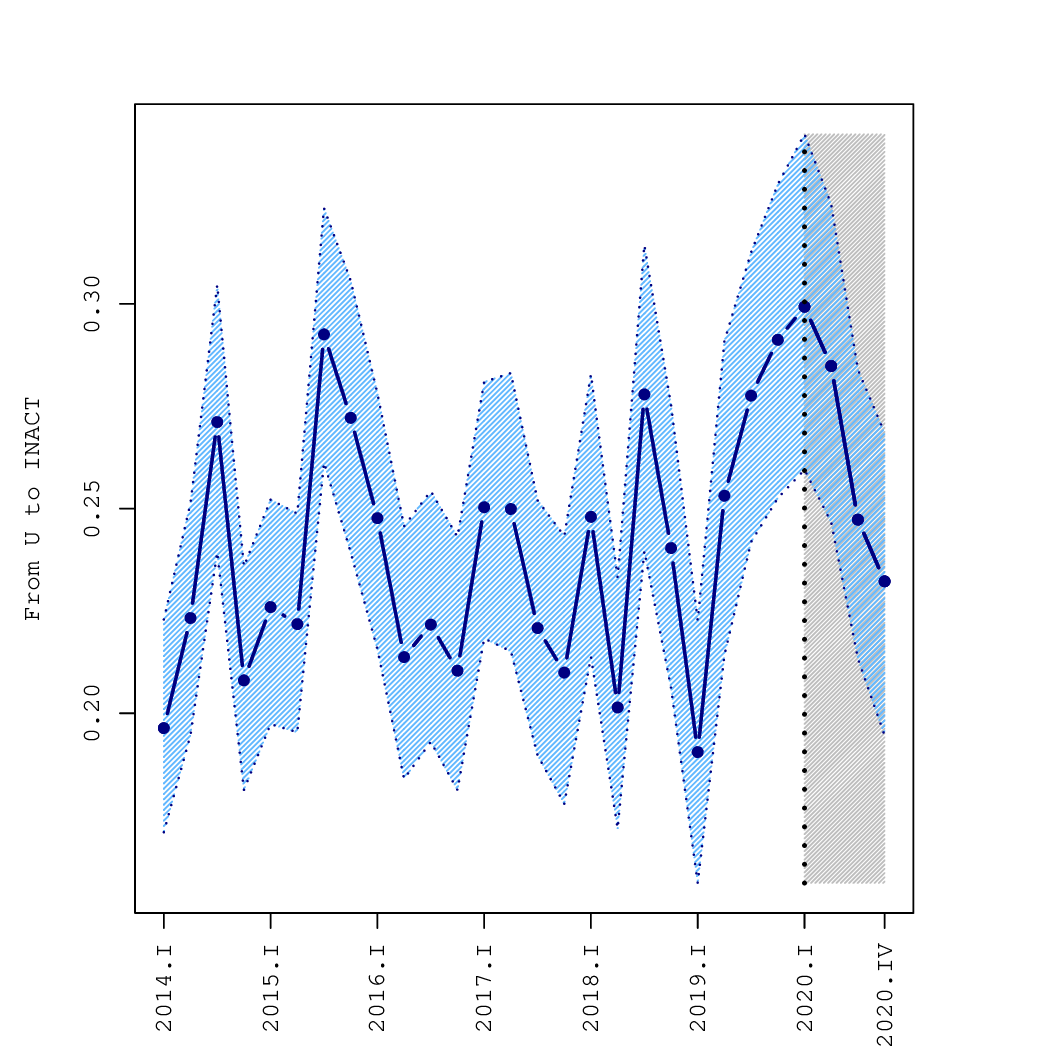}
		\caption{U.}
		\label{fig:transProbFromEDUtoPE_IIIIIIIIIIIIIIIIIIIIIIIIIIII}
	\end{subfigure}
	\vspace{0.1cm}
	\caption*{\scriptsize{\textbf{South}}.}
	\begin{subfigure}[t]{0.24\textwidth}
		\centering
		\includegraphics[width=\linewidth]{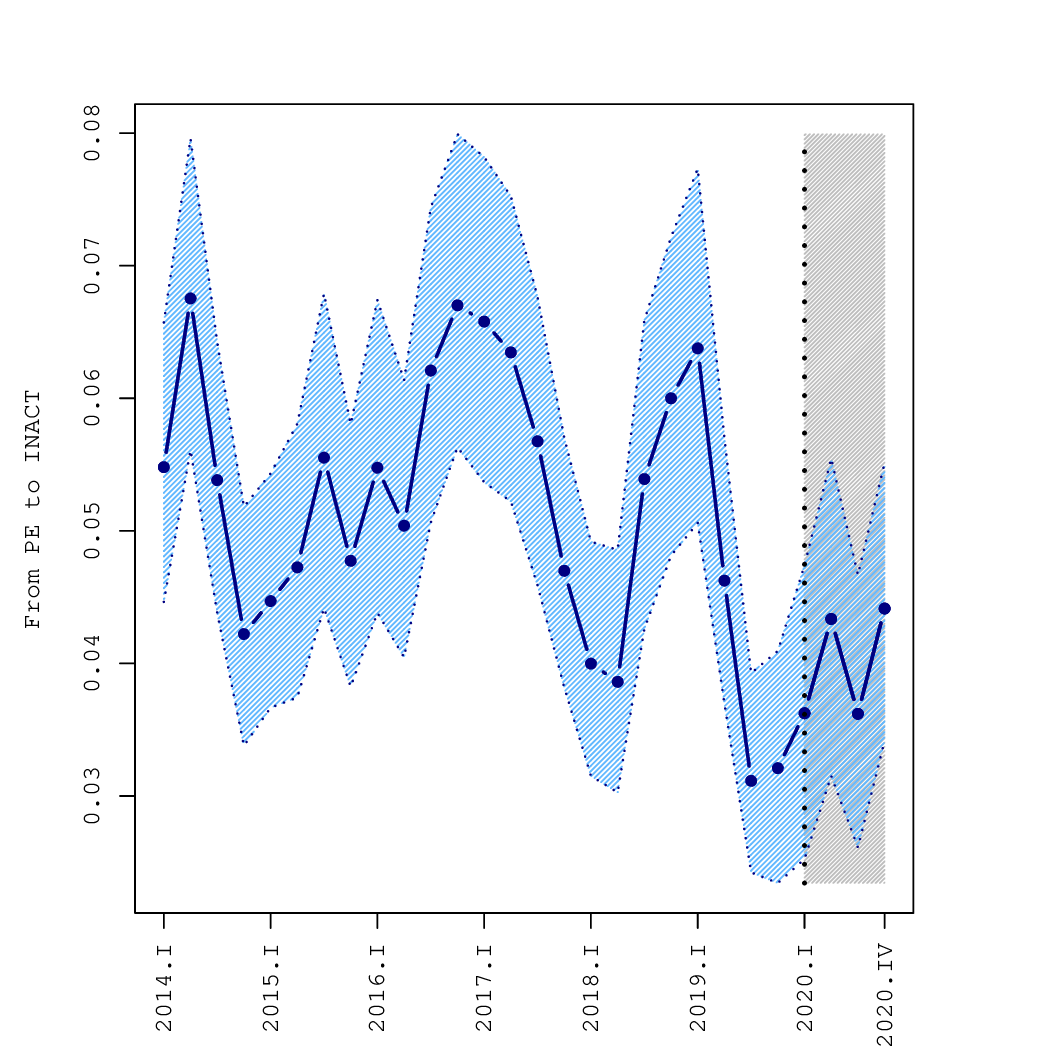}
		\caption{PE.}
		\label{fig:transProbFromEDUtoTE_IIIIIII}
	\end{subfigure}
	\begin{subfigure}[t]{0.24\textwidth}
		\centering
		\includegraphics[width=\linewidth]{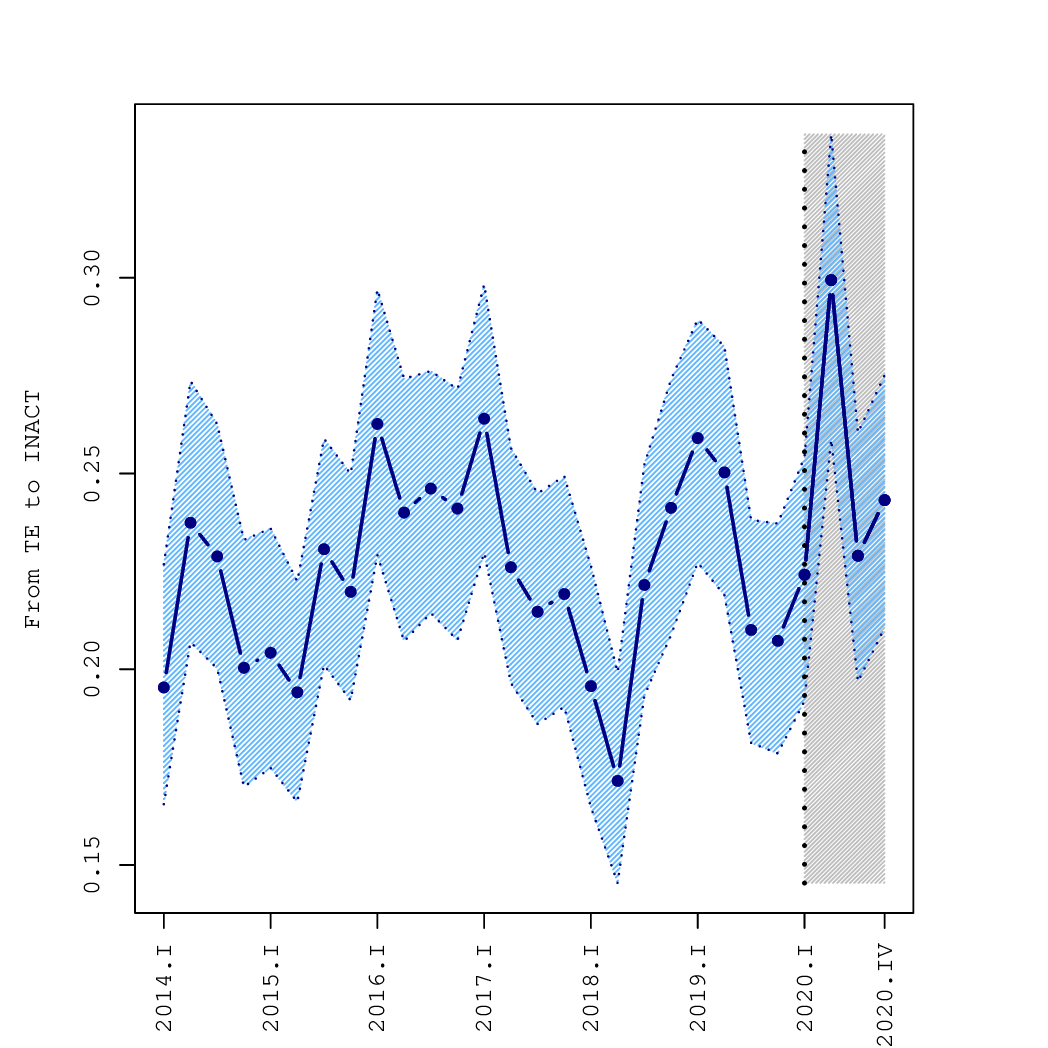}
		\caption{TE.}
		\label{fig:transProbFromEDUtoU_IIIIIIIIIIIIIIIIIIIIIII}
	\end{subfigure}
	\begin{subfigure}[t]{0.24\textwidth}
		\centering
		\includegraphics[width=\linewidth]{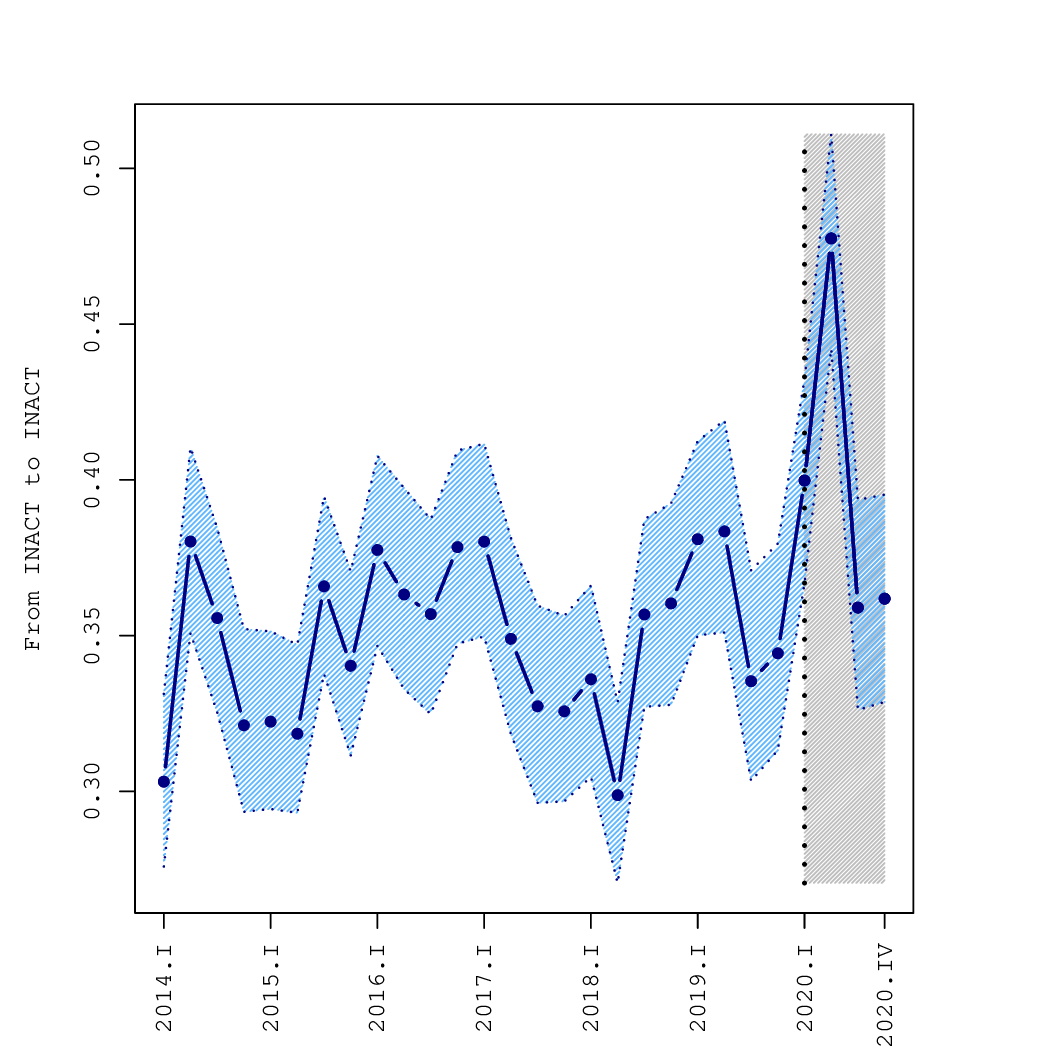}
		\caption{INACT.}
		\label{fig:transProbFromEDUtoU_IIIIIIIIIIIIIIIIIIIIIIII}
	\end{subfigure}
	\begin{subfigure}[t]{0.24\textwidth}
		\centering
		\includegraphics[width=\linewidth]{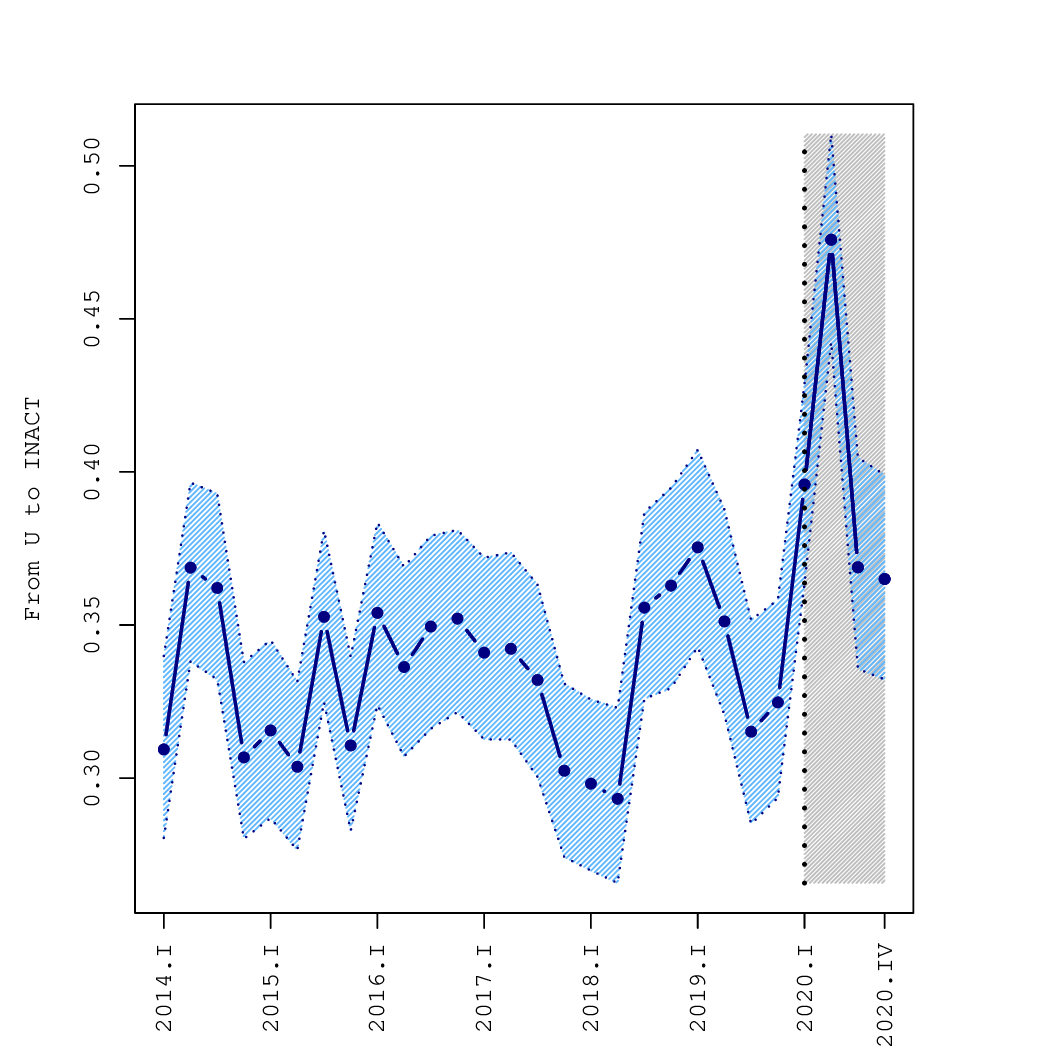}
		\caption{U.}
		\label{fig:transProbFromEDUtoU_IIIIIIIIIIIIIIIIIIIIIIIII}
	\end{subfigure}
	\vspace{0.2cm}
	\caption*{\scriptsize{\textit{Note}: Confidence intervals at 90\% are computed using 1000 bootstraps. The gray area identifies the COVID period. North includes regions in the North and the Center. \textit{Source}: LFS 3-month longitudinal data as provided by the Italian Institute of Statistics (ISTAT).}}
\end{figure}

\clearpage

\begin{figure}[!htbp]
	\caption{Annual transition probabilities of individuals aged 40-49 from  temporary employment, unemployment, inactive, and permanent employment to the inactive  state in the North and South of Italy.}
	\label{fig:transprob4049females}
	\caption*{\scriptsize{\textbf{Females}}.}
	\centering
	\begin{subfigure}[t]{0.24\textwidth}
		\centering
		\includegraphics[width=\linewidth]{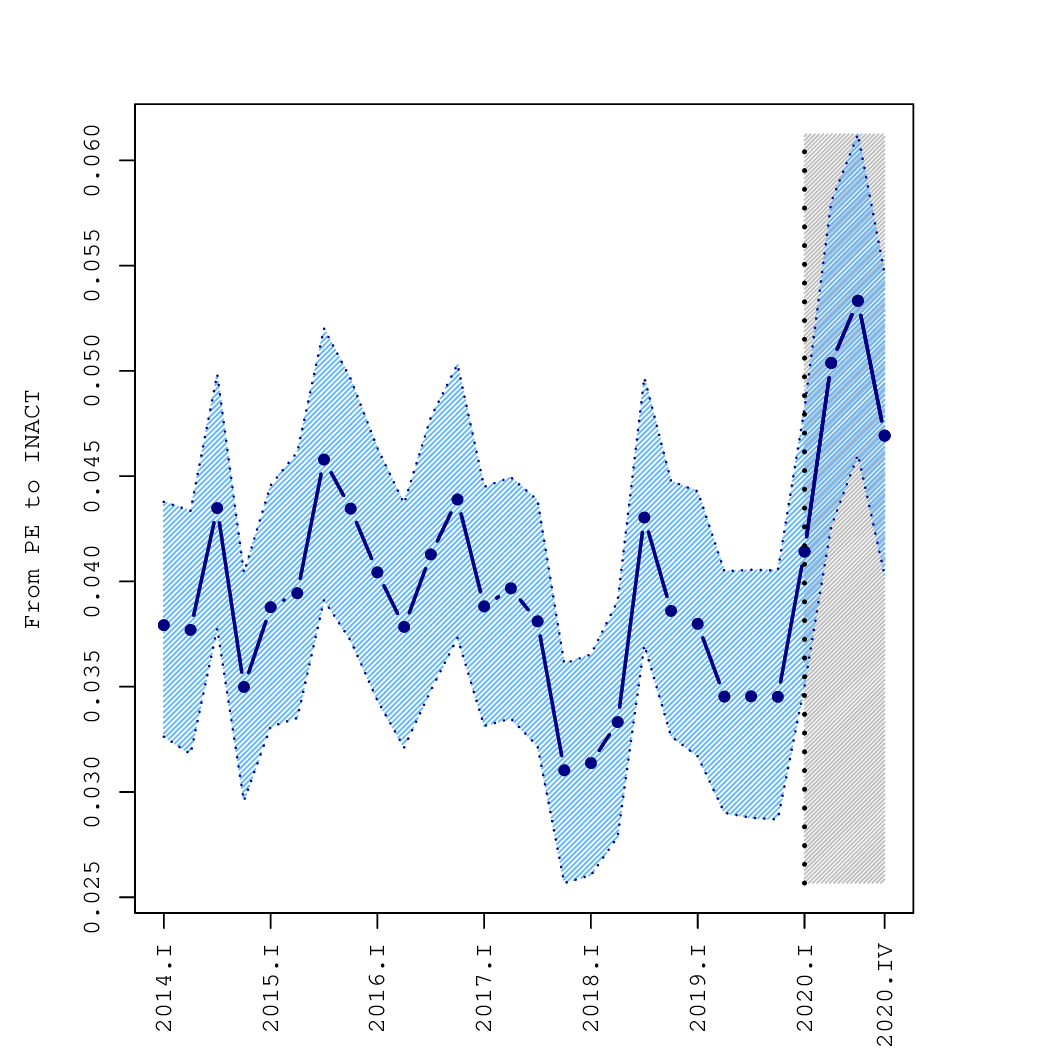}
		\caption{PE - North.}
		\label{fig:transProbFromEDUtoSE_IIIIIIIII}
		\vspace{0.2cm}
	\end{subfigure}
	\begin{subfigure}[t]{0.24\textwidth}
		\centering
		\includegraphics[width=\linewidth]{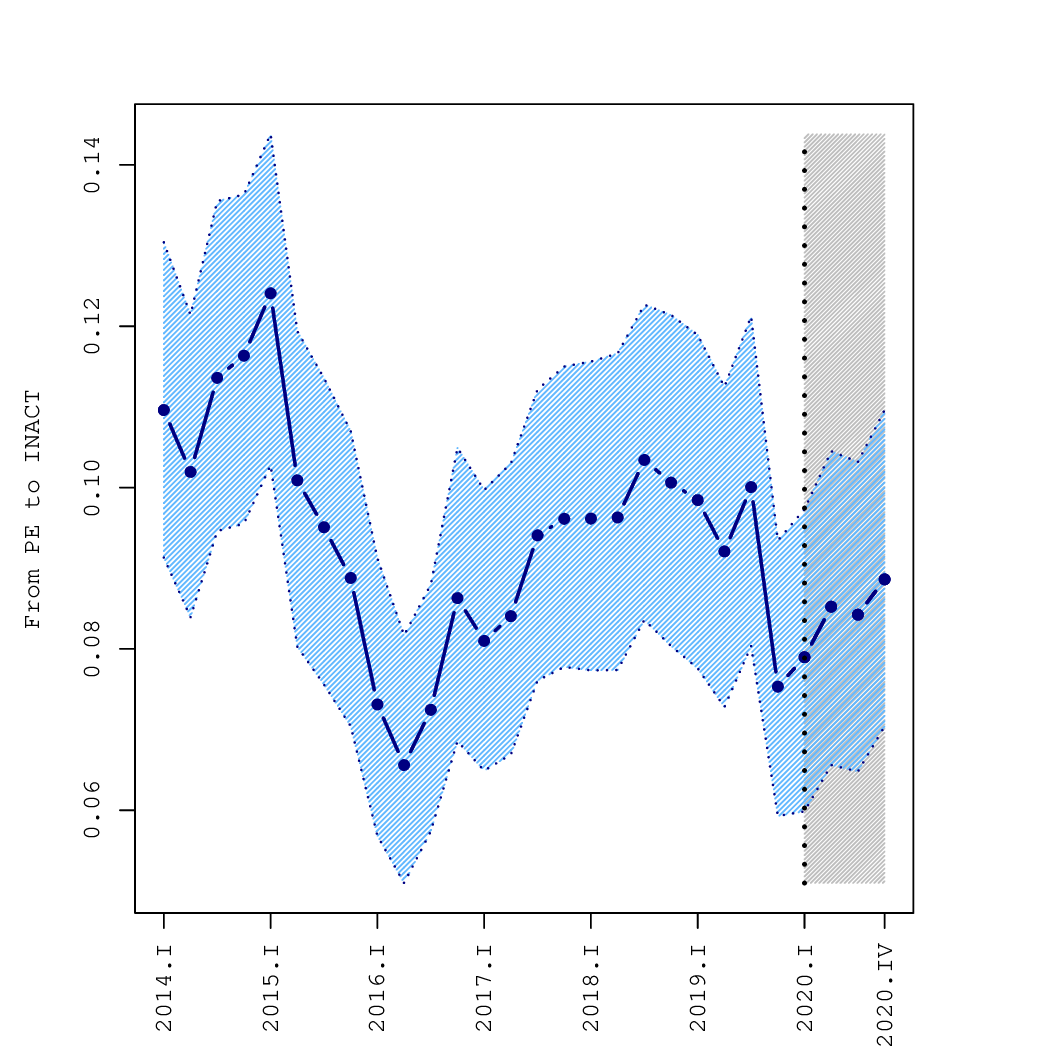}
		\caption{PE - South.}
		\label{fig:transProbFromEDUtoTE_IIIIIIII}
	\end{subfigure}
	\begin{subfigure}[t]{0.24\textwidth}
		\centering
		\includegraphics[width=\linewidth]{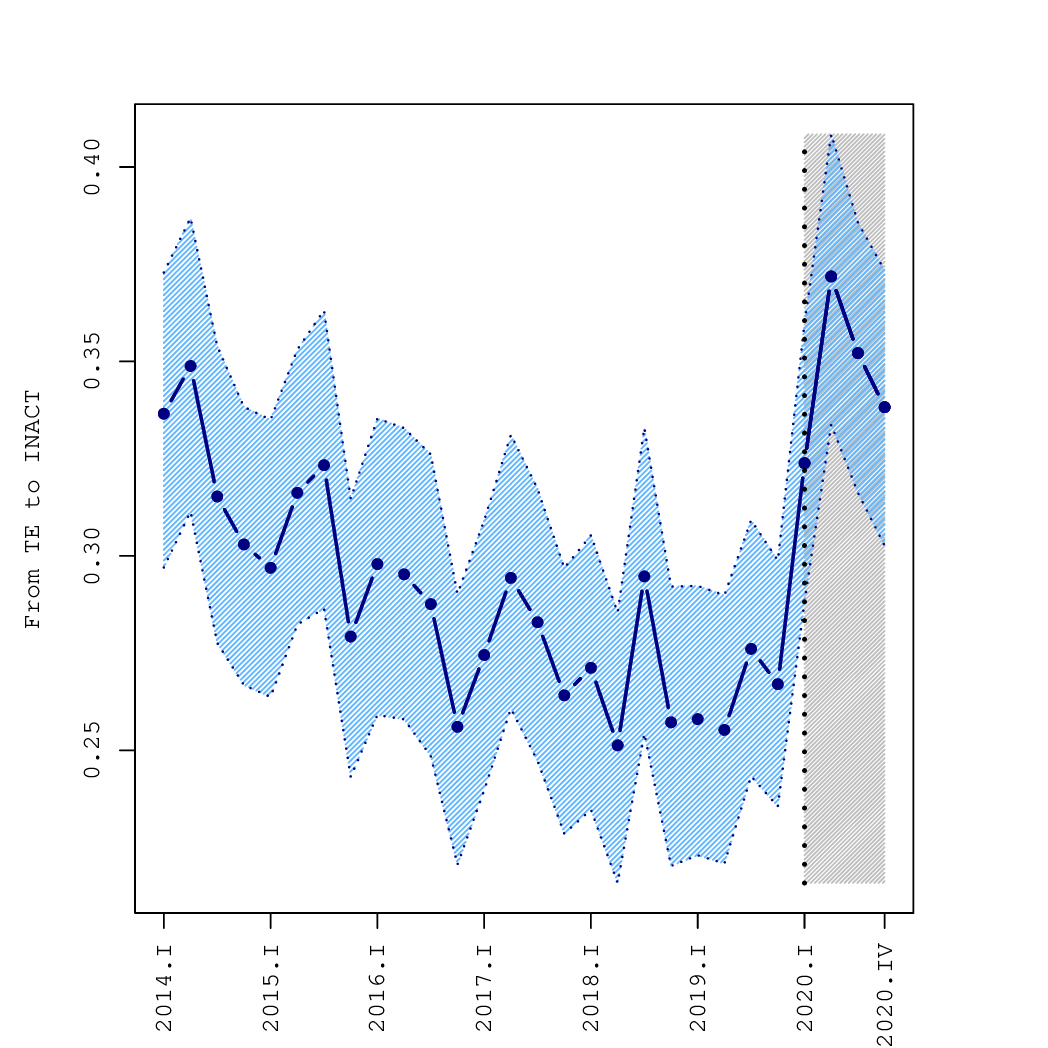}
		\caption{TE - North.}
		\label{fig:transProbFromEDUtoPE_IIIIIIIIIIIIIIIIIIIIIIIIIIIII}
	\end{subfigure}
	\begin{subfigure}[t]{0.24\textwidth}
		\centering
		\includegraphics[width=\linewidth]{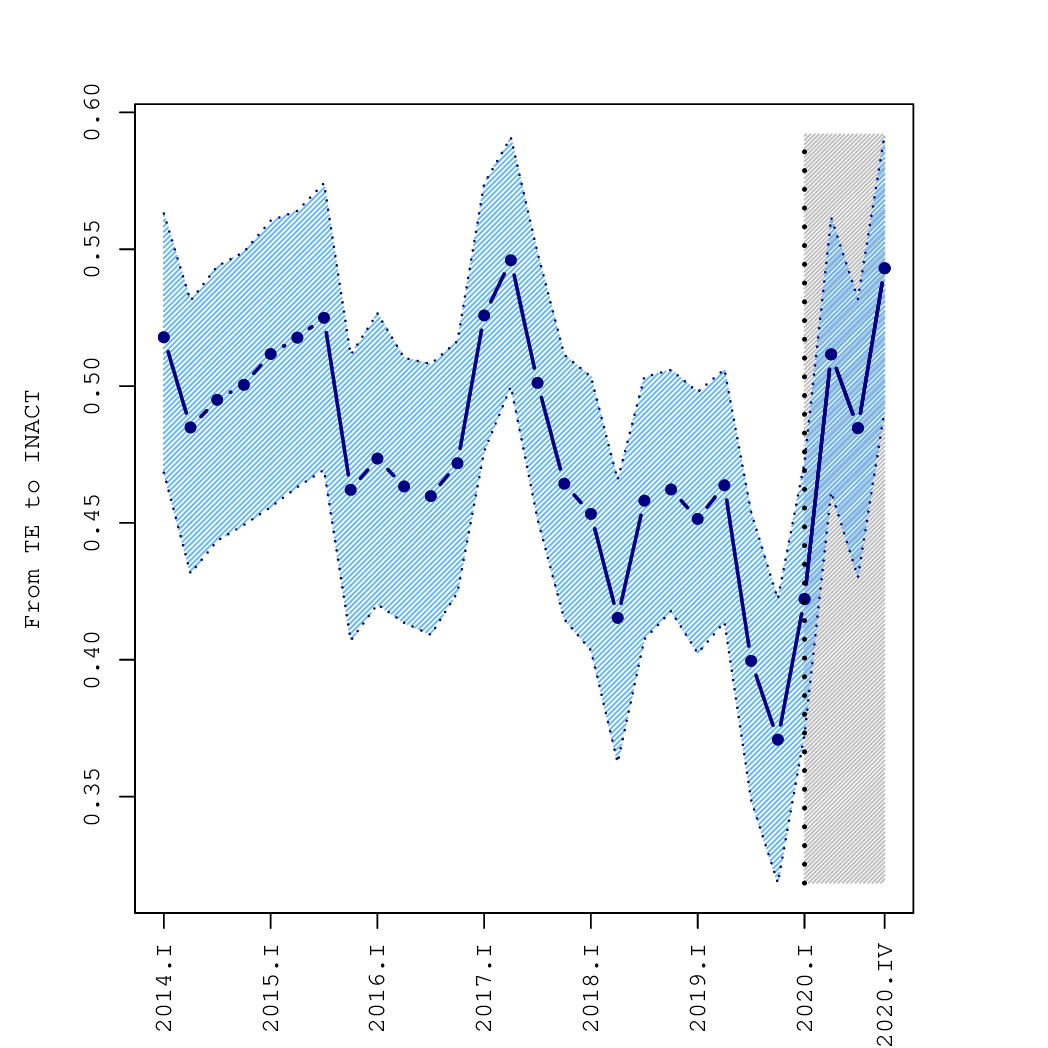}
		\caption{TE - South.}
		\label{fig:transProbFromEDUtoU_IIIIIIIIIIIIIIIIIIIIIIIIII}
	\end{subfigure}
	\begin{subfigure}[t]{0.24\textwidth}
		\centering
		\includegraphics[width=\linewidth]{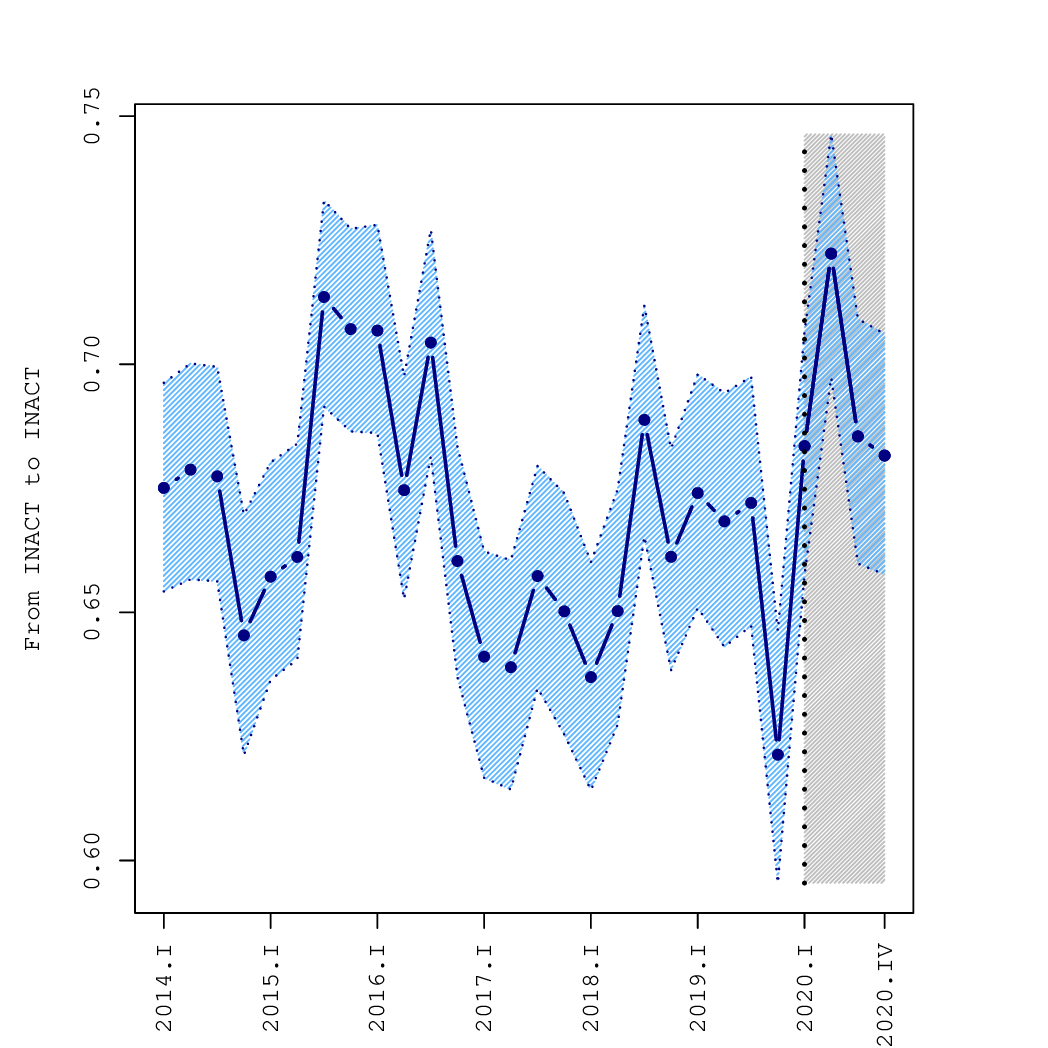}
		\caption{INACT - North.}
		\label{fig:transProbFromEDUtoPE_IIIIIIIIIIIIIIIIIIIIIIIIIIIIII}
	\end{subfigure}
	\begin{subfigure}[t]{0.24\textwidth}
		\centering
		\includegraphics[width=\linewidth]{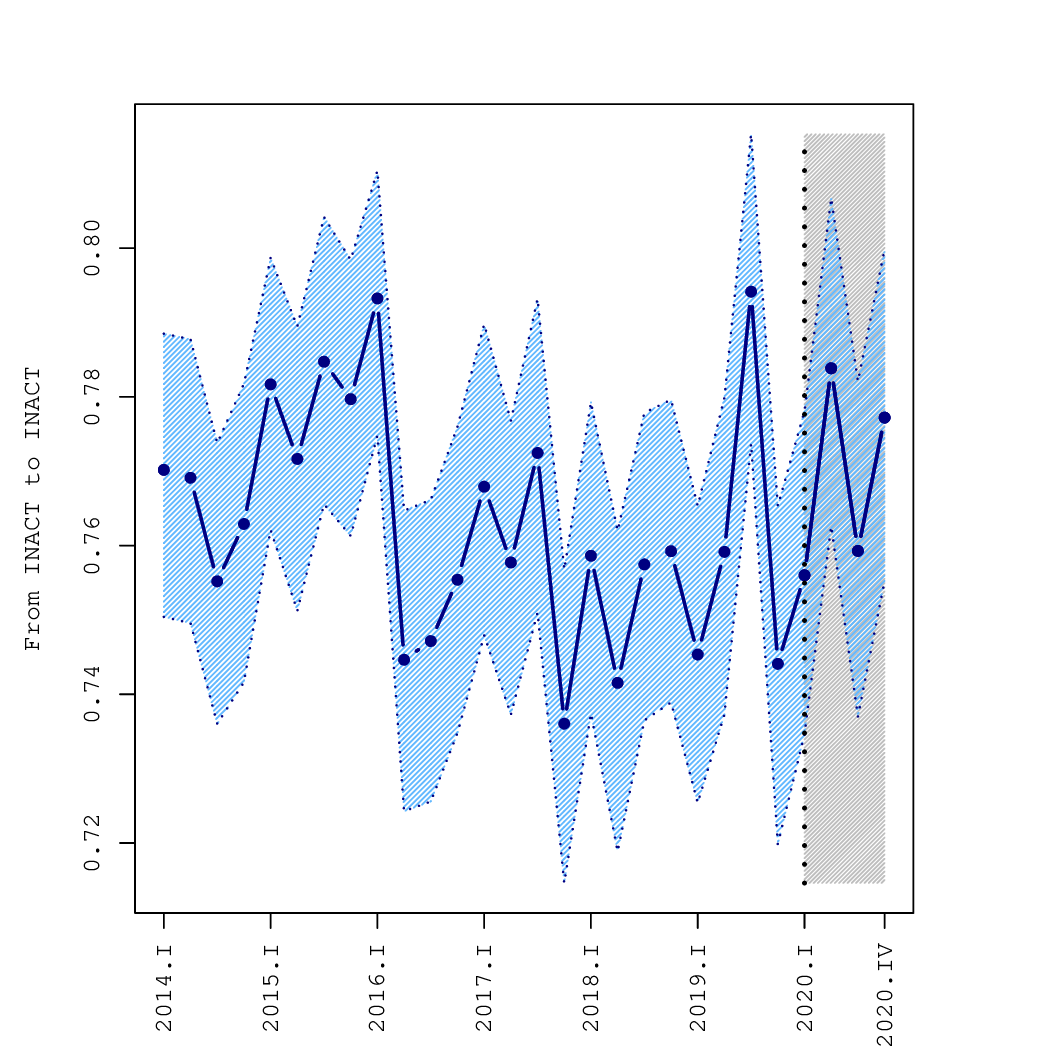}
		\caption{INACT - South.}
		\label{fig:transProbFromEDUtoU_IIIIIIIIIIIIIIIIIIIIIIIIIII}
	\end{subfigure}
	\begin{subfigure}[t]{0.24\textwidth}
		\centering
		\includegraphics[width=\linewidth]{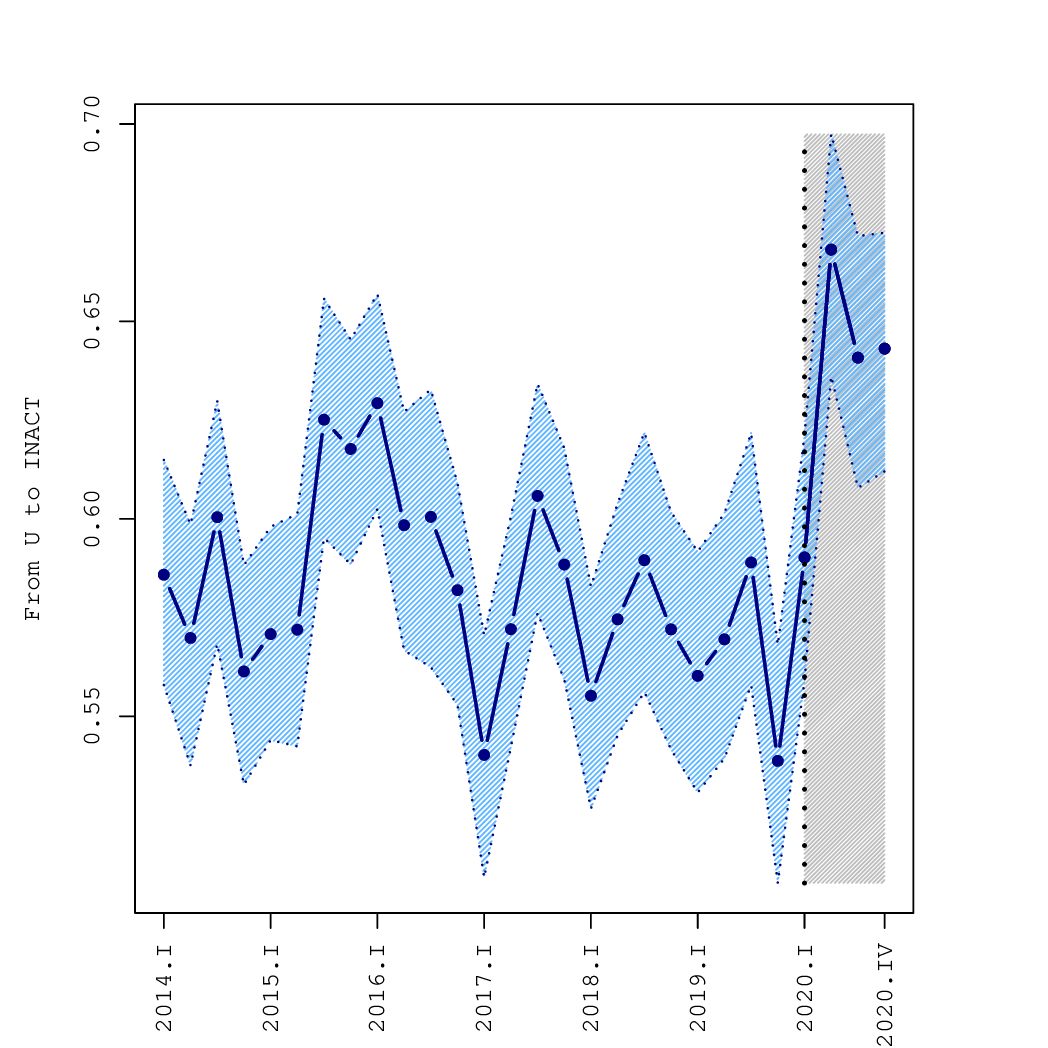}
		\caption{U - North.}
		\label{fig:transProbFromEDUtoPE_IIIIIIIIIIIIIIIIIIIIIIIIIIIIIII}
	\end{subfigure}
	\begin{subfigure}[t]{0.24\textwidth}
		\centering
		\includegraphics[width=\linewidth]{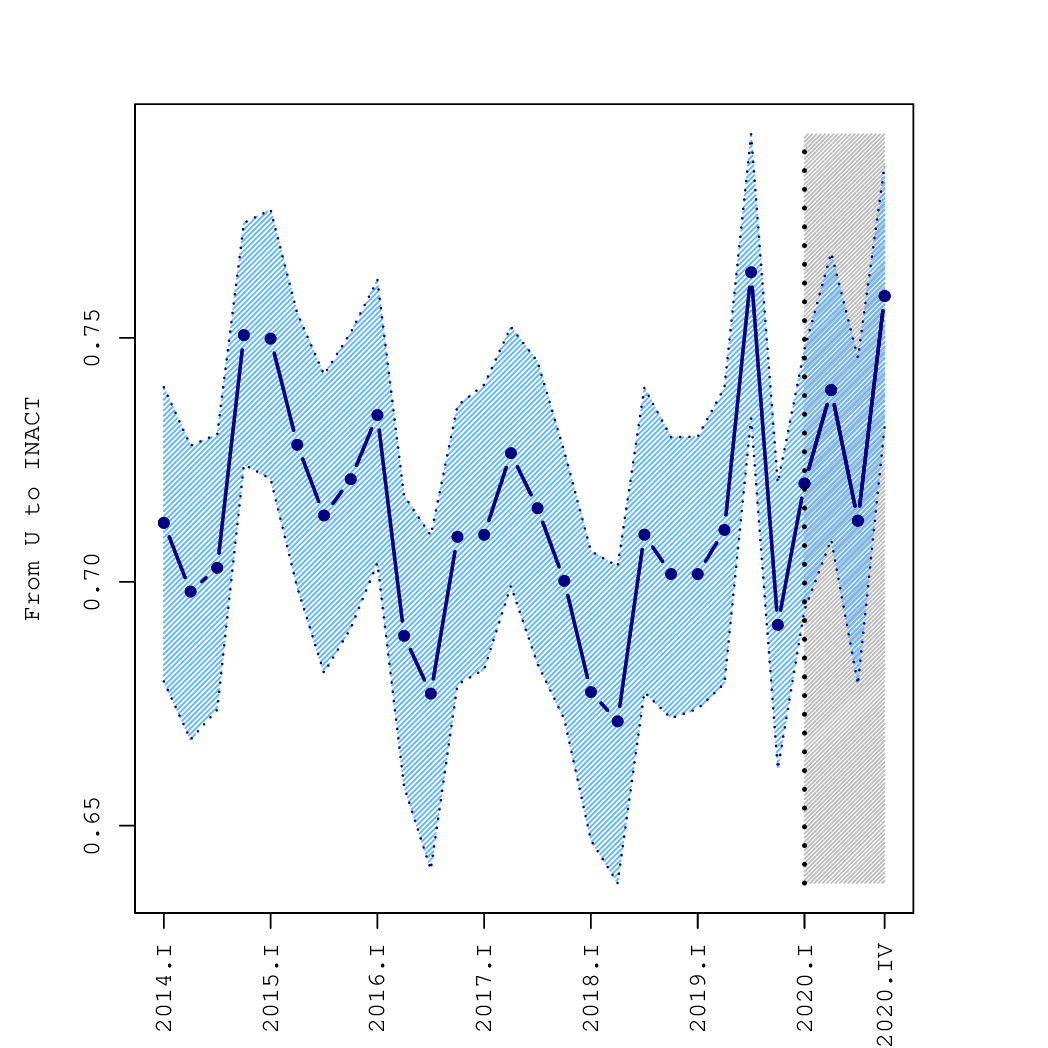}
		\caption{U  - South.}
		\label{fig:transProbFromEDUtoU_IIIIIIIIIIIIIIIIIIIIIIIIIIIIIII}
	\end{subfigure}
	\vspace{0.5cm}
	\caption*{\scriptsize{\textbf{Males}}.}
	\centering
	\begin{subfigure}[t]{0.24\textwidth}
		\centering
		\includegraphics[width=\linewidth]{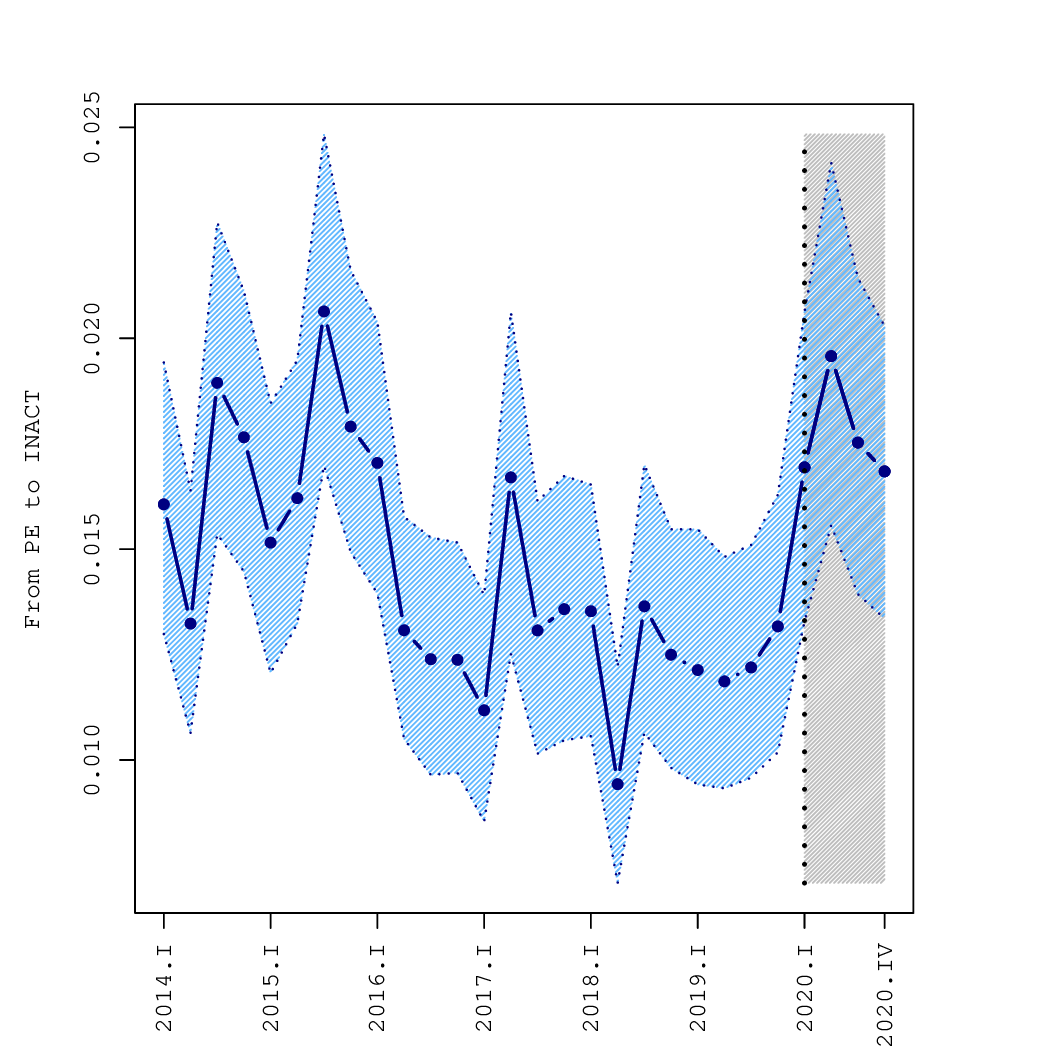}
		\caption{PE - North.}
		\label{}
		\vspace{0.2cm}
	\end{subfigure}
	\begin{subfigure}[t]{0.24\textwidth}
		\centering
		\includegraphics[width=\linewidth]{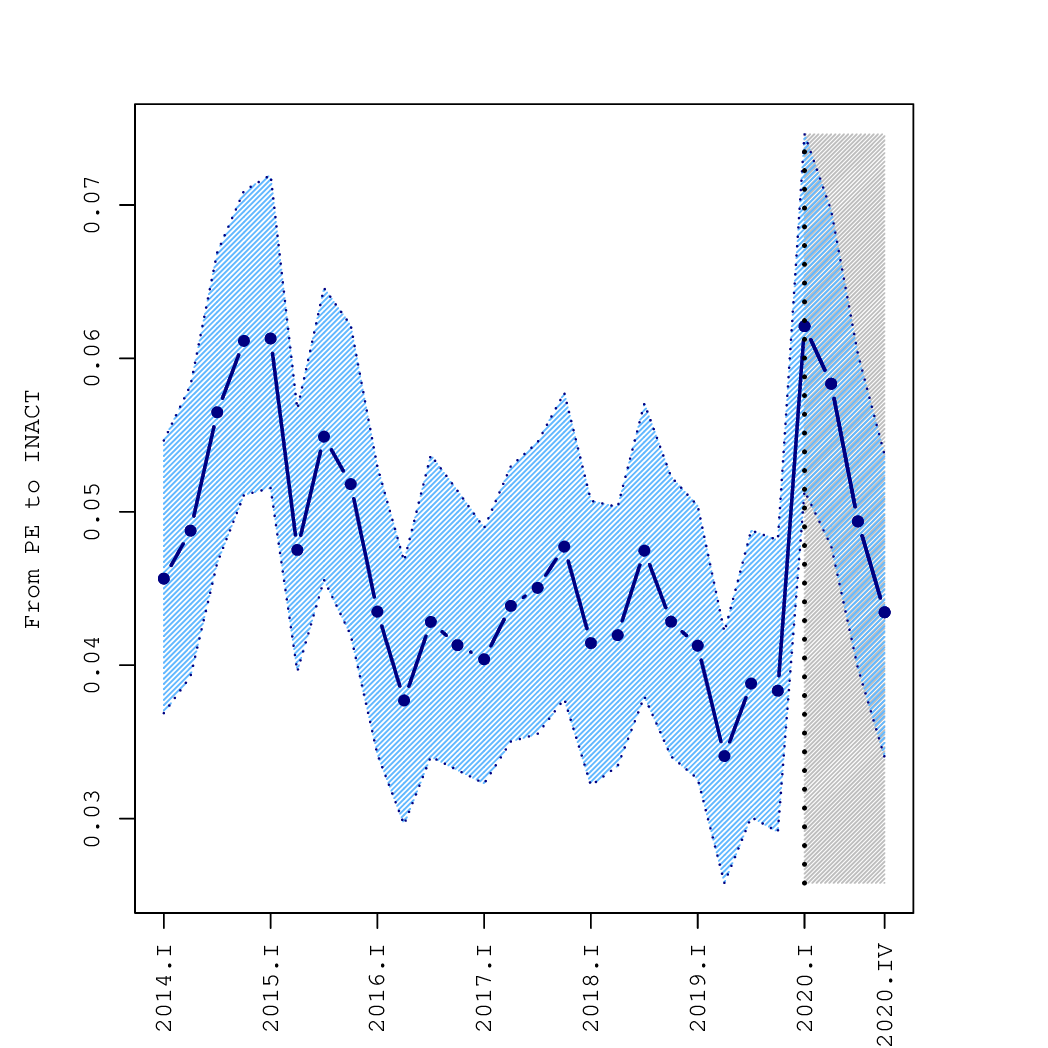}
		\caption{PE - South.}
		\label{fig:transProbFromEDUtoTE}
	\end{subfigure}
	\begin{subfigure}[t]{0.24\textwidth}
		\centering
		\includegraphics[width=\linewidth]{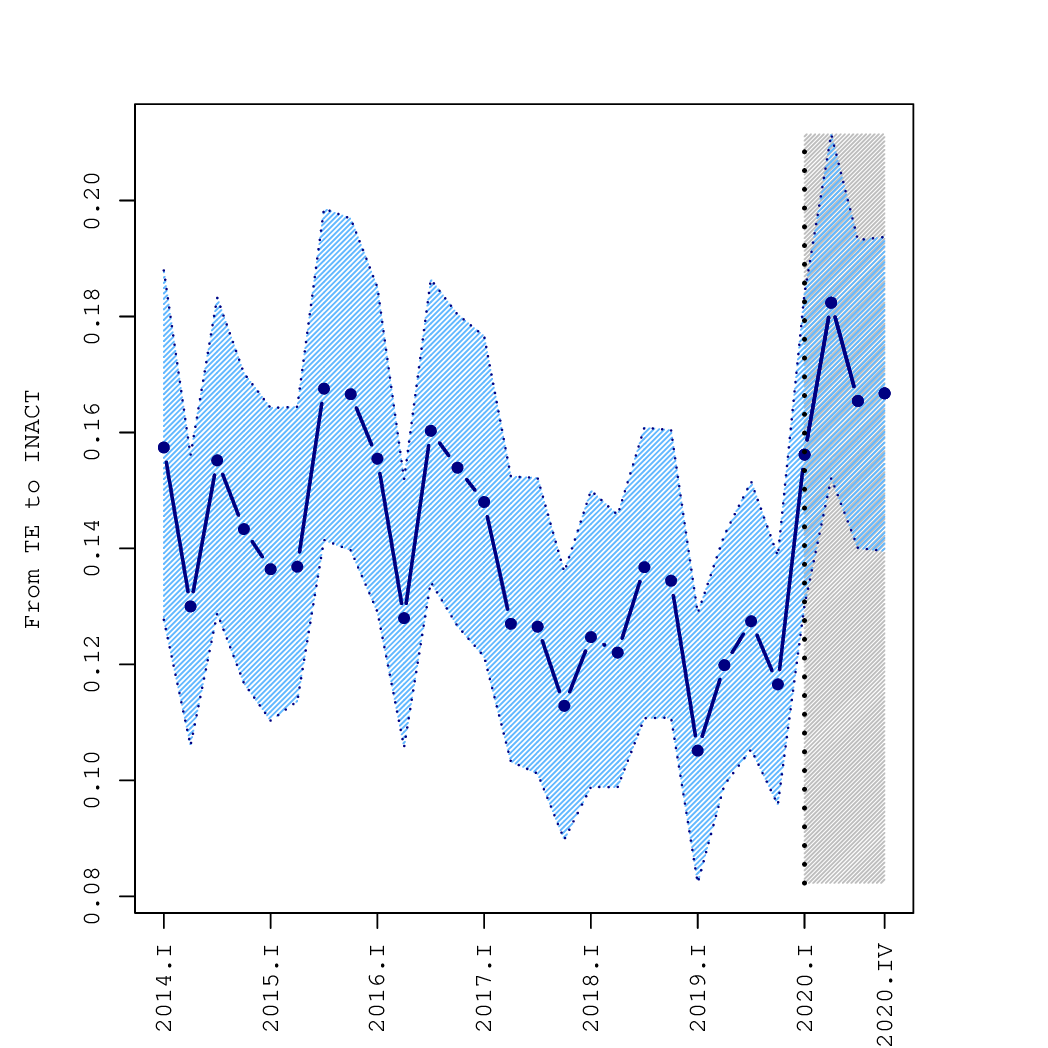}
		\caption{TE - North.}
		\label{fig:transProbFromEDUtoPE_IIIIIIIIIIIIIIIIIIIIIIIIIIIIIIII}
	\end{subfigure}
	\begin{subfigure}[t]{0.24\textwidth}
		\centering
		\includegraphics[width=\linewidth]{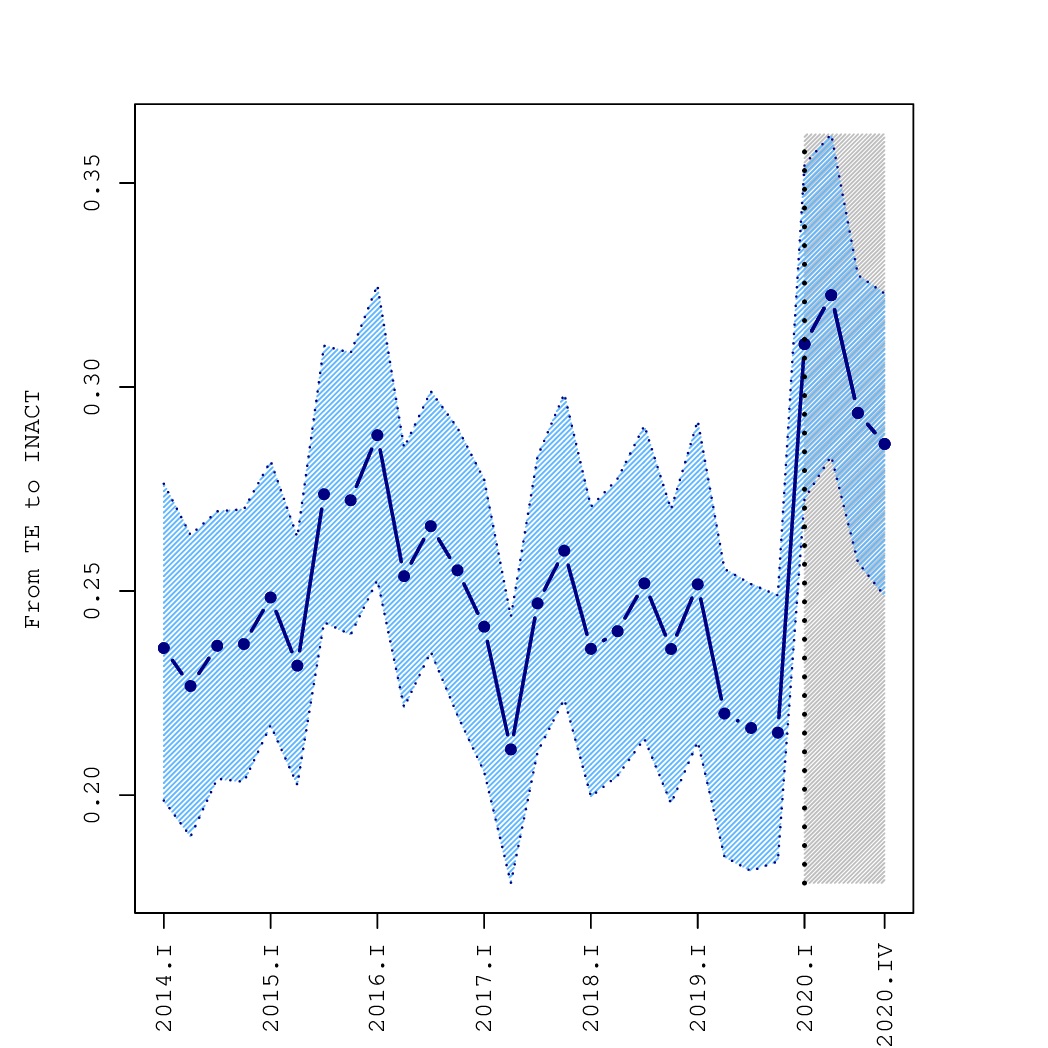}
		\caption{TE - South.}
		\label{fig:transProbFromEDUtoU_IIIIIIIIIIIIIIIIIIIIIIIIIIIIIIII}
	\end{subfigure}
	\begin{subfigure}[t]{0.24\textwidth}
		\centering
		\includegraphics[width=\linewidth]{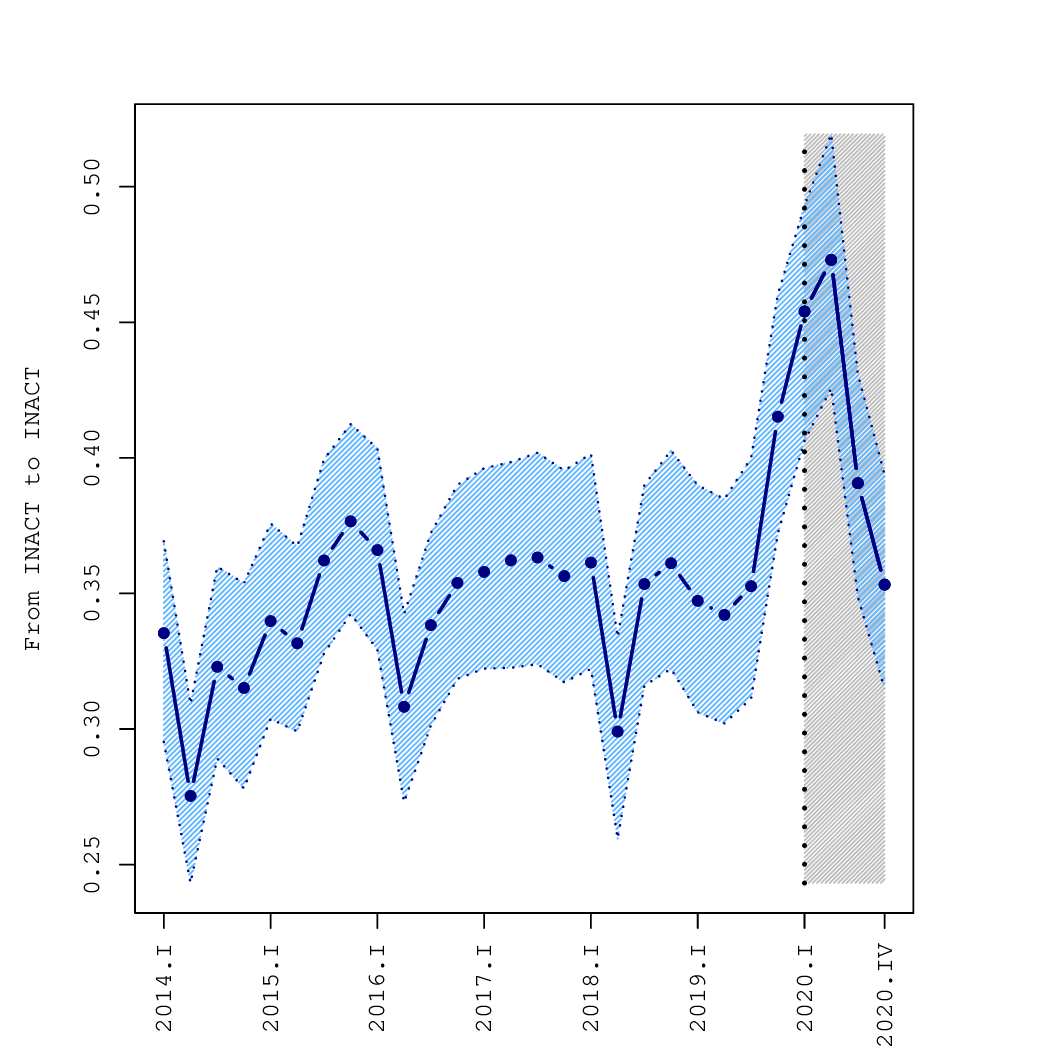}
		\caption{INACT - North.}
		\label{fig:transProbFromEDUtoPE_IIIIIIIIIIIIIIIIIIIIIIIIIIIIIIIII}
	\end{subfigure}
	\begin{subfigure}[t]{0.24\textwidth}
		\centering
		\includegraphics[width=\linewidth]{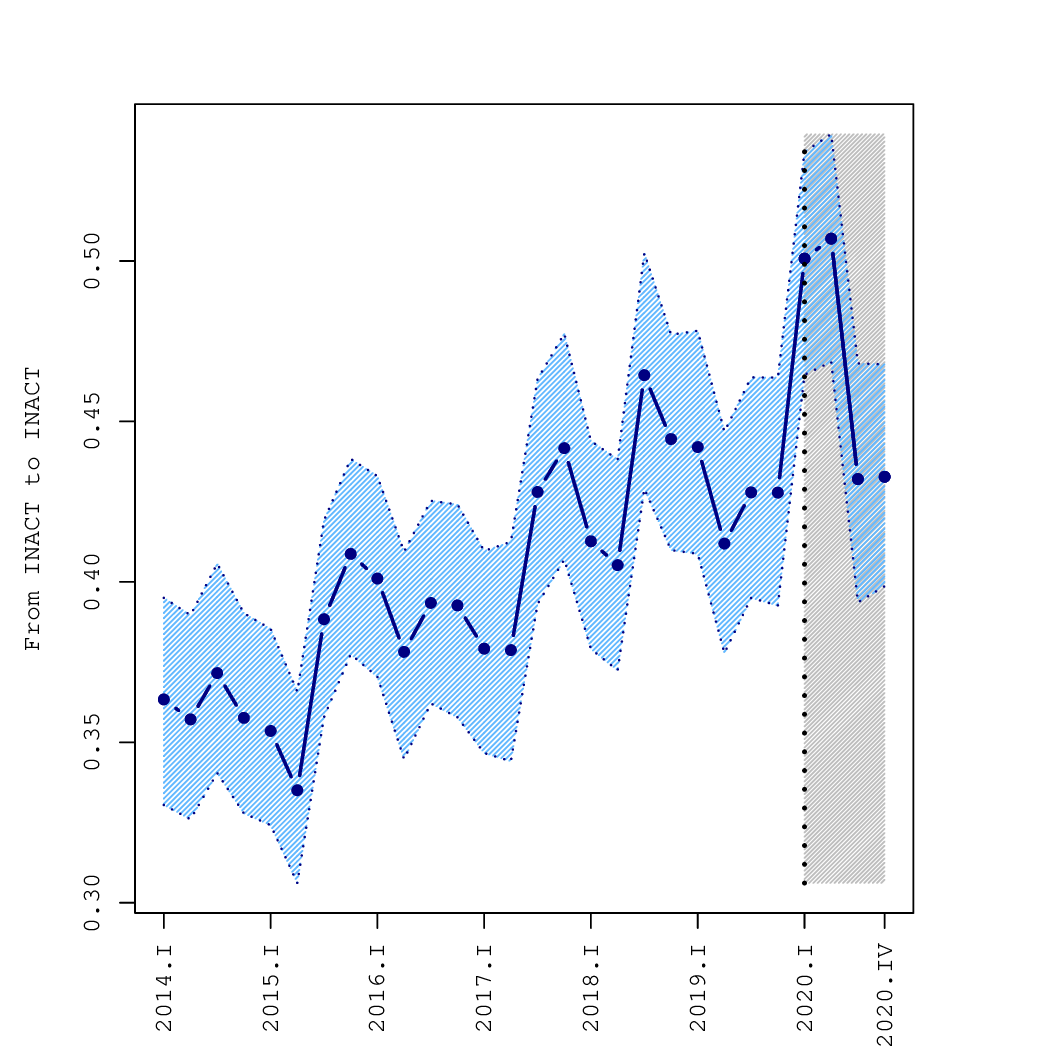}
		\caption{INACT  - South.}
		\label{fig:transProbFromEDUtoU_IIIIIIIIIIIIIIIIIIIIIIIIIIIIIIIII}
	\end{subfigure}
	\begin{subfigure}[t]{0.24\textwidth}
		\centering
		\includegraphics[width=\linewidth]{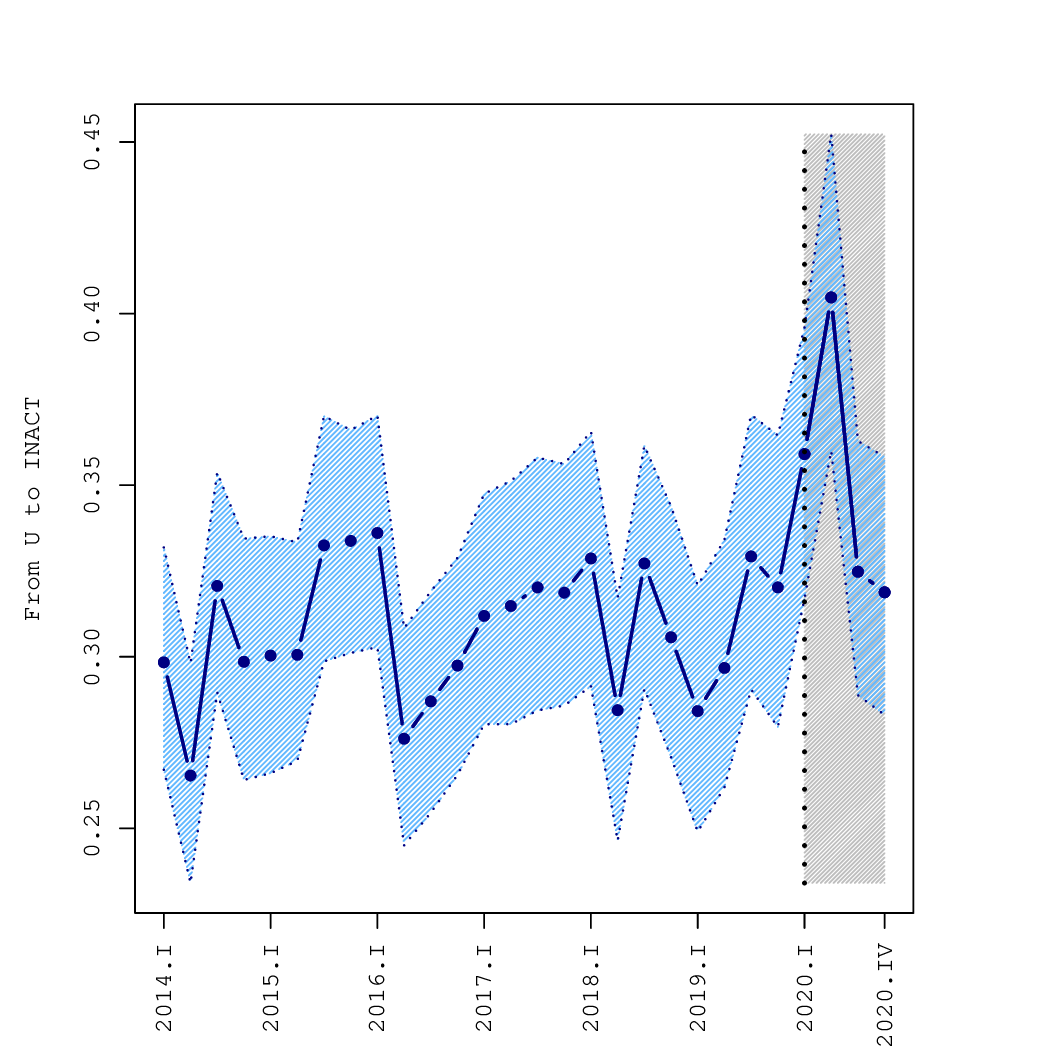}
		\caption{U - North.}
		\label{fig:transProbFromEDUtoPE_IIIIIIIIIIIIIIIIIIIIIIIIIIIIIIIIII}
	\end{subfigure}
	\begin{subfigure}[t]{0.24\textwidth}
		\centering
		\includegraphics[width=\linewidth]{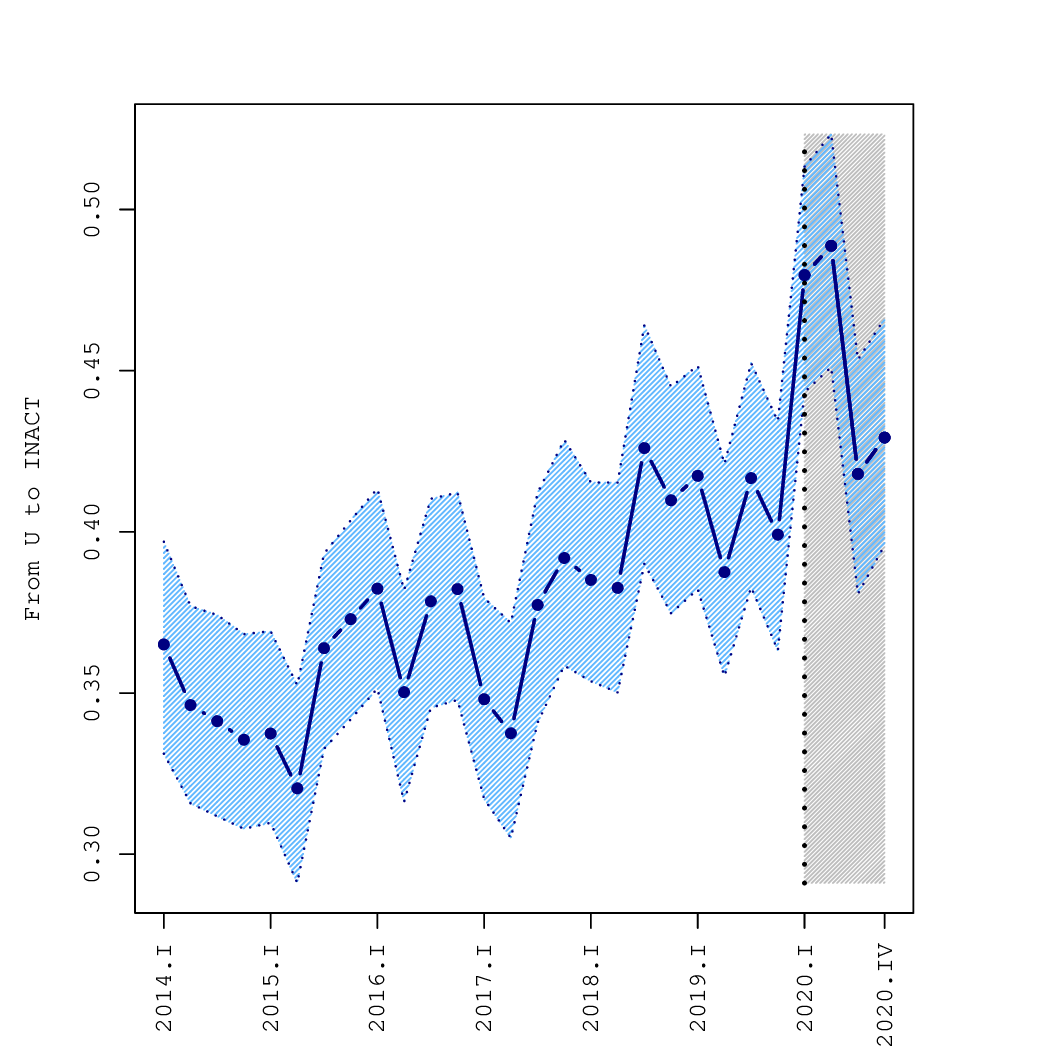}
		\caption{U  - South.}
		\label{fig:transProbFromEDUtoU}
	\end{subfigure}
	\vspace{0.2cm}
	\caption*{\scriptsize{\textit{Note}: Confidence intervals at 90\% are computed using 1000 bootstraps. The gray area identifies the COVID period. North includes regions in the North and the Center. \textit{Source}: LFS 3-month longitudinal data as provided by the Italian Institute of Statistics (ISTAT).}}
\end{figure}

\clearpage

\section{The case of 25-29 female workers.}\label{app:2529women}

\begin{figure}[!htbp]
	\caption{Transition probabilities of females from temporary employment to the inactive state by age groups.}
	\label{appfig:transProbTEtoinactiveFemales}
	\vspace{0.1cm}
	\caption*{\scriptsize{\textbf{Age 25-29}}.}
	\vspace{0.1cm}
	\centering
	\begin{subfigure}[t]{0.3\textwidth}
		\centering
		\includegraphics[width=\linewidth]{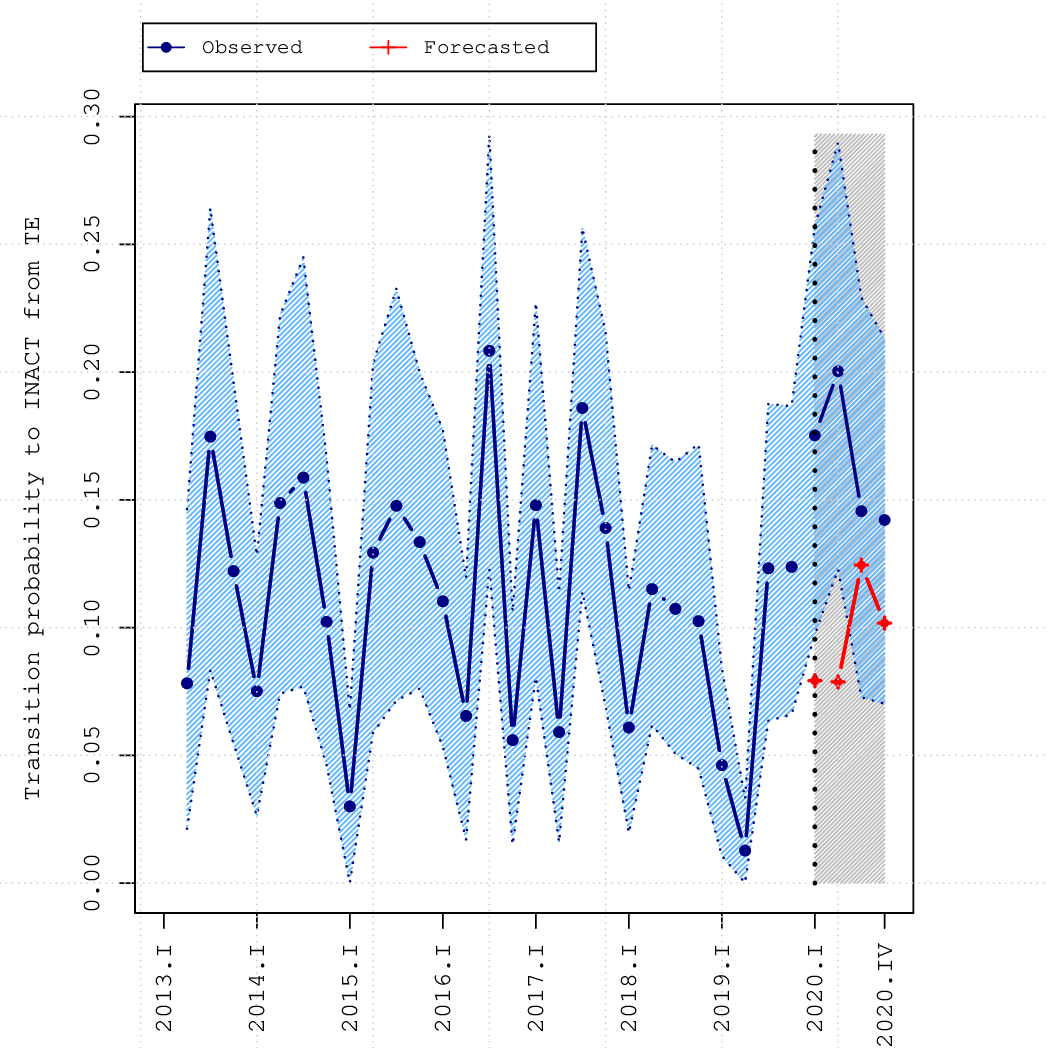}
		\caption{South.}
		\label{fig:transProbFromEDUtoSEApp_IIIIIIIIIIIIIIIIIIIIIIIIIIIIIII}
		\vspace{0.2cm}
	\end{subfigure}
	\begin{subfigure}[t]{0.3\textwidth}
		\centering
		\includegraphics[width=\linewidth]{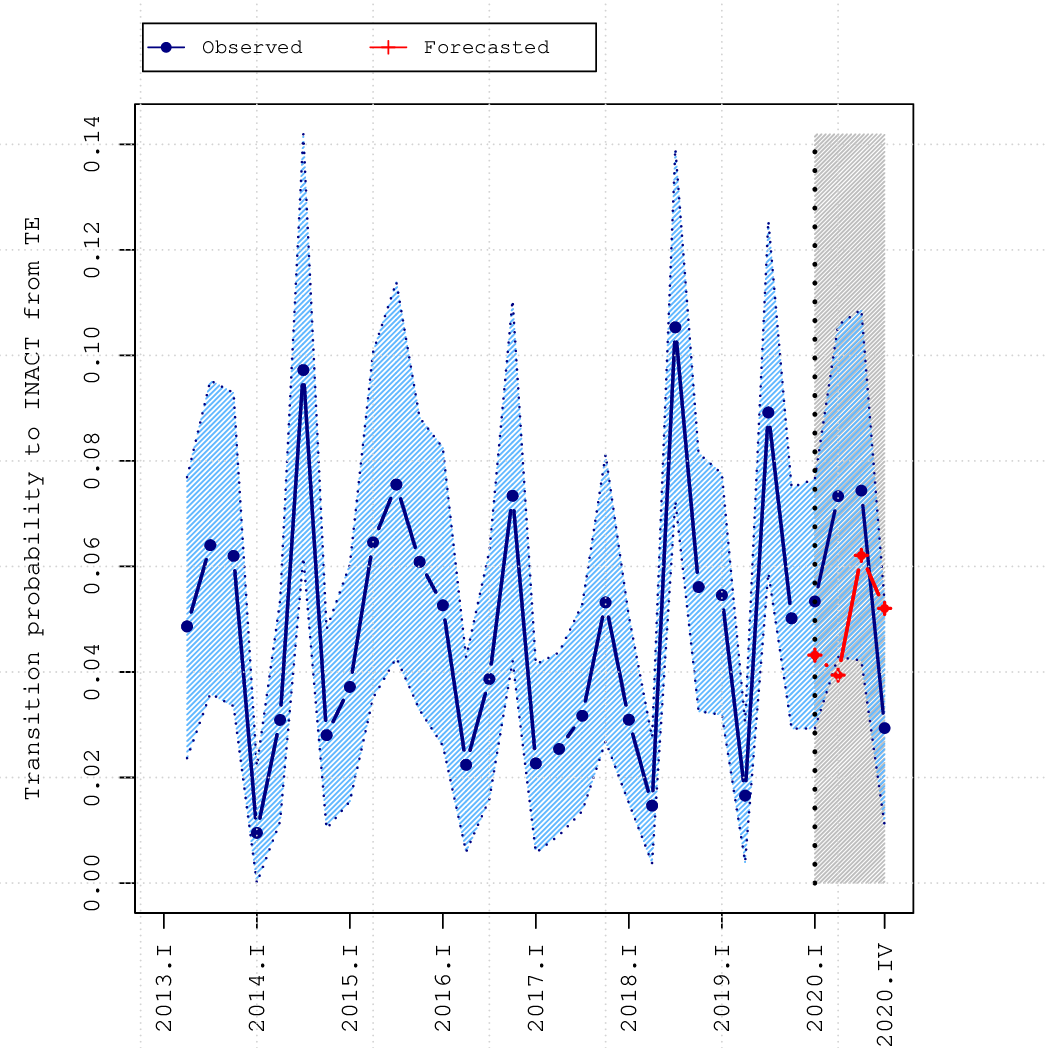}
		\caption{North.}
		\label{fig:transProbFromEDUtoTEApp_IIIIIIIIIIIIIIIIIIIIIIIIIIIIIII}
	\end{subfigure}
	\vspace{0.1cm}
	\caption*{\scriptsize{\textbf{Age 30-39}}.}
	\vspace{0.1cm}
	\centering
	\begin{subfigure}[t]{0.3\textwidth}
		\centering
		\includegraphics[width=\linewidth]{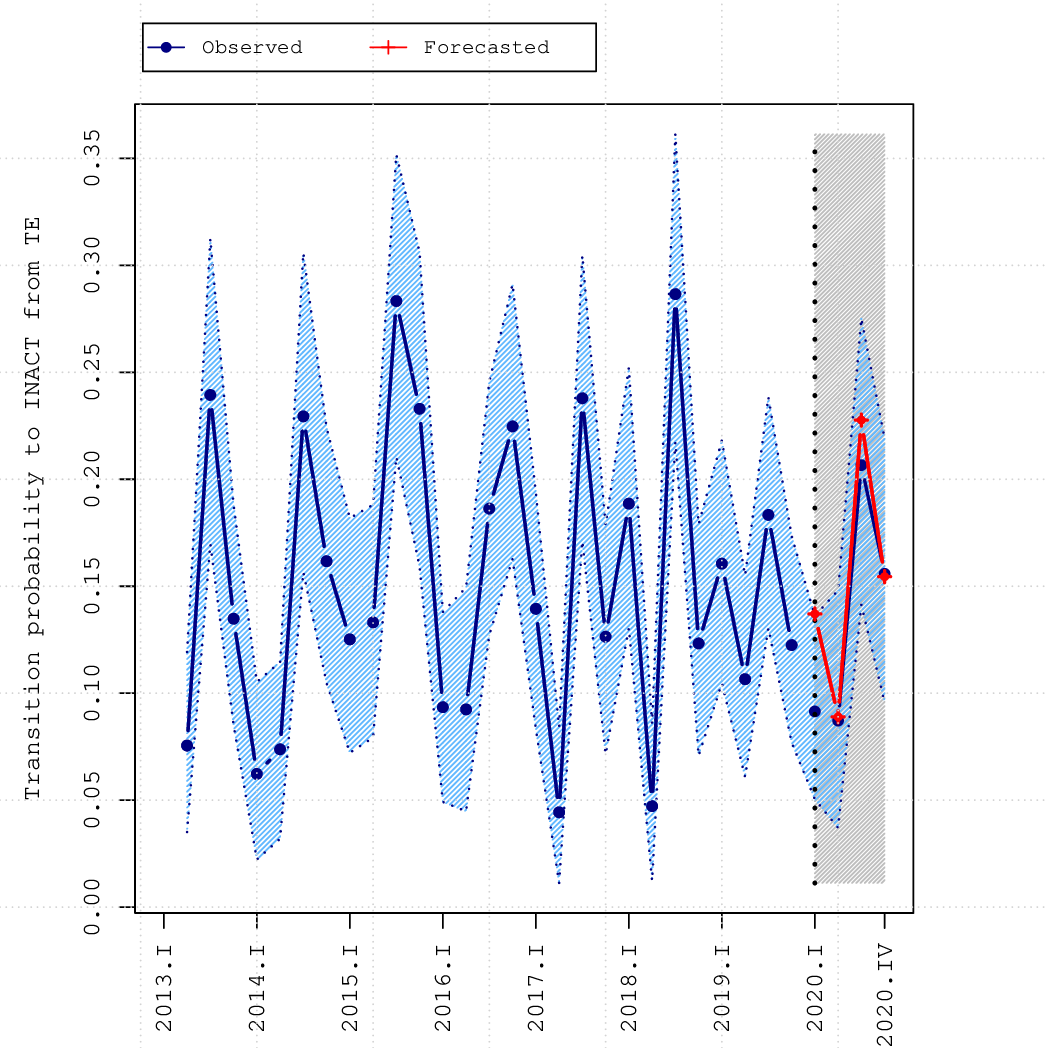}
		\caption{South.}
		\label{fig:transProbFromEDUtoSEApp_IIIIIIIIIIIIIIIIIIIIIIIIIIIIIIII}
		\vspace{0.2cm}
	\end{subfigure}
	\begin{subfigure}[t]{0.3\textwidth}
		\centering
		\includegraphics[width=\linewidth]{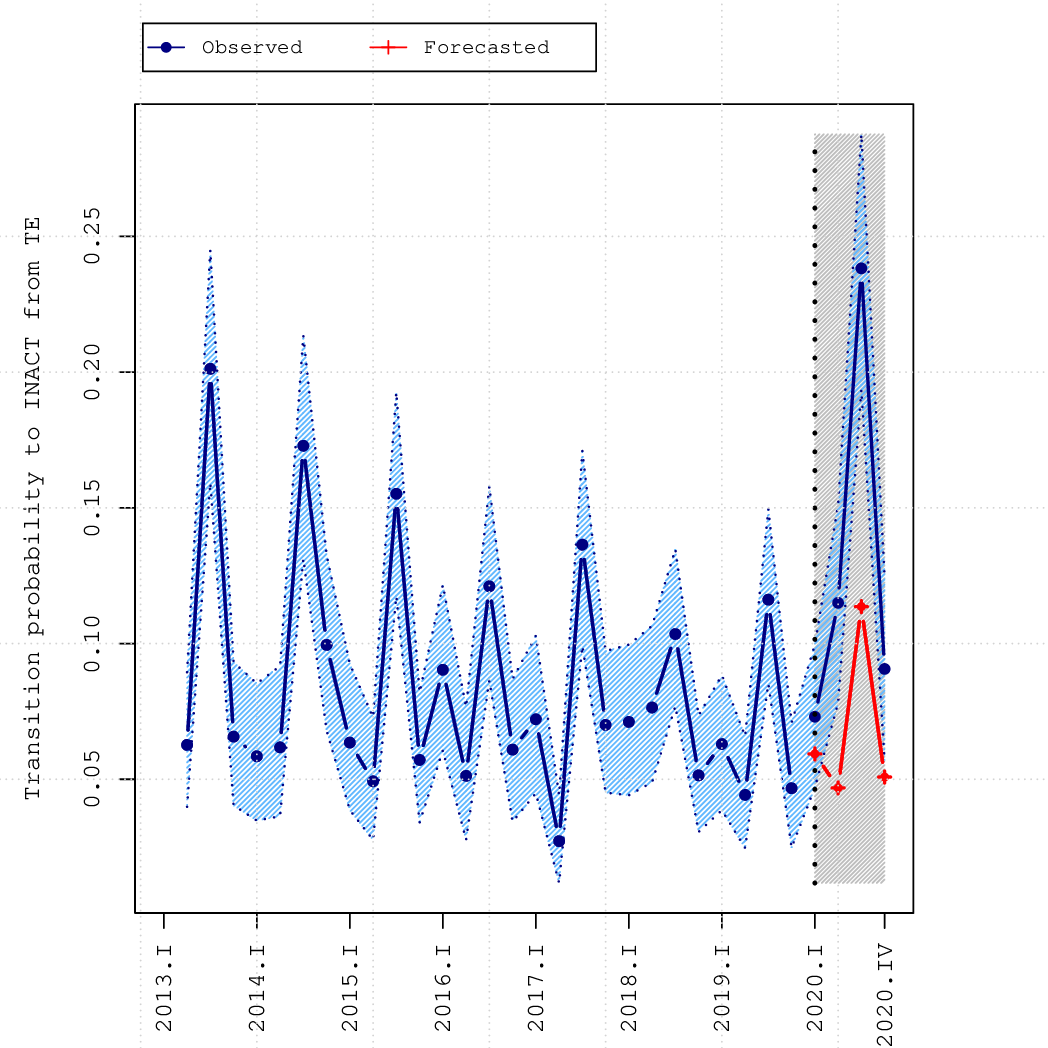}
		\caption{North.}
		\label{fig:transProbFromEDUtoTEApp_IIIIIIIIIIIIIIIIIIIIIIIIIIIIIIII}
	\end{subfigure}
	\vspace{0.1cm}
	\caption*{\scriptsize{\textbf{Age 40-49}}.}
	\vspace{0.1cm}
	\centering
	\begin{subfigure}[t]{0.3\textwidth}
		\centering
		\includegraphics[width=\linewidth]{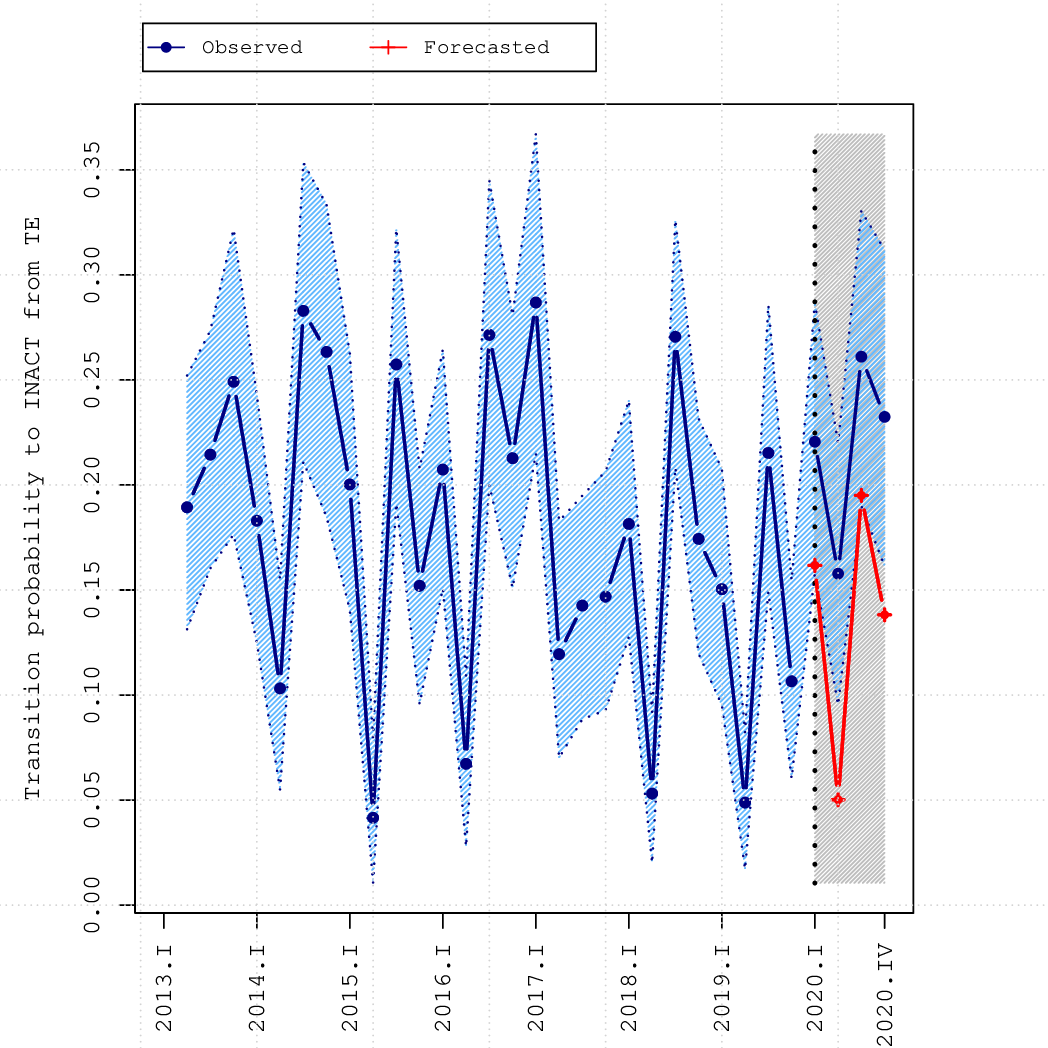}
		\caption{South.}
		\label{fig:transProbFromEDUtoSEApp_III}
		\vspace{0.2cm}
	\end{subfigure}
	\begin{subfigure}[t]{0.3\textwidth}
		\centering
		\includegraphics[width=\linewidth]{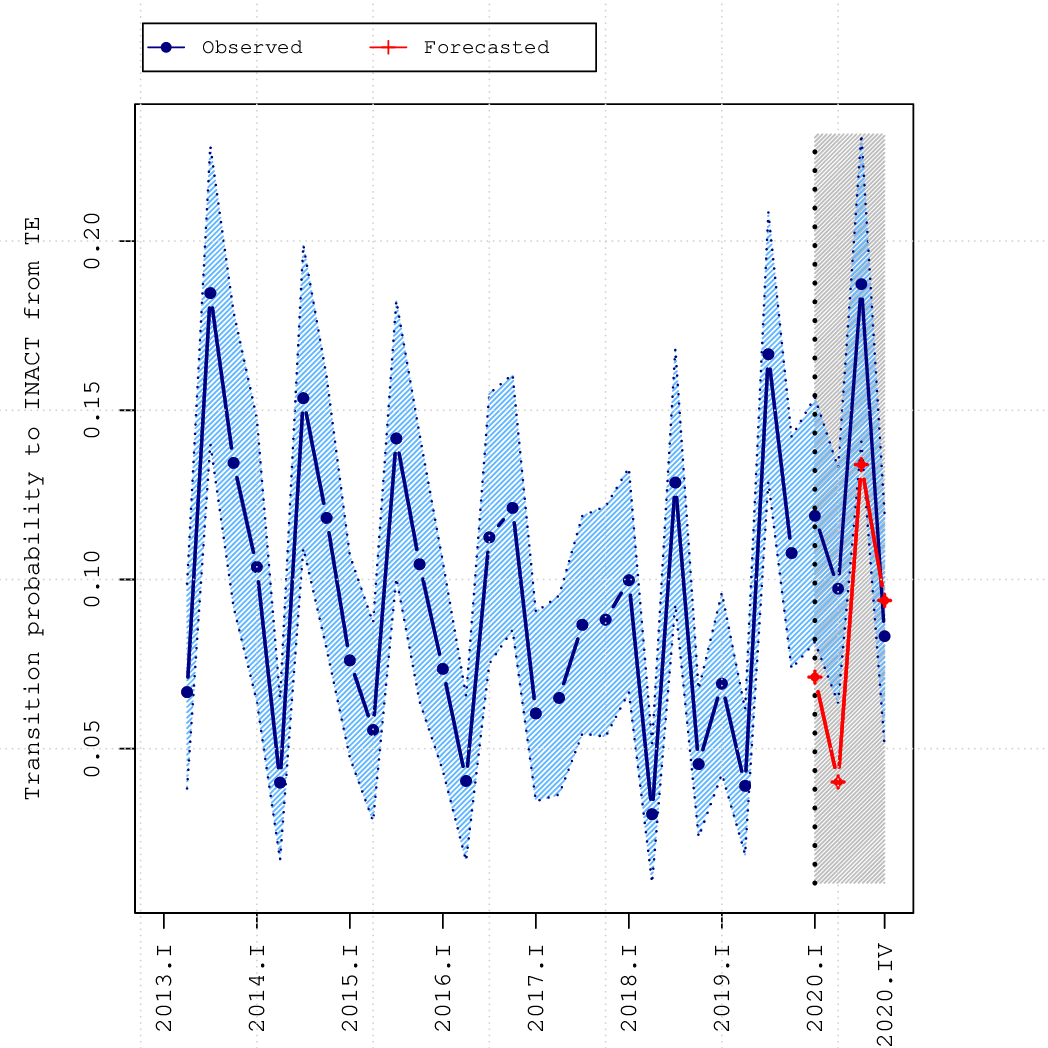}
		\caption{North.}
		\label{fig:transProbFromEDUtoTEApp_IIIIIIIIIIIIIIIIIIIIIIIIIIIIIIIII}
	\end{subfigure}
	\vspace{0.2cm}
	\caption*{ \scriptsize{\textit{Note}: The forecasted transition probabilities are computed using a combination of four forecasting models (ETS, TSLM, THETAF, and ARIMA) \citep{HyndmanAthanasopoulosforecasting2021} in the period 2013 (quarter I)- 2019 (quarter IV). Confidence intervals at 90\% are computed using 1000 bootstraps and reported in parenthesis. The gray area identifies the COVID period. North includes regions in the North and the Center. \textit{Source}: LFS 3-month longitudinal data as provided by the Italian Institute of Statistics (ISTAT).}}
\end{figure}

Although not the focus of this paper, we provide some discussion on the Covid effect for females in the 25-29 age category. 
Table \ref{tab:sharechangesqIVtext}  indicates that the share of 25-29 females in the inactive state has increased everywhere in Italy, but more in the South. We also observe from Figures \ref{fig:transprob2529females} and \ref{appfig:transProbTEtoinactiveFemales} that the transition probabilities for females in the 25-29 age category increased from unemployment and temporary employment to inactivity mainly in the South, and only marginally in the North. We do not observe any other significant increase in the transition probabilities from the other states to inactivity. 

%
%
%
%
%
%

To dig further into this phenomenon, we look at the distribution of females in the 25-29 age category by employment status and household size, which we use as a proxy for the presence of children, across the two geographical areas. Table \ref{apptab:women2529unemHH} shows that in the South the percentage of unemployed women living in a household with more than two members is much higher compared to the North. Similarly, in the South the percentage of women in temporary employment living in a household with more than two members is much higher compared to the North.

These findings provide further support to the main results of the paper which point to the presence of children as the main driver behind the decision of women to leave the labour force during the Covid pandemic.

\begin{table}[!htbp] \centering 
	\caption{Distribution of female aged 25-29  workers by employment status (unemployment and temporary employment), household size and geographical location.}  
	\label{apptab:women2529unemHH} 
	\footnotesize{
		\begin{tabular}{@{\extracolsep{5pt}} lcc} 
			\hline \hline \\[-1.8ex] 
	\multicolumn{3}{c}{Unemployed}	 \\
	\hline \\[-1.8ex]
			&South &	North
			\\
			\hline \\[-1.8ex]
		HH $<=2$	& 18.6&	26.4
			\\
				HH $>2$	& 81.4&	73.6
			\\
		
	\hline
	
	\\[-1.8ex]
\multicolumn{3}{c}{In temporary employment}\\	 
\hline \\[-1.8ex]
	
	&South &	North
	\\
	\hline \\[-1.8ex]
	HH $<=2$	& 21.8&	31.6
	\\
	HH $>2$	& 78.2&	68.4
	\\
	\hline
		\multicolumn{3}{l}{\tiny{\textit{Note}: North includes regions in the North and }}\\
	\multicolumn{3}{l}{\tiny{the Center. \textit{Source}: LFS 3-month longitudinal }}\\
	\multicolumn{3}{l}{\tiny{data as provided  by the Italian Institute of}}\\ \multicolumn{3}{l}{\tiny{ Statistics (ISTAT).}}
\end{tabular} }
\end{table}

\clearpage

\section{Regional distribution of median age of the mother at birth.}

\vspace{-0.5cm}
\begin{table}[!htbp] \centering 
	\caption{Regional distribution of median age of the mother at first child by citizenship.}  
	\label{apptab:birthagebycitiz} 
	\scriptsize{
		\begin{tabular}{@{\extracolsep{5pt}} lccc} 
			\hline \hline \\[-1.8ex] 
			Region& Macro-area&	\multicolumn{2}{c}{Citizenship}\\ 
			\cline{3-4}
			&&Italian &	Foreign
			\\
			\hline \\[-1.8ex]
			Piemonte&North&	32.11&	28.68
			\\
			Valle d'Aosta&North	&32.08	&28.42
			\\
			Lombardia&North&	32.60&	28.54
			\\
			Prov. Auton. Bolzano&North&	30.72&	29.33
			\\
			Prov. Auton. Trento	&North&31.90	&28.33
			\\
			Veneto	&North&32.27&	28.21
			\\
			Friuli Venezia Giulia&North&	32.40&	28.30
			\\
			Liguria	&North&32.71&	28.38
			\\
			Emilia Romagna&North&	32.55&	28.56
			\\
			Toscana	&Center&32.86&	28.39
			\\
			Umbria&Center&	32.44&	29.76
			\\
			Marche&Center	&32.51&	29.15
			\\
			Lazio&Center&	32.96&	29.92
			\\
			Abruzzo&Center	&32.33&	28.74
			\\
			Molise&South&	31.61	&-
			\\
			Campania&South&	30.59&	28.73
			\\
			Puglia&South&	31.11&	28.50
			\\
			Basilicata&South&	31.66&	24.67
			\\
			Calabria&South&	31.02&	28.90
			\\
			Sicilia	&South&30.59&	28.56
			\\
			Sardegna&South&	32.60&	30.03
			\\
			\hline \\[-1.8ex] 
			Italy &&	31.92&	28.73\\
			\hline \hline \\[-1.8ex] 
			\multicolumn{3}{l}{\tiny{Source: Italian Ministry of Health, 2019.}}
	\end{tabular} }
\end{table} 

\vspace{-0.5cm}
\begin{table}[!htbp]
	\centering 
	\caption{Regional distribution of births by age of the mother.}  
	\label{apptab:birthbyage} 
	\scriptsize{
		\begin{tabular}{@{\extracolsep{5pt}} ccccccc} 		
			\hline \hline
			\\[-1.8ex] 
			Region&Macro-area&	\multicolumn{4}{c}{Age groups}& $\#$ births\\
			\cline{2-5} \\[-1.8ex]
			&&$<$20&	20-29&	30-39&	$>40$& \\
			\hline \\[-1.8ex]		
			Piemonte&North&	0.74&	28.16&	60.76&	10.33&	27,296	\\
			Valle d'Aosta&North&	0.49&	27.53&	60.99&	10.99&	810
			\\
			Lombardia&North&	0.73&	26.56&	62.18&	10.52&	72,702	\\
			P.A. Bolzano&North&	0.73&	32.64	&58.93&	7.70&	5,218
			\\
			P.A. Trento&North&	0.47&	27.82&	62.01&	9.64&	4,004
			\\
			Veneto&North&	0.61&	26.91&	62.17&	10.30&	32,845
			\\
			Friuli Venezia Giulia&North&	0.63&	27.99&	60.35&	11.00&	7,892
			\\
			Liguria	&North&0.87&	27.71&	60.36&	11.04&	8,375	\\
			Emilia Romagna	&North&0.58&	27.97&	60.69&	10.63&	31,123
			\\
			Toscana	&Center&0.64&	25.72&	61.78&	11.81&	23,626	\\
			Umbria&Center&	0.70&	26.58&	62.25&	10.47&	6,016\\
			Marche&Center&	0.66&	25.03&	62.67&	11.34&	9,358
			\\
			Lazio&Center&	0.77&	24.64&	61.41&	13.06&	38,388	
			\\
			Abruzzo&Center&	0.59&	25.99&	62.78&	10.63&	8,272
			\\
			Molise&South&	0.96&	27.15&	60.77&	11.00&	1,672
			\\
			Campania&South&	1.61&	31.55	&58.72&	8.09&	46,833	\\
			Puglia&South	&1.53&	28.86&	60.12&	9.48&	27,539
			\\
			Basilicata&South&	0.42&	26.65&	61.82&	11.09&	3,824	
			\\
			Calabria&South&	1.29&	30.34&	59.74&	8.62&	12,674	\\
			Sicilia&South	&2.24&	33.14&	56.40&	8.21&	38,047
			\\
			Sardegna&South&	0.86&	23.18&	61.01&	14.94&	8,556	\\
			\hline \\[-1.8ex] 
			Total&&	1.01&	28.04&	60.62&	10.29&415,070	\\
			
			\hline \hline \\[-1.8ex] 
			\multicolumn{6}{l}{\tiny{Source: Italian Ministry of Health, 2019.}}
	\end{tabular} }
\end{table}

\clearpage

\section{Labour market dynamics  during COVID-19}\label{app:dynamicsduring}

\subsection{Shares of individuals by age, gender and geographical location}\label{app:dynamicssharesduring}


\begin{table}[!htbp] \centering 
	\caption{Changes in the shares in different labour market states from quarter IV of 2019 to quarter IV of 2020 by category of individuals.}  
	\label{sharechangesqIVtext} 
	\scriptsize{
		\begin{tabular}{@{\extracolsep{5pt}} cccccccc} 
			\hline \hline \\[-1.8ex] 
			\multicolumn{8}{c}{\textbf{Females - North}}\\\\[-1.8ex] 
			\hline \\[-1.8ex] 
			& SE & TE & PE & U & inactive  & EDU & FS \\ 
			\hline \\[-1.8ex] 
			20-24 & $$-$0.002$ & $$-$0.041^{***}$ & $$-$0.011^{*}$ & $0.013^{***}$ & $\textbf{0.034}^{***}$ & $$-$0.001$ & $0.008^{***}$ \\
			& $(0.305)$ & $(0.000)$ & $(0.050)$ & $(0.009)$ & $(0.000)$ & $(0.467)$ & $(0.000)$ \\ 
			25-29 & $$-$0.0001$ & $$-$0.021^{***}$ & $$-$0.027^{***}$ & $0.009^{*}$ & $\textbf{0.017}^{**}$ & $$-$0.005$ & $0.027^{***}$ \\ 
			& $(0.485)$ & $(0.005)$ & $(0.002)$ & $(0.077)$ & $(0.022)$ & $(0.244)$ & $(0.000)$ \\ 
			30-39 & $$-$0.009^{**}$ & $$-$0.007^{*}$ & $$-$0.029^{***}$ & $$-$0.015^{***}$ & $\textbf{0.034}^{***}$ & $$-$0.001^{}$ & $0.027^{***}$ \\ 
			& $(0.012)$ & $(0.052)$ & $(0.000)$ & $(0.000)$ & $(0.000)$ & $(0.225)$ & $(0.000)$ \\ 
			40-49 & $$-$0.009^{***}$ & $$-$0.010^{***}$ & $$-$0.020^{***}$ & $$-$0.007^{***}$ & $\textbf{0.020}^{***}$ & $0.0004$ & $0.024^{***}$ \\ 
			& $(0.008)$ & $(0.000)$ & $(0.000)$ & $(0.001)$ & $(0.000)$ & $(0.247)$ & $(0.000)$ \\
			\hline \\[-1.8ex] 
			\multicolumn{8}{c}{\textbf{Females - South}}\\\\[-1.8ex] 
			\hline \\[-1.8ex] 
			& SE & TE & PE & U & inactive  & EDU & FS \\ 
			\hline \\[-1.8ex] 
			20-24 & $$-$0.014^{***}$ & $$-$0.024^{***}$ & $$-$0.018^{***}$ & $$-$0.019^{**}$ & $0.013$ & $\textbf{0.050}^{***}$ & $0.013^{***}$ \\
			& $(0.000)$ & $(0.000)$ & $(0.001)$ & $(0.017)$ & $(0.143)$ & $(0.000)$ & $(0.000)$ \\ 
			25-29 & $$-$0.006$ & $$-$0.020^{***}$ & $$-$0.016^{**}$ & $$-$0.037^{***}$ & $\textbf{0.058}^{***}$ & $0.007$ & $0.014^{***}$ \\
			& $(0.166)$ & $(0.003)$ & $(0.041)$ & $(0.000)$ & $(0.000)$ & $(0.254)$ & $(0.000)$ \\ 
			30-39 & $0.001$ & $0.007$ & $$-$0.013^{**}$ & $$-$0.022^{***}$ & $0.010$ & $$-$0.001$ & $0.017^{***}$ \\ 
			& $(0.418)$ & $(0.109)$ & $(0.042)$ & $(0.000)$ & $(0.154)$ & $(0.424)$ & $(0.000)$ \\
			40-49 & $0.003$ & $$-$0.001$ & $$-$0.023^{***}$ & $$-$0.005$ & $0.011$ & $0.002^{*}$ & $0.013^{***}$ \\
			& $(0.304)$ & $(0.461)$ & $(0.000)$ & $(0.148)$ & $(0.116)$ & $(0.086)$ & $(0.000)$ \\ 
			\\[-1.8ex]\hline \\[-1.8ex] 
			\multicolumn{8}{c}{ \textbf{Males - North}}\\\\[-1.8ex] 
			\hline \\[-1.8ex] 
			& SE & TE & PE & U & inactive  & EDU & FS \\ 
			\hline \\[-1.8ex] 
			20-24 & $0.002$ & $$-$0.011$ & $$-$0.011^{*}$ & $0.005$ & $\textbf{0.018}^{***}$ & $$-$0.020^{**}$ & $0.016^{***}$ \\ 
			& $(0.331)$ & $(0.131)$ & $(0.088)$ & $(0.212)$ & $(0.000)$ & $(0.041)$ & $(0.000)$ \\ 
			25-29 & $$-$0.007$ & $$-$0.015^{**}$ & $$-$0.024^{***}$ & $0.003$ & $\textbf{0.020}^{***}$ & $$-$0.003$ & $0.026^{***}$ \\ 
			& $(0.197)$ & $(0.041)$ & $(0.002)$ & $(0.300)$ & $(0.000)$ & $(0.330)$ & $(0.000)$ \\ 
			30-39 & $$-$0.004$ & $$-$0.010^{***}$ & $$-$0.016^{***}$ & $$-$0.002$ & $0.003$ & $0.002$ & $0.027^{***}$ \\ 
			& $(0.222)$ & $(0.002)$ & $(0.006)$ & $(0.232)$ & $(0.127)$ & $(0.144)$ & $(0.000)$ \\
			40-49 & $$-$0.006$ & $$-$0.008^{***}$ & $$-$0.020^{***}$ & $$-$0.004^{**}$ & $\textbf{0.013}^{***}$ & $$-$0.0004$ & $0.026^{***}$ \\ 
			& $(0.121)$ & $(0.000)$ & $(0.000)$ & $(0.047)$ & $(0.000)$ & $(0.146)$ & $(0.000)$ \\
			\hline \\[-1.8ex] 
			\multicolumn{8}{c}{\textbf{Males - South}}\\\\[-1.8ex] 
			\hline \\[-1.8ex] 
			& SE & TE & PE & U & inactive  & EDU & FS \\ 
			\hline \\[-1.8ex] 
			20-24 & $$-$0.009^{**}$ & $$-$0.020^{**}$ & $0.008$ & $$-$0.016^{**}$ & $\textbf{0.016}^{*}$ & $0.010$ & $0.011^{***}$ \\
			& $(0.038)$ & $(0.014)$ & $(0.161)$ & $(0.042)$ & $(0.079)$ & $(0.229)$ & $(0.000)$ \\ 
			25-29 & $0.002$ & $$-$0.024^{***}$ & $$-$0.018$ & $$-$0.022^{**}$ & $\textbf{0.017}^{*}$ & $\textbf{0.034}^{***}$ & $0.012^{***}$ \\ 
			& $(0.409)$ & $(0.002)$ & $(0.046)$ & $(0.011)$ & $(0.065)$ & $(0.000)$ & $(0.000)$ \\ 
			30-39 & $$-$0.004$ & $$-$0.014^{***}$ & $$-$0.019^{**}$ & $$-$0.021^{***}$ & $\textbf{0.027}^{***}$ & $0.004$ & $0.027^{***}$ \\
			& $(0.285)$ & $(0.002)$ & $(0.017)$ & $(0.000)$ & $(0.000)$ & $(0.110)$ & $(0.000)$ \\ 
			40-49 & $0.005$ & $$-$0.012^{***}$ & $$-$0.017^{**}$ & $$-$0.016^{***}$ & $\textbf{0.016}^{***}$ & $0.0005$ & $0.025^{***}$ \\
			& $(0.236)$ & $(0.001)$ & $(0.018)$ & $(0.000)$ & $(0.003)$ & $(0.227)$ & $(0.000)$ \\ 
			\hline \hline \\[-1.8ex] 
			\multicolumn{8}{l}{\textit{Note}: The attained significance levels (ASL)  of the null hypothesis of equality between the  shares}\\  \multicolumn{8}{l}{ in the two periods computed using 1000 bootstraps are reported in parenthesis}\\  \multicolumn{8}{l}{\citep[p.220]{efron1994introduction}; North includes regions in the North and the Center.} \\  \multicolumn{8}{l}{ $^{*}$ASL$<$0.1; $^{**}$ASL$<$0.05; $^{***}$ASL$<$0.01.}
	\end{tabular} }
\end{table} 

\clearpage

\begin{table}[!htbp]
	\centering 
	\caption{Changes in the shares of different categories of individuals in different labour market states from quarter III of 2019 to quarter III of 2020, by age groups.}  
	\label{sharechangesMalesNQIII} 
	\scriptsize{
		\begin{tabular}{@{\extracolsep{5pt}} cccccccc} 
			\\[-1.8ex]\hline  \hline \\[-1.8	ex]
			\multicolumn{8}{c}{\textbf{Males - North}}\\\\[-1.8ex] 
			\hline \\[-1.8ex] 
			& SE & TE & PE & U & inactive  & EDU & FS \\ 
			\hline \\[-1.8ex] 
			20-24 & $0.002$ & $$-$0.011$ & $$-$0.004$ & $$-$0.008$ & $\textbf{0.007}^{*}$ & $$-$0.0001$ & $0.013^{***}$ \\ 
			& $(0.284)$ & $(0.117)$ & $(0.307)$ & $(0.108)$ & $(0.098)$ & $(0.464)$ & $(0.000)$ \\ 
			25-29 & $0.011$ & $0.006$ & $$-$0.033^{***}$ & $0.002$ & $$-$0.001$ & $$-$0.006$ & $0.020^{***}$ \\ 
			& $(0.060)$ & $(0.219)$ & $(0.000)$ & $(0.390)$ & $(0.437)$ & $(0.215)$ & $(0.000)$ \\ 
			30-39 & $$-$0.009$ & $$-$0.005$ & $$-$0.008$ & $$-$0.011^{***}$ & $\textbf{0.008}^{***}$ & $0.001$ & $0.023^{***}$ \\ 
			& $(0.047)$ & $(0.102)$ & $(0.122)$ & $(0.000)$ & $(0.000)$ & $(0.176)$ & $(0.000)$ \\ 
			40-49 & $$-$0.001$ & $$-$0.006^{**}$ & $$-$0.022^{***}$ & $$-$0.003^{*}$ & $\textbf{0.011}^{***}$ & $0.0002$ & $0.021^{***}$ \\ 
			& $(0.428)$ & $(0.024)$ & $(0.000)$ & $(0.074)$ & $(0.000)$ & $(0.274)$ & $(0.000)$ \\ 
			\hline \\[-1.8ex] 
			
		\end{tabular}
		\begin{tabular}{@{\extracolsep{5pt}} cccccccc} 
			\\[-1.8ex]
			\multicolumn{8}{c}{\textbf{Males - South}}\\\\[-1.8ex] 
			\hline \\[-1.8ex] 
			& SE & TE & PE & U & inactive  & EDU & FS \\ 
			\hline \\[-1.8ex] 
			20-24 & $$-$0.003$ & $$-$0.025^{***}$ & $0.012^{*}$ & $$-$0.013$ & $0.004$ & $0.017$ & $0.009^{***}$ \\ 
			& $(0.289)$ & $(0.000)$ & $(0.080)$ & $(0.102)$ & $(0.351)$ & $(0.114)$ & $(0.000)$ \\ 
			25-29 & $0.005$ & $$-$0.003$ & $$-$0.016^{*}$ & $$-$0.033^{***}$ & $0.013$ & $\textbf{0.024}^{***}$ & $0.012^{***}$ \\ 
			& $(0.266)$ & $(0.385)$ & $(0.067)$ & $(0.000)$ & $(0.116)$ & $(0.002)$ & $(0.000)$ \\
			30-39 & $$-$0.007$ & $$-$0.003$ & $$-$0.027^{***}$ & $$-$0.015^{***}$ & $\textbf{0.023}^{***}$ & $\textbf{0.006}^{**}$ & $0.023^{***}$ \\
			& $(0.172)$ & $(0.315)$ & $(0.001)$ & $(0.002)$ & $(0.000)$ & $(0.010)$ & $(0.000)$ \\  
			40-49 & $$-$0.005$ & $$-$0.002$ & $$-$0.011^{*}$ & $$-$0.015^{***}$ & $\textbf{0.012}^{**}$ & $0.0002$ & $0.022^{***}$ \\ 
			& $(0.214)$ & $(0.312)$ & $(0.095)$ & $(0.000)$ & $(0.031)$ & $(0.340)$ & $(0.000)$ \\
			\hline \\[-1.8ex]
		\end{tabular} 
		
		\begin{tabular}{@{\extracolsep{5pt}} cccccccc} 
			\\[-1.8ex]
			\multicolumn{8}{c}{\textbf{Females - North}}\\\\[-1.8ex] 
			\hline \\[-1.8ex] 
			& SE & TE & PE & U & inactive  & EDU & FS \\ 
			\hline \\[-1.8ex] 
			20-24 & $$-$0.004$ & $$-$0.024^{***}$ & $$-$0.00000$ & $$-$0.001$ & $\textbf{0.021}^{***}$ & $0.002$ & $0.007^{***}$ \\ 
			& $(0.153)$ & $(0.000)$ & $(0.482)$ & $(0.417)$ & $(0.000)$ & $(0.443)$ & $(0.000)$ \\ 
			25-29 & $$-$0.002$ & $$-$0.023^{***}$ & $$-$0.017^{*}$ & $0.003$ & $\textbf{0.024}^{***}$ & $$-$0.010^{*}$ & $0.025^{***}$ \\
			& $(0.355)$ & $(0.001)$ & $(0.055)$ & $(0.307)$ & $(0.001)$ & $(0.093)$ & $(0.000)$ \\  
			30-39 & $$-$0.005^{}$ & $$-$0.012^{***}$ & $$-$0.025^{***}$ & $$-$0.010^{***}$ & $\textbf{0.031}^{***}$ & $$-$0.001$ & $0.022^{***}$ \\ 
			& $(0.131)$ & $(0.000)$ & $(0.000)$ & $(0.000)$ & $(0.000)$ & $(0.311)$ & $(0.000)$ \\ 
			40-49 & $$-$0.005$ & $$-$0.012^{***}$ & $$-$0.009^{*}$ & $$-$0.007^{***}$ & $\textbf{0.012}^{***}$ & $$-$0.0004$ & $0.021^{***}$ \\ 
			& $(0.135)$ & $(0.000)$ & $(0.064)$ & $(0.001)$ & $(0.000)$ & $(0.215)$ & $(0.000)$ \\
			\hline \\[-1.8ex]
		\end{tabular} 
		
		\begin{tabular}{@{\extracolsep{5pt}} cccccccc} 
			\\[-1.8ex]
			\multicolumn{8}{c}{\textbf{Female - South}}\\\\[-1.8ex] 
			\hline \\[-1.8ex] 
			& SE & TE & PE & U & inactive  & EDU & FS \\ 
			\hline \\[-1.8ex] 
			20-24 & $$-$0.005$ & $$-$0.024^{***}$ & $$-$0.009^{*}$ & $$-$0.010$ & $$-$0.005$ & $\textbf{0.041}^{***}$ & $0.011^{***}$ \\ 
			& $(0.159)$ & $(0.000)$ & $(0.088)$ & $(0.157)$ & $(0.340)$ & $(0.000)$ & $(0.000)$ \\ 
			25-29 & $$-$0.012^{*}$ & $$-$0.021^{***}$ & $0.004$ & $$-$0.033^{***}$ & $\textbf{0.042}^{**}$ & $0.008$ & $0.011$ \\ 
			& $(0.016)$ & $(0.000)$ & $(0.327)$ & $(0.000)$ & $(0.000)$ & $(0.230)$ & $(0.000)$ \\ 
			30-39 & $0.002$ & $0.003$ & $$-$0.024^{***}$ & $$-$0.016^{***}$ & $\textbf{0.016}^{**}$ & $0.003$ & $0.016^{***}$ \\ 
			& $(0.363)$ & $(0.260)$ & $(0.000)$ & $(0.001)$ & $(0.045)$ & $(0.147)$ & $(0.000)$ \\ 
			40-49 & $0.005$ & $0.001$ & $$-$0.011^{*}$ & $0.005$ & $$-$0.010$ & $0.0004$ & $0.011^{***}$ \\ 
			& $(0.176)$ & $(0.400)$ & $(0.060)$ & $(0.162)$ & $(0.125)$ & $(0.355)$ & $(0.000)$ \\ 
			\hline \hline \\[-1.8ex] 
			\multicolumn{8}{l}{\textit{Note}: The attained significance levels (ASL)  of the null hypothesis of equality between the  shares}\\  \multicolumn{8}{l}{ in the two periods computed using 1000 bootstraps are reported in parenthesis}\\  \multicolumn{8}{l}{\citep[p.220]{efron1994introduction}; North includes regions in the North and the Center.}\\  \multicolumn{8}{l}{ $^{*}$ASL$<$0.1; $^{**}$ASL$<$0.05; $^{***}$ASL$<$0.01.}\\
		\end{tabular} 
	}
\end{table} 

\clearpage

\subsection{Transition probabilities of females by age and geographical location}

\clearpage

\begin{figure}[!htbp]
	\caption{Transition probabilities of females aged 30-39 from temporary employment to the inactive in the North and South of Italy by household size.}
	\label{appfig:transProbHHsizefromTE}
	\centering
	\begin{subfigure}[t]{0.3\textwidth}
		\centering
		\includegraphics[width=\linewidth]{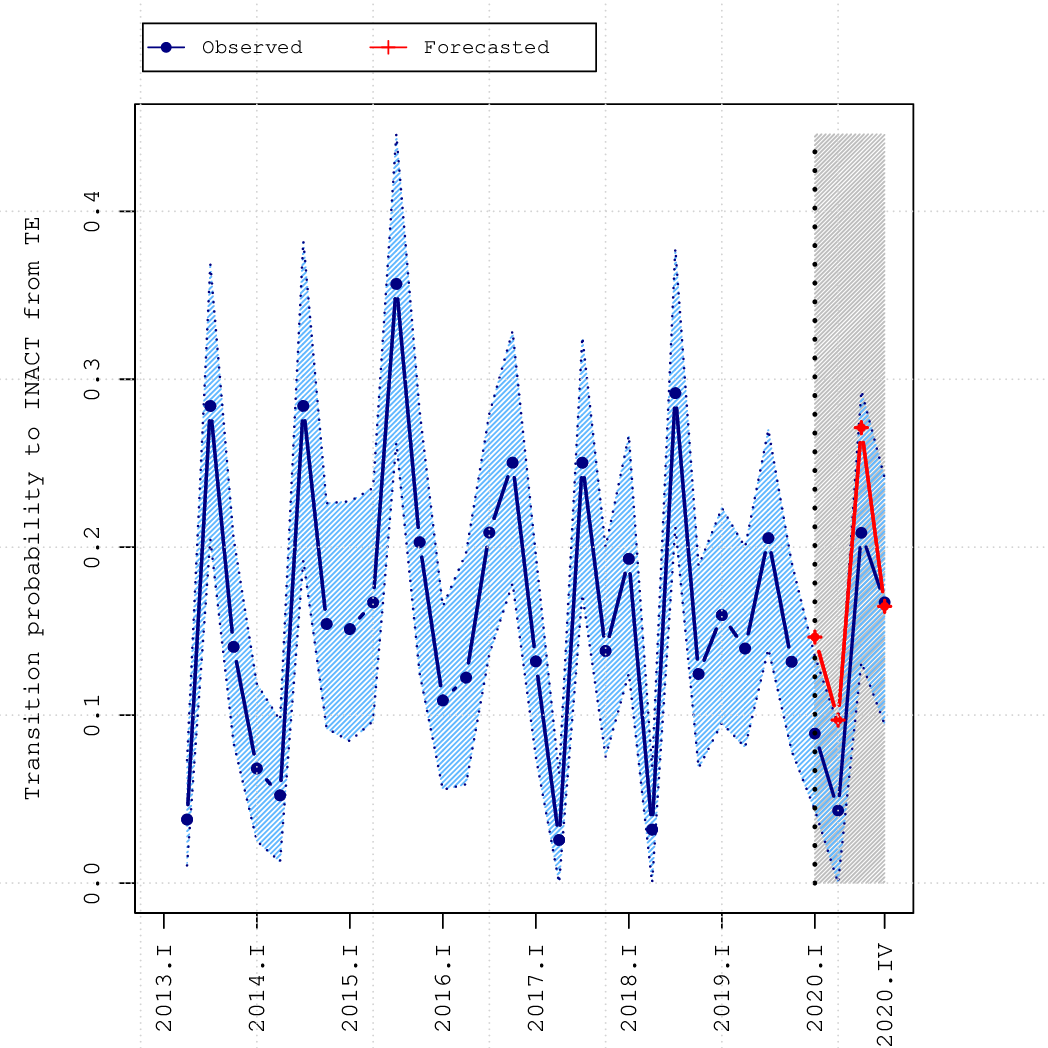}
		\caption{South - HH $>$ 2.}
		\label{fig:transProbFromEDUtoPE_IIIIIIIIIIIIIIIIIIIIIIIIIIIIIIIIIII}
	\end{subfigure}
	\begin{subfigure}[t]{0.3\textwidth}
		\centering
		\includegraphics[width=\linewidth]{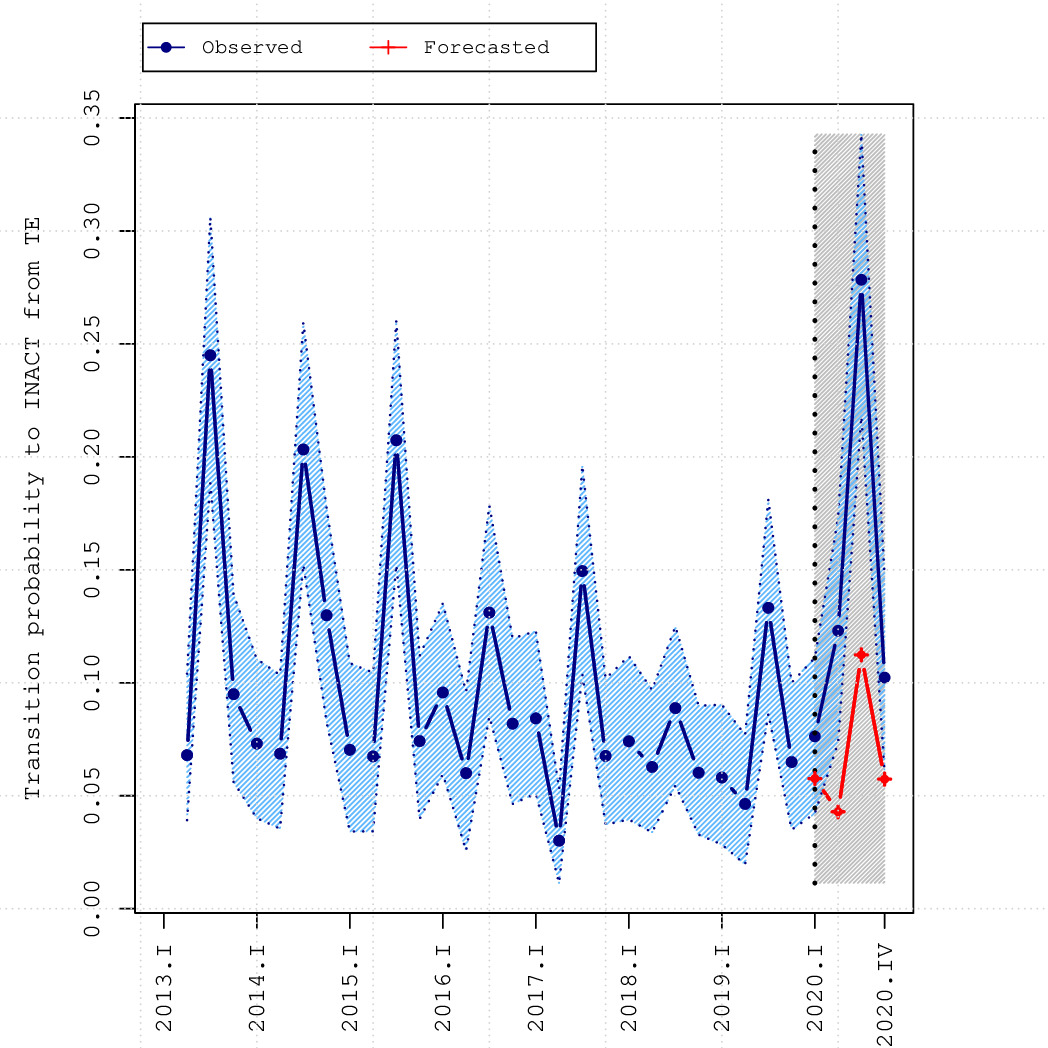}
		\caption{North - HH $>$ 2.}
		\label{fig:transProbFromEDUtoSE}
		\vspace{0.1cm}
	\end{subfigure}\\
	\begin{subfigure}[t]{0.3\textwidth}
		\centering
		\includegraphics[width=\linewidth]{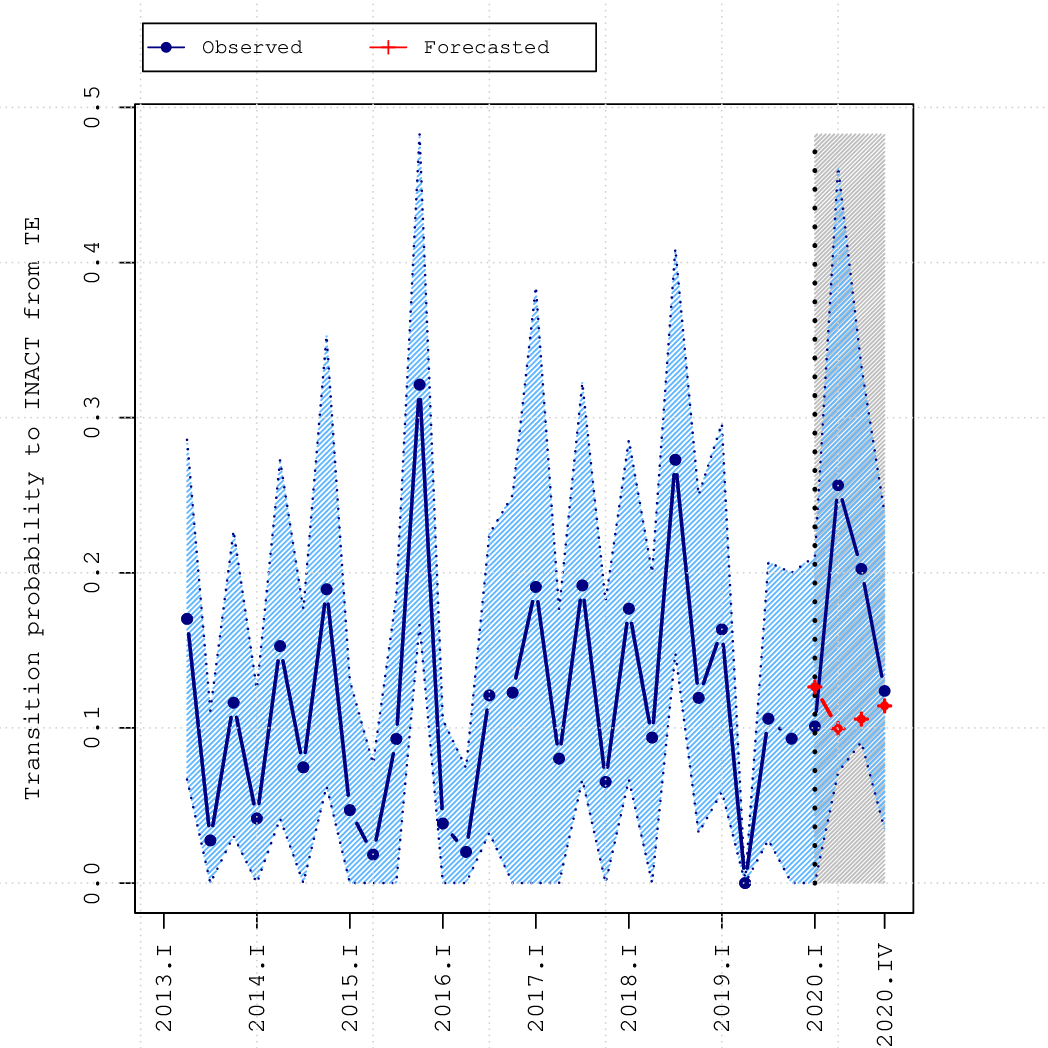}
		\caption{South - HH $\leq$ 2.}
		\label{fig:transProbFromEDUtoPE_IIIIIIIIIIIIIIIIIIIIIIIIIIIIIIIIIIII}
	\end{subfigure}
	\begin{subfigure}[t]{0.3\textwidth}
		\centering
		\includegraphics[width=\linewidth]{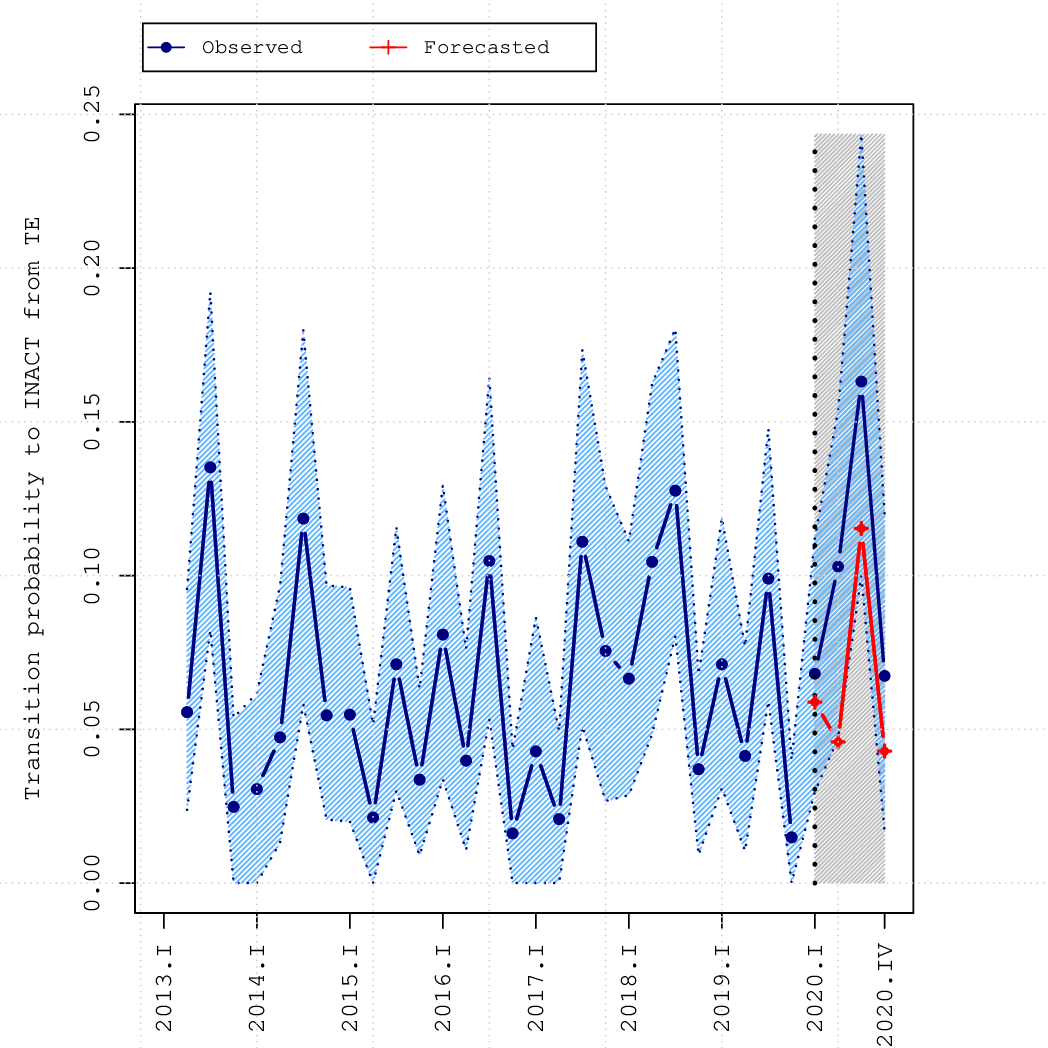}
		\caption{North - HH $\leq$ 2.}
		\label{fig:transProbFromEDUtoPE_IIIIIIIIIIIIIIIIIIIIIIIIIIIIIIIIIIIII}
	\end{subfigure}
	\vspace{0.2cm}
	\caption*{\scriptsize{\textit{Note}: Confidence intervals at 90\% are computed using 1000 bootstraps. The gray area identifies the COVID period. North includes regions in the North and the Center. \textit{Source}: LFS 3-month longitudinal data as provided by the Italian Institute of Statistics (ISTAT).}}
\end{figure}

\clearpage

\subsection{Transition probabilities of males by age and geographical location}

\begin{figure}[!htbp]
	\caption{Transition probabilities of males  from unemployment to the inactive state by age groups.}
	\label{appfig:tranProbUtoinactiveMales}
	\centering
	\vspace{0.1cm}
	\caption*{\scriptsize{\textbf{Age 30-39}}.}
	\vspace{0.1cm}
	\begin{subfigure}[t]{0.3\textwidth}
		\centering
		\includegraphics[width=\linewidth]{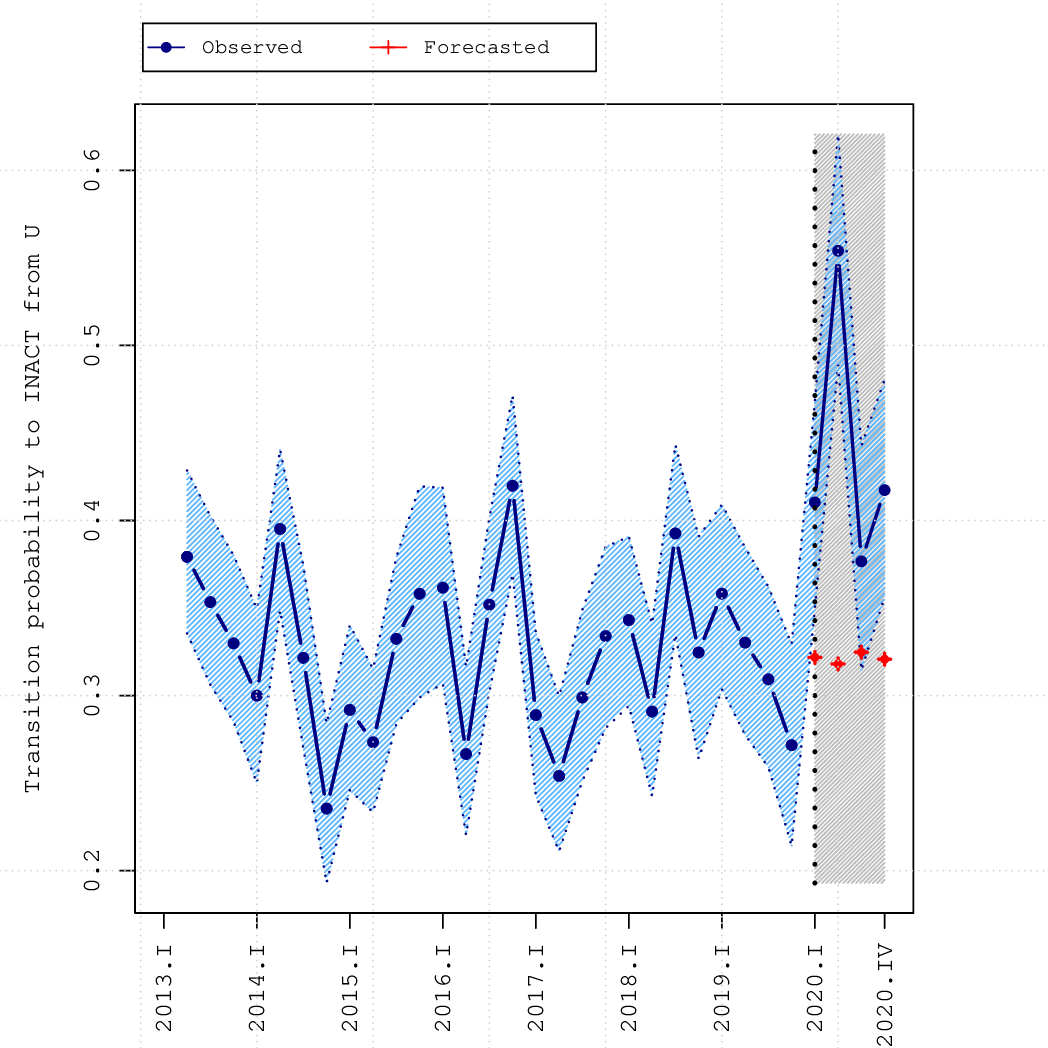}
		\caption{South.}
		\label{fig:transProbFromEDUtoPEApp_IIIIIIIIIIIIIIIIIIIIIIIIIIIIIII}
	\end{subfigure}
	\begin{subfigure}[t]{0.3\textwidth}
		\centering
		\includegraphics[width=\linewidth]{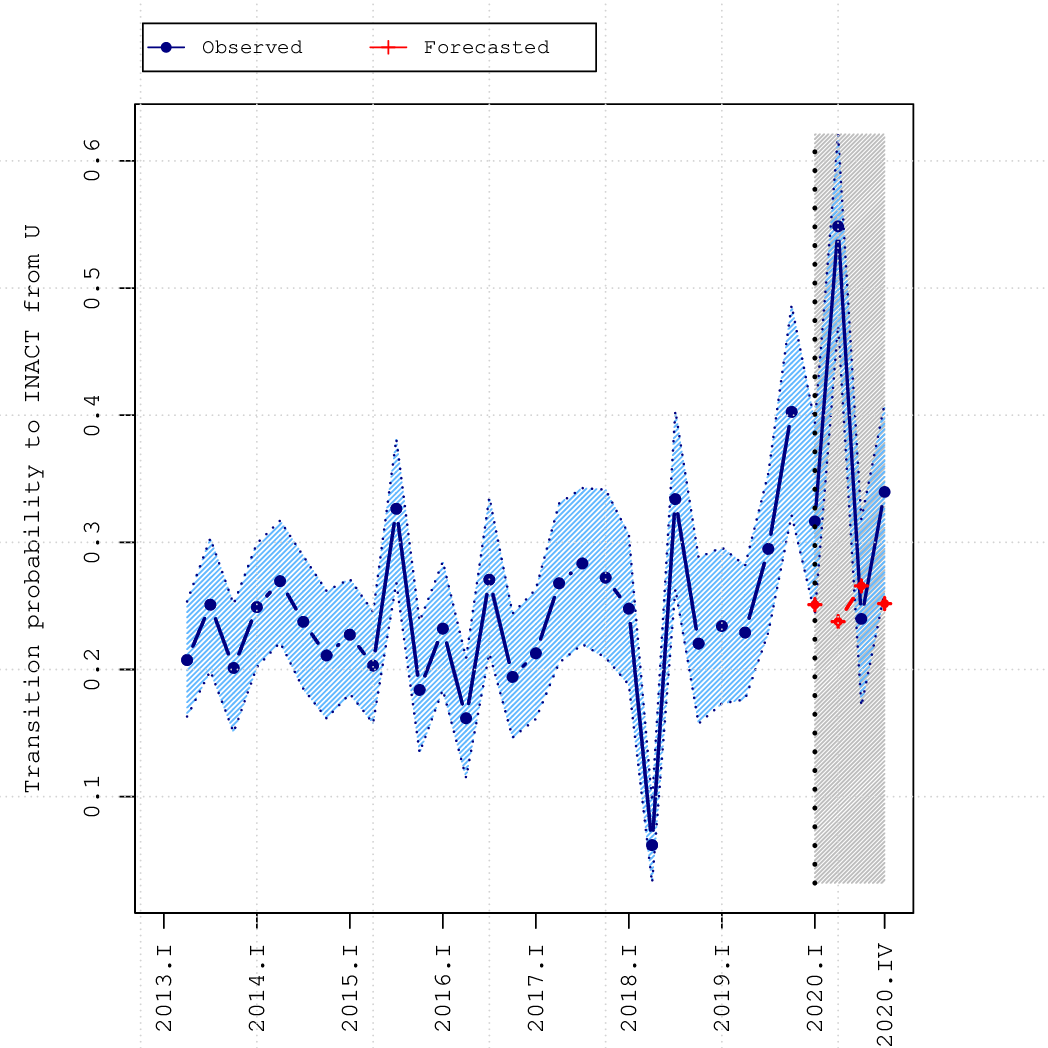}
		\caption{North.}
		\label{fig:transProbFromEDUtoUApp_IIIIIIIIIIIIIIIIIIIIIIIIIIIIIII}
	\end{subfigure}
	\vspace{0.1cm}
	\caption*{\scriptsize{\textbf{Age 40-49}}.}
	\vspace{0.1cm}
	\begin{subfigure}[t]{0.3\textwidth}
		\centering
		\includegraphics[width=\linewidth]{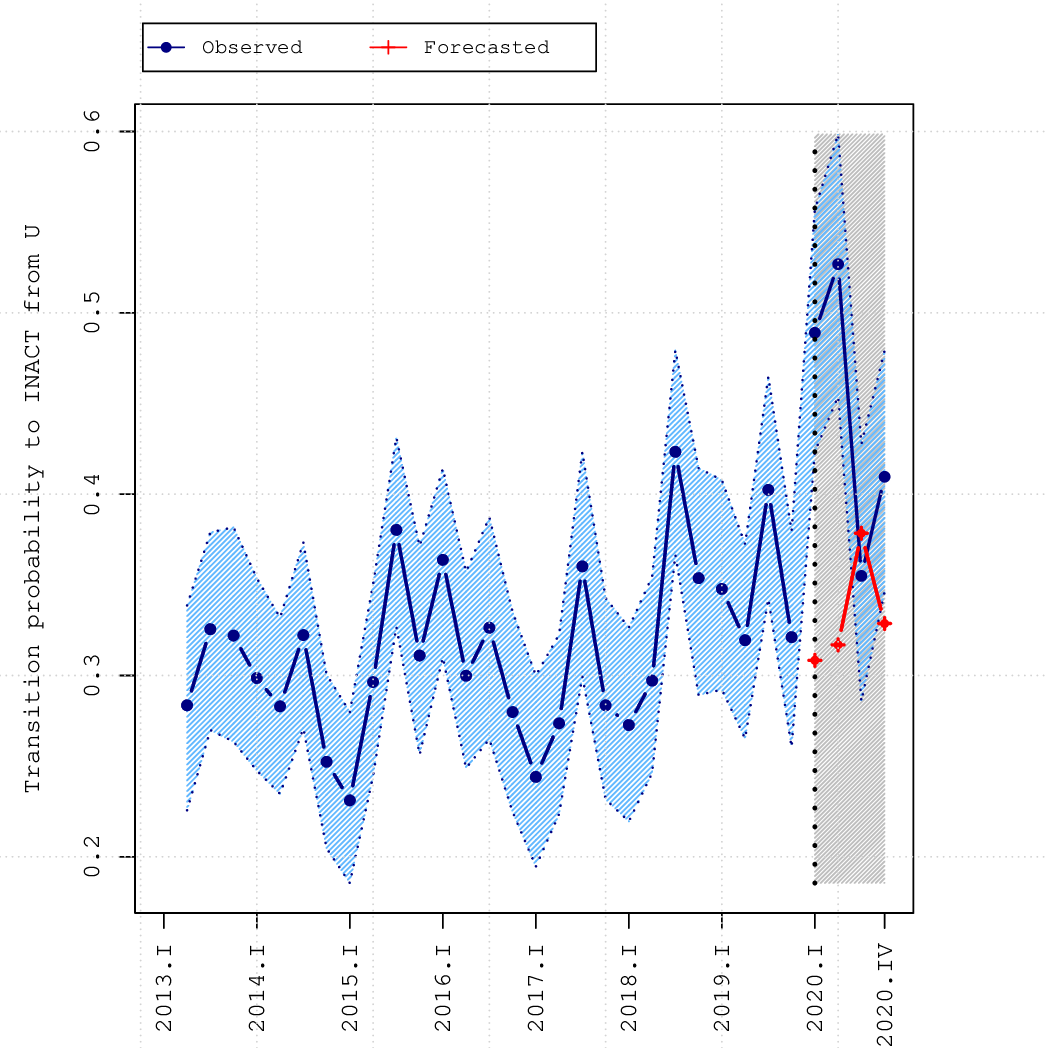}
		\caption{South.}
		\label{fig:transProbFromEDUtoPEApp_IIIIIIIIIIIIIIIIIIIIIIIIIIIIIIII}
	\end{subfigure}
	\begin{subfigure}[t]{0.3\textwidth}
		\centering
		\includegraphics[width=\linewidth]{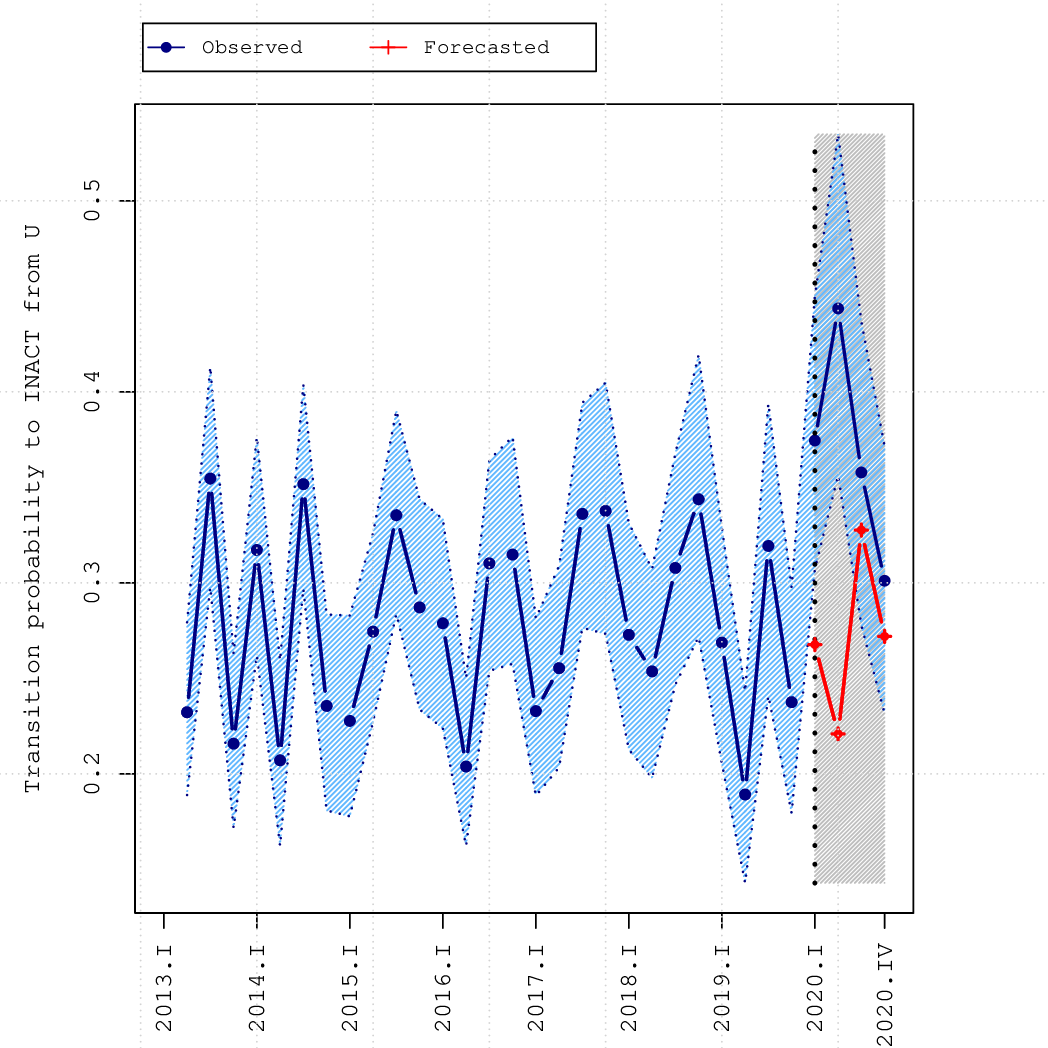}
		\caption{North.}
		\label{fig:transProbFromEDUtoUApp_IIIIIIIIIIIIIIIIIIIIIIIIIIIIIIII}
	\end{subfigure}
	\vspace{0.2cm}
	\caption*{ \scriptsize{\textit{Note}: The forecasted transition probabilities are computed using a combination of four forecasting models (ETS, TSLM, THETAF, and ARIMA) \citep{HyndmanAthanasopoulosforecasting2021} in the period 2013 (quarter I)- 2019 (quarter IV). Confidence intervals at 90\% are computed using 1000 bootstraps and reported in parenthesis. The gray area identifies the COVID period. North includes regions in the North and the Center.  \textit{Source}: LFS 3-month longitudinal data as provided by the Italian Institute of Statistics (ISTAT).}}
\end{figure}

\clearpage

\begin{figure}[!htbp]
	\caption{Transition probabilities of males from temporary employment to the inactive  state by age groups.}
	\label{appfig:transProbTEtoinactiveMales}
	\vspace{0.1cm}
	\caption*{\scriptsize{\textbf{Age 30-39}}.}
	\vspace{0.1cm}
	\centering
	\begin{subfigure}[t]{0.3\textwidth}
		\centering
		\includegraphics[width=\linewidth]{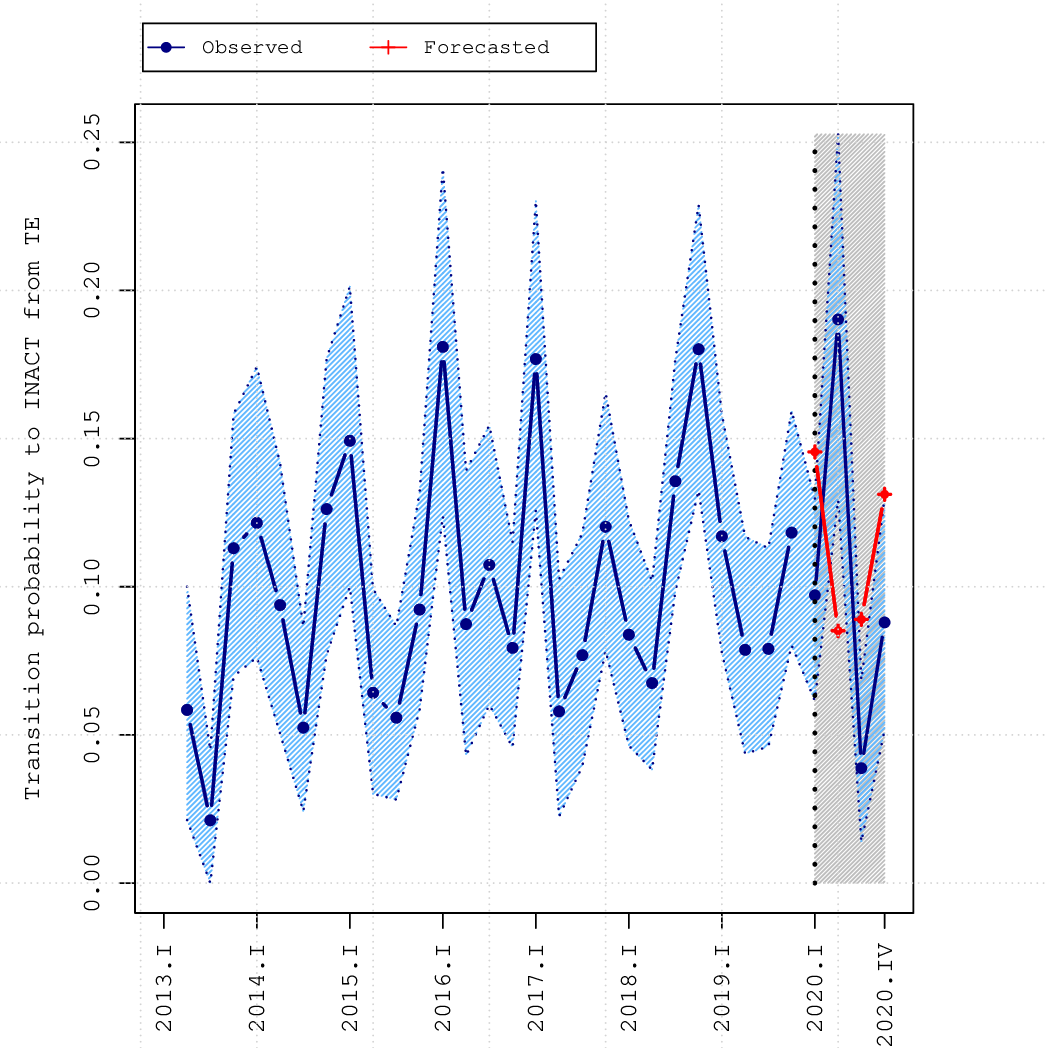}
		\caption{South.}
		\label{fig:transProbFromEDUtoPEApp_IIIIIIIIIIIIIIIIIIIIIIIIIIIIIIIII}
	\end{subfigure}
	\begin{subfigure}[t]{0.3\textwidth}
		\centering
		\includegraphics[width=\linewidth]{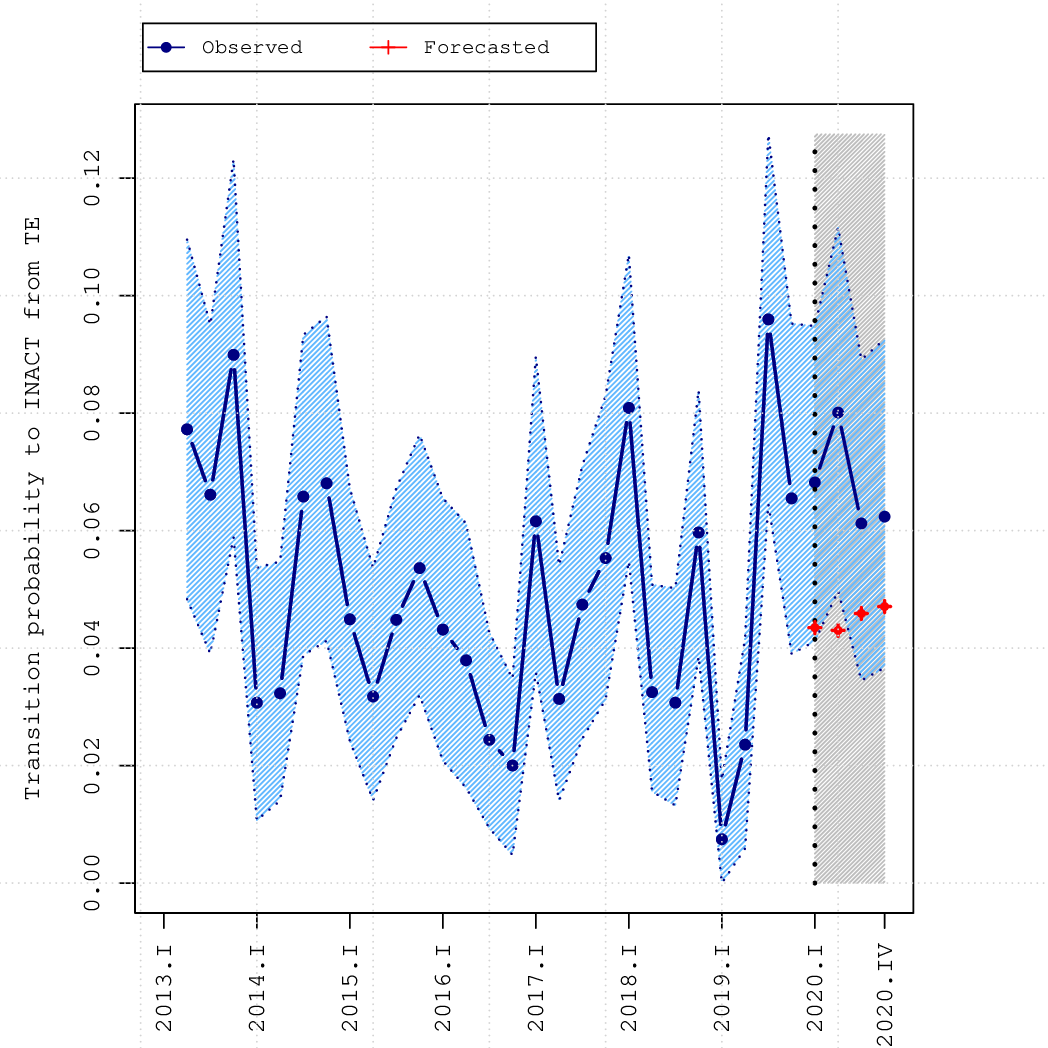}
		\caption{North.}
		\label{fig:transProbFromEDUtoUApp_IIIIIIIIIIIIIIIIIIIIIIIIIIIIIIIII}
	\end{subfigure}
	\vspace{0.1cm}
	\caption*{\scriptsize{\textbf{Age 40-49}}.}
	\vspace{0.1cm}
	\begin{subfigure}[t]{0.3\textwidth}
		\centering
		\includegraphics[width=\linewidth]{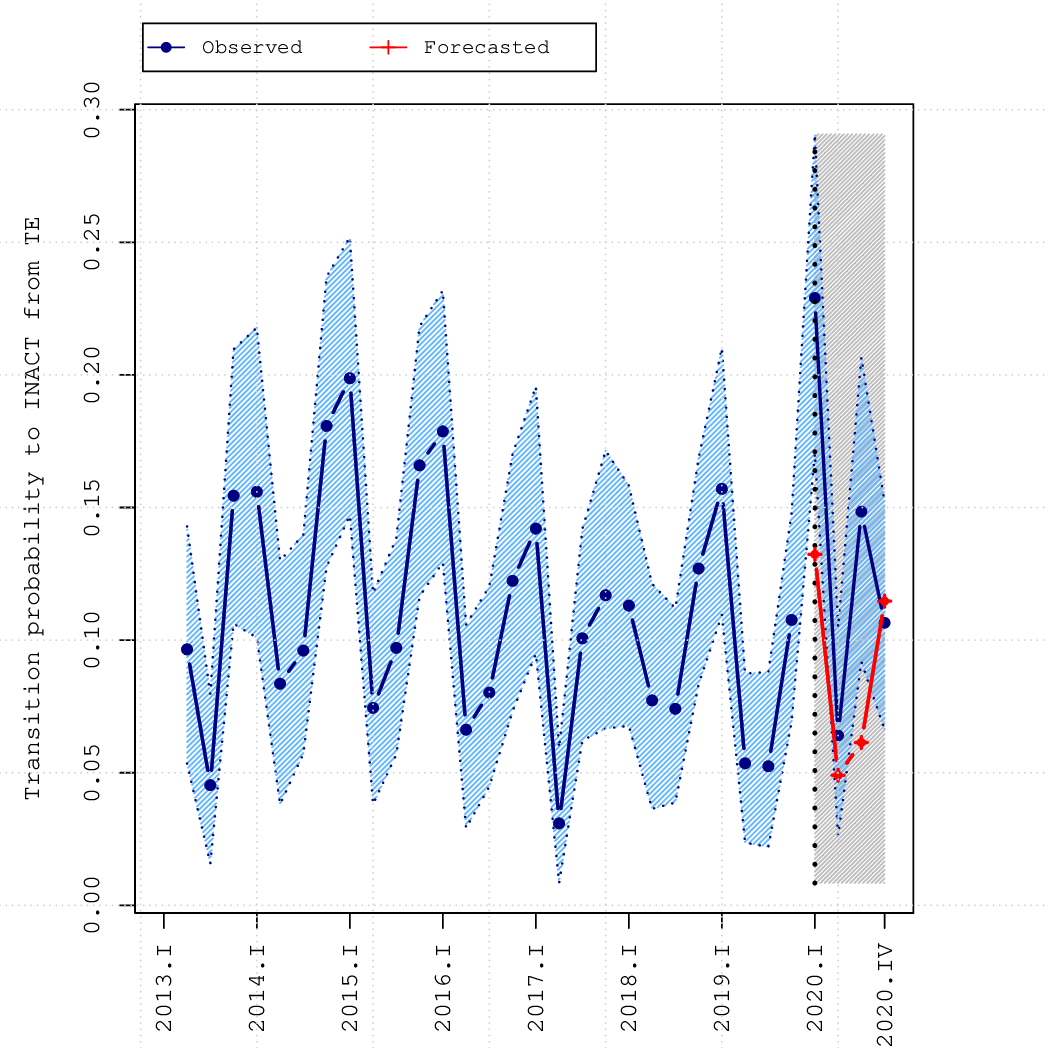}
		\caption{South.}
		\label{fig:transProbFromEDUtoPEApp_IIIIIIIIIIIIIIIIIIIIIIIIIIIIIIIIII}
	\end{subfigure}
	\begin{subfigure}[t]{0.3\textwidth}
		\centering
		\includegraphics[width=\linewidth]{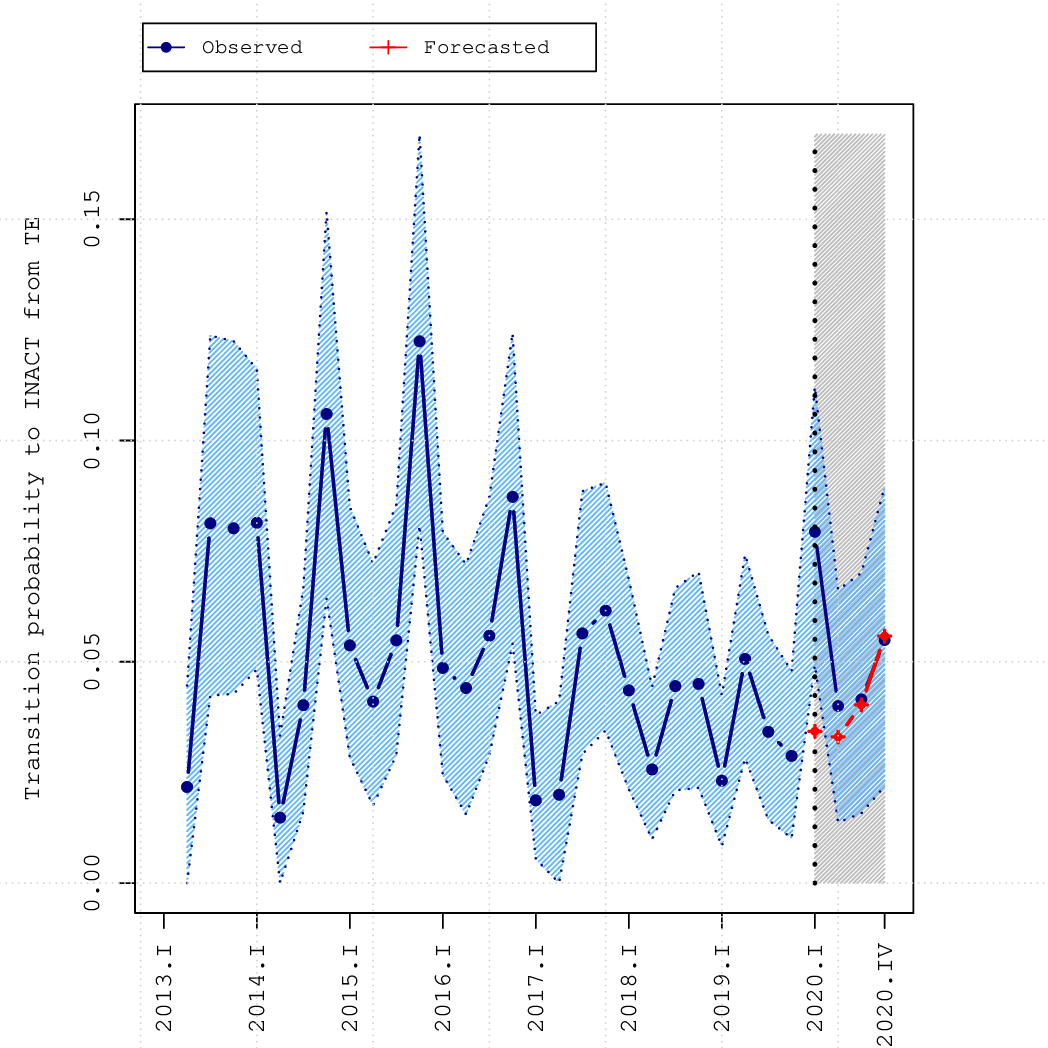}
		\caption{North.}
		\label{fig:transProbFromEDUtoUApp_IIIIIIIIIIIIIIIIIIIIIIIIIIIIIIIIII}
	\end{subfigure}
	\vspace{0.2cm}
	\caption*{ \scriptsize{\textit{Note}: The forecasted transition probabilities are computed using a combination of four forecasting models (ETS, TSLM, THETAF, and ARIMA) \citep{HyndmanAthanasopoulosforecasting2021} in the period 2013 (quarter I)- 2019 (quarter IV). Confidence intervals at 90\% are computed using 1000 bootstraps and reported in parenthesis. The gray area identifies the COVID period. North includes regions in the North and the Center. \textit{Source}: LFS 3-month longitudinal data as provided by the Italian Institute of Statistics (ISTAT).}}
\end{figure}

\clearpage

\begin{figure}[!htbp]
	\caption{Transition probabilities of males from permanent employment to the inactive  state by age groups.}
	\label{appfig:transProbPEtoinactiveMales}
	\vspace{0.1cm}
	\caption*{\scriptsize{\textbf{Age 30-39}}.}
	\vspace{0.1cm}
	\centering
	\begin{subfigure}[t]{0.3\textwidth}
		\centering
		\includegraphics[width=\linewidth]{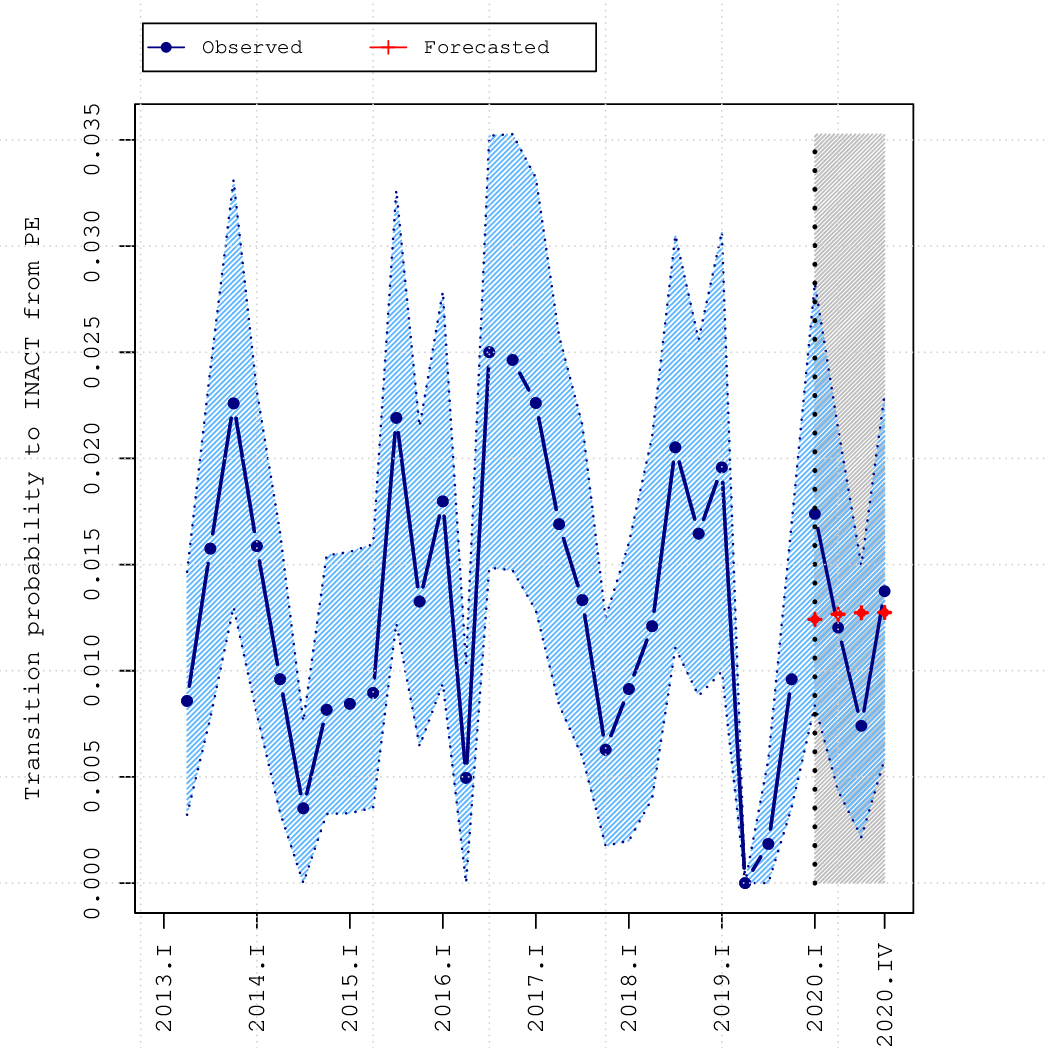}
		\caption{Male South.}
		\label{fig:transProbFromEDUtoPEApp_IIIIIIIIIIIIIIIIIIIIIIIIIIIIIIIIIII}
	\end{subfigure}
	\begin{subfigure}[t]{0.3\textwidth}
		\centering
		\includegraphics[width=\linewidth]{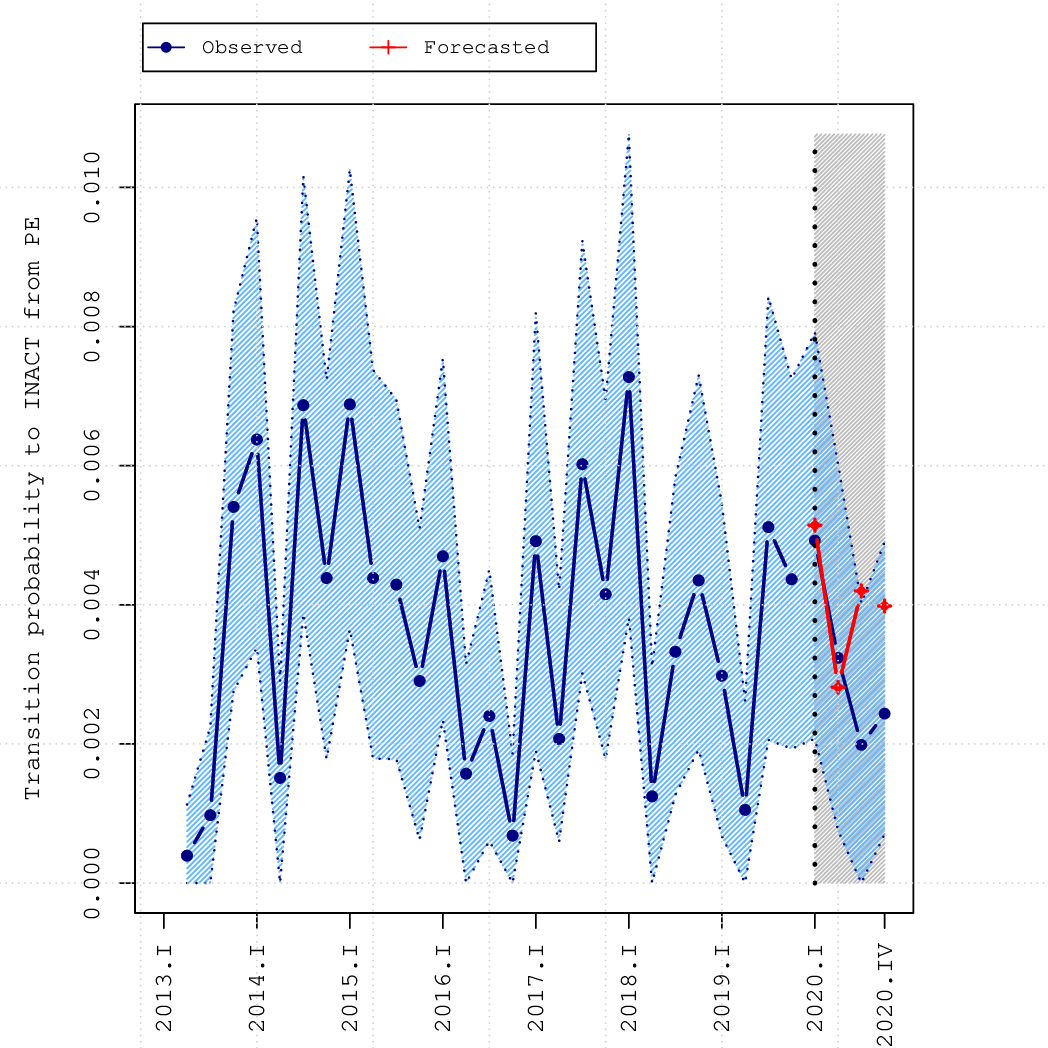}
		\caption{Male North.}
		\label{fig:transProbFromEDUtoUApp_IIIIIIIIIIIIIIIIIIIIIIIIIIIIIIIIIII}
	\end{subfigure}
	\vspace{0.1cm}
	\caption*{\scriptsize{\textbf{Age 40-49}}.}
	\vspace{0.1cm}
	\begin{subfigure}[t]{0.3\textwidth}
		\centering
		\includegraphics[width=\linewidth]{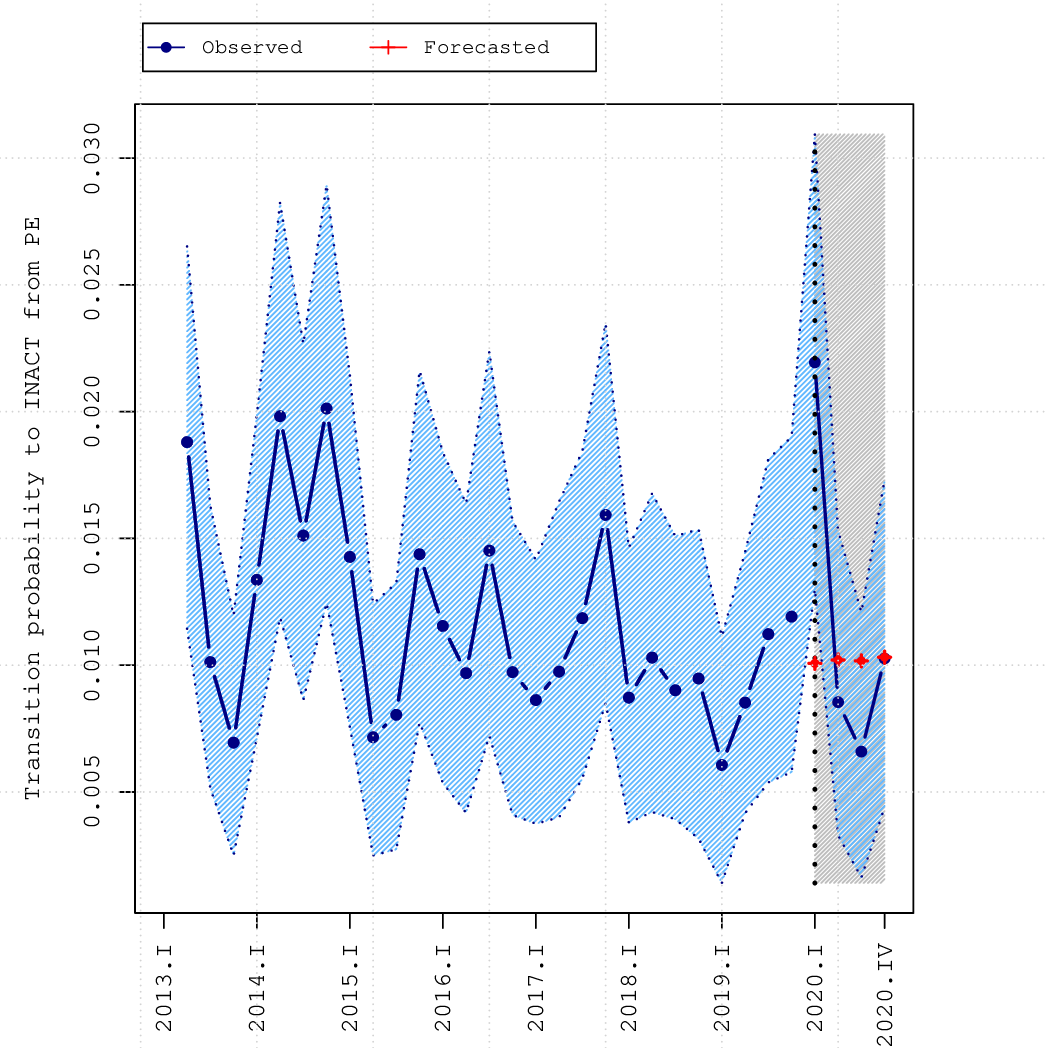}
		\caption{Male South.}
		\label{fig:transProbFromEDUtoPEApp}
	\end{subfigure}
	\begin{subfigure}[t]{0.3\textwidth}
		\centering
		\includegraphics[width=\linewidth]{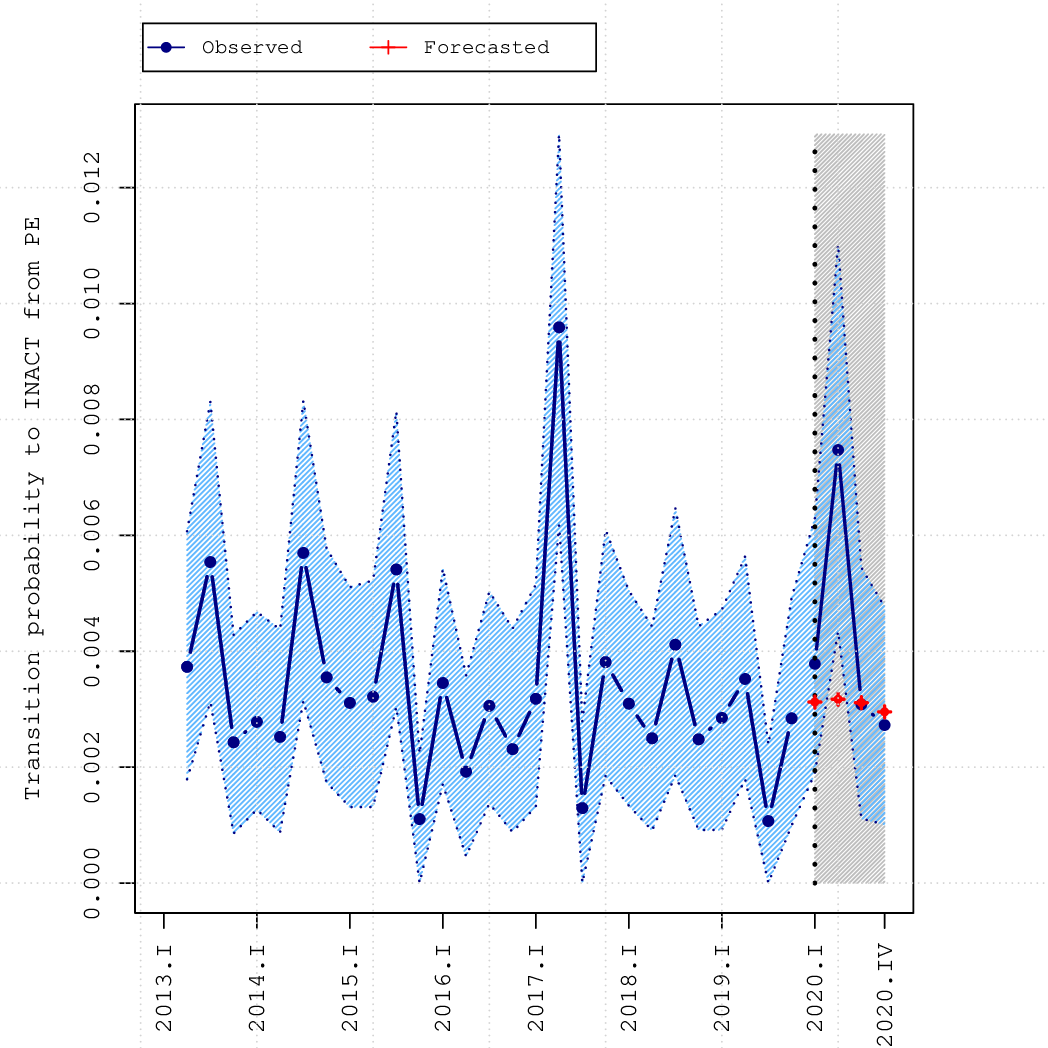}
		\caption{Male North.}
		\label{fig:transProbFromEDUtoUApp}
	\end{subfigure}
	\vspace{0.2cm}
	\caption*{ \scriptsize{\textit{Note}: The forecasted transition probabilities are computed using a combination of four forecasting models (ETS, TSLM, THETAF, and ARIMA) \citep{HyndmanAthanasopoulosforecasting2021} in the period 2013 (quarter I)- 2019 (quarter IV). Confidence intervals at 90\% are computed using 1000 bootstraps and reported in parenthesis. The gray area identifies the COVID period. North includes regions in the North and the Center.  \textit{Source}: LFS 3-month longitudinal data as provided by the Italian Institute of Statistics (ISTAT).}}
\end{figure}

\clearpage

\section{Household size by categories of individuals \label{app:householdSize}}

\begin{table}[!htbp]
	\centering
	\footnotesize 
	\caption{Percentage of females with at least one child below the age of 11 by geographical area and employment status in 2020.}  
	\label{tab:percentagewomenkids} 
	\begin{tabular}{crrrr} 
		\hline 
		\hline
		\\[-1.8ex] 
		&\multicolumn{2}{c}{Age 30-39}&\multicolumn{2}{c}{Age 40-49}\\	
		\hline \\[-1.8ex]
		&North & South & North & South\\
		\hline \\[-1.8ex]
		Permanent  & 46.1&20.5&56.1&29.0\\
		Temporary &7.9&6.4&5.6&5.8\\
		Self-employed & 8.1&6.5&11.1&9.4\\
		Unemployed & 5.4 & 8.5&4.1&6.9\\
		\textbf{Inactive} &	31.8&57.1&22.3&48.3\\
		\hline 
		\\[-1.8ex]
		Total (in 000s)& 1261&715&1188&557\\
		\hline 
		\hline
		\multicolumn{5}{l}{\scriptsize{\textit{Note}: North includes regions in the North and the Center.}}\\\multicolumn{5}{l}{\scriptsize{\textit{Source}: ELFS data.}}
	\end{tabular}	
\end{table}	

\clearpage

\section{Distribution of workers by sector}\label{appsec:distributionWorkersSector}

\begin{table}[!htbp]
	\caption{Distribution of workers by sector in 2019.}
	\label{apptab:workersBySector}
	\centering
	\begin{footnotesize}
		\begin{tabular}{lcccc}
			\hline
			\hline
			\\[-1.8ex]
			& \multicolumn{2}{c}{Females 30-39}   & \multicolumn{2}{c}{Males 30-39} \\ \\[-1.8ex]
			\hline
			\\[-1.8ex]
			& South & North & South &North \\  
			\hline
			Agriculture & 22944 & 9879 & 72711 & 57287 \\ 
			Industry & 28505 & 245513 & 151164 & 561974 \\ 
			Constructions & 3652 & 25569 & 96395 & 207688 \\ 
			Commerce & 89911 & 247142 & 168699 & 251949 \\ 
			Hotels and restaurants & 38826 & 111932 & 47009 & 113679 \\ 
			Transport & 7061 & 50319 & 62694 & 135119 \\ 
			Communications & 4708 & 25661 & 23436 & 104134 \\ 
			Finance & 13598 & 69859 & 11318 & 54810 \\ 
			Real estate & 82127 & 274795 & 86161 & 251191 \\ 
			Public administration & 7797 & 14500 & 32488 & 69909 \\ 
			Education and health & 110635 & 368974 & 51503 & 118223 \\ 
			Others & 73041 & 151200 & 46293 & 80096 \\ 
			\hline
			\hline
			\multicolumn{5}{l}{\scriptsize{ \textit{Source}: LFS 3-month longitudinal data as provided by the Italian Institute}}\\ \multicolumn{5}{l}{\scriptsize{ of Statistics (ISTAT).}}
		\end{tabular}
	\end{footnotesize}
\end{table}

\begin{table}[!htbp]
	\caption{Distribution of workers by sector in 2019 ($\%$).}
	\label{apptab:workersBySectorPercentage}
	\centering
	\begin{footnotesize}
		\begin{tabular}{lcccc}
			\hline
			\hline
			\\[-1.8ex]
			& \multicolumn{2}{c}{Females 30-39}   & \multicolumn{2}{c}{Males 30-39} \\ \\[-1.8ex]
			\hline
			\\[-1.8ex]
			& South & North & South &North \\  
			\hline
			Agriculture  & 0.05 & 0.01 & 0.09 & 0.03 \\ 
			Industry & 0.06 & 0.15 & 0.18 & 0.28 \\ 
			Constructions & 0.01 & 0.02 & 0.11 & 0.10 \\ 
			Commerce & 0.19 & 0.15 & 0.20 & 0.13 \\ 
			Hotels and restaurants & 0.08 & 0.07 & 0.06 & 0.06 \\ 
			Transport & 0.01 & 0.03 & 0.07 & 0.07 \\ 
			Communications & 0.01 & 0.02 & 0.03 & 0.05 \\ 
			Finance & 0.03 & 0.04 & 0.01 & 0.03 \\ 
			Real estate & 0.17 & 0.17 & 0.10 & 0.13 \\ 
			Public administration & 0.02 & 0.01 & 0.04 & 0.03 \\ 
			Education and health & 0.23 & 0.23 & 0.06 & 0.06 \\ 
			Others & 0.15 & 0.09 & 0.05 & 0.04 \\ 
			\hline
			All  & 1 & 1 & 1 & 1 \\
			\hline
			\hline
			\multicolumn{5}{l}{\scriptsize{ \textit{Source}: LFS 3-month longitudinal data as provided by the Italian Institute}}\\ \multicolumn{5}{l}{\scriptsize{ of Statistics (ISTAT).}}
		\end{tabular}
	\end{footnotesize}
\end{table}

\clearpage

\section{Logit estimates} \label{app:logit}

\begin{table}[!htbp] \centering 
	\caption{Odds-ratios of being active next quarter for an individual \textbf{currently active} in the labour market (\textbf{females age 30-39}).} 
	\label{apptab:logit3039femalepart} 
	\begin{footnotesize}
		\begin{tabular}{lccccc}
			\hline \hline
			\\[-1.8ex]
			& C.I. & C.I.  & Mean & C.I.  & C.I.  \\ 
			&  2.5\% &  5\% &   bootstrap &  95\% &  97.5\% \\ 
			\\[-1.8ex]
			\hline
			\\[-1.8ex]
			\textbf{	COVID $\times$ North $\times$}  &&&&& \\
			\textbf{ $\times$ Household members$>$2} & 0.445 & 0.470 & \textbf{0.690} & 0.964 & 1.035 \\
			\hline
			\\[-1.8ex]
			Constant & 5.726 & 5.960 & \textbf{7.562} & 9.600 & 10.040 \\ 
			2014 & 0.887 & 0.907 & 1.025 & 1.147 & 1.171 \\ 
			2015 & 0.967 & 0.989 & 1.118 & 1.251 & 1.283 \\ 
			2016 & 0.926 & 0.948 & 1.069 & 1.193 & 1.220 \\ 
			2017 & 1.016 & 1.043 & 1.177 & 1.316 & 1.346 \\ 
			2018 & 1.072 & 1.097 & 1.244 & 1.405 & 1.443 \\ 
			2019 & 0.691 & 0.720 & 0.887 & 1.067 & 1.112 \\ 
			2020 & 0.475 & 0.509 & 0.886 & 1.436 & 1.574 \\ 
			Quarter II & 0.814 & 0.827 & 0.908 & 0.991 & 1.009 \\ 
			Quarter III & 0.667 & 0.680 & \textbf{0.746} & 0.819 & 0.829 \\ 
			Quarter IV & 0.865 & 0.880 & 0.966 & 1.057 & 1.070 \\ 
			EU citizen & 0.939 & 0.960 & 1.109 & 1.290 & 1.327 \\ 
			No EU citizen & 0.789 & 0.805 & 0.915 & 1.040 & 1.066 \\ 
			Industry & 2.379 & 2.450 & \textbf{2.984} & 3.542 & 3.676 \\ 
			Constructions & 1.915 & 2.015 & \textbf{2.886} & 3.991 & 4.294 \\ 
			Commerce & 2.667 & 2.753 & \textbf{3.303} & 3.923 & 4.036 \\ 
			Hotels and restaurants & 1.130 & 1.170 & \textbf{1.400} & 1.672 & 1.701 \\ 
			Transport & 2.382 & 2.486 & \textbf{3.377} & 4.555 & 4.792 \\ 
			Communications & 2.210 & 2.319 & \textbf{3.226} & 4.394 & 4.700 \\ 
			Finance & 3.385 & 3.619 & \textbf{4.917} & 6.608 & 7.105 \\ 
			Real estate & 2.450 & 2.539 & \textbf{3.053} & 3.631 & 3.749 \\ 
			Public administration & 3.276 & 3.492 & \textbf{5.111} & 7.299 & 7.907 \\ 
			Education and health & 1.709 & 1.760 & \textbf{2.077} & 2.457 & 2.537 \\ 
			Others & 1.574 & 1.624 & \textbf{1.958} & 2.320 & 2.388 \\ 
			Primary education & 0.551 & 0.564 & \textbf{0.629} & 0.694 & 0.705 \\ 
			Secondary education & 0.772 & 0.785 & \textbf{0.862 }& 0.943 & 0.959 \\ 
			North & 1.553 & 1.595 & \textbf{1.836} & 2.094 & 2.139 \\ 
			Household members$>$2 & 0.592 & 0.610 & \textbf{0.705} & 0.804 & 0.819 \\ 
			North  x Household members$>$2 & 0.833 & 0.856 & 1.009 & 1.181 & 1.222 \\
			COVID x EU citizen & 0.724 & 0.767 & 1.100 & 1.511 & 1.643 \\ 
			COVID x No EU citizen & 0.775 & 0.804 & 1.081 & 1.408 & 1.486 \\ 
			COVID x Industry & 0.358 & 0.393 & 0.674 & 1.014 & 1.098 \\ 
			COVID x Constructions & 0.175 & 0.204 & 0.548 & 1.194 & 1.412 \\ 
			COVID x Commerce & 0.295 & 0.344 & \textbf{0.572} & 0.861 & 0.936 \\ 
			COVID x Hotels and restaurants & 0.218 & 0.247 & \textbf{0.411} & 0.619 & 0.665 \\ 
			COVID x Transport & 0.161 & 0.179 & \textbf{0.362} & 0.624 & 0.700 \\ 
			COVID x Communications & 0.307 & 0.358 & 0.800 & 1.488 & 1.745 \\ 
			COVID x Finance & 0.381 & 0.442 & 1.076 & 2.068 & 2.614 \\ 
			COVID x Real estate & 0.514 & 0.560 & 0.974 & 1.458 & 1.565 \\ 
			COVID x Public administration & 0.594 & 0.668 & 2.422 & 5.852 & 6.747 \\ 
			COVID x Education and health & 0.372 & 0.413 & 0.678 & 1.010 & 1.091 \\ 
			COVID x Others & 0.425 & 0.458 & 0.777 & 1.168 & 1.265 \\ 
			COVID x Primary education & 0.853 & 0.891 & 1.162 & 1.461 & 1.543 \\ 
			COVID x Secondary education & 0.700 & 0.725 & 0.878 & 1.045 & 1.073 \\ 
			COVID x North & 0.688 & 0.730 & 1.014 & 1.356 & 1.420 \\ 
			COVID x Household members$>$2 & 0.939 & 0.993 & 1.379 & 1.811 & 1.912 \\ 
			\hline
			Observations&\multicolumn{5}{c}{57264}\\
			\hline
			\hline
			\multicolumn{6}{l}{\scriptsize{\textit{Note}: North includes regions in the North and the Center.  \textit{Source}: LFS 3-month longitudinal data }}\\\multicolumn{6}{l}{\scriptsize{as provided by the Italian Institute of Statistics (ISTAT).}}
		\end{tabular}
	\end{footnotesize}
\end{table}

\clearpage

\begin{table}[!htbp] \centering 
	\caption{Odds-ratios of being inactive next quarter for an individual \textbf{currently inactive} in the labour market (\textbf{females age 30-39}).} 
	\label{apptab:logit3039femalenonpart} 
	\begin{footnotesize}
		\begin{tabular}{lccccc}
			\hline \hline
			\\[-1.8ex]
			& C.I. & C.I.  & Mean & C.I.  & C.I.  \\ 
			&  2.5\% &  5\% &   bootstrap &  95\% &  97.5\% \\ 
			\\[-1.8ex]
			\hline
			\\[-1.8ex]
			\textbf{	COVID $\times$ North $\times$}  &&&&& \\
			\textbf{$\times$ Household members$>$2} & 0.729 & 0.774 & 1.025 & 1.345 & 1.415 \\ 
			\hline
			Constant & 0.945 & 0.968 & 1.074 & 1.189 & 1.215 \\ 
			2014 & 0.857 & 0.870 & 0.938 & 1.011 & 1.026 \\ 
			2015 & 0.931 & 0.948 & 1.022 & 1.101 & 1.119 \\ 
			2016 & 0.895 & 0.906 & 0.981 & 1.054 & 1.073 \\ 
			2017 & 0.908 & 0.918 & 0.993 & 1.076 & 1.092 \\ 
			2018 & 0.962 & 0.974 & 1.052 & 1.135 & 1.153 \\ 
			2019 & 0.960 & 0.971 & 1.054 & 1.140 & 1.156 \\ 
			2020 & 0.763 & 0.807 & 1.007 & 1.234 & 1.285 \\ 
			Quarter II & 0.983 & 0.993 & 1.050 & 1.106 & 1.118 \\ 
			Quarter III & 1.064 & 1.076 & \textbf{1.137} & 1.201 & 1.211 \\ 
			Quarter IV & 0.858 & 0.868 & \textbf{0.915} & 0.962 & 0.974 \\ 
			EU citizen & 0.975 & 0.988 & 1.074 & 1.165 & 1.180 \\ 
			No EU citizen & 1.333 & 1.348 & \textbf{1.434} & 1.527 & 1.545 \\ 
			Primary education & 2.246 & 2.265 & \textbf{2.398} & 2.531 & 2.556 \\ 
			Secondary education & 1.545 & 1.556 & \textbf{1.644} & 1.731 & 1.752 \\ 
			North  & 0.569 & 0.577 & \textbf{0.630} & 0.683 & 0.694 \\ 
			Household members$>$2 & 1.676 & 1.699 & \textbf{1.827} & 1.961 & 1.984 \\
			North  x Household members$>$2 & 1.030 & 1.047 & \textbf{1.156} & 1.270 & 1.289\\ 
			COVID x EU citizen & 0.647 & 0.669 & 0.840 & 1.045 & 1.091 \\ 
			COVID x No EU citizen & 0.949 & 0.981 & 1.189 & 1.425 & 1.469 \\ 
			COVID x Primary education & 0.833 & 0.858 & 1.021 & 1.195 & 1.226 \\ 
			COVID x Secondary education & 0.767 & 0.789 & 0.914 & 1.055 & 1.089 \\ 
			COVID x North  & 0.833 & 0.866 & 1.111 & 1.406 & 1.470 \\ 
			COVID x Household members$>$2 & 0.854 & 0.892 & 1.098 & 1.331 & 1.366  \\ 
			\hline
			Observations&\multicolumn{5}{c}{44428}\\
			\hline \hline
			\multicolumn{6}{l}{\scriptsize{\textit{Note}: North includes regions in the North and the Center. \textit{Source}: LFS 3-month longitudinal }}\\	\multicolumn{6}{l}{\scriptsize{data as provided by the Italian Institute of Statistics (ISTAT).}}
		\end{tabular}
	\end{footnotesize}
\end{table}

\clearpage

\begin{table}[!htbp] \centering 
	\caption{Odds-ratios of being active next quarter for an individual \textbf{currently active} in the labour market (\textbf{males aged 30-39}).} 
	\label{apptab:logit3039malepart} 
	\begin{footnotesize}
		\begin{tabular}{lccccc}
			\hline \hline
			\\[-1.8ex]
			& C.I. & C.I.  & Mean & C.I.  & C.I.  \\ 
			&  2.5\% &  5\% &   bootstrap &  95\% &  97.5\% \\ 
			\\[-1.8ex]
			\hline
			\\[-1.8ex]
			\textbf{	COVID $\times$ North or Center $\times$}  &&&&& \\
			\textbf{	$\times$ Household members$>$2} & 0.465 & 0.499 & 0.721 & 0.999 & 1.080 \\ 
			\\[-1.8ex]
			\hline
			\\[-1.8ex]
			Constant & 10.539 & 10.948 & \textbf{14.275} & 18.155 & 18.864 \\ 
			2014 & 0.913 & 0.933 & 1.070 & 1.212 & 1.237 \\ 
			2015 & 0.877 & 0.901 & 1.034 & 1.173 & 1.198 \\ 
			2016 & 1.046 & 1.070 & \textbf{1.228} & 1.406 & 1.444 \\ 
			2017 & 1.064 & 1.086 & \textbf{1.251} & 1.432 & 1.469 \\ 
			2018 & 0.971 & 0.993 & 1.147 & 1.305 & 1.332 \\ 
			2019 & 1.080 & 1.125 & \textbf{1.398 }& 1.729 & 1.813 \\ 
			2020 & 0.892 & 0.972 & 1.690 & 2.814 & 3.014 \\ 
			Quarter II & 0.933 & 0.960 & 1.058 & 1.163 & 1.180 \\ 
			Quarter III & 1.241 & 1.269 & \textbf{1.408} & 1.558 & 1.588 \\ 
			Quarter IV & 0.935 & 0.952 & 1.050 & 1.155 & 1.172 \\ 
			EU citizen & 0.586 & 0.599 & \textbf{0.694} & 0.809 & 0.832 \\ 
			No EU citizen & 0.833 & 0.852 & 0.961 & 1.080 & 1.111 \\ 
			Industry & 1.960 & 2.018 & \textbf{2.363} & 2.731 & 2.828 \\ 
			Constructions & 0.927 & 0.956 & 1.116 & 1.286 & 1.311 \\ 
			Commerce & 2.063 & 2.135 & \textbf{2.545} & 3.003 & 3.104 \\ 
			Hotels and restaurants & 0.593 & 0.610 & \textbf{0.723} & 0.838 & 0.861 \\ 
			Transport & 1.829 & 1.892 & \textbf{2.352} & 2.881 & 2.999 \\ 
			Communications & 1.865 & 1.965 & \textbf{2.730} & 3.714 & 3.956 \\ 
			Finance & 3.046 & 3.268 & \textbf{5.576} & 9.073 & 10.117 \\ 
			Real estate & 1.543 & 1.589 & \textbf{1.934} & 2.306 & 2.412 \\ 
			Public administration & 3.965 & 4.168 & \textbf{5.869} & 8.097 & 8.713 \\ 
			Education and health & 1.028 & 1.071 & \textbf{1.336} & 1.670 & 1.731 \\ 
			Others & 0.884 & 0.918 & 1.112 & 1.348 & 1.400 \\ 
			Primary education & 0.433 & 0.446 & \textbf{0.516} & 0.595 & 0.609 \\ 
			Secondary education & 0.641 & 0.660 & \textbf{0.753} & 0.862 & 0.877 \\ 
			North  & 1.632 & 1.670 & \textbf{1.935} & 2.204 & 2.258 \\ 
			Household members$>$2 & 0.703 & 0.721 & \textbf{0.822} & 0.931 & 0.949 \\ 
			North  x Household members$>$2 & 1.024 & 1.062 & \textbf{1.264} & 1.482 & 1.524\\
			COVID x EU citizen & 0.512 & 0.552 & 0.781 & 1.075 & 1.125 \\ 
			COVID x No EU citizen & 0.621 & 0.643 & 0.846 & 1.070 & 1.131 \\ 
			COVID x Industry & 0.160 & 0.177 & \textbf{0.288} & 0.415 & 0.443 \\ 
			COVID x Constructions & 0.259 & 0.294 & \textbf{0.504} & 0.749 & 0.802 \\ 
			COVID x Commerce & 0.152 & 0.165 & \textbf{0.275} & 0.402 & 0.425 \\ 
			COVID x Hotels and restaurants & 0.144 & 0.154 & \textbf{0.259} & 0.383 & 0.423 \\ 
			COVID x Transport & 0.169 & 0.184 & \textbf{0.329} & 0.528 & 0.565 \\ 
			COVID x Communications & 0.133 & 0.144 & \textbf{0.306} & 0.552 & 0.645 \\ 
			COVID x Finance & 0.159 & 0.190 & 0.731 & 1.792 & 2.155 \\ 
			COVID x Real estate & 0.203 & 0.228 & \textbf{0.390} & 0.597 & 0.648 \\ 
			COVID x Public administration & 0.239 & 0.287 & 0.915 & 2.437 & 2.894 \\ 
			COVID x Education and health & 0.195 & 0.224 & \textbf{0.413} & 0.663 & 0.725 \\ 
			COVID x Others & 0.257 & 0.288 & \textbf{0.525} & 0.840 & 0.908 \\ 
			COVID x Primary education & 0.914 & 0.970 & 1.352 & 1.763 & 1.875 \\ 
			COVID x Secondary education & 0.710 & 0.745 & 0.996 & 1.276 & 1.330 \\ 
			COVID x North & 0.651 & 0.688 & 0.941 & 1.232 & 1.304 \\ 
			COVID x Household members$>$2 & 0.848 & 0.895 & 1.211 & 1.559 & 1.618 \\ 
			\hline
			Observations&\multicolumn{5}{c}{71126}\\
			\hline \hline
		\multicolumn{6}{l}{\scriptsize{\textit{Note}: North includes regions in the North and the Center.  \textit{Source}: LFS 3-month longitudinal data }}\\\multicolumn{6}{l}{\scriptsize{as provided by the Italian Institute of Statistics (ISTAT).}}
		\end{tabular}
	\end{footnotesize}
\end{table}

\clearpage

\begin{table}[!htbp] \centering 
	\caption{Odds-ratios of being inactive next quarter for an individual \textbf{currently inactive} in the labour market (\textbf{males aged 30-39}).} 
	\label{apptab:logit3039malenonpart} 
	\begin{footnotesize}
		\begin{tabular}{lccccc}
			\hline \hline
			\\[-1.8ex]
			& C.I. & C.I.  & Mean & C.I.  & C.I.  \\ 
			&  2.5\% &  5\% &   bootstrap &  95\% &  97.5\% \\ 
			\\[-1.8ex]
			\hline
			\\[-1.8ex]
			\textbf{	COVID $\times$ North  $\times$}  &&&&& \\ 
			\textbf{ $\times$ Household members$>$2 }& 0.548 & 0.578 & 0.807 & 1.097 & 1.183 \\ 
			\hline \\[-1.8ex]
			Constant & 0.665 & 0.679 & \textbf{0.781} & 0.890 & 0.905 \\ 
			2014 & 1.019 & 1.036 & \textbf{1.142} & 1.253 & 1.270 \\ 
			2015 & 1.077 & 1.097 & \textbf{1.212} & 1.335 & 1.361 \\ 
			2016 & 1.111 & 1.132 & \textbf{1.246} & 1.370 & 1.396 \\ 
			2017 & 0.976 & 0.995 & 1.102 & 1.210 & 1.232 \\ 
			2018 & 1.165 & 1.189 & \textbf{1.323} & 1.460 & 1.485 \\ 
			2019 & 1.189 & 1.218 & \textbf{1.348} & 1.493 & 1.522 \\ 
			2020 & 1.334 & 1.411 & \textbf{1.916} & 2.521 & 2.647 \\ 
			Quarter II & 0.869 & 0.880 & 0.945 & 1.010 & 1.025 \\ 
			Quarter III & 0.926 & 0.939 & 1.005 & 1.076 & 1.091 \\ 
			Quarter IV & 0.910 & 0.922 & 0.991 & 1.068 & 1.083 \\ 
			EU citizen & 0.593 & 0.608 & \textbf{0.699} & 0.799 & 0.819 \\ 
			No EU citizen & 0.611 & 0.622 & \textbf{0.683} & 0.744 & 0.753 \\ 
			Primary education & 1.071 & 1.093 & \textbf{1.187} & 1.278 & 1.293 \\ 
			Secondary education & 0.960 & 0.976 & 1.064 & 1.151 & 1.169 \\ 
			North  & 0.665 & 0.677 & \textbf{0.744} & 0.815 & 0.831 \\ 
			Household members$>$2 & 0.853 & 0.864 & 0.934 & 1.005 & 1.017 \\ 
			North  x Household members$>$2 & 1.021 & 1.048 &\textbf{ 1.175} & 1.314 & 1.339 \\ 
			COVID x EU citizen & 0.911 & 0.971 & 1.526 & 2.191 & 2.343 \\ 
			COVID x No EU citizen & 0.671 & 0.705 & 0.917 & 1.161 & 1.215 \\ 
			COVID x Primary education & 0.502 & 0.527 & \textbf{0.677} & 0.842 & 0.883 \\ 
			COVID x Secondary education & 0.510 & 0.532 & \textbf{0.679} & 0.848 & 0.887 \\ 
			COVID x North & 0.664 & 0.696 & 0.917 & 1.177 & 1.239 \\ 
			COVID x Household members$>$2 & 0.967 & 1.021 & 1.278 & 1.568 & 1.641 \\
				\hline
			Observations&\multicolumn{5}{c}{20358}\\
			\hline \hline
			\multicolumn{6}{l}{\scriptsize{\textit{Note}: North includes regions in the North and the Center.  \textit{Source}: LFS 3-month longitudinal data }}\\\multicolumn{6}{l}{\scriptsize{as provided by the Italian Institute of Statistics (ISTAT).}}
		\end{tabular}
	\end{footnotesize}
\end{table}

\clearpage

\begin{table}[!htbp] \centering 
	\caption{Odds-ratios of being active next quarter for an individual \textbf{currently active} in the labour market (\textbf{females aged 40-49}).} 
	\label{apptab:logit4049femalepart} 
	\begin{footnotesize}
		\begin{tabular}{lccccc}
			\hline \hline
			\\[-1.8ex]
			& C.I. & C.I.  & Mean & C.I.  & C.I.  \\ 
			&  2.5\% &  5\% &   bootstrap &  95\% &  97.5\% \\ 
			\\[-1.8ex]
			\hline
			\\[-1.8ex]
			\textbf{	COVID $\times$ North  $\times$}  &&&&& \\
			\textbf{	$\times$ Household members$>$2} & 0.554 & 0.601 & 0.880 & 1.234 & 1.312 \\ 
			\hline
			\\[-1.8ex]	
			Constant & 4.605 & 4.788 & \textbf{5.982} & 7.370 & 7.703 \\ 
			2014 & 0.990 & 1.012 & 1.136 & 1.270 & 1.298 \\ 
			2015 & 0.989 & 1.009 & 1.144 & 1.282 & 1.312 \\ 
			2016 & 1.025 & 1.048 &\textbf{ 1.176} & 1.319 & 1.354 \\ 
			2017 & 1.073 & 1.100 & \textbf{1.240} & 1.395 & 1.433 \\ 
			2018 & 1.036 & 1.065 & \textbf{1.205 }& 1.353 & 1.381 \\ 
			2019 & 0.919 & 0.950 & 1.147 & 1.375 & 1.433 \\ 
			2020 & 0.855 & 0.936 & 1.520 & 2.309 & 2.506 \\ 
			Quarter II & 0.944 & 0.954 & 1.037 & 1.121 & 1.141 \\ 
			Quarter III & 0.815 & 0.831 & \textbf{0.906} & 0.981 & 0.998 \\ 
			Quarter IV & 0.985 & 1.003 & 1.092 & 1.187 & 1.205 \\ 
			EU citizen & 0.724 & 0.742 & 0.859 & 0.985 & 1.021 \\ 
			No EU citizen & 0.736 & 0.752 & \textbf{0.854} & 0.966 & 0.999 \\ 
			Industry & 3.128 & 3.246 &\textbf{ 3.766} & 4.330 & 4.456 \\ 
			Constructions & 1.405 & 1.470 & \textbf{1.934} & 2.477 & 2.660 \\ 
			Commerce & 3.873 & 4.010 & \textbf{4.658} & 5.375 & 5.538 \\ 
			Hotels and restaurants & 1.273 & 1.300 & \textbf{1.497} & 1.705 & 1.753 \\ 
			Transport & 3.142 & 3.268 & \textbf{4.311} & 5.622 & 5.900 \\ 
			Communications & 5.029 & 5.303 & \textbf{7.679} & 11.013 & 12.127 \\ 
			Finance & 7.846 & 8.254 & \textbf{11.981} & 17.406 & 18.679 \\ 
			Real estate & 2.964 & 3.049 & \textbf{3.527 }& 4.038 & 4.169 \\ 
			Public administration & 11.861 & 12.661 & \textbf{18.641} & 27.033 & 29.429 \\ 
			Education and health & 3.861 & 3.925 & \textbf{4.535} & 5.182 & 5.306 \\ 
			Others & 2.120 & 2.166 & \textbf{2.503} & 2.878 & 2.952 \\ 
			Primary education & 0.558 & 0.567 &\textbf{ 0.631} & 0.698 & 0.716 \\ 
			Secondary education & 0.810 & 0.826 & 0.910 & 0.999 & 1.015 \\ 
			North  & 1.386 & 1.429 &\textbf{ 1.652} & 1.875 & 1.926 \\ 
			Household members$>$2 & 0.628 & 0.650 & \textbf{0.745} & 0.849 & 0.870 \\ 
			North  x Household members$>$2 & 1.189 & 1.232 & \textbf{1.438} & 1.676 & 1.749 \\ 
			COVID x EU citizen & 0.898 & 0.947 & 1.319 & 1.822 & 1.946 \\ 
			COVID x No EU citizen & 0.814 & 0.856 & 1.120 & 1.457 & 1.507 \\ 
			COVID x Industry & 0.476 & 0.517 & 0.737 & 1.008 & 1.064 \\ 
			COVID x Constructions & 0.518 & 0.567 & 1.105 & 1.976 & 2.341 \\ 
			COVID x Commerce & 0.363 & 0.388 &\textbf{ 0.554} & 0.753 & 0.793 \\ 
			COVID x Hotels and restaurants & 0.260 & 0.279 & \textbf{0.389} & 0.530 & 0.568 \\ 
			COVID x Transport & 0.329 & 0.368 &\textbf{ 0.677} & 1.148 & 1.289 \\ 
			COVID x Communications & 0.230 & 0.276 & 0.660 & 1.343 & 1.644 \\ 
			COVID x Finance & 0.164 & 0.183 & 0.366 & 0.644 & 0.735 \\ 
			COVID x Real estate & 0.432 & 0.467 & 0.673 & 0.931 & 0.992 \\ 
			COVID x Public administration & 0.259 & 0.295 & 0.829 & 2.006 & 2.849 \\ 
			COVID x Education and health & 0.368 & 0.398 & \textbf{0.569} & 0.775 & 0.823 \\ 
			COVID x Others & 0.515 & 0.547 & 0.778 & 1.053 & 1.119 \\ 
			COVID x Primary education & 0.614 & 0.643 & 0.809 & 1.001 & 1.045 \\ 
			COVID x Secondary education & 0.495 & 0.514 & \textbf{0.634} & 0.761 & 0.787 \\ 
			COVID x North  & 0.496 & 0.537 & 0.740 & 0.989 & 1.059 \\ 
			COVID x Household members$>$2 & 0.773 & 0.820 & 1.144 & 1.519 & 1.616 \\ 
				\hline
			Observations&\multicolumn{5}{c}{87060}\\
			\hline \hline
			\multicolumn{6}{l}{\scriptsize{\textit{Note}: North includes regions in the North and the Center.  \textit{Source}: LFS 3-month longitudinal data }}\\\multicolumn{6}{l}{\scriptsize{as provided by the Italian Institute of Statistics (ISTAT).}}
		\end{tabular}
	\end{footnotesize}
\end{table}

\clearpage

\begin{table}[!htbp] \centering 
	\caption{Odds-ratios of being active next quarter for an individual \textbf{currently active} in the labour market (\textbf{males aged 40-49}).} 
	\label{apptab:logit4049malepart} 
	\begin{footnotesize}
		\begin{tabular}{lccccc}
			\hline \hline
			\\[-1.8ex]
			& C.I. & C.I.  & Mean & C.I.  & C.I.  \\ 
			&  2.5\% &  5\% &   bootstrap &  95\% &  97.5\% \\ 
			\\[-1.8ex]
			\hline
			\\[-1.8ex]
			\textbf{COVID $\times$ North  $\times$}  &&&&& \\
			\textbf{$\times$ Household members$>$2} & 0.729 & 0.775 & 1.112 & 1.530 & 1.657 \\ 
			\hline
			\\[-1.8ex]
			Constant & 11.186 & 11.632 & \textbf{15.073} & 18.974 & 19.979 \\ 
			2014 & 0.790 & 0.809 & 0.916 & 1.033 & 1.058 \\ 
			2015 & 0.866 & 0.895 & 1.020 & 1.153 & 1.189 \\ 
			2016 & 1.043 & 1.072 &\textbf{ 1.217} & 1.384 & 1.412 \\ 
			2017 & 1.057 & 1.090 & \textbf{1.246} & 1.423 & 1.456 \\ 
			2018 & 1.170 & 1.205 & \textbf{1.375} & 1.557 & 1.594 \\ 
			2019 & 1.089 & 1.145 & \textbf{1.432} & 1.780 & 1.862 \\ 
			2020 & 0.757 & 0.812 & 1.325 & 2.034 & 2.245 \\ 
			Quarter II & 0.923 & 0.939 & 1.028 & 1.115 & 1.133 \\ 
			Quarter III & 1.317 & 1.344 & \textbf{1.482} & 1.632 & 1.662 \\ 
			Quarter IV & 1.015 & 1.032 & \textbf{1.122} & 1.220 & 1.242 \\ 
			EU citizen & 0.432 & 0.443 & \textbf{0.520} & 0.614 & 0.637 \\ 
			No EU citizen & 0.525 & 0.533 & \textbf{0.599} & 0.673 & 0.685 \\ 
			Industry & 2.474 & 2.540 & \textbf{2.914} & 3.329 & 3.427 \\ 
			Constructions & 1.061 & 1.087 & \textbf{1.240} & 1.404 & 1.436 \\ 
			Commerce & 2.875 & 2.957 & \textbf{3.498} & 4.073 & 4.179 \\ 
			Hotels and restaurants & 0.877 & 0.902 & 1.069 & 1.248 & 1.290 \\ 
			Transport & 2.203 & 2.278 & \textbf{2.749} & 3.308 & 3.414 \\ 
			Communications & 2.610 & 2.808 & \textbf{3.999} & 5.543 & 6.038 \\ 
			Finance & 5.107 & 5.595 & \textbf{9.654} & 17.332 & 19.133 \\ 
			Real estate & 1.799 & 1.870 & \textbf{2.245} & 2.691 & 2.776 \\ 
			Public administration & 6.663 & 6.947 & \textbf{9.523} & 12.830 & 13.696 \\ 
			Education and health & 1.999 & 2.076 & \textbf{2.591} & 3.213 & 3.354 \\ 
			Others & 1.199 & 1.245 & \textbf{1.492 }& 1.786 & 1.846 \\ 
			Primary education & 0.336 & 0.346 & \textbf{0.409} & 0.479 & 0.491 \\ 
			Secondary education & 0.536 & 0.559 & \textbf{0.657} & 0.765 & 0.787 \\ 
			North  & 1.878 & 1.919 & \textbf{2.234} & 2.569 & 2.635 \\ 
			Household members$>$2 & 0.974 & 1.001 & 1.144 & 1.298 & 1.328 \\ 
			North  x Household members$>$2 & 0.886 & 0.914 & 1.088 & 1.272 & 1.306 \\ 
			COVID x EU citizen & 0.973 & 1.042 & 1.483 & 2.087 & 2.251 \\ 
			COVID x No EU citizen & 1.041 & 1.102 & \textbf{1.440} & 1.824 & 1.934 \\ 
			COVID x Industry & 0.271 & 0.289 & \textbf{0.397} & 0.528 & 0.549 \\ 
			COVID x Constructions & 0.471 & 0.503 & 0.713 & 0.957 & 1.039 \\ 
			COVID x Commerce & 0.343 & 0.369 & \textbf{0.551} & 0.766 & 0.818 \\ 
			COVID x Hotels and restaurants & 0.211 & 0.226 & \textbf{0.333} & 0.456 & 0.490 \\ 
			COVID x Transport & 0.238 & 0.263 & \textbf{0.400} & 0.572 & 0.612 \\ 
			COVID x Communications & 0.145 & 0.163 & \textbf{0.317} & 0.532 & 0.595 \\ 
			COVID x Finance & 0.138 & 0.158 & 0.551 & 1.458 & 1.830 \\ 
			COVID x Real estate & 0.467 & 0.506 & 0.794 & 1.157 & 1.269 \\ 
			COVID x Public administration & 0.686 & 0.829 & 2.564 & 6.475 & 7.480 \\ 
			COVID x Education and health & 0.328 & 0.367 & 0.616 & 0.960 & 1.058 \\ 
			COVID x Others & 0.382 & 0.411 & 0.638 & 0.932 & 1.006 \\ 
			COVID x Primary education & 0.671 & 0.716 & 1.004 & 1.338 & 1.422 \\ 
			COVID x Secondary education & 0.571 & 0.613 & 0.851 & 1.128 & 1.220 \\ 
			COVID x North  & 0.523 & 0.551 & 0.755 & 1.002 & 1.067 \\ 
			COVID x Household members$>$2 & 0.607 & 0.639 & 0.855 & 1.092 & 1.156 \\ 
			\hline
			Observations&\multicolumn{5}{c}{109226}\\
			\hline
			\hline
			\multicolumn{6}{l}{\scriptsize{\textit{Note}: North includes regions in the North and the Center.  \textit{Source}: LFS 3-month longitudinal data }}\\\multicolumn{6}{l}{\scriptsize{as provided by the Italian Institute of Statistics (ISTAT).}}
		\end{tabular}
	\end{footnotesize}
\end{table}

\end{document}